\documentclass[a4paper,11pt]{article}
\usepackage{jheppub}
\usepackage[T1]{fontenc}
\usepackage{lmodern}

\usepackage{amstext}
\usepackage{amsfonts}
\usepackage{mathrsfs}
\usepackage{amsmath}
\usepackage{amssymb}
\usepackage{cases}
\usepackage{bm}
\usepackage{extarrows}
\usepackage{tensor}
\usepackage{xcolor}
\usepackage{float}
\usepackage{bbm, dsfont}
\usepackage{stmaryrd}
\usepackage{hyperref}
\usepackage[low-sup]{subdepth}
\usepackage{verbatim}

\definecolor{Green}{RGB}{199,238,206}
\makeatletter
\newsavebox\myboxA
\newsavebox\myboxB
\newlength\mylenA
\newcommand*\xoverline[2][0.82]{%
\sbox{\myboxA}{$\m@th#2$}%
\setbox\myboxB\null
\ht\myboxB=\ht\myboxA%
\dp\myboxB=\dp\myboxA%
\wd\myboxB=#1\wd\myboxA
\sbox\myboxB{$\m@th\overline{\copy\myboxB}$}
\setlength\mylenA{\the\wd\myboxA}
\addtolength\mylenA{-\the\wd\myboxB}%
\ifdim\wd\myboxB<\wd\myboxA%
\rlap{\hskip 1.2\mylenA\usebox\myboxB}{\usebox\myboxA}%
\else
\hskip -0.5\mylenA\rlap{\usebox\myboxA}{\hskip 0.5\mylenA\usebox\myboxB}%
\fi}

\newcommand{\beq}{\begin{equation}}
\newcommand{\eeq}{\end{equation}}
\newcommand{\ba}{\begin{array}}
\newcommand{\ea}{\end{array}}
\newcommand{\baa}{\begin{array}}
\newcommand{\eaa}{\end{array}}
\newcommand{\beqa}{\begin{eqnarray}}
\newcommand{\eeqa}{\end{eqnarray}}
\newcommand{\beqs}{\begin{subequations}}
\newcommand{\eeqs}{\end{subequations}}
\def\dis{\displaystyle}

\newcommand{\fr}[2]{\frac{{#1}}{#2}}
\newcommand{\Fr}[2]{\mbox{$\frac{\,{#1}\,}{#2}$}}
\renewcommand{\rm}{\mathrm}
\newcommand{\Dbrack}[1]{\llbracket#1\rrbracket}

\def\under{\underline}
\def\phih{\hat{\phi}}
\def\psih{\hat{\psi}}
\def\AP{A_{\rm{P}}}
\def\AT{A_{\rm{T}}}
\def\PhiB{\boldsymbol{\Phi}}

\def\tr{\text{tr}}

\def\geqq{\geqslant}
\def\({\left(}
\def\){\right)}
\def\[{\left[}
\def\]{\right]}
\def\LB{\left\{}
\def\RB{\right\}}
\def\nn{\nonumber}
\def\pd{\partial}
\def\d{\rm{d}}

\def\to{\rightarrow}
\def\ito{\!\rightarrow\!}

\def\under{\underline}

\def\sp{\mathfrak{s}}
\def\qB{\bar{q}}
\def\pB{\bar{p}}

\def\FF{\mathcal{F}}

\def\FFd{\underline{\mathcal{F}}}
\def\FFt{\widetilde{\mathcal{F}}}

\def\FFB{\underline{\mathbb{F}}}
\def\hP{h_{\rm{P}}^{}}
\def\hS{h_{\rm{S}}^{}}
\def\hTT{h_{\rm{T}}^{}}
\def\htd{\tilde{h}}
\def\MM{\mathcal{M}}
\def\MMT{\widetilde{\mathcal{M}}}

\def\QQ{\mathbb{Q}}

\def\ba{\bar{a}}

\def\CC{\mathcal{C}}

\def\D{\mathcal{D}}

\def\EE{\mathcal{E}}

\def\hbar{\bar{h}}
\def\ii{\text{i}}

\def\La{\mathcal{L}}

\def\NN{\mathcal{N}}

\def\SS{\mathcal{S}}

\def\TT{\mathcal{T}}
\def\tT{\widetilde{\mathcal{T}}}

\def\vt{\tilde{v}}
\def\VV{\mathcal{V}}

\def\Zh{\widehat{\mathbb{Z}}}

\def\SS{\mathbb{S}}

\def\YY{\mathbb{Y}}
\def\YT{\widetilde{\mathbb{Y}}}

\def\al{\alpha}

\def\be{\beta}

\def\ka{\kappa}
\def\ab{\alpha\beta}
\def\mn{\mu\nu}

\def\PP{\rm{P}}
\def\ep{\epsilon}
\def\epP{\epsilon_{\rm{P}}}
\def\epS{\epsilon_{\rm{S}}}
\def\epL{\epsilon_{\rm{L}}}
\def\epT{\epsilon_{\rm{T}}}
\def\bepT{\bar{\epsilon}_{\rm{T}}}
\def\tepT{\widetilde{\epsilon}_{\rm{T}}}
\def\epX{\epsilon_{\rm{X}}}
\def\ct{c_\theta^{}}
\def\st{s_\theta^{}}

\def\ctt{c_{2\theta}^{}}

\def\stt{s_{2\theta}^{}}

\def\MP{M_{\text{Pl}}^{}}

\def\DD{\boldsymbol{\mathcal{D}}}

\def\pB{\bar{p}}
\def\qB{\bar{q}}
\def\sB{\bar{s}}
\def\hf{\frac{1}{2}}
\def\hs{\hspace*{0.3mm}}
\def\hsx{\hspace*{0.5mm}}
\def\hsm{\hspace*{-0.3mm}}
\def\hsmx{\hspace*{-0.5mm}}

\def\vs{\vspace*{1mm}}
\def\ssh{\hat{\tt s}}
\def\mt{\widetilde{m}}
\def\etaB{\bar{\eta}}
\def\ct{c_\theta^{}}
\def\st{s_\theta^{}}
\def\cht{c_{\theta/2}^{}}
\def\sht{s_{\theta/2}^{}}

\def\ctt{c_{2\theta}^{}}

\def\stt{s_{2\theta}^{}}

\def\pu{\underline{\,p}}
\allowdisplaybreaks[4]
\def\End{\end{document}}

\renewcommand{\thefootnote}{\fnsymbol{footnote}}

\title{Structure of Chern-Simons Graviton Scattering\\ 
Amplitudes from Topological Graviton Equivalence\\ Theorem and Double Copy}

\author{Hong-Xu Liu\,$^a$, ~Zi-Xuan Yi\,$^a$, 
       ~Hong-Jian He\,$^{a,b,c,}$\footnote{corresponding author.}}

\affiliation[a]{Tsung-Dao Lee Institute \& School of Physics and Astronomy, \\
	Shanghai Jiao Tong University, Shanghai, China}

\affiliation[b]
{Department of Physics, Tsinghua University, Beijing, China}
\affiliation[c]
{Center for High Energy Physics, Peking University, Beijing, China}

\emailAdd{hongxu-liu@sjtu.edu.cn,\ zixuanyi@sjtu.edu.cn,\ hjhe@sjtu.edu.cn}

\abstract{\\	
Gravitons naturally acquire topological masses in the 3d topologically massive gravity (TMG) theory
that includes the gravitational Chern-Simons term.\ We present a Weyl-transformed 
TMG (WTMG) formulation by introducing an unphysical dilaton field through the Weyl transformation.\ 
We perform the BRST quantization of the WTMG, which reduces to the conventional
TMG in the unitary gauge.\ 
We demonstrate that this WTMG theory conserves the physical degrees of freedom (DoF)
in the massless limit, under which the physical massive graviton 
becomes an unphysical massless graviton
and its physical DoF is converted to the massless dilaton.\
With these, we newly establish 
a Topological Graviton Equivalence Theorem (TGRET),
which connects each scattering amplitude of physical gravitons
to the corresponding dilaton scattering amplitude in the high energy limit.\
The TGRET provides a general mechanism to guarantee all the large energy cancellations 
in any massive graviton scattering amplitudes.\
Applying the TGRET and using the generalized gravitational power counting rule, we prove that
the $N$-point massive graviton amplitudes ($N\!\!\geqq\!4$) have striking energy cancellations 
by powers proportional to $\frac{5}{2}N$ ($\frac{7}{2}N$) in the Landau (unitary) gauge.\   
For four graviton scattering amplitudes, this explains their large energy cancellations of 
$E^{11}\!\!\to\hsm\!E^1$ (Landau gauge) and $E^{12}\!\!\to\hsm\!E^1$ (unitary gauge).\ 
We compute the four-point graviton (dilaton) amplitudes 
and explicitly demonstrate the TGRET and these large energy cancellations.\
With the extended massive double-copy approach,
we systematically construct the three- and four-point graviton (dilaton) scattering amplitudes
in the WTMG theory from the corresponding gauge boson (adjoint scalar) amplitudes 
in the topologically massive Yang-Mills theory.\
\\[2mm]
{[\,arXiv:2512.10870 [hep-th]{\hs}]}
}

\begin{document}
\maketitle
\flushbottom

\renewcommand{\thefootnote}{\arabic{footnote}}
\setcounter{page}{2}

\vspace*{3mm}
\section{\hspace*{-2mm}Introduction}
\label{sec:1}

The (2+1)-dimensional (3d) spacetime has distinctive features characterized by the
gauge and gravitational Chern-Simons terms\,\cite{Deser:1981wh}-\cite{Dunne:1998},
which generate topological masses for gauge bosons and gravitons,
realize fractional statistics and predict the existence of
anyon-like quasiparticles\,\cite{Wilczek}.\
Chern-Simons (CS) terms are topological invariants in mathematics\,\cite{CS}
and serve as the theoretical key ingredients
of a wide range of applications in modern physics, including
fractional quantum Hall effect\,\cite{Tong:2016kpv},
models of the high-temperature superconductivity and strongly correlated systems\,\cite{Qi},
and topological quantum computing\,\cite{TQC}.\
On the other hand, studying the structure of scattering amplitudes of massive gauge bosons
and gravitons in the Chern-Simons theories provides an important means for understanding
the mechanism of topological mass-generations\,\cite{Hang:2021oso}\cite{Hang:2023fkk} 
and for realizing the deep gauge-gravity duality connection, 
(Gravity)\,=\,(Gauge\,Theory)$^2$, in the topologically massive theories 
that nontrivially extends the original massless double-copy approach\,\cite{KLT}-\cite{BCJ-rev}.\ 

\vs

In the 3d Chern-Simons gauge and gravity theories\,\cite{Deser:1981wh}\cite{Deser-CS1982PRL},
gauge bosons and gravitons acquire gauge-invariant topological mass terms
without invoking the conventional Higgs mechanism\cite{Higgs} in the 4d standard model (SM).\
The previous work\,\cite{Hang:2021oso}\cite{Hang:2023fkk} formulated
this 3d topological mass-generation mechanism
at the $S$-matrix level for both the Abelian and non-Abelian Chern-Simons gauge theories.\
For this, a Topological Gauge-boson Equivalence Theorem (TGAET) 
was proposed\,\cite{Hang:2021oso}\cite{Hang:2023fkk},
which connects the $N$-point scattering amplitudes of the
physical polarization states of massive gauge bosons ($\AP$)
to the scattering amplitudes of the corresponding
transversely polarized gauge boson states ($\AT$)
at the leading order of high energy expansion.\
(It was also shown that
in the massless limit, the physical gauge boson $\AP$ reduces to
the transversely polarized massless state $\AT\hs$.)
For the 3d Topologically Massive Yang-Mills (TMYM) theory
that includes the non-Abelian Chern-Simons term,
it was proved 
that the TGAET provides an elegant mechanism to generally ensure
large nontrivial energy cancellations in the $N$-point scattering amplitude ($N\!\!\geqq\! 4$)
on-shell physical gauge bosons at tree level:
$\!E^4\hsm\ito\hsm E^{4-N}\hsm$,$\hs$
without invoking any conventional Higgs boson.\
This means that for the case of four-point physical gauge boson scattering,
its tree-level amplitude has large energy cancellations of
$\,E^4\!\to\! E^0\,$ under high energy expansion.\
In contrast, for the 3d Topologically Massive Gravity (TMG) theory
that includes the gravitational Chern-Simons term\,\cite{Deser:1981wh}\cite{Deser-CS1982PRL},
it was found\,\cite{Hang:2021oso}\cite{Hang:2023fkk}
that the four-point graviton scattering amplitude at tree level
has even more striking energy cancellations of $E^{12}\!\ito\hsm E^1$,
which is hard to understand.\
By extending the conventional double-copy method for massless gauge/gravity
theories\,\cite{BCJ}-\cite{BCJ-rev} to the 3d massive gauge/gravity theories
of TMYM and TMG, the double-copy constructions of the four-point physical
graviton amplitudes were given in
Refs.\,\cite{TMG-DCx} and \cite{Hang:2021oso} independently.\ 
The double-copy analysis\,\cite{Hang:2021oso} found that at tree level
the reconstructed four-point graviton scattering amplitude could have
leading energy-power dependence of $E^4$ at most, suggesting large
energy cancellations of $E^{12}\!\ito E^4$.\ Beyond this, further 
energy cancellations of $E^{4}\ito E^1$ were found\,\cite{Hang:2021oso} 
by explicit calculations, which remain puzzling.\

Based on what was achieved\,\cite{Hang:2021oso}\cite{Hang:2023fkk}
for formulating the topological mass generation of gauge bosons through the TGAET 
in the TMYM theory as discussed above, it is natural and highly motivated to study
why the massive graviton scattering amplitudes in the TMG theory have even more striking
energy cancellations and whether a topological graviton equivalence theorem (TGRET) 
can be established for the TMG theory 
to fully explain such large energy cancellations.\
However, extending the above formulation for the TMYM theory
to the case of the TMG theory is much more difficult and highly nontrivial due to the complications
of the gravitational gauge-fixing term and of the graviton polarization tensor.\
Furthermore, in the conventional TMG theory, there is no proper
scalar-field degree of freedom whose scattering amplitudes
could quantitatively mimic the high-energy behavior of the corresponding
graviton scattering amplitudes.\ We will demonstrate in this work that  
a Weyl-transformed TMG theory 
(including an unphysical dilaton field $\phi\hs$)
is essential for the correct formulation of the TGRET.\  
For instance, we find that the four-point scattering amplitude 
of massive gravitons ($\hP$) does not equal the corresponding
amplitude of the transversely polarized gravitons ($\hTT$) in the high energy limit; 
this situation differs from the TGAET 
of the TMYM gauge theory.\footnote{%
This situation of the TMG theory is also contrary to 
our naive expectation based on
the double-copy approach established 
for the massive Kaluza-Klein (KK) gauge/gravity theories 
in the literature\,\cite{Hang:2021fmp}-\cite{Hang:2025},  
where the KK gravitational equivalence theorem (KK~GRET)\,\cite{Hang:2021fmp}\cite{Hang:2022rjp}\cite{Hang:2024uny}-\cite{chivukula:2024}  
can be obtained from the KK gauge-boson equivalence theorem 
(KK~GAET)\,\cite{KKET0}\cite{KKET1}\cite{KKET2} 
by using the double-copy construction
and the KK GRET ensures\,\cite{Hang:2021fmp}\cite{Hang:2022rjp}
a {\it different type} of large energy cancellations 
for the $N$-point (helicity-zero) longitudinal KK graviton 
scattering amplitudes, 
including the energy cancellations $E^{10}\!\ito\hsm E^2$
in the four-point amplitudes.}\
Hence,  a major task of the present work is to overcome this obstacle.

\vs

In this work, we present a new formulation of the TMG theory
by introducing an unphysical dilaton field through the Weyl transformation.\
We perform the BRST quantization
of this Weyl-transformed TMG (WTMG) theory by constructing proper gauge-fixing terms 
and ghost terms.\
For this, we construct proper 
gauge-fixing terms with a class of gauge choices, 
including the Landau gauge ($\xi \!=\!\zeta \!=\!0$), 
the Feynman gauge ($\xi \!=\!\zeta \!=\!1$), 
and the unitary gauge ($\zeta \!=\!\infty$) 
that corresponds to the conventional TMG theory.\ 
We demonstrate that this WTMG theory conserves the physical degrees of freedom (DoF)
in the massless limit, under which the physical massive graviton
becomes an unphysical massless graviton
and its physical DoF is converted to the massless dilaton.\
With this setup, we newly establish 
a Topological Graviton Equivalence Theorem (TGRET),
which connects each scattering amplitude of physical gravitons 
to the corresponding scattering amplitude of	 dilatons 
in the high energy limit.\
The TGRET provides a general mechanism that ensures all the nontrivial
large energy cancellations in any massive graviton scattering amplitudes.\
In this way, the leading energy-power dependence of $N$-point massive 
graviton amplitudes is fully determined by the leading energy-power 
dependence of the corresponding $N$-point dilaton amplitudes.\ 
Applying the generalized gravitational power counting rule 
and using the TGRET, 
we will demonstrate that the $N$-point massive graviton scattering amplitudes 
($N\!\!\geqq\!4$) have striking 
energy cancellations by energy powers proportional to 
$\frac{5}{2}N$ ($\frac{7}{2}N$) in the Landau (unitary) gauge.\   
For the four-point scattering, 
the leading energy dependence of the dilaton amplitudes 
scales as $E^{1}$, which implies the large energy cancellations of $E^{11}\!\!\to\hsm\!E^1$ (Landau gauge) and $E^{12}\!\!\to\hsm\!E^1$ (unitary gauge) in the four graviton scattering amplitudes.\  
Then, we compute the four-point scattering amplitudes of gravitons and of dilatons,  
and explicitly demonstrate these large energy cancellations
and prove the validity of the TGRET.\

Finally, using the extended massive double-copy approach,
we construct the scattering amplitudes of gravitons
and of dilatons (of physical scalars) in the WTMG theory 
(coupled to a physical scalar) 
from the corresponding scattering amplitudes of gauge bosons 
and of adjoint scalars in the TMYM theory 
(coupled to adjoint scalars).\footnote{%
There are some studies in the literature 
that studied double-copies of three- and two-algebra gauge theories 
for the 3d supersymmetric 
theories\,\cite{other1a-3d-CS}-\cite{other1c-3d-CS}, 
and gave certain double-copy analyses for the amplitudes with
matter fields or with topological modes in 3d CS gauge theory\,\cite{other2a-3d-CS}-\cite{Emond:2025nxa}.\ 
%
Other extended massive double-copy studies include 
the compactified flat or warped 5d KK gauge/gravity 
theories\,\cite{Hang:2021fmp}-\cite{Hang:2024uny},
the 4d massive Yang-Mills (YM) theory versus Fierz-Pauli-like
gravity\,\cite{dRGT}\cite{dRGT2},
the spontaneously broken YM-Einstein supergravity models
with adjoint Higgs fields\,\cite{DC-4dx1}, and 
the KK-inspired effective gauge theory with extra global
U(1)\,\cite{DC-5dx}.}\ 
With the extended massive double-copy, we can 
connect the TGAET in the TMYM gauge theory (including adjoint scalar fields)
to the TGRET in the WTMG-Scalar theory (including dilaton field and
a physical scalar field).\

\begin{figure}[t]
\centering
\includegraphics[width=0.8\textwidth]{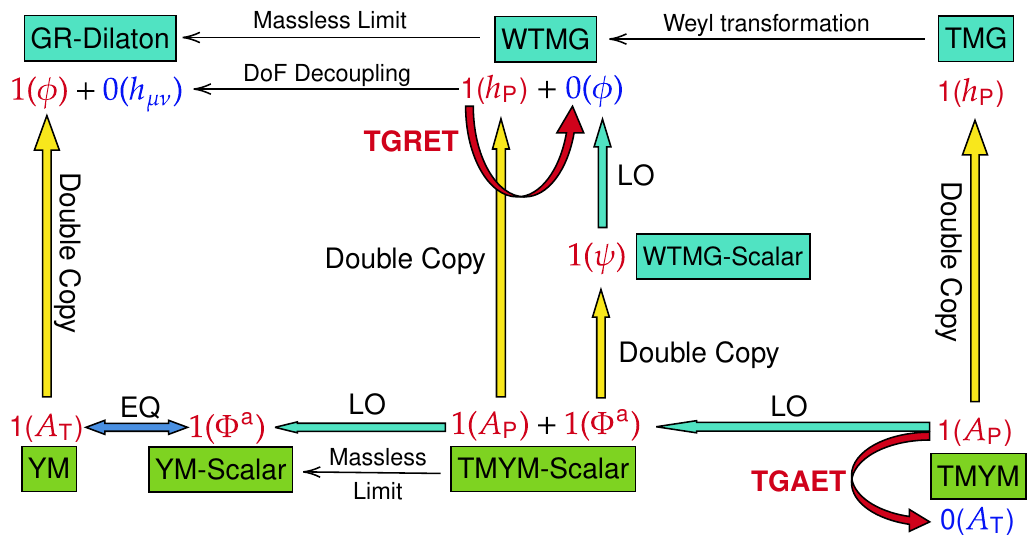}
\vspace*{-4mm}
\caption{\hspace*{-1.5mm} 
Schematic outline and summary of the present study:\
1).\,the topological graviton equivalence theorem (TGRET) 
in the WTMG theory, in comparison with the 
topological gauge-boson equivalence theorem (TGAET) 
in the TMYM theory;
2).\,the massive double-copy constructions of both the graviton 
$\hP\!$ amplitudes and the physical-scalar $\psi$ amplitudes
(dilaton $\phi$ amplitudes) 
in the WTMG-Scalar theory (coupled to physical scalar $\psi$)
from the gauge boson amplitudes and the adjoint scalar amplitudes in
the TMYM-Scalar theory (coupled to adjoint scalars $\Phi^a$).\
Further explanations are given in the text.}
\label{fig:1}
\end{figure}

\vspace*{2mm}
\noindent
{\bf Outline and Summary:}
\\[1mm]
To make the presentation more transparent for readers, we provide an outline and summary of our
main logic and key results as illustrated in the schematic plot of Fig.\,\ref{fig:1}.\
\vspace*{-1mm}
\begin{itemize}
	
\item
In the top line of Fig.\,\ref{fig:1}, we start from the 3d TMG theory
(containing the massive graviton field $h_{\mn}^{}\!$
with 1 physical degree of freedom denoted as $\hP$)
and make a Weyl transformation to introduce an unphysical
dilaton field $\phi$, which results in the Weyl-transformed TMG (WTMG) theory
(containing 1 physical graviton $\hP$ and 1 unphysical dilaton $\phi$).\ 
Then, taking the massless limit, the WTMG
theory reduces to the 3d massless general relativity (GR) plus dilaton $\phi\hs$,
where the physical degree of freedom is converted from the graviton field to the dilaton field,
resulting in an unphysical massless graviton field $h_{\mn}^{}$
and a physical massless dilaton field $\phi\hs$.\
These will be discussed in Section\,\ref{sec:4.3}.\

\item
In the middle of the top line of Fig.\,\ref{fig:1},
in the WTMG theory, we will generally prove the
Topological Graviton Equivalence Theorem (TGRET) 
(marked in red color) which states that
each scattering amplitude of gravitons ($\hP$) is connected to
the corresponding scattering amplitude of dilatons ($\phi$)
in the high energy limit, as presented in Section\,\ref{sec:3}
and explicitly demonstrated up to the four-point 
amplitudes in Sections\,\ref{sec:4.1}.\

\end{itemize}
\begin{itemize}
\vspace*{-3mm}	
	
\item
In the bottom line of Fig.\,\ref{fig:1}, 
we start from the 3d TMYM gauge theory
(containing the massive gauge field $A_\mu^a$ 
with 1 physical degree of freedom
denoted as $\AP$) at the right end of this line,
and consider the TMYM theory coupled to 
an adjoint scalar field $\Phi^a$
(called the TMYM-Scalar theory, with the abbreviation TMYMS) 
in the middle of this line.\
In the TMYM theory, it was proved\,\cite{Hang:2021oso} that
the physical $\AP$ scattering amplitude equals the corresponding
$\AT$ scattering amplitude in the high energy limit, called
the Topological Gauge-boson Equivalence Theorem (TGAET), 
as marked in red color around the right end of the bottom line.\

\item
Furthermore, along the bottom line, from the right to the left,
we will show that the four-point $\AP$ scattering amplitude
in the TMYM theory equals the corresponding scattering amplitude of adjoint scalar $\Phi^a$
in the TMYM-Scalar (TMYMS) theory
at the leading order (LO) of high energy expansion,
as studied in Sections\,\ref{sec:4.3}.\
Moreover, the $\Phi^a$ scattering amplitude in the TMYM-Scalar
theory equals the corresponding $\Phi^a$ amplitude in
the massless YM-Scalar theory at the LO of high energy expansion.\
In addition, the four-point $\Phi^a$ scattering amplitude in the massless YM-Scalar theory
equals the corresponding $\AT^a$ amplitude in the massless YM theory
(as shown in Section\,\ref{sec:4.3}), where the abbreviation ``EQ'' 
around the lower left corner of Fig.\,\ref{fig:1} 
stands for this equality.

\item
From the bottom line to the top line,
we will demonstrate that through the extended double-copy approach,
the graviton (dilaton) scattering amplitudes of the gravitational theories in the top line
can be constructed from the gauge boson (adjoint scalar) scattering
amplitudes of the gauge theories in the bottom line,
as presented in Section\,\ref{sec:4.2} for the three-point and
four-point amplitudes.\
Specifically, from the lower right corner to the upper right corner,
we use massive gauge boson ($\AP$) scattering amplitudes in the TMYM theory to construct
the corresponding massive graviton ($\hP$) scattering amplitudes in the WTMG (or TMG) theory
via the double-copy approach (corresponding to the 
unitary gauge construction of the new WTMG formulation, 
as done in the literature\,\cite{TMG-DCx}\cite{Hang:2021oso}).\
From the middle of the bottom line to the middle of the top line, 
we use the double-copy method to construct the graviton amplitudes 
and the physical-scalar (dilaton) amplitudes  
in the WTMG-Scalar theory (WTMG coupled to a physical scalar field)
from the corresponding gauge-boson amplitudes 
and adjoint-scalar amplitudes in the TMYM-Scalar theory
(TMYM coupled to adjoint scalars),
where the double copy of the scattering amplitudes 
of adjoint scalars ($\Phi^a$) gives the physical scalar amplitudes
(which equal the dilaton $\phi$ amplitudes at the LO of 
high energy expansion) in the WTMG-Scalar theory.\
Finally, from the lower left corner to the upper left corner,
the 3d massless dilaton scattering amplitudes 
in the GR-dilaton theory can be constructed
from the gauge boson scattering amplitudes 
in the massless YM theory.

\end{itemize}

The rest of this paper is organized as follows.\
In Section\,\ref{sec:2.1}, present a new formulation of the TMG theory by introducing a massless
unphysical dilaton field through the Weyl transformation and study its BRST quantization.\ 
This Weyl-transformed TMG theory (WTMG) 
(including the dilaton field) reduces to the conventional TMG theory 
by taking the gauge-fixing parameter be infinity $\zeta\ito\infty$ 
(the unitary gauge of the WTMG)
under which the dilaton field decouples.\ 
In the gravitational Landau gauge ($\xi \!=\!\zeta \!=\!0$) 
of the WTMG theory,
we show that the graviton propagator has better high-energy
behavior $m/p^3$ than the $1/p^2$ behavior 
in the Feynman gauge ($\xi \!=\!\zeta \!=\!1$) and 
in the unitary gauge ($\zeta\!=\!\infty$).\   
In Section\,\ref{sec:2.2new}, 
we analyze the pure dilaton (scalar) self-interactions
under field redefinition in the WTMG (WTMG-Scalar) theory
and will prove that in the WTMG (WTMG-Scalar) theory the 
pure dilaton (scalar) self-interactions do not contribute 
to the on-shell leading-order dilaton (scalar) amplitudes 
under high energy expansion.\
We further prove that for the WTMG theory coupled to a physical scalar $\psi$ (called WTMGS),
the tree-level $N$-point dilaton amplitudes and physical scalar amplitudes are equal
at the leading order of high energy expansion. 
Then, for Section\,\ref{sec:3}, we derive in Section\,\ref{sec:3.1new}
some key relations for the 3d massive polarization vectors and tensors 
that will be used for the proof of the TGRET in Section\,\ref{sec:3.2new}
and for the explicit calculations of Section\,\ref{sec:4.1}.\
In Section\,\ref{sec:3.2new}, we formulate the Topological Graviton Equivalence Theorem (TGRET)
in the WTMG theory, which connects the physical graviton scattering amplitudes 
to the corresponding dilaton scattering amplitudes
at the leading order of high energy expansion.\ 
In Sections\,\ref{sec:3.3new}, we present a generalized gravitational power counting method 
to extract the leading high-energy behaviors of 
general $N$-point graviton scattering amplitudes and dilaton scattering amplitudes.\ 
This can be applied to analyze large energy cancellations of the massive graviton 
scattering amplitudes because the TGRET connects the graviton scattering amplitudes 
to the corresponding scalar dilaton scattering amplitudes whose energy-dependence is
manifest without invoking any nontrivial energy cancellations.\   
In this way, we prove that these large energy cancellations 
in the physical graviton scattering amplitudes are guaranteed by the TGRET.\  
In Section\,\ref{sec:4}, using the new WTMG formulation 
we explicitly analyze the three-point and 
four-point scattering amplitudes of massive gravitons
and of massless dilatons in the Landau gauge in comparison with 
the corresponding graviton amplitudes 
in the unitary gauge of the WTMG theory 
(which is the conventional TMG theory).\
We explicitly demonstrate that the physical graviton scattering amplitudes as computed in
both the Landau gauge and unitary gauge take the same form and are thus gauge-invariant.\ 
With these, we explicitly prove in Section\,\ref{sec:4.1}
that the TGRET holds for the three-point and four-point scattering amplitudes of gravitons (dilatons).\
We further compute the four-point physical scalar amplitude in
the WTMG-Scalar theory and explicitly prove its equivalence to
four-point dilaton amplitude 
at the leading order of high energy expansion.\
Then, in Section\,\ref{sec:4.2}, 
we use the extended double-copy approach to further construct the three-point and four-point scattering amplitudes of gravitons and 
of physical scalars (dilatons) in the WTMG-Scalar theory 
from the corresponding scattering amplitudes of gauge bosons 
and of adjoint scalars in the TMYM-Scalar theory.\
In Section\,\ref{sec:4.3}, we discuss the massless limit of the graviton scattering amplitudes 
in the TMG theory and their relations to the corresponding amplitudes in the 3d massless theories 
as well as their double-copy constructions.\
With these, we conclude in Section\,\ref{sec:5}.\
Finally, in Appendices\,\ref{app:A}-\ref{app:C}, 
we present some technical derivations and results 
used for the analyses in the main text.\
Among them, Appendix\,\ref{app:A} provides the formulation 
of massive spinor-helicity basis in the 3d spacetime.\ 
Appendix\,\ref{app:B} derives all the relevant cubic and 
quartic Feynman rules for the WTMG theory, 
as well as some gravitational scattering amplitudes used
for discussions of the main text.\ 
Appendix\,\ref{app:Cnew} provides additional formulas 
for the four-point physical scalar amplitudes 
in the WTMGS and TMGS theories.\ 
Appendix\,\ref{app:D} presents the BRST quantization 
of the WTMG theory and 
the relevant identities as used for the LSZ (Lehmann–Symanzik–Zimmermann) reduction 
analysis in the main text.

\section{\hspace*{-2mm}Topologically Massive Gravity and Quantization}
\label{sec:2}

In this section, we study the quantization of the Topologically Massive Gravity (TMG)
and its extension by including a physical scalar as matter field.\ 
In Section\,\ref{sec:2.1}, we present a new formulation of the TMG theory 
by introducing an unphysical dilaton field ($\phi$) through the
Weyl transformation and study its quantization
\`{a} la BRST (Becchi-Rouet-Stora-Tyutin) method\,\cite{BRST}\cite{BRST2}.\
In this Weyl-transformed TMG (WTMG) formulation, 
we construct the proper gauge-fixing terms 
with a class of gauge choices, 
including the Landau gauge ($\xi \!=\!\zeta \!=\!0$), 
the Feynman gauge ($\xi \!=\!\zeta \!=\!1$), 
and the unitary gauge ($\zeta \!=\!\infty$) 
that corresponds to the conventional TMG theory.\ 
Furthermore, we show that in the Landau gauge 
($\xi \!=\!\zeta \!=\!0$) of the WTMG theory,
the graviton propagator has better high-energy
behavior of $1/p^3$, 
rather than the usual $1/p^2$ behavior of the
graviton propagator in the unitary gauge
(corresponding to the conventional TMG theory).\ 
In Section\,\ref{sec:2.2new}, we analyze the pure dilaton self-interactions
under field redefinition
and will prove that in the WTMG theory the 
pure dilaton self-interactions 
do not contribute to on-shell leading-order dilaton amplitudes.\
We further prove that for the WTMG theory coupled to a physical scalar $\psi$ (called WTMGS),
the tree-level $N$-point dilaton amplitudes and physical scalar amplitudes are equal
at the leading order of high energy expansion.

\vspace*{1mm}
\subsection{\hspace*{-2mm}Quantization of Weyl-Transformed Topologically Massive Gravity}
\label{sec:2.1}
\vspace*{1mm}
\vs 

For the three-dimensional (3d) spacetime, we may construct the asymptotic states in the gravitational theory from the tensor product of the asymptotic states in the gauge theory.\
But, we note that the 3d Yang-Mills (YM) gauge field $A^{a}_{\mu}$ has only one physical degree of freedom and is a pseudo-vector field.\
By making the tensor product of two YM gauge fields, 
we expect the correspondence:
\begin{equation}
	\label{eq:AxA=hmunu+phi-0}
	A_{\mu}^a \otimes A_{\nu}^b \,\Longrightarrow\, h_{\mu \nu}^{} \oplus \phi \,,
\end{equation}
where $h_{\mu \nu}^{}$ denotes the graviton field of
the 3d gravitational theory,
and $\phi$ is the massless scalar field known as dilaton.\

\vs

We note that the graviton field in the 3d massless GR has
no local physical degrees of freedom, whereas in the 
3d topologically massive gravity (TMG) theory 
the massive graviton has one physical degree of freedom.\
(Here the abbreviation GR stands for General Relativity.)
But, as we will show, 
the degrees of freedom of the massless scalar dilaton field have
a different feature, with one physical degree of freedom
(when the dilaton is coupled to the 3d massless GR);
whereas in the TMG theory, the dilaton is introduced as
an unphysical field and has no physical degree of freedom.\
Thus, for the 3d massless YM gauge theories
and the 3d topologically massive gauge field theory (TMYM),
we may deduce the following relations for
counting the physical degrees of freedom:
\beqs
\label{eq:AxA=hmunu+phi}
\begin{align}
\label{eq:m0-AxA=0.hmunu+1.phi}
\text{3d Massless YM:} &  \hspace*{7mm}
1(A^a_{\mu}) \otimes 1(A^b_{\nu}) =\hs 0(h_{\mu \nu}) + 1(\phi) \hs,
\\
\label{eq:TMG-AxA=1.hmunu+0.phi}
\text{3d TMYM (Massive):} & \hspace*{7mm} 
1(A^a_{\mu}) \otimes 1(A^b_{\nu}) =\hs 1(h_{\mu \nu}^{}) + 0(\phi) \hs .
\end{align}
\eeqs
This illustration of counting the physical degrees of freedom
does not imply whether the double copy for the graviton scattering amplitudes
generally holds for 3d theories,
which is a highly nontrivial issue and will be further
studied in Section\,\ref{sec:4.2} via explicit analyses.\footnote{%
The 3d classical double copy was shown for massless YM 
solutions\,\cite{classic-DC1}-\cite{classic-DC3}.\ 
For scattering amplitudes, the massive double copy for 
the 3d TMYM and TMG theories has been verified 
at tree level and only up to
four-point amplitudes analytically\,\cite{TMG-DCx}\cite{Hang:2021oso} 
and up to five-point amplitudes numerically\,\cite{TMG-DCx}.\  
Since there is no general proof of the BCFW recursions\,\cite{BCFW} in 3d spacetime for non-supersymmetric theories (except for very special supersymmetric theories 
such as the 3d $\mathcal{N}\hsm\!=\!6$ ABJM theory\,\cite{ABJM}), 
extensions of the 3d double-copy to higher-point amplitudes 
are much harder and could be studied only case by case.}

\vs 

It is known that by adding the topological Chern-Simons (CS) term 
to the 3d YM Lagrangian and to the 3d Einstein-Hilbert action 
of general relativity (GR) respectively, 
the corresponding YM gauge field
and graviton field can acquire topological masses
without breaking the gauge symmetry or diffeomorphism invariance.\
The Lagrangian of the TMYM theory takes the following 
form\,\cite{Deser:1981wh}\cite{Deser-CS1982PRL}:
%
\begin{align}
\label{eq:TMYM-L}
\mathcal{L}_{\text{TMYM}}^{}
&= -\frac{1}{2}\text{tr}\hs\mathbf{F}_{\mu \nu}^2 \!
+ \mt\hs\varepsilon^{\mu \nu \rho}
{\tr}\Big({\hsm\mathbf{A}_{\mu}\partial_{\nu}\mathbf{A}_{\rho}
\!-\! \frac{\,\ii\hs 2g\,}{3}\mathbf{A}_{\mu}\mathbf{A}_{\nu}\mathbf{A}_{\rho}}\Big) ,
\end{align}
%
where $\mathbf{A}_{\mu}\!\!=\!\!A_{\mu}^{a}T^{a}$ and
$\mathbf{F}_{\mu \nu}\!\!=\!F_{\mu \nu}^{a}T^{a}$,
with $\mathbf{F}_{\mu \nu}$ being the YM field strength defined as
$\mathbf{F}_{\mu \nu}^{}\hsm\!=\!
\partial_{\mu}\mathbf{A}_{\nu}
\!-\hsm \partial_{\nu}\mathbf{A}_{\mu}\!-\!\ii\hs g
[\mathbf{A}_{\mu},\mathbf{A}_{\nu}]\,$
and with $T^{a}$ being the generators of the non-Abelian group SU($N$).\
Then, the parameter $m\!=\!|\mt |$
of the CS term provides a topological mass for the YM gauge field,
and the ratio
$\,\mathfrak{s}={\mt}/{m}=\pm 1$\,
corresponds to its spin projection\,\cite{Deser:1981wh}\cite{Dunne:1998}.
With a proper choice of the 
quantization condition\,\cite{Tong},
$4\pi\mt /g^{2} \!\in\! \mathbb{Z}\,$,
it can be shown that under the gauge transformation
$A_{\mu}\!\ito U^{-1}\hsm A_{\mu}U \!+\!\frac{\ii}{\,g\,}U^{-1}\partial_{\mu}U\hs$,
the action $\int \!\d^3x\, \mathcal{L}_{\text{TMYM}}$
remains gauge-invariant.\

Given the TMYM Lagrangian \eqref{eq:TMYM-L}, 
we can properly construct the following 
covariant gauge-fixing term and
its corresponding Faddeev-Popov ghost term:
%
\label{eq:LGF-FP}
\begin{align}
	\label{eq:LGF-LFP}
	\La_{\rm{GF}}^{} = -\frac{1}{\,2\xi\,}(\pd^\mu \!A_\mu^a)^2 ,
	\hspace*{8mm}
	\La_{\rm{FP}}^{} =
	\bar{c}^a\pd^\mu \hsm\big(\delta^{ab} \pd_\mu \!-\! gf^{abc}A_\mu^c\big)c^b,
\end{align}	
%
where $f^{abc}$ is the structure constant of the non-Abelian gauge group.\
Thus, we can derive the complete gauge boson propagator in the general $\xi$-gauge as follows:
\beq
\mathcal{D}_{\mn}^{ab} = \frac{~\ii\hs\delta^{ab}\hsm\Delta^{}_{\mn}~}{p^2\!+\hsm m^2}
=  -\hsm\ii\hs\delta^{ab}\!\left[\hsmx\frac{\,1\,}{~p^2\!+\hsm m^2\,}
\hsm\!\(\! \eta^{}_{\mu \nu} \!-\!\frac{\,p^{}_{\mu}p^{}_{\nu}\,}{p^{2}} \!-\!
\frac{\,\ii\hs m\hs\varepsilon^{\mu \nu \rho}p_{\rho}^{}\,}{p^{2}}\hsm\!\)
\!+ \xi\frac{\,p^\mu p^\nu\,}{p^4}
\hsm\right]\!,
\eeq
where $\xi\!=\!0$ corresponds to the Landau gauge.\

\vs

Next, we consider the following Lagrangian of the
3d TMG theory\,\cite{Deser:1981wh}\cite{Deser-CS1982PRL}:
%
\begin{align}
\mathcal{L}_{\text{TMG}} &= \frac{1}{\,\kappa^2\,}
\!\left[ -\sqrt{-g\,}R 
+\!\frac{1}{\,2\hs\mt\,}
\varepsilon^{\mu \nu \rho}\Gamma_{\ \rho \beta}^{\alpha}
\Big(\partial_{\mu}\Gamma_{\ \alpha \nu}^{\beta}
\!+\!\frac{2}{\hs 3\hs}\Gamma_{\ \mu \gamma }^{\beta}\Gamma_{\ \nu \alpha}^{\gamma}\Big)\!\right] \!,
\label{eq:TMG-L}
\end{align}
%
where the 3d gravitational coupling constant $\ka$ is connected to the Newton constant $G$
by the relation  $\,\ka \!=\!\sqrt{16\pi G\,}\,$
with the reduced Planck mass $\,\MP\!=\!1/(8\pi G)\,$.\
The parameter $\,\mt\,$ of Eq.\eqref{eq:TMG-L} provides the graviton mass $\,m=|\mt|$\,,\,
as will be shown later.\
Using the weak field expansion
$g_{\mu \nu}^{}\!=\!\eta_{\mu \nu}\!+\!\kappa\hs h_{\mu \nu}$
and the linearized diffeomorphism transformation
$\,h_{\mn} \!\to h_{\mn}'\!= h_{\mn}^{} \!+\! \pd_\mu^{}\xi_\nu^{}
+  \pd_\nu^{}\xi_\mu^{}\hsx$,
the gravitational Chern-Simons term in Eq.\eqref{eq:TMG-L} changes only by a total derivative,
so the TMG action is diffeomorphism invariant by the path integral formulation.

\vspace*{0.5mm}

In Eqs.\eqref{eq:TMYM-L} and \eqref{eq:TMG-L},
the parameters $g$ and $\kappa$ denote 
the coupling strengths of the YM gauge fields
and of the gravitational fields respectively,
where the gauge coupling $g$ has mass-dimension $\!\frac{1}{\,2\,}$
and the gravitational coupling $\kappa$ has mass-dimension $\!-\frac{1}{\,2\,}$.\
Moreover, the spin angular momentum in 3d spacetime
is a pseudoscalar\,\cite{Deser:1981wh}\cite{Deser-CS1982PRL}\cite{Jackiw:1991} and
each gauge boson has only one physical degree of freedom with helicity either $-1$ or $+1$,
whereas the 3d massive graviton just has
one physical degree of freedom with helicity either $-2$ or $+2\hs$.\

\vs

We note that for the TMYM gauge theory,
the Chern-Simons term is topological and does not introduce any additional degrees of freedom to the system.\
Thus, when the TMYM theory takes the $m\ito 0$ limit, it naturally returns to the case of the pure massless YM gauge theory,
where the gauge field still has one physical degree of freedom.\
But for the TMG theory, when we take the $m\ito 0$ limit, the Lagrangian 
\eqref{eq:TMG-L} becomes singular
because the 3d massless graviton contains 
no physical degree of freedom\,\cite{Witten1988}-\cite{Hinterbichler:2011tt}
and this massless limit does not conserve the physical degree of freedom.\
We will resolve later this problem of the nonconservation of physical degrees of freedom in the massless limit.

\vs 

For the present analysis, we reparametrize the metric tensor as follows:
\beq
\label{eq:CT-gmunu}
g_{\mn}^{} = \bar{g}_{\mn}^{}\hs e^{{-}\kappa\phi},
\eeq
where the conformal factor $e^{-\kappa\phi}$
contains a scalar dilaton field $\phi\,$.\footnote{%
We note that Ref.\,\cite{Deser1990} introduced 
a different conformal factor
$\Phi^4\!\!=\hsm\!(1\hsm\!+\hsm\!\frac{1}{4}\kappa\phi)^4$
and also a different gauge-fixing term  
for analyzing the renormalizability issue of the TMG theory.}\ 
Under the weak field expansion
$\bar{g}_{\mn}^{}\!=\!\eta_{\mn}^{}\!+\hsm\kappa \bar{h}_{\mn}^{}$,
we derive the relation between the graviton fields $h_{\mn}^{}$ and $\bar{h}_{\mn}^{}$
up to $O(\kappa^2)$,
\begin{equation}
\label{eq:conformaltransformation}
h_{\mu \nu}^{} = \bar{h}_{\mu\nu}^{}
\!-\! \bigg(\!\phi\hsm -\!
\frac{\hs\kappa\hs}{2}\phi^2
\!+\!\frac{\,\kappa^2}{6}\phi^3\!\bigg)\eta_{\mu\nu}^{}
\!-\!\bigg(\!\kappa\phi\hsm
-\!\frac{\,\kappa^2}{2}\phi^2\!\bigg)\bar{h}_{\mu\nu}^{} \hsm +\hsm O(\kappa^3)
\hs.
\hspace*{8mm}
\end{equation}
Thus, substituting \eqref{eq:CT-gmunu}
into the original TMG Lagrangian \eqref{eq:TMG-L},
we derive the following Weyl-transformed TMG (WTMG) Lagrangian
including dilaton field:
\begin{align}
\label{eq:LTMG-phi}
\hspace*{-4mm}
\mathcal{L}_{\rm{WTMG}}^{} = \frac{\,-1\,}{\,\kappa^2\,}\hsm\!
\left\{\!\!\sqrt{\hsm -\bar{g}~}\hs e^{{-}\kappa\phi/2}\hsm\bigg[\hsm R\!+\!\fr{\,\kappa^{2}}{2}
\bar{g}_{\mn}^{}\partial^{\mu}\hsm\phi\hs\partial^{\nu}\!\phi\bigg]
\hsm\!-\!\frac{~\varepsilon^{\mu\nu\rho}\,}{\,2\hs\tilde{m}\,}
\Gamma_{\ \rho \beta}^{\alpha}
\Big(\!\partial_{\mu}\Gamma_{\ \alpha \nu}^{\beta}
\!+\!\frac{2}{3}\Gamma_{\ \mu \gamma }^{\beta}\Gamma_{\ \nu \alpha}^{\gamma}\hsm\Big)
\!\hsm\right\} \!,
\end{align}
where
$\bar{g}_{\mn}^{}\hsm =\hsm\eta_{\mn}^{}+\hs\kappa\hs\bar{h}_{\mn}^{}$.\
Since the gravitational Chern-Simons term is conformally invariant, 
it is not affected by the reparametrization of the metric tensor in Eq.\eqref{eq:CT-gmunu}.\
We note that besides the diffeomorphism invariance, 
the WTMG Lagrangian \eqref{eq:LTMG-phi} has a parametrization-induced Weyl-type local redundancy, namely, it is invariant under the 
following local transformation:
\begin{align}
\label{eq:Weyl-localtransf}
\phi ~\to~ \phi' = \phi +\omega(x)\hs ,
\hspace*{10mm}
\bar{g}_{\mn}^{} \,\to~ 
\bar{g}_{\mn}' = e^{-\ka\hs\omega(x)}\bar{g}_{\mn}^{}\,,
\end{align}
and this invariance can be explicitly verified.

\vs

Expanding the WTMG Lagrangian \eqref{eq:LTMG-phi},
we derive the quadratic part in the following form:
\\[-7mm]
\begin{align}
\label{eq:linearTMG}
\hspace*{-7mm}
\mathcal{S}_{\text{WTMG}}^{(2)}  =& \int\!\! \d^{3}x\,
\frac{1}{\,2\,}\!\bigg[ \frac{1}{\,2\,}\partial_{\lambda}^{}
\hbar_{\mu\nu}\partial^{\lambda}\hbar^{\mu \nu}
\!\!-\hsm\partial_{\mu}^{}\hbar_{\nu\lambda}^{}\partial^{\nu}
\hbar^{\mu\lambda} \!+\hsm \partial_{\mu}^{}\hbar^{\mu \nu}\partial_{\nu}\hbar
-\!\frac{1}{\,2\,}\partial_{\lambda}^{}\hbar\partial^{\lambda}\hbar
\\
\hspace*{-7mm}
& \hspace*{15mm}
+\hsm\partial_{\mu}\phi\partial^{\mu}\phi\hsm +\hsm
\partial_{\mu}\phi\partial^{\mu}\hbar
\hsm -\hsm \partial_{\mu}^{}\phi\partial_{\nu}\hbar^{\mu\nu}
{-\frac{1}{\,2m\,}{\varepsilon_{\lambda\mu\nu}\partial_{\rho}^{}\hbar^{\lambda \sigma}\partial_{\sigma}^{}\partial^{\mu}\hbar^{\nu\rho}}}
\nn\\
\hspace*{-7mm}
& \hspace*{16mm}
{+\frac{1}{\,2m\,}{\varepsilon_{\lambda\mu\nu}^{}\partial_{\rho}^{}
\hbar^{\lambda\sigma}\partial^{\rho}\partial^{\mu}\hbar^{\nu}_{\,\sigma}}}
\bigg]  .
\nn
\end{align}
Then, we need two gauge-fixing terms 
for the WTMG Lagrangian \eqref{eq:LTMG-phi}: 
one is to fix the gauge for the diffeomorphism invariance and
another is to fix the Weyl-type local-redundancy invariance 
\eqref{eq:Weyl-localtransf} due to the unphysical dilaton field.\ 
Thus, we construct the following Lorentz-covariant gauge-fixing terms 
for the WTMG Lagrangian, 
\\[-7mm]
\beqs
\label{eq:gaugefix}
\begin{align}
\label{eq:gaugefix-1}
\mathcal{L}_{\text{WTMG}}^{\rm{GF1}}
& =  
\frac{1}{\,2\hs\xi\,}\!
\(\mathcal{F}_{\rm{GF1}}^{\mu}\hsm\)^2\!,
\hspace*{-11mm}
&&\mathcal{F}_{\text{GF1}}^{\mu}\!= \partial_{\nu}\hbar^{\mu\nu}
\hsm\!-\! \frac{1}{\,2\,}\partial^{\mu}\hsm
(\hbar\hsm -\hsm {\xi}\phi)\hs,
\\
\label{eq:gaugefix-2}
\mathcal{L}_{\rm{WTMG}}^{\rm{GF2}}
& =  
\frac{1}{\,2\hs\zeta\,}\hsm
\(\mathcal{F}_{\text{GF2}}^{\mu}\hsm\)^2\!,
\hspace*{-11mm}
&&\mathcal{F}_{\text{GF2}}^{\mu}\!=
\frac{1}{\,2\,}\partial^{\mu}\hsm
(\hbar\hsm -\hsm {\zeta}\phi)\hs,
\end{align}
\eeqs
where $(\xi,\hs \zeta)$ are gauge-fixing parameters.\
Since the reparametrization \eqref{eq:CT-gmunu} separates $\phi$
from the original metric tensor $g_{\mn}^{}$
as an unphysical degree of freedom, we need the second gauge-fixing term
\eqref{eq:gaugefix-2} to impose the constraint between
the trace of the graviton field
$\,\hbar\!=\!\hbar^{\mu}_{\mu}\,$ and the dilaton field $\phi\hs$,
where both $(\hbar,\hs\phi)$ represent unphysical degrees of freedom
in the metric tensor $g_{\mn}^{}\hs$.\  
We note that our gauge-fixing terms in Eq.\eqref{eq:gaugefix}
are in analogy with the $R_\xi^{}$ gauge-fixing\,\cite{Rxi} 
of the electroweak standard model (SM) 
that contains a family of Lorentz-covariant gauges 
with general gauge-fixing parameter $\xi$ 
[including the Feynman gauge ($\xi\!=\!1$), the Landau gauge ($\xi\!=\!0$), 
and the unitary gauge ($\xi\!=\!\infty$, with unphysical Goldstone bosons decoupled)].\ 
As we will show shortly, our covariant gauge-fixing \eqref{eq:gaugefix} 
for the WTMG theory contains the gravitational Feynman gauge ($\xi\!=\!\zeta\!=\!1$),
Landau gauge ($\xi\!=\!\zeta\!=\!0$), and unitary gauge  
($\zeta\!=\!\infty$, with the unphysical dilaton decoupled).\ 
From the above gauge-fixing terms,
we further derive the Faddeev-Popov ghost terms
$\La_{\rm{WTMG}}^{\rm{FP}}$ in Appendix\,\ref{app:C}.\ 
Thus, the quantized full Lagrangian
$\La_{\rm{WTMG}}^{\phi}\hsm +\hsm\La_{\rm{WTMG}}^{\rm{GF1}}
\hsm +\hsm\La_{\rm{WTMG}}^{\rm{GF2}}\hsm +\hsm\La_{\rm{WTMG}}^{\rm{FP}}$
is invariant under the BRST transformations.

\vspace*{0.4mm}

Including the gauge-fixing terms \eqref{eq:gaugefix-1}-\eqref{eq:gaugefix-2},
we can diagonalize the quadratic part of the action
\eqref{eq:linearTMG} as follows:
\begin{align}
& \mathcal{S}_{\rm{WTMG}}^{[2]} =\,
 \mathcal{S}_{\rm{WTMG}}^{(2)} \hsm
 +\mathcal{S}_{\rm{WTMG}}^{\rm{GF1}}
\hsm +\mathcal{S}_{\text{WTMG}}^{\rm{GF2}}
\nn\\
&=\int\!\! \d^{3}x
\hs\frac{1}{\,2\,}\!\hsmx\left[\frac{1}{\,2\,}\partial_{\lambda}h_{\mu \nu}\partial^{\lambda}h^{\mu \nu}
\!-\!\bigg(\!1\!-\!\frac{1}{\,\xi\,}\!\bigg)\partial_{\mu}^{}h_{\nu\lambda}^{}
\partial^{\nu}h^{\mu \lambda} \!+\!\bigg(\!1\!-\!\hsm\frac{1}{\,\xi\,}\!\bigg)
\partial_{\mu}h^{\mu \nu}\partial_{\nu}h \right.
\nn\\
& \hspace*{5mm}
-\!\frac{1}{\,2\,}\bigg(\!1\!-\!\frac{1}{\,2\xi\,}\!-\!\frac{1}{\,2\zeta\,}\!\bigg)
\partial_{\lambda}h\partial^{\lambda}h\!-\!
\bigg(\!1\!-\!\frac{\xi}{\,4\,}\!-\!\frac{\zeta}{\,4\,}\!\bigg)
\partial_{\mu}\phi\partial^{\mu}\phi
{\,-\frac{1}{\,2m\,}{\varepsilon_{\lambda \mu \nu}\partial_{\rho}h^{\lambda \sigma}\partial_{\sigma}\partial^{\mu}h^{\nu \rho}}}
\nn\\
& \hspace*{5mm}
\left.
{+\frac{1}{\,2m\,}{\varepsilon_{\lambda \mu \nu}\partial_{\rho}h^{\lambda \sigma}\partial^{\rho}\partial^{\mu}h^{\nu}_{\,\sigma}}}
\right]\!.  \hspace*{7mm}
%
\end{align}
In the above formula and hereafter, we suppress the bar of the graviton field
$\bar{h}_{\mn}^{}$ for notational simplicity unless specified otherwise.\
For the convenience of deriving the graviton propagator, we reexpress the above  quadratic action
in the following form:
\begin{equation}
\mathcal{S}_{\rm{WTMG}}^{[2]}
= \int\!\! \d^{3}x\, \frac{1}{\,2\,}h^{\mu \nu}D_{\mu \nu \alpha \beta}^{-1}h^{\alpha \beta} \,, 
\end{equation}
where the inverse propagator takes the form:
\begin{align}
\label{eq:D-inv}
\hspace*{-5mm}
D_{\mu \nu \alpha \beta}^{-1} =\,&
(\eta_{\mu \alpha}\eta_{\nu \beta} \!+\! \eta_{\mu \beta}
\eta_{\nu \alpha})\partial^{2} \!-\!\hsm
\bigg(\!2\!-\!\frac{1}{\xi} \!-\! \frac{1}{\zeta}\hsm\bigg)
\eta_{\mu\nu}^{}\eta_{\alpha\beta}^{}\partial^{2}
\!+\!2\bigg(\hsm\!1\hsm\!-\hsm\!\frac{1}{\xi}\hsm\bigg)\!
(\eta_{\mu \nu}^{}\partial_{\alpha}^{}\partial_{\beta}^{} \!+\hsm
\eta_{\alpha\beta}\partial_{\mu}\partial_{\nu})
\nn\\
&-\!\bigg(\!1\!-\!\frac{1}{\,\xi\,}\!\bigg)\!
\big(\eta_{\mu\alpha}\partial_{\nu}\partial_{\beta}\!+\!\eta_{\mu \beta}
\partial_{\nu}\partial_{\alpha}^{}\!+\eta_{\nu \alpha}\partial_{\mu}\partial_{\beta}
\!+\hsm \eta_{\nu \beta}\partial_{\mu}\partial_{\alpha}\big)
\nn\\
&+\frac{1}{\,m\,}\hsm\bigg[\varepsilon_{\mu\rho\alpha}^{}
(\partial_{\nu}\partial_{\beta}\partial^{\rho} \!-
\eta_{\nu \beta}^{}\partial^{2}\partial^{\rho})
\!+\varepsilon_{\nu \rho \alpha}^{}
(\partial_{\mu}\partial_{\beta}\partial^{\rho}
\!-\!\eta_{\mu \beta}\partial^{2}\partial^{\rho})\hsm
\bigg] .
\end{align}
Then, using Eq.\eqref{eq:D-inv} and going to the momentum space,
we derive the following massive graviton propagator
and dilaton propagator:
\beqs
\label{eq:DhDphi-xi-zeta}
\begin{align}
D_{\mu\nu\alpha\beta}^{h}(p) \!=\, &
\frac{\,-\ii\hs (\etaB_{\mu \alpha}^{}\etaB_{\nu \beta}^{}\!+\!\etaB_{\mu \beta}^{}\etaB_{\nu \alpha}^{}
	\!-\!\etaB_{\mu \nu}^{}\etaB_{\alpha \beta}^{})\,}{~p^2\!+\!m^2\,}
\hsm\! + \!\frac{\ii\hs\xi}{\,p^4\,}\hsm 
(\eta_{\mu \alpha}^{}p_{\nu}^{}p_{\beta}^{}\!+\!\eta_{\mu\beta}^{}p_{\nu}^{}p_{\alpha}^{}\!+\!
\eta_{\nu \alpha}^{}p_{\mu}^{}p_{\beta}^{}\!+\!\eta_{\nu \beta}^{}p_{\mu}^{}p_{\alpha}^{} )
\nn\\
& -\!\frac{m\hs p^{\rho}}{\,2\hs p^{2}(p^{2}\!+\!m^{2})\,}\!\hsm
\(\varepsilon_{\rho\mu\alpha}^{}\etaB_{\nu \beta}^{} \hsmx +\hsmx\varepsilon_{\rho\mu\beta}^{}
\etaB_{\nu\alpha}^{} \! +\hsm\varepsilon_{\rho\nu\alpha}^{}\etaB_{\mu \beta}^{}
\!+\hsm\varepsilon_{\rho \nu \beta}\etaB_{\mu \alpha}\)
\nn\\
& +\! \frac{\ii}{\,p^2\,}\!
\(\!\etaB_{\mu \alpha}^{}\etaB_{\nu\beta}^{} \!+\! \etaB_{\mu \beta}\etaB_{\nu \alpha}
\!-\! 2\etaB_{\mu \nu}\etaB_{\alpha \beta}\)
\!-\!\frac{\ii\hs 2}{\,p^4\,}
({\eta_{\mu\nu}^{}p_{\alpha}^{}p_{\beta}^{}}\hsm +\hsm \eta_{\alpha\beta}^{}p_{\mu}^{}p_{\nu}^{})
\label{eq:Dh-xi-zeta}
\\
& -\!\frac{\ii\hs 4}{\,(\xi\!+\!\zeta\!-\!4)\hs p^{2}\,}\!\!
\left[\eta_{\mu \nu}^{}\eta_{\alpha \beta}^{}\!+\!\frac{(1\!-\!\xi)}{p^{2}}
(\eta_{\mu \nu}p_{\alpha}p_{\beta} +\eta_{\alpha \beta}p_{\mu}p_{\nu})
\!+\!\frac{(1\!-\!\xi)^2}{p^4}p_{\mu}^{}p_{\nu}^{}p_{\alpha}p_{\beta}
\right]\!,
\nn 
%
\\
D^{}_{\phi}(p) \!=\, &
\frac{\ii\hs 4}{~(\xi\!+\!\zeta\!-\!4)\hs p^2~} \,,
\label{eq:Dphi-xi-zeta}
\end{align}
\eeqs
where we have defined the notation
$\hs\bar{\eta}_{\mu \nu}^{}\! \equiv\! \eta_{\mu \nu}^{}\hsm\! -\! p_{\mu}^{}p_{\nu}^{}/p^{2}$.\
Note that $p^{2}\hsm\!=\hsm\!0$ poles disappear in the above formula
when the graviton propagator is contracted with the conserved sources\,\cite{Deser:1981wh}.\
The last line of \eqref{eq:Dh-xi-zeta}
is mainly generated by the second gauge-fixing term \eqref{eq:gaugefix-2} (due to the existence
of the dilaton field $\phi$ in our new WTMG formulation) and can be decoupled
by taking $\zeta\ito \infty\,$.\  
The choice $\zeta\!=\!\infty$ is called unitary gauge of our WTMG theory
and just corresponds to the conventional TMG theory\,\cite{Deser:1981wh}
that contains no unphysical dilaton field $\phi\hs$.\
Moreover, choosing the unitary gauge with $\zeta\hsm\!=\!\infty\hs$ and $\xi$ finite,
we find that, as expected,
the dilaton propagator $D^{}_{\phi}(p)$ vanishes (implying that the dilaton $\phi$ decouples)
and the graviton propagator
$D_{\mu\nu\alpha\beta}^{h}(p)$ reduces to the following form:
\begin{align}
D_{\mu\nu\alpha\beta}^{h(\rm{U})}(p) \!=\, &
\frac{\,-\ii\hs (\etaB_{\mu \alpha}^{}\etaB_{\nu \beta}^{}\!+\!\etaB_{\mu \beta}^{}\etaB_{\nu \alpha}^{}
\!-\!\etaB_{\mu \nu}^{}\etaB_{\alpha \beta}^{})\,}{~p^2\!+\!m^2\,}
\hsm\! + \!\frac{\ii\hs\xi}{\,p^4\,}\hsm 
(\eta_{\mu \alpha}^{}p_{\nu}^{}p_{\beta}^{}\!+\!\eta_{\mu\beta}^{}p_{\nu}^{}p_{\alpha}^{}\!+\!
\eta_{\nu \alpha}^{}p_{\mu}^{}p_{\beta}^{}\!+\!\eta_{\nu \beta}^{}p_{\mu}^{}p_{\alpha}^{} )
\nn\\
& -\!\frac{m\hs p^{\rho}}{\,2\hs p^{2}(p^{2}\!+\!m^{2})\,}\!\hsm
\(\varepsilon_{\rho\mu\alpha}^{}\etaB_{\nu \beta}^{} \hsmx +\hsmx\varepsilon_{\rho\mu\beta}^{}
\etaB_{\nu\alpha}^{} \! +\hsm\varepsilon_{\rho\nu\alpha}^{}\etaB_{\mu \beta}^{}
\!+\hsm\varepsilon_{\rho \nu \beta}\etaB_{\mu \alpha}\)
\nn\\
& +\! \frac{\ii}{\,p^2\,}\!
\(\!\etaB_{\mu \alpha}^{}\etaB_{\nu\beta}^{} \!+\! \etaB_{\mu \beta}\etaB_{\nu \alpha}
\!-\! 2\etaB_{\mu \nu}\etaB_{\alpha \beta}\)
\!-\!\frac{\ii\hs 2}{\,p^4\,}
({\eta_{\mu\nu}^{}p_{\alpha}^{}p_{\beta}^{}}\hsm +\hsm \eta_{\alpha\beta}^{}p_{\mu}^{}p_{\nu}^{})\hs.
\label{eq:Dh-unitary} 
\end{align} 
In this case the WTMG theory reduces the original TMG Lagrangian \eqref{eq:TMG-L} 
with the gauge-fixing term \eqref{eq:gaugefix-1} (by setting $\phi\!=\!0$).\ 
The unitary-gauge graviton propagator \eqref{eq:Dh-unitary} 
also agrees with Eq.(E.6) of Ref.\,\cite{Hang:2021oso}
that is derived for the conventional TMG theory\,\cite{Deser:1981wh}.\

\vs   

Then, we consider the Landau gauge with the gauge-fixing parameters $\xi\!=\hsm\zeta\!=\hsm 0\hs$.\
In this gauge, we can derive the following graviton and dilaton propagators:
\beqs
\label{eq:DhDphi-Landau}  
\begin{align}
\mathcal{D}_{\mu\nu\alpha\beta}^{h}(p)
=\, & -\!\frac{m\hs p^{\rho}}{\,2\hs p^{2}(p^{2}\!+\!m^{2})\,}
\!\hsm\(\varepsilon_{\rho \mu \alpha}^{}\etaB_{\nu \beta}^{}\!+\hsm \varepsilon_{\rho \mu \beta}^{}
\etaB_{\nu \alpha}^{}\!+\hsm\varepsilon_{\rho \nu \alpha}^{}\etaB_{\mu \beta}^{}\hsm +\hsm
\varepsilon_{\rho \nu \beta}^{}\etaB_{\mu \alpha}^{}\)
\nn\\
& +\!\frac{\ii\hs m^2}{\,p^2(p^{2}\!+\!m^{2})\,}\!
\(\etaB_{\mu \alpha}^{}\etaB_{\nu \beta}^{}\!+\!\etaB_{\mu \beta}^{}\etaB_{\nu \alpha}^{}
\!-\!\etaB_{\mu \nu}^{}\etaB_{\alpha \beta}^{}\)\!,
\label{eq:Dh-Landau}  
\\
D_{}^{\phi}(p) \!=\, &
-\!\frac{\ii}{~p^2~} \,,
\label{eq:Dphi-Landau}
\end{align}
\eeqs
with the gauge parameters ${\xi\hsm =\hsm \zeta\! =\hsm 0}\hs$.\
We find that the above graviton propagator \eqref{eq:Dh-Landau} has the leading high-energy
scaling of $m\hsm /p^3$ (contributed by the first line of $\mathcal{D}_{\mu\nu\alpha\beta}^{h}$
and arising from the gravitational Chern-Simons term)\footnote{%
This leading high-energy behavior $m\hsm /p^3$ of the Landau-gauge graviton propagator
\eqref{eq:Dh-Landau} agrees with Eq.(15) of Ref.\,\cite{Deser1990} 
which used a different gauge-fixing term and BRST quantization of the TMG with dilaton field.}, 
whereas the rest of the graviton propagator
$\mathcal{D}_{\mu\nu\alpha\beta}^{h}$ scales as $m^2\hsm /p^4$ which contributes the sub-leading
high energy behavior.\
This leading high-energy behavior ($m\hsm /p^3$) of the Landau-gauge graviton propagator 
\eqref{eq:Dh-Landau} is better than that of the leading high-energy behavior ($1/p^2$) of the
unitary-gauge graviton propagator \eqref{eq:Dh-unitary}.\

\vs

From Eq.\eqref{eq:Dh-xi-zeta}, we may choose the gauge-fixing parameters  
$\xi\!=\!\zeta\!=\!1$ to define the Feynman gauge.\ 
Thus, we derive the following graviton and dilaton propagators 
in the Feynman gauge:
\beqs
\begin{align}
\mathcal{D}_{\mu \nu \alpha \beta}^{h}(p) =\,
& -\!\frac{\ii}{\,(p^{2}\!+\!m^{2})\,}\!\hsm
\(\!\etaB_{\mu \alpha}^{}\etaB_{\nu \beta}^{}\!+\!\etaB_{\mu \beta}^{}\etaB_{\nu \alpha}^{}
\!-\!\etaB_{\mu \nu}^{}\etaB_{\alpha \beta}^{}\)
\!+\! \frac{\ii}{\,p^{2}\,}\!\hsm
\(\!\etaB_{\mu \alpha}\etaB_{\nu \beta} \!+\! \etaB_{\mu \beta}^{}\etaB_{\nu \alpha}^{}
\! -\hsm 2\hs\etaB_{\mu \nu}^{}\etaB_{\alpha \beta}^{}\)
\nn\\
& -\!\frac{m\hs p^{\rho}}{\,2\hs p^{2}(p^{2}\!+\!m^{2})\,}
\!\hsm\(\varepsilon_{\rho \mu \alpha}^{}\etaB_{\nu \beta}^{}\!+\hsm \varepsilon_{\rho \mu \beta}^{}
\etaB_{\nu \alpha}^{}\!+\!\varepsilon_{\rho \nu \alpha}^{}\etaB_{\mu \beta}^{}\hsm +\hsm
\varepsilon_{\rho \nu \beta}^{}\etaB_{\mu \alpha}^{}\)
\nn\\
& +\!\frac{\ii}{\,p^4\,}(\eta_{\mu\alpha}^{}p_{\nu}^{}p_{\beta}^{}\!+\!
\eta_{\mu \beta}^{}p_{\nu}^{}p_{\alpha}^{}\!+\!\eta_{\nu \alpha}^{}p_{\mu}^{}p_{\beta}^{}
+\eta_{\nu \beta}^{}p_{\mu}^{}p_{\alpha}^{})
\nn\\
& +\!\frac{\ii\hs 2}{\,p^4\,}\big(\eta_{\mu \nu}^{}\eta_{\alpha \beta}^{}\hs p^2
\hsm -\hsm \eta_{\mu \nu}^{}\hs p_{\alpha}^{}p_{\beta}^{}
\!-\hsm\eta_{\alpha \beta}^{}\hs p_{\mu}^{}p_{\nu}^{}\big) \,,
\label{eq:Dh-xi=zeta=1}
\\
D_{}^{\phi}(p) \!=\, &
-\frac{~\ii\hs 2~}{~p^2~} \,,
\label{eq:Dphi-xi=zeta=1}
\end{align}
\eeqs
where we choose the gauge-fixing parameters
$\xi\hsm =\hsm \zeta\hsm =\hsm 1\hs$.\ 

\vs 

Then, we take the massless limit $m\hsm\ito 0$ for the graviton propagator \eqref{eq:Dh-Landau}
in Landau gauge and find that the graviton propagator \eqref{eq:Dh-Landau} vanishes in the massless limit:
\beq
\label{eq:Dh-m=0-Landau}
\lim_{m\rightarrow 0}^{}\! \mathcal{D}_{\mu \nu \alpha \beta}^{h}(p) = 0 \,,
\hspace*{10mm}
({\xi\hsm =\hsm \zeta\hsm =\hsm 0}).
\eeq
We can also take the massless limit for the graviton propagator \eqref{eq:Dh-xi=zeta=1} in Feynman gauge
and derive the following form:
\beq
\begin{split}
\lim_{m\rightarrow 0}^{}\!\!
\mathcal{D}_{\mu \nu \alpha \beta}^{h}(p) \!=\,
& \frac{\ii}{\,p^{2}\,}\!\hsm
\( \etaB_{\mu \nu}^{} \etaB_{\alpha \beta}^{}
\hsm-\hsm \frac{\,2\hs p_{\mu}^{}p_{\nu}^{}p_{\al}^{}p_{\be}^{}\,}{p^4}\!
\)
\\
& +\!\frac{\ii}{\,p^4\,}(\eta_{\mu\alpha}^{}p_{\nu}^{}p_{\beta}^{}\!+\!
\eta_{\mu \beta}^{}p_{\nu}^{}p_{\alpha}^{}\!+\!\eta_{\nu \alpha}^{}p_{\mu}^{}p_{\beta}^{}
\!+\hsm\eta_{\nu \beta}^{}p_{\mu}^{}p_{\alpha}^{})
\hs,
\hspace*{6mm}
({\xi\hsm =\hsm \zeta\hsm =\hsm 1}).~~
\label{eq:Dh-m=0-Feynman}
\end{split}
\eeq
Eq.\eqref{eq:Dh-m=0-Landau} shows that 
in the massless limit the graviton propagator
$\mathcal{D}_{\mu \nu \alpha \beta}^{h}(p)$
vanishes in the Landau gauge.\
This is expected since the 3d massless graviton field $h_{\mn}^{}$
has no physical degree of freedom and
could only have a gauge-dependent unphysical propagator.\
Hence it is no surprise that the gauge-dependent unphysical massless graviton propagator 
simply vanishes under a certain gauge choice.\
Eq.\eqref{eq:Dh-m=0-Feynman} shows that in the massless limit, 
the graviton propagator in the Feynman gauge contains 
no physical component, and this reflects the fact that 
the 3d massless graviton has no physical degree of freedom.

\vs

Taking the massless limit we have a direct correspondence: 
the unphysical scalar field $\phi$ becomes
a physical massless scalar field known as dilaton 
(having one physical degree of freedom),
and $h_{\mu\nu}^{}$ becomes the massless graviton field 
of a pure 3d GR theory (denoted as GR$_3$) 
that has no physical degree of freedom 
and simply does not propagate in the Landau gauge.\
We have seen explicitly that the correct massless limit 
of the TMG is not just a pure GR$_3$,
but instead it contains the pure GR$_3$ theory plus 
an extra massless scalar degree of freedom
(known as the dilaton $\phi\hs$).\
This is because the massless pure GR$_3$ theory 
has no physical degree of freedom, and the one physical degree
of freedom of the 3d massive graviton 
in the WTMG theory is converted to the one physical degree
of freedom of the scalar dilaton field in the massless limit.\
This massless dilaton field $\phi$ couples to 
the trace of energy momentum tensor $T_{\mu\nu}^{}\hs$.\
Taking out this extra scalar dilaton field leads 
to a discontinuity between the 3d pure gravity theory and
the massless limit of the conventional 3d TMG theory.\footnote{%
In passing, the appearance of a dilaton-like scalar 
contribution in the massless limit of graviton-exchange was shown for a 4-point 
matter scattering amplitude\,\cite{other2a-3d-CS}.\ 
However, our current WTMG formula has essential difference 
from \cite{other2a-3d-CS} because we introduce the 
dilaton field $\phi$ through the Weyl transformation 
\eqref{eq:CT-gmunu} and explicitly determines dilation-graviton
interactions at the action (Lagrangian) level.\ 
The constraint between the dilaton $\phi$ and the 
trace of graviton field ($h$) is imposed through our 
second gauge-fixing term \eqref{eq:gaugefix-2}.}\ 
On the other hand, our new formulation of 
the WTMG theory (including the unphysical dilaton field $\phi$) can provide a consistent realization 
of the massless limit of the TMG theory, because 
our WTMG formulation naturally conserves the physical 
degrees of freedom of the TMG theory.

Next, for comparison, we consider the 3d massless GR with the harmonic gauge-fixing:
\beq
\mathcal{L}_{\rm{GR}}^{} =\,
\frac{1}{\,\kappa^2\,}\sqrt{-g\,}R -
\frac{1}{\,{2}\hs\xi\,}
\Big(\hsm\partial_{\nu}^{}h^{\mn}\!-\!\frac{1}{\hs 2\hs}\partial_{\mu}^{} h\Big)^{\hsm\!2},
\eeq
where we flip an overall sign such that the Einstein-Hilbert term has the correct sign
for the massless graviton.\
Then, we derive the massless graviton propagator:
\beqs
\begin{align}
\label{eq:Dh=m0-3d}
\mathcal{D}_{\mu \nu \alpha \beta}^{h}(p) =\,&
\frac{\,-\ii~}{\,p^2\,}\hsm\Big[\hsm
\big(\eta_{\mu \alpha}^{}\eta_{\nu \beta}^{}\!+\hsm \eta_{\mu \beta}^{}\eta_{\nu \alpha}^{}
\!-\!2\eta_{\mu \nu}\eta_{\alpha \beta}\big)
\nn\\
&\hspace*{7mm} +\!\frac{\,(\xi\!-\!1)\,}{\,p^2\,}(\eta_{\mu\alpha}^{}p_{\nu}^{}p_{\beta}^{}\!+\!
\eta_{\mu \beta}^{}p_{\nu}^{}p_{\alpha}^{}\!+\!\eta_{\nu \alpha}^{}p_{\mu}^{}p_{\beta}^{}
\!+\hsm\eta_{\nu \beta}^{}p_{\mu}^{}p_{\alpha}^{}\big)\hsm\Big]
\hspace*{8mm}
\\
=\,&  \frac{-\ii}{~p^2~}
\big(\eta_{\mu \alpha}^{}\eta_{\nu \beta}^{}\!+\hsm \eta_{\mu \beta}^{}\eta_{\nu \alpha}^{}
\!-\!2\hs\eta_{\mu \nu}\eta_{\alpha \beta}\big) ,
\hspace*{7mm} (\xi\!=\!1)\hs.
\label{eq:Dh=m0-3d-Feynman}
\end{align}
\eeqs
This also differs from the \eqref{eq:Dh-m=0-Landau} or \eqref{eq:Dh-m=0-Feynman} that we derived
from the massive graviton propagator of our WTMG theory under the massless limit.\
Since the 3d massless graviton has no propagating physical degree of freedom, its massless propagator
is a gauge artefact, so it is expected that different ways of obtaining this 3d massless graviton
propagator may give different answers.\ A key point is that the 3d equation of motion (Einstein equation)
for the on-shell massless graviton field has zero solution (corresponding to the flat Minkowski metric).

As another comparison, we consider
the Fierz-Pauli gravity\,\cite{PF}-\cite{Hinterbichler:2011tt}
defined in 3d spacetime 
for massive graviton fields $h^{\mn}$ with mass $m\,$.\ 
It has the following Lagrangian,
\begin{equation}
	\La_{\rm{FP}}=
	\fr{1}{2}(\pd_\mu h)^2 \!-\! \fr{1}{2}(\pd_\al h_{\mn})^2
	\!-\! \pd_\mu h^{\mn}\pd_\nu h + \pd_\mu h^{\mu\al}\pd^\nu h_{\nu\al}
	\!+\! \fr{1}{2}{m^2}(h^2 \!-\! h_{\mn}^2) \,,
\end{equation}
and gives the massive graviton propagator, 
\begin{equation}
\label{eq:Dh-PF}
\D_{\mn\ab}^{\text{FP}}(p) \,=\,	
-\frac{\ii}{\,2\,}\frac{~{\hat\eta}_{\mu\al}^{}{\hat\eta}^{\nu\be} \!+\hsm \hat{\eta}^{\mu\be}\hat{\eta}^{\nu\al}
\!-\hsm \hat{\eta}^{\mu\nu}\hat{\eta}^{\al\be}~}
{\,p^{2} \hsm + m^{2}\,}\,,
\end{equation}
where we have denoted $\hat{\eta}_{\mn}^{}\!\equiv\hsm\eta_{\mn}^{}\!+p^{\mu}p^{\nu}\hsmx /m^2$.\
The massive graviton of the 3d Fierz-Pauli gravity has spin-2 with 2 physical degrees of freedom.\
Taking the massless limit $m\ito 0\hs$, we find that the three terms in the
numerator of the Fierz-Pauli graviton propagator \eqref{eq:Dh-PF} have coefficients $(1,\,1,\,-1)$,
which should be compared to the coefficients $(1,1,-2)$ of the three terms in the numerator of
the 3d massless graviton propagator \eqref{eq:Dh=m0-3d-Feynman}.\
The comparison between the two types of numerator coefficients  $(1,\,1,\,-1)$ versus $(1,1,-2)$
shows a discontinuity in taking the massless limit of the 3d Fierz-Pauli graviton propagator,
which originates from the non-conservation of the physical degrees of freedom of the
3d Fierz-Pauli graviton in the massless limit, namely, $2\neq 0\hs$.\
This discontinuity is also evident from comparing the 3d Fierz-Pauli graviton propagator with
the massless limit of the massive graviton propagator of the WTMG theory 
(whose massive graviton has 1 physical degree of freedom), 
as shown in Eq.\eqref{eq:Dh-m=0-Landau} or \eqref{eq:Dh-m=0-Feynman}.\
For comparison, 
we note that the 4d Fierz-Pauli massive graviton propagator has a different discontinuity 
in the massless limit,
known as the van\,Dam-Veltman-Zakharov (vDVZ) discontinuity\,\cite{vDVZ}\cite{vDVZ2},
where the massive and massless graviton propagators have numerator coefficients  
$\(\!1,1,-\frac{2}{3}\)$ versus $(1,\,1,\,-1)$.\ 
This is because the physical degrees of freedom of the 4d Fierz-Pauli graviton are not conserved 
in the massless limit, namely, $5\neq 2\hs$, (instead of $2\neq 0\hs$ in the case of taking massless
limits for the 3d Fierz-Pauli graviton).\ 

\vs

From the above, we see that by taking the massless limit,  
our new WTMG formulation thus has 
a massless graviton field of the 3d GR theory (called GR$_3$, 
having no physical degree of freedom)
and a massless scalar field dilaton (having one physical degree of freedom).\
Hence, the physical degree of freedom is conserved before and 
after taking the massless limit of the TMG theory:
\beq
\label{eq:hmunu0Mlimit1=0+1} 
1\Dbrack{h_{\mu\nu}^{}|\rm{TMG}} \xLongrightarrow{\rm{Massless\,Limit~}}
0\Dbrack{{h_{\mu\nu}^{}}|\rm{GR}_3} + 1\Dbrack{\phi|\rm{Dilaton}} ,
\eeq
where the number ``1'' stands for one physical degree of freedom and the number ``0'' stands for zero physical degree of freedom.\
To take massless limit of the WTMG theory, we remove the Chern-Simons term of the 
WTMG Lagrangian \eqref{eq:LTMG-phi} and obtain: 
%
\begin{equation}
\label{eq:limitTMG}
\mathcal{L}^{\rm{massless}}_{\rm{WTMG}} =
-\frac{1}{\,\kappa^2\,}\sqrt{-g\,}\hs e^{-\kappa\phi/2}
\!\(\!R +
\frac{1}{2}g_{\mn}^{}\partial^{\mu}\hsm\phi\hs\partial^{\nu}\hsm\phi \!\)\!,
\end{equation}
which we call GRD$_3$ for notational convenience.\
The Lagrangian \eqref{eq:limitTMG} shows that the dilaton $\phi$ becomes a physical
scalar field in the massless limit.\footnote{%
We will further elaborate this point below Eq.\eqref{eq:gauge-fixingLSZ-2}.\ 
We note that the on-shell gauge-fixing function vanishes 
in the massless limit 
$m\!=\!0\hs$, so the combination $(h\hsm -\hsm\phi)$ 
is no longer constrained by this gauge-fixing in the massless limit  
and the dilaton $\phi$ is released to be a physical massless scalar.\ 
Moreover, scattering amplitude of the physical dilatons ($\phi$) 
is gauge-independent, and corresponds to the double copy 
of the scattering amplitude of the physical gauge bosons 
($\AT^a$) in 3d massless YM theory 
(cf.\ Sec.\,\ref{sec:4.3}).}\ 
This clearly demonstrates that the massless limit of the 3d WTMG theory contains
the massless GR$_3$ (with an unphysical massless graviton field) 
coupled to a physical massless dilaton field.\
In this massless limit, the one physical degree of freedom 
of the massive graviton decouples and converts to 
a physical scalar boson (corresponding to the transverse 
trace mode of the massive graviton, known as the dilaton).\

\vspace*{0.5mm}

We can gain some insight into the origin of the above massless
effective Lagrangian \eqref{eq:limitTMG} from the string theory.\
Consider the following low-energy effective Lagrangian of
the bosonic string
in $D$-dimension\,\cite{Gross:1986mw}\cite{Horowitz:1993jc}:
\begin{equation}
\label{eq:L-string-3d}
\mathcal{L}_{\rm{string}}^{}= \frac{1}{\,\kappa^2\,}\sqrt{-g\,}\!
\left[\hsm R\hsm -\hsm\frac{1}{\,D\!-\!2\,}(\nabla\phi)^{2} \!-\!
\frac{1}{12}e^{-4\phi/(D-2)}H_{\mu \nu \rho}H^{\mu \nu \rho} \right]\!.
\end{equation}
We note that the dilaton kinetic term has the correct sign (when $D\!\geqq\!3$) 
and makes the dilaton a physical field.\
In the 3d spacetime, the anti-symmetric tensor field $H_{\mu\nu\rho}^{}$ 
is proportional to the volume form $\varepsilon_{\mu\nu\rho}^{}\hs$.\  
By choosing $D\!=\!3$ and
$H_{\mu\nu\rho}^{}\!=\!(\ii\hs 2e^{2\phi}/\ell)\varepsilon_{\mu \nu \rho}^{}$
with $\hs\ell\,$ being a constant of dimensions of length,
we derive the Lagrangian \eqref{eq:L-string-3d}	as follows:
\begin{equation}
\label{eq:L-string2-3d}
\mathcal{L}_{\text{string}}^{} = \frac{1}{\,\kappa^2\,}
\sqrt{-g\,}\!\left[\hsm R \!-\!(\nabla\phi)^{2} \!-\!\frac{2}{\,\ell^2\,}\right] \!.
\end{equation}
In the flat spacetime, the cosmological constant 
$\Lambda\hsm\!=\!-1/\ell^{2}$ vanishes, corresponding to
$\ell\!=\!\infty\,$.\
In this case, we reproduce the effective Lagrangian of the WTMG theory 
to the lowest order of $\phi$ and in the massless limit \eqref{eq:limitTMG}
with rescaling $\,\phi\rightarrow\phi/\sqrt{2\,}\,$.

\vspace*{0.5mm}

In passing, Ref.\,\cite{other2a-3d-CS} showed a surviving term 
as a hint of dilaton-like contribution   
of the four-point scalar amplitude in the massless graviton limit 
for the conventional TMG coupled to a massive scalar, 
but it did not introduce the dilaton field at Lagrangian level
and it remains unclear whether this surviving term
is the real contribution of a dilaton field.\
We will explicitly clarify the origin of this surviving 
contribution by using our new WTMG formulation (coupled to
massive scalar field) in a separate paragraph below 
Eq.\eqref{eq:Amp4S|m=0|U=L(phi)}.\  
We also note that Ref.\,\cite{1311.4736}
considered an extended 3d TMG Lagrangian 
by adding a cosmological constant ($\Lambda$) and 
a Fierz-Pauli term (with a different mass $m$ 
from the Chern-Simons mass $\mu\,$); 
it claimed its graviton has 3 physical degrees of freedom.\ 
Then, it considered around its Eq.(48) the anyon scattering 
and got the related potential energy for the TMG without 
the Fierz-Pauli mass.\ It took the limit 
$\Lambda,\hs m\ito 0\hs$ to avoid the vDVZ discontinuity; 
but it did not explain how the 3 physical DoF could be 
conserved in this massless limit.\  
Moreover, as Ref.\,\cite{Deser-2002} pointed out, 
such mixed models (combining the TMG theory with the
Fierz-Pauli mass term) suffer from inconsistencies 
causing the complex masses and ghost excitations.\ 

\vspace*{1.5mm}
\subsection{\hspace*{-2mm}Pure Scalar Interactions in the WTMG and WTMGS Theories}
\label{sec:2.2new}
\vspace*{1mm}
\vs  

We note that WTMG Lagrangian \eqref{eq:LTMG-phi} contains the following
dilaton kinematic term and pure self-interactions:
\\[-7mm]
\begin{align}
\label{eq:WTMG-Lphi}
\mathcal{L}_{\rm{WTMG}}^{\phi} = 
-\fr{1}{\hs 2\hs}e^{{-}\kappa\phi/2}
{\eta}_{\mn}^{}\partial^{\mu}\hsm\phi\hs\partial^{\nu}\!\phi \,,
\end{align}
where the exponential factor $e^{-\kappa\phi/2}$ will induce infinity number of
dilaton self-interactions of the form 
$\,\phi^n(\partial\phi)^2\,$ (with $n\!=\!1,2,3,\cdots\hsm$).\ 
The naive power counting would suggests that based on the pure dilaton Lagrangian 
\eqref{eq:WTMG-Lphi}, a general $N$-point dilaton amplitude has the high-energy-power dependence
of $E^2$.\ But our explicit calculations in Sec.\,\ref{sec:4.1.2} find 
that the sum of contributions from pure dilaton self-interactions to 
the on-shell four-dilaton amplitude actually vanishes.\   

A key observation here is to show that the pure dilaton Lagrangian 
\eqref{eq:WTMG-Lphi} is equivalent to a free massless scalar theory 
under a local and invertible field redefinition.\ 
We can prove that the above pure dilaton self-interactions
can be removed by a nonlinear field redefinition,
\beq 
\label{eq:transf-phi-phihat}
\phi ~\to~ \phih = 
\frac{4}{\hs\kappa\hs}\!\(1\!-\hsm e^{-\kappa\phi/4}\) \hsm ,
\eeq 
which gives 
$\hs \phi\!=\!-\frac{4}{\hs\kappa\hs}
 \ln\!\big(1\!-\!\frac{\,\kappa\hs}{4}\phih\big)$.\ 
Substituting this into Eq.\eqref{eq:WTMG-Lphi}, 
we deduce the following pure dilaton Lagrangian:
\beq 
\label{eq:LphiHat}
\mathcal{L}_{\rm{WTMG}}^{\phih} = 
-\fr{1}{\hs 2\hs}
{\eta}_{\mn}^{}\partial^{\mu}\hsm\phih\hs\partial^{\nu}\!\phih \,,
\eeq
which describes a free massless scalar without any self-interactions.\
Since the a nonlinear field redefinition such as Eq.\eqref{eq:transf-phi-phihat}
will leave the on-shell $S$-matrix element invariant, 
the two Lagrangians \eqref{eq:WTMG-Lphi}
and \eqref{eq:LphiHat} are physically equivalent.\  
This invariance is based on 
a general theorem\,\cite{FieldRedef-1}-\cite{FieldRedef-6} 
stating that the on-shell $S$-matrix elements remain invariant  
under local (with no derivatives) or almost-local (with a finite 
number of derivatives) field redefinitions 
(where appropriate wave function renormalizations will be introduced
at loop level).\ 
(The present study in Section\,\ref{sec:4}
is confined to the tree-level analyses.)\
This invariance means that an on-shell $N$-point dilaton amplitude ($N\!\!\!\geqq\!\!3$) 
as computed by using the the pure dilaton Lagrangian \eqref{eq:WTMG-Lphi} must vanish.\ 
Hence, the pure dilaton self-interactions \eqref{eq:WTMG-Lphi} 
cannot have net contribution to the leading high-energy behavior 
of dilaton scattering amplitudes.\ 
Based on this conclusion, we can ignore the pure dilaton self-interactions \eqref{eq:WTMG-Lphi} 
in our power counting analysis of the $N$-point dilaton amplitudes
in Section\,\ref{sec:3.3new}.\ 

\vs 

Next, we couple the TMG Lagrangian \eqref{eq:TMG-L} to a physical massive 
real scalar field $\psi$ with mass $m_s^{}$ (which we denote as TMGS theory):
\vspace*{-2.5mm}
\begin{align}
\mathcal{L}_{\rm{TMGS}}^{} =&\,
\!\sqrt{-g\,}\!\left(\!\hsm -\kappa^{-2}R
\hsm -\!\fr{1}{\,2\,}g^{}_{\mn}
\partial^\mu\hsm\psi\hs\partial^\nu\hsm\psi
	\!-\!\fr{\,m_{\hsm s}^{2}}{2}\psi^2
	\!-\frac{\,m_s^2\kappa^2\hs}{4!}\psi^4\!\right) 
\nn\\
& 
+\!\frac{\,\varepsilon^{\mu\nu\rho}\,}{~2\hs\tilde{m}\hs\kappa^2~}
	\Gamma_{\ \rho\beta}^{\alpha}
	\Big(\partial_{\mu}\Gamma_{\ \alpha \nu}^{\beta}
	\!+\!\frac{\hs 2\hs}{3}\Gamma_{\ \mu \gamma }^{\beta}\Gamma_{\ \nu \alpha}^{\gamma}\Big) . 
\label{eq:L-TMGS-UG} 
\end{align}
Then, we make the Weyl transformation \eqref{eq:CT-gmunu} 
for the TMGS Lagrangian \eqref{eq:L-TMGS-UG} 
and derive the following WTMG-Scalar (WTMGS) Lagrangian 
that couples the massive scalar field $\psi$ to the WTMG Lagrangian: 
\begin{align}
\label{eq:L-WTMGS}
\mathcal{L}_{\text{WTMGS}}= 
& \sqrt{\hsm -{g}\,}\hs e^{{-}\kappa\phi/2}\hsm
\bigg(\!\!\! -\!\kappa^{-2}R \!-\!\fr{1}{\hs 2\hs}
{g}_{\mn}^{}\partial^{\mu}\hsm\phi\hs\partial^{\nu}\!\phi
\!-\!\fr{1}{\hs 2\hs}{g}_{\mn}^{}
\partial^{\mu}\hsm\psi\hs\partial^{\nu}\!\psi
\!-\!\fr{\,m_s^2\,}{2}e^{{-}\kappa\phi}
\psi^2
\nn\\
&~
{-\frac{\,m_s^2 \kappa^2}{4!}e^{{-}\kappa\phi}\psi^4\!\bigg)\!}
+\!\frac{~\varepsilon^{\mu\nu\rho}\,}{\,2\hs\mt\hs\kappa^2\,}
\Gamma_{\ \rho \beta}^{\alpha}
\Big(\!\partial_{\mu}\Gamma_{\ \alpha \nu}^{\beta}
\!+\!\frac{2}{3}\Gamma_{\ \mu \gamma }^{\beta}\Gamma_{\ \nu \alpha}^{\gamma}\hsm\Big) .
\end{align}
In the above we include a quartic scalar self-interaction term
$\psi^4$ (whose coupling has mass-dimension $+1$ in 3d spacetime
and is super-renormalizable), and it is needed for the double-copy
analysis of Section\,\ref{sec:4.2.2}.\  
The above WTMGS Lagrangian contains 
the following scalar kinematic terms and 
pure scalar interactions for $(\phi,\hs \psi)\hs$:
\begin{align}
\label{eq:WTMG-Lphi-psi}
\hspace*{-4mm}
\mathcal{L}_{\rm{WTMGS}}^{\phi\psi} = 
-\hsm\fr{1}{\hs 2\hs}e^{{-}\kappa\phi/2}{\eta}_{\mn}^{}
\big(\partial^{\mu}\hsm\phi\hs\partial^{\nu}\!\phi 
\hsm +\hsm\partial^{\mu}\hsm\psi\hs\partial^{\nu}\!\psi\big)
\!-\hsm e^{-3\kappa\phi/2}\hsm\bigg(\!\hsm 
\frac{~m_s^2\hs}{\hs 2\hs}\psi^2
\!+\!\frac{m_s^2 \kappa^2}{4!}\psi^4\!\bigg) .
\end{align}
We make the following field redefinitions for $(\phi,\hs \psi)$:
\begin{equation}
\label{eq:transf-phi-psi}
\begin{aligned}
\phi ~\to~ \phih &= \frac{\,4\,}{\hs\kappa\hs}\! 
\(\!1\!-\!e^{-\kappa\phi/4}\hsm\cos\!\frac{\,\kappa\psi\,}{4}\!\) \!,
\\
\psi ~\to~ \psih &= \frac{\,4\,}{\kappa}\hs e^{-\kappa\phi/4}\hsm 
\sin\!\frac{\hs\kappa\psi\hs}{4} \,.
\end{aligned}
\end{equation}
Substituting the above field redefinitions into the scalar Lagrangian \eqref{eq:WTMG-Lphi-psi}, 
we derive its first term (associated with the partial derivatives) as follows: 
\begin{align}
\label{eq:L-phih/psih-kin}
\mathcal{L}_{\rm{WTMGS}}^{\phi\psi,\rm{kin}} = -\frac{1}{2}{\eta}_{\mn}^{}\partial^{\mu}\hsm\phih\hs\partial^{\nu}\!\phih-\frac{1}{2}{\eta}_{\mn}^{}\partial^{\mu}\hsm\psih\hs\partial^{\nu}\!\psih \,.
\end{align}
This describes two free massless scalar fields $\phih$ and $\psih\hs$,
and the associated nonlinear dilaton factor $e^{-\kappa\phi/2}$ disappears.\  
The on-shell scattering amplitudes of $\phi$ and $\psi$
should be invariant under field redefinitions \eqref{eq:transf-phi-psi}.\ 
Thus, the only nontrivial pure scalar-interactions of the Lagrangian \eqref{eq:WTMG-Lphi-psi} arise from its second parentheses 
(containing $e^{-3\kappa\phi/2}\psi^2$ and $e^{-3\kappa\phi/2}\psi^4$ terms) after the field redefinitions \eqref{eq:transf-phi-psi},  
but they do not contain any partial derivative.\ 
This means that for the $N$-point pure scalar scattering amplitudes 
of $\phi$ and/or $\psi\hs$, the contribution from the derivative interactions of the first parentheses of 
Eq.\eqref{eq:WTMG-Lphi-psi} must vanish, 
whereas the contribution from the non-derivative interactions
of the second parentheses of Eq.\eqref{eq:WTMG-Lphi-psi} should scale 
as no more than $O(E^0)$ in the high energy limit. 

Moreover, we observe that the only nontrivial scalar derivative interactions are given by
the following interactions of $(\phih,\psih)$ with gravitons:
\begin{align}
\label{eq:L-phih/psih-h}
\mathcal{L}_{\rm{WTMGS}}^{\phih\psih h,\rm{kin}} = -\frac{1}{2}\sqrt{-g\,}\hs{g}_{\mn}^{}\hsm\!\(\!\partial^{\mu}\hsm\phih\hs\partial^{\nu}\!\phih
+\partial^{\mu}\hsm\psih\hs\partial^{\nu}\!\psih\) \!.
\end{align}
This Lagrangian is invariant under exchange of the two scalar fields, 
$\phih \!\leftrightarrow\!\psih\hs$.\ 
Hence, $\phih$ and $\psih$ interact with the graviton field
$h_{\mn}^{}$ in precisely the same way.\ 
We note that the derivative interactions of Eq.\eqref{eq:L-phih/psih-h} 
contribute to the leading-order amplitudes of the $N$-point $\phih$ scattering and 
of the $N$-point $\psih$ scattering under high energy expansion.\  
Hence, the leading-order amplitudes of the $N$-point $\phih$ scattering 
and of the $N$-point $\psih$ scattering are contributed by the scalar-graviton vertices
$h^n(\partial\phih)^2$ and $h^n(\partial\psih)^2$, 
and the two leading-order amplitudes should be equal to each other,
\beq 
\label{eq:M0[Nphi]=M0[Npsi]}
\MM_0^{}[N\phih] =\! \MM_0^{}[N\psih],
~~\Longrightarrow~~ \MM_0^{}[N\phi] =\! \MM_0^{}[N\psi],
\eeq 
where the subscript ``$_0$'' denotes the leading order amplitude 
under high energy expansion.\  
(It might be possible that the LO amplitude happens to vanish
due to certain accidental cancellations, under which Eq.\eqref{eq:M0[Nphi]=M0[Npsi]}
would reduces a trivial identity $0\hsm =\hsm 0\hs$.)
The second equality of Eq.\eqref{eq:M0[Nphi]=M0[Npsi]} holds 
because each $N$-point on-shell scattering amplitude is
invariant under the field redefinitions \eqref{eq:transf-phi-psi}.\ 
We will use the second equality of Eq.\eqref{eq:M0[Nphi]=M0[Npsi]} 
for the tree-level double-copy construction of the 
four-point leading-order dilaton amplitudes in Section\,\ref{sec:4.2.2}.\

\section{\hspace*{-2mm}%
TGRET for Topological Graviton Mass Generation}
\label{sec:3}
\label{sec:3new}
\vspace*{1mm}

In this section, we formulate the mechanism of topological
mass-generation of gravitons in the Weyl-transformed TMG (WTMG) theory 
at the $S$-matrix level by newly proposing and 
proving a Topological Graviton Equivalence Theorem (TGRET).\
In Section\,\ref{sec:3.1new} we derive some key relations for the 3d massive
polarization vectors and tensors that will be used for the proof of the TGRET
in the following subsection.\  
In Section\,\ref{sec:3.1}, we prove the TGRET which quantitatively connects the 
$N$-point scattering amplitudes of massive gravitons ($\hP$)  
to the corresponding scattering amplitudes of dilatons ($\phi$)
in the high energy limit.\ 
Different from the Topological Gauge-boson Equivalence Theorem
(TGAET)\,\cite{Hang:2021oso}\cite{Hang:2023fkk} formulated 
previously for the 3d TMYM theory, this is highly nontrivial 
because in the conventional TMG theory there is no proper
scalar-field degree of freedom whose scattering amplitudes
could quantitatively mimic the high-energy behavior of the corresponding
graviton scattering amplitudes.\  Hence, the new WTMG theory with dilaton field 
(Section\,\ref{sec:2}) will be essential 
for the correct formulation of the TGRET.\   
In Section\,\ref{sec:3.2}, 
we further develop a generalized gravitational power counting method
for the graviton (dilaton) scattering amplitudes 
in the WTMG theory
and demonstrate that the TGRET can provide a general mechanism
to ensure the nontrivial large energy cancellations in the
$N$-point massive graviton scattering amplitudes 
(with $N\hsm\!\geqq\! 4$).\

\subsection{\hspace*{-2mm}Relations for 3d Massive Polarization Vectors and Tensors}
\label{sec:2.2}
\label{sec:3.1new}
\vspace*{1mm}

We consider the gauge boson $A^{a\mu}$ of the 3d TMYM theory, having a general momentum
$p^{\mu}\!=\! E(1,\beta s_{\theta}^{},\beta c_{\theta}^{})$
with $\beta\!=\hsm\!\sqrt{1\!-\!m^2/E^2\,}$. We solve the equation of motion \eqref{eq.EOM} and
explicitly construct the physical polarization vector of 
the gauge boson $A^{a\mu}$ as follows:
\begin{equation}
\label{eq.PVfromEOM}
\epP^{\mu}
=\frac{1}{\sqrt{2\,}\,}\hsm\big(\bar{E}\beta,\bar{E}s_{\theta}^{}\!+\!\ii\hs \sp\hs c_{\theta}^{},
\bar{E}c_{\theta}^{}\!-\!\ii\hs\sp\hs s_{\theta}^{}\big) \hs,
\end{equation}
where we have defined a dimensionless energy parameter
$\bar{E}\!=\!E/m\,$ for convenience.\ 
In the above, $\sp\!=\!\mt/m\!=\!\pm 1\hs$ 
denotes the helicity of the gauge boson and
only $-1$ or $+1$ (rather than both values) is physically independent in the 3d spacetime,
so we will just choose $\sp\!=\!-1\hs$ for the following analysis unless specified otherwise.\

\vs

We further construct the longitudinal and transverse polarizations of $A^{a\mu}$ as follows:
\begin{equation}
\label{eq:epL-epT}  
\epL^{\mu}=\bar{E}\hs (\beta,s_{\theta}^{},c_{\theta}^{})\hs,
\hspace*{5mm}
\epT^{\mu} =  (0,-c_{\theta}^{},s_{\theta}^{}) \hs,
\end{equation}
where under high energy expansion we have,
$\epL^{\mu}\!=\hsm\epS^{\mu} + v^{\mu}$, with
$\epS^{\mu}\!=\!p^{\mu}/m$ and $v^{\mu}\!=\!O(m/E)\hs$.\
Thus, we can express the physical polarization vector $\epP^{\mu}$
into a sum of the longitudinal and transverse polarizations\,\cite{Hang:2021oso}:
\beq
\label{eq:epP=epL+epT}
\epP^{\mu} = \frac{1}{\sqrt{2\,}\,}
\big(\epsilon^{\mu}_{\rm{L}}+\bepT^{\,\mu}\big)\hs ,
\eeq
where we denote $\,\bepT^{\,\mu}\!\equiv\! -\ii\hs\sp\hs\epT^{\,\mu}\hs$
for notational convenience.\
We note that the other orthogonal combination
$\epX^{\mu}\!=\!\big(\epsilon^{\mu}_{\rm{L}}\hsm\!-\!
\bepT^{\,\mu}\big)/\sqrt{2\,}$
corresponds to the polarization of an unphysical degree of freedom.\

\vs

For the polarization tensor of graviton field $h_{\mn}^{}$
in the TMG theory, it can be expressed as the tensor product
of the polarization vectors of the YM gauge fields, namely, $\epP^{\mn} \!\!=\! \epP^{\mu}\epP^{\nu}\hs$.\
This is equivalent to decomposing the tensor product of two vector representations
of the little group into an irreducible representation\,\cite{Rumbutis:2022gqh}.\
Then, using the formula \eqref{eq:epP=epL+epT} and the relation
$\epL^{\mu}\!=\hsm\epS^{\mu}\!+\hsm v^{\mu}$,
we further derive the physical polarization tensor $\epP^{\mu\nu}$
of the graviton field as follows:
\begin{equation}
\label{eq:gravitonPol}
\epP^{\mn} = \epP^{\mu}\epP^{\nu}
= \frac{1}{\hs 2\hs}\Big[\epS^{\mn} \!+\hsm\bepT^{\,\mn} \!+\!\big(\epS^\mu\tepT^{\,\nu} \!+\!\epS^\nu\tepT^{\,\mu}\big)
\!+\!\big(v^\mu\bepT^{\,\nu}\!+\!v^\nu\bepT^{\,\mu} \!+\! v^\mu v^\nu\big)\!
\Big],
\end{equation}
where we denote $\tepT^{\,\mu}\!\equiv\!\bepT^{\,\mu}\!+\! v^\mu$ and
$v^\mu\!=\!\epL^\mu\!-\!\epS^\mu\!=\!O(m/E)$
is an energy-suppressed quantity
under high energy expansion.\
In the above, we have defined the scalar polarization tensor
and transverse polarization tensor:
\beq
\label{eq:epS-epT-munu}
\epS^{\mn} = \epS^\mu\epS^\nu \,,   \hspace*{6mm}
\epT^{\mn} = \epT^{\,\mu}\epT^{\,\nu}\,,  \hspace*{6mm}
\bepT^{\,\mn}
= \bepT^{\,\mu}\bepT^{\,\nu}
= -\epT^{\mn} \hs,
\eeq
which are the unphysical polarizations of the graviton field in the TMG theory.

\vs 

According to the definition of the transverse polarization vector $\epT^\mu$
in Eq.\eqref{eq:epL-epT},
we can further derive the form of the transverse polarization tensor $\epT^{\mn}$ as follows:
\\[-4mm]
\beq
\label{eq:epT-munu}
\epT^{\mn} = \epT^{\,\mu}\epT^{\,\nu} =
\begin{pmatrix}
0 & 0 \!& 0\! \\
0 & c_{\theta}^2 \!& -s_{\theta}^{}c_{\theta}^{} \!\\
0 & -s_{\theta}^{}c_{\theta}^{} \!& s_{\theta}^2\!
\end{pmatrix}
= \eta^{\mn} \!+\hsm
\frac{~p_0^\mu q_0^\nu \!+\hsm p_0^\nu q_0^\mu\,}
{p_0^{\alpha} p_0^{\beta}\delta_{\alpha\beta}^{}}\,,
\eeq
where the massless momenta $(p_0^\mu,\hs q_0^\mu)$ are given by
$\hs p_0^\mu\!=\!E(1,\hs s_{\theta}^{},\hs c_{\theta}^{})\hs$
and $\hs q_0^\mu\!=\!E(1, -s_{\theta}^{}, -c_{\theta}^{})\hs$,
with $p_0^2\!=\! q_0^2\!=\! 0\,$.\
We note that the above formula \eqref{eq:epT-munu} coincides with that of Eq.\eqref{eq:epT-munu-sp}
which is derived independently by using the spinor formulation.
Under high energy expansion, we can rewrite the last equality of Eq.\eqref{eq:epT-munu}
as follows:
\begin{align}
\epT^{\mn} &= \epT^{\,\mu}\epT^{\,\nu}
= \eta^{\mn} \!+\hsm
\frac{~p^\mu q^\nu \!+\hsm p^\nu q^\mu\,}{2E^2}
+O\!\hsm\(\hsm\!\frac{m^2}{E^2}\!\)
\nn\\
& 
= \eta^{\mn} \!+\hsm
\frac{~\epS^\mu \bar{q}^\nu \!+\hsm \epS^\nu \bar{q}^\mu\,}{2\bar{E}^2}
+O\!\hsm\(\hsm\!\frac{m^2}{E^2}\!\) \!,
\label{eq:epT-munu-expd}
\end{align}
where $\epS^\mu \!=\!p^{\mu}\hsm /m\hs$,
$\bar{q}^\mu\!=\!q^\mu\hsm /m\hs$, and $\bar{E}\!=\!E/m\hs$.\
In the above, the momenta $(p^\mu\!,\hs q^\mu )$ are given by
$\hs p^\mu\!=\!E(1,\hs \be s_{\theta}^{},\hs \be c_{\theta}^{})\hs$
and $\hs q^\mu\!=\!E(1, -\be s_{\theta}^{}, -\be c_{\theta}^{})\hs$
with $p_\mu^2 \!=\hsm q_\mu^2 \!=\! -m^2$.\

\vs

Finally, we substitute Eq.\eqref{eq:epT-munu-expd} into the $\epP^{\mn}$ formula
\eqref{eq:gravitonPol} and derive the following:
\begin{equation}
\label{eq:gravitonPol-2}
\epP^{\mn}
= \frac{1}{\hs 2\hs}\Big[\epS^{\mn} \!-\hsm\eta^{\,\mn}
\!+\!\big(\epS^\mu\chi^{\nu} \!+\!\epS^\nu\chi^{\mu}\big)
\!+\! \vt^{\mn}
\Big],
\end{equation}
where the quantities $\chi^\mu$ and $\vt^{\mn}$ are given by
\beqs
\begin{align}
\label{eq:chi}
\chi^\mu &= \tepT^{\,\mu}-\!\frac{\bar{q}^{\hs\mu}}{\,2\bar{E}^2\,}\,,
\\
\label{eq:vt-munnu}
\vt^{\mn} &= \big(v^\mu\bepT^{\,\nu}\!+\!v^\nu\bepT^{\,\mu} \!+\! v^\mu v^\nu\big)
\!+\hsm O\!\hsm\(\!\frac{m^2}{E^2}\!\) = O\!\hsm\(\hsm\frac{m}{E\,}\hsm\) \!.
\end{align}
\eeqs
In the above, the precise expression of $\vt^{\mn}$ is not needed for the following analysis except that
we know its order of magnitude, $\vt^{\mn}\!=\hsm O(m/E)$.\

\vs

Given the TMYM Lagrangian \eqref{eq:TMYM-L}, we can add the covariant
gauge-fixing term and its corresponding Faddeev-Popov ghost term:
%
\label{eq:LGF-FP}
\begin{align}
\label{eq:LGF-LFP}
\La_{\rm{GF}}^{} = -\frac{1}{\,2\xi\,}(\pd^\mu \!A_\mu^a)^2 ,
\hspace*{8mm}
\La_{\rm{FP}}^{} =
\bar{c}^a\pd^\mu \hsm\big(\delta^{ab} \pd_\mu \!-\! gf^{abc}A_\mu^c\big)c^b,
\end{align}	
%
where $f^{abc}$ is the structure constant of the non-Abelian gauge group.\
Thus, we can derive the complete gauge boson propagator in the general $\xi$-gauge as follows:
\beq
\mathcal{D}_{\mn}^{ab} = \frac{~\ii\hs\delta^{ab}\hsm\Delta^{}_{\mn}~}{p^2\!+\hsm m^2}
=  -\hsm\ii\hs\delta^{ab}\!\left[\hsmx\frac{\,1\,}{~p^2\!+\hsm m^2\,}
\hsm\!\(\! \eta^{}_{\mu \nu} \!-\!\frac{\,p^{}_{\mu}p^{}_{\nu}\,}{p^{2}} \!-\!
\frac{\,\ii\hs m\hs\varepsilon^{\mu \nu \rho}p_{\rho}^{}\,}{p^{2}}\hsm\!\)
\!+ \xi\frac{\,p^\mu p^\nu\,}{p^4}
\hsm\right]\!,
\eeq
where $\xi\!=\!0$ corresponds to the Landau gauge.\

\vspace*{2.5mm}
\subsection{\hspace*{-2mm}Formulation of Topological Graviton Equivalence Theorem}
\label{sec:3.1}
\label{sec:3.2new}
\vspace*{1.5mm}

In this subsection, we newly propose and prove the TGRET in the
WTMG theory, which formulates the topological graviton 
mass-generation at the $S$-matrix level.\ 
The TGRET connects the $N$-point scattering amplitudes 
of massive gravitons ($\hP$)  
to the corresponding scattering amplitudes of dilatons ($\phi$)
in the high energy limit.\ 

\vs 

According to the gauge-fixing terms \eqref{eq:gaugefix} of the WTMG theory,
we have the gauge-fixing functions:
\\[-7mm]
\beqs
\begin{align}
\label{eq:gaugefixl1}
\mathcal{F}_{\rm{GF1}}^{\mu}
&= \partial_{\nu}h^{\mu \nu}
\!-\!\Fr{1}{\hs 2\hs}\hs\partial^{\mu}\hsm
(h\hsm -\hsm\xi\phi)\hs,
\\[1mm]
\label{eq:gaugefixl2}
\mathcal{F}_{\rm{GF2}}^{\mu}
&= \Fr{1}{\hs 2\hs}\hs\partial^{\mu}\hsm
(h\hsm -\hsm\zeta\phi)\hs.
\end{align}
\eeqs
In the above and hereafter, we suppress the bar of the graviton field
$\bar{h}_{\mn}^{}$ [introduced in Eq.\eqref{eq:CT-gmunu}]
for notational simplicity unless specified otherwise.\
For the present analysis, we will choose $\hs\xi\hsm =\hsm\zeta\,$
for convenience.\footnote{%
A further choice of $\xi\!=\hsm\zeta\!=\!1$ 
may be called the gravitational Feynman gauge.}\
Thus, we sum up \eqref{eq:gaugefixl1} and \eqref{eq:gaugefixl2}
under the choice $\hs\xi\hsm =\hsm\zeta\,$
and obtain the following function:
\beq
\label{eq:gaugefixl3}
\FF_{\rm{GF3}}^{\mu}
= \partial_{\nu}h^{\mu\nu}\hs.
\eeq

Then, we derive the corresponding Faddeev-Popov ghost terms
and the BRST  transformations
in Appendix\,\ref{app:C}.\
With these, we can derive the following Slavnov-Taylor-type identities
for the gauge-fixing functions in the momentum space:
\beqs
\label{eq:gf}
\begin{align}
\label{eq:gf1}
\left\langle 0\!\left\lvert \FF_{\text{GF1}}^{\mu_1}(p_1^{})\FF_{\text{GF1}}^{\mu_2}(p_2^{})\cdots
\FF_{\text{GF1}}^{\mu_N^{}}(p_N^{}) \Phi \right\rvert\! 0\right\rangle & = 0 \,,
\\
\label{eq:gf2}
\left\langle 0\!\left\lvert \FF_{\text{GF2}}^{\mu_1}(p_1^{})\FF_{\text{GF2}}^{\mu_2}(p_2^{})\cdots
\FF_{\text{GF2}}^{\mu_N^{}}(p_N^{}) \Phi \right\rvert\! 0\right\rangle & = 0 \,,
\\
\label{eq:gf3}
\left\langle 0\!\left\lvert \FF_{\text{GF3}}^{\mu_1}(p_1^{})\FF_{\text{GF3}}^{\mu_2}(p_2^{})\cdots
\FF_{\text{GF3}}^{\mu_N^{}}(p_N^{})\Phi \right\rvert\! 0\right\rangle & = 0 \,,
\end{align}
\eeqs
where the symbol $\Phi$ denotes any other on-shell physical fields after the Lehmann-Symanzik-Zimmermann (LSZ) reduction.\
In the above, we have set each external momentum to be on-shell
according to the mass of the corresponding physical graviton $h^{\mu \nu}\!$,$\hs$
$p_{j}^{2}\!=\!-m^2$ (with $j\!=\!1,2,3,\cdots$).\
We can also derive a mixed identity similar to Eqs.\eqref{eq:gf1}-\eqref{eq:gf3},
but with its left-hand side containing
any number of the gauge-fixing functions
${\FF}_{\rm{GF1}}^{\mu}$, ${\FF}_{\rm{GF2}}^{\mu}$, and ${\FF}_{\rm{GF3}}^{\mu}$
in the external states.\
Furthermore, we note that the above 3d Slavnov-Taylor-type identity \eqref{eq:gf} is a direct consequence of
the diffeomorphism (gauge) symmetry of the TMG theory,
which is similar to the gauge-fixing-function identities 
we derived for the 4d SM\,\cite{ET-SM}\cite{ET-SM-Rev}
and for the compactified 5d Kaluza-Klein (KK) gravity theories
with either flat or warped extra dimensions\,\cite{Hang:2021fmp}-\cite{Hang:2024uny}.\

\vs

By contracting the external momentum $p^{\mu}$ with the corresponding gauge-fixing functions
${\FF}_{\rm{GF1}}^{\mu}$ and ${\FF}_{\rm{GF2}}^{\mu}$
of Eqs.\eqref{eq:gaugefixl1}-\eqref{eq:gaugefixl2}
under the choice $\xi\hsm =\hsm\zeta\hs$,
we drive the following formulas in the momentum space and with on-shell external fields:
\beqs
\label{eq:gauge-fixingLSZ}
\begin{align}
\label{eq:gauge-fixingLSZ-1}
\mathcal{F}_{\text{GF1}}^{} & \equiv p_{\mu} \mathcal{F}_{\text{GF1}}^{\mu}
= \ii\hs m^{2} \!\left[ \epS^{\mn}h_{\mu\nu}^{} \!+\!
\frac{1}{\,2\,}(h\hsm -\hsm\xi\phi) \hsm\right]\!
= \ii\hs m^{2}\mathcal{F}_{\rm{1}}^{}\,,
\\
\label{eq:gauge-fixingLSZ-2}
\mathcal{F}_{\text{GF2}}^{} & \equiv p_{\mu} \mathcal{F}_{\text{GF2}}^{\mu}
= -\frac{\,\ii\hs m^2\hs}{2} (h\hsm -\hsm\xi\phi)
= -\frac{\,\ii\hs m^2\hs}{2} \mathcal{F}_{\rm{2}}^{}\,,
\end{align}
\eeqs
where 
$\epS^{\mn}\!=\!\epS^{\mu}\epS^{\nu}\!=\!p^\mu p^\nu\!/m^2\hs$
is the scalar polarization tensor of graviton.\
We note that when taking the massless limit 
for the above gauge-fixing functions,
the term $m^2(h\hsm -\hsm\phi)$ 
vanishes due to $m\ito 0\hs$, and thus
$\mathcal{F}_{\text{GF1}}^{}\hsm\!=\hsm\ii\hs p^\mu p^\nu h_{\mn}^{}$
and
$\mathcal{F}_{\text{GF2}}^{}\!=\hsm 0\hs$.\
This means that the field-combination $(h\hsm -\hsm\phi)$ 
is no longer constrained
in the massless limit $m\!=\! 0\hs$ and the dilaton $\phi$
can become a physical massless scalar.\
This fact also supports our discussion around Eqs.\eqref{eq:hmunu0Mlimit1=0+1}-\eqref{eq:limitTMG}.\ 

\vs 

From Eq.\eqref{eq:gaugefixl3}, we may define another contracted function:
\beqs
\label{eq:gauge-fixing-3}
\begin{align}
\label{eq:gauge-fixing-31}
\FF_{\rm{GF3}}^{\mu}
& = \ii\hs m\hs \ep_{\rm{S}\hs \nu}^{}h^{\mu\nu}
= \ii\hs m\hs \widetilde{\mathcal{F}}_{\rm{3}}^{\mu}\,,
\\
\FF_{\rm{GF3}}^{} &\equiv p_\mu^{}\FF_{\rm{GF3}}^{\mu}
= \ii\hs m^2 \epS^{\mn}h_{\mu\nu}^{}
= \ii\hs m^2 \mathcal{F}_{\rm{3}}^{}\,.
\label{eq:gauge-fixing-32}
\end{align}
\eeqs
In Eqs.\eqref{eq:gauge-fixingLSZ} and \eqref{eq:gauge-fixing-3},
the functions $\mathcal{F}_1^{}$, $\mathcal{F}_2^{}$, $\mathcal{F}_3^{}$,
and $\widetilde{\mathcal{F}}_{\rm{3}}^{\mu}$ are defined as follows:
\beq
\label{eq:F1-F3}
\FF_{1}^{} \!= \hS \!+\!\frac{1}{\,2\,}(h\hsm - \hsm\xi\phi)\,,
\hspace*{4mm}
\FF_{2}^{} \!=  h\hsm -\hsm \xi\phi \,,
\hspace*{4mm}
\FF_3^{} \!=\hS \,,
\hspace*{4mm}
\widetilde{\mathcal{F}}_{\rm{3}\mu}^{} \!=
\epS^{\nu}h^{}_{\mu\nu} \,,
\eeq
where $\hS=\epS^{\mn}h_{\mu\nu}$ is the graviton's scalar polarization and
$\hs h\!=\!h^{\mu}_{\mu}\hs$ is the trace of graviton field.\
With the definitions of Eqs.\eqref{eq:gauge-fixingLSZ}-\eqref{eq:F1-F3} and for nonzero graviton mass
$m\!\neq\!0\hs$, we can reexpress the identities \eqref{eq:gf1}-\eqref{eq:gf3} as follows:
\beqs
\label{eq:GF-ID}
\begin{align}
\label{eq:GF-ID1}
\left\langle 0\!\left\lvert \FF_1^{}(p_1^{})\FF_1^{}(p_2^{})\cdots
\FF_1^{}(p_N^{}) \Phi \right\rvert\! 0\right\rangle & = 0 \,,
\\
\label{eq:GF-ID2}
\left\langle 0\!\left\lvert \FF_2^{}(p_1^{})\FF_2^{}(p_2^{})\cdots
\FF_2^{}(p_N^{}) \Phi \right\rvert\! 0\right\rangle & = 0 \,,
\\
\label{eq:GF-ID3}
\left\langle 0\!\left\lvert \FF_3^{}(p_1^{})\FF_3^{}(p_2^{})\cdots
\FF_3^{}(p_N^{}) \Phi \right\rvert\! 0\right\rangle & = 0 \,,
\\
\label{eq:GF-ID4}
\langle 0\lvert \FFt_3^{\mu_{1}}\hsmx (p_1^{})\FFt_3^{\mu_{2}}\hsm (p_2^{})\cdots
\FFt_3^{\mu_N^{}}\hsmx (p_N^{}) \Phi \rvert 0\rangle & = 0 \,,
\end{align}
\eeqs
where we set each external momentum to be on-shell
according to the mass of the corresponding physical graviton $h^{\mu \nu}$,
$p_{j}^{2}\!=\!-m^2$ (with $j\!=\!1,2,3,\cdots$).\
We can also derive a mixed identity similar to Eqs.\eqref{eq:GF-ID1}-\eqref{eq:GF-ID4},
but with its left-hand side containing
any number of the gauge-fixing functions
${\FF}_1^{}$, ${\FF}_2^{}$, ${\FF}_3^{}$, and ${\FFt}_3^{\mu}$
in the external states.\

\vs

Then, for the $\FF_2^{}$ formula in Eq.\eqref{eq:F1-F3},
we can reexpress the left-hand side of the
identity \eqref{eq:GF-ID2} as the Green's function
with $N$ external states of ${\FF}_{2}^{}\hs$:
\begin{equation}
\label{eq:F2-GID}
\mathcal{G} \big[\FF_2^{}(p_1^{}),\FF_2^{}(p_2^{}),
\cdots \!,\FF_2^{}(p_N^{}), \Phi \big] = 0 \,.
\end{equation}
For each external state $\FF_2^{}(p)$ in Eq.\eqref{eq:F2-GID},
we first perform
the LSZ reduction at tree level and derive the following:
\begin{align}
&\mathcal{G} \big[\FF_2^{}(p),\cdots\big]
= \eta^{\alpha\beta}
\mathcal{D}_{\alpha \beta}^{h~~\mu\nu}\hsm (p)\hs\mathcal{M}\big[h_{\mu\nu}^{}(p),\cdots\big]
\!-\hsm\xi\hs\mathcal{D}^{\phi}(p)\mathcal{M}\big[\phi(p),\cdots\big]
\nn\\
&= \frac{\ii\hs 2\hs\xi\,}{\,(2-\xi)\hs p^{2}\,}\Big\{\hsm
\MM\hsm\big[h(p),\cdots\big]
\!+\MM\hsm\big[\phi(p),\cdots\big]\!\Big\}
= \frac{\ii\hs 2\hs\xi\,}{\,(2-\xi)\hs p^{2}\,}\MM\hsm
\big[\FFd_2^{}(p),\cdots\big]\hs,
\label{eq:F2-LSZ}
\end{align}
where we have used the graviton propagator 
\eqref{eq:Dh-xi-zeta} and 
dilaton propagator \eqref{eq:Dphi-xi-zeta}.\ 
In the above formula, the function $\FFd_2^{}$ is given by
%
\begin{align}
\label{eq:F2-under}
\FFd_2^{} = h + \phi \,,
\end{align}
%
with $\hs h\hsm =\hsm\eta^{\mn}h_{\mn}^{}\hs$
being the trace of the graviton field.\
Moreover, using a general BRST identity, we can perform the LSZ reduction up to all loop orders,
which induces a radiative modification factor
$\hat{C}\!=\!1\!+\hsm O(\rm{loop})$
in the above function $\FFd_2^{}$ such that
\beq
\label{eq:F2-under-C}
\FFd_2^{} = h + \hat{C}\hs\phi \,,
\eeq
where $\hs\hat{C}\,$ is derived in the formula right below Eq.\eqref{appeq:F2-Q}
and in Eq.\eqref{appeq:C}.\
This general LSZ reduction together with the loop-induced modification factor  $\hat{C}$ are presented in Appendix\,\ref{app:C}.\
Thus, applying the LSZ reduction \eqref{eq:F2-LSZ} to all external
$\FF_2^{}$ states of the left-hand side of Eq.\eqref{eq:F2-GID},
we derive the following LSZ-amputated identity:
\begin{equation}
\label{eq:F2-MID}
\MM \big[\FFd_2^{}(p_1^{}),\FFd_2^{}(p_2^{}),
	\cdots \!,\FFd_2^{}(p_N^{}), \Phi \big] = 0 \,.
\end{equation}

Next, we contract the two sides of Eq.\eqref{eq:gravitonPol-2} with the graviton field $h_{\mn}^{}$
and derive a formula for the on-shell physical graviton field:
\beq
\hP = \frac{1}{\hs 2\hs} \!\left[
\hS \!-\! h \!+\! \big(\epS^\mu\hs\chi^\nu\hsm\!+\hsm\epS^\nu\hs\chi^\mu\big)h_{\mn}^{}\!+\hsm \tilde{h}_v
\right]\!,
\eeq
where $\chi^{\mu}$ is given by Eq.\eqref{eq:chi} and
$\htd_v^{}\hsm =\hsm\vt^{\mn}h_{\mn}^{}\!=\hsm O(m/E)\hs$
with $\vt^{\mn}$ defined in Eq.\eqref{eq:vt-munnu}.\
Thus, we derive the trace of the graviton field $\hs h\hs$ as follows:
\beq
h = -2\hP \!+\! \FF _3^{}\!+\! (\FFt_{3\nu}^{}\chi^\nu\hsm\!+\hsm\FFt_{3\mu}^{}\chi^\mu) +\htd_v^{}\,.
\eeq
We substitute the above formula of $h$ into
Eq.\eqref{eq:F2-under-C} and derive the following:
\beq
\FFd_2^{}  = -2\hP\!+\hsm \hat{C}\hs\phi \hsm +\!
\big(\FF_3^{}\!+\hsm\FFt_{3\nu}^{}\chi^\nu\hsm\!+\hsm\FFt_{3\mu}^{}\chi^\mu\big) \!+\hsm \htd_v^{}\,.
\eeq
Thus, we can further derive the $S$-matrix identity \eqref{eq:F2-MID}
as follows:
\begin{equation}
\label{eq:F2-MID2}
\MM \big[\FFB_2^{}(p_1^{}),\FFB_2^{}(p_2^{}),
\cdots \!,\FFB_2^{}(p_N^{}), \Phi \big] = 0 \,.
\end{equation}
where $\FFB_2^{}$ is given by
\beqs
\label{eq:F2b}
\begin{align}
\label{eq:F2b1}
\FFB_2^{} & = \hP\!-\hsm \frac{1}{\hs 2\hs}
\big(\hat{C}\hs \phi  +\hsm \htd_v^{}\big)
= \hP \! -\hsm \QQ\,,
\\
\label{eq:F2b2}
\QQ & = \frac{1}{\hs 2\hs}
\big(\hat{C}\hs \phi \hsm +\hsm \htd_v^{}\big) \hs,
\end{align}
\eeqs
in which we have removed an overall factor $-2\hs$ and
have dropped the terms of $\FF_3^{}$, $\FFt_{3\nu}^{}$ and $\FFt_{3\mu}^{}$ based on
the identities \eqref{eq:GF-ID3} and \eqref{eq:GF-ID4}.\
In Eq.\eqref{eq:F2b}, $\,\hP\!=\!\epP^{\mn}h_{\mn}^{}\,$
is the physical graviton field,
$\phi$ is the dilaton field, and
$\htd_v^{}\! =\!\vt^{\mn}h_{\mn}^{}$ is suppressed by
the tensor factor $\hs \vt^{\mn}\!=\hsm O(m/E)\hs$.\

\vs

Then, we can compute the amplitude:
\begin{align}
\label{eq:Q-hP-ID}
\MM[\QQ(p_1^{}),\cdots\!,\Phi] = \MM[\hP(p_1^{})\!-\!\FFB_2^{}(p_1^{}),\cdots\!,\Phi]
= \MM[\hP(p_1^{}),\cdots\!,\Phi]\hs,
\end{align}
where in the last step we have made use of the identity \eqref{eq:F2-MID2}
which states that any physical amplitude with at least one external state replaced by
the $\FFB_2^{}$ state must vanish.\
Thus, from Eq.\eqref{eq:Q-hP-ID} we derive the following identity to connect the
$N$-point graviton scattering amplitude to the $\QQ$-amplitude:
\begin{align}
\label{eq:TGET-ID}
\MM[\hP(p_1^{}),\cdots\!,\hP(p_N^{}),\Phi] =
\MM[\QQ(p_1^{}),\cdots\!,\QQ(p_N^{}),\Phi]
\hs,
\end{align}
where the quantity $\QQ \!=\! \frac{1}{2}\big(\hat{C}\hs\phi \hsm +\hsm  \htd_v^{}\big)$
is given by Eq.\eqref{eq:F2b2}
with the radiative modification factor $\hat{C}\!=\!1\!+\hsm O(\rm{loop})$ and the residual term
$\htd_v^{}\!=\!v^{\mn}h_{\mn}^{}$ [suppressed by the tensor factor $\hs v^{\mn}\!=\!{O}({m}/{E})\hs$].\
Hence, under high energy expansion, we derive the
Topological Graviton Equivalence Theorem (TGRET) for the WTMG (TGM) theory
that connects the $N$-point graviton scattering amplitude to the corresponding dilaton amplitude:
\begin{equation}
\label{eq:TGRET}
{\MM}\hsm\big[\hP(p_1^{}),\cdots\!,h_{\text{P}}(p_N^{}), \Phi\big]
= \(\!\Fr{1}{2}\!\)^{\hsm\!N}\!\!C_{\rm{mod}}^{}{\MM}\hsm
\big[\phi(p_{1}),\cdots\!,\phi(p_N^{}), \Phi\big]
\!+ {O}\Big(\hsm\frac{m}{\hs E\,}\!\Big) \hs,
\end{equation}
where each external momentum obeys the condition $p_j^2\!=\!-m^2$
and the radiative modification factor
$C_{\rm{mod}}^{}\!=\!1\hsm + O(\rm{loop})\hs$.\
We note that the physical graviton scattering amplitude on the left-hand side of
the TGRET \eqref{eq:TGRET} is gauge-invariant (diffeomorphism-invariant) and thus can be computed 
in either the WTMG theory or the TMG theory (corresponding to 
the unitary gauge of the WTMG).\ 
In contrast, the dilaton scattering amplitude 
on the righ-hand side of the TGRET \eqref{eq:TGRET}
is defined only in the WTMG theory, 
and it can mimic the high-energy behavior 
of the corresponding physical graviton scattering 
on the left-hand side of Eq.\eqref{eq:TGRET}.\ 

\vs

We further note that the scattering amplitude of transverse gravitons $\hTT$ cannot reproduce the
left-hand side of Eq.\eqref{eq:TGRET} in the high energy limit.\    
This differs from the TGAET (topological gauge-boson equivalence 
theorem)\,\cite{Hang:2021oso}\cite{Hang:2023fkk} 
for the TMYM theory,
\begin{equation}
\label{eq:TGAET}
\TT [\AP^{a_1^{}}\!(p_1^{}),\!\cdots\!,\AP^{a_N^{}}\!(p_N^{}),\Phi] 
=\! \(\!\Fr{1}{\sqrt{2\hs}\,}\!\)^{\hsm\!N}\!\hsm 
\TT [\AT^{a_1^{}}(p_1^{}),\!\cdots\!,\AT^{a_N^{}}(p_N^{}),\Phi]
+O\!\(\frac{m}{E}\)\hsm ,
\end{equation}
which connects the $N$-point physical gauge-boson ($\AP^a$) scattering amplitude 
to the corresponding unphysical transverse gauge-boson ($\AT^a$) scattering amplitude.\ 
As a supplemental analysis of Section\,\ref{sec:4.1.2}, 
we have explicitly computed the
tree-level $4\hTT$ scattering amplitudes in Eq.\eqref{eq:Amp-4hT} of Appendix\,\ref{app:B2}
and find that its leading-order amplitude \eqref{eq:Amp-4hT-exp} does not equal
that of the physical $4\hP$ amplitude in the high energy limit.\  
In contrast, we will show in Eq.\eqref{eq:TGRET-4pt} of Section\,\ref{sec:4.1.2} that
the $4\hs\phi$ dilaton scattering amplitude equals 
the physical $4\hP$ amplitude in the high energy limit.\  
This also demonstrates that our formulation of the WTMG theory
(including the dilaton field) is essential 
for the successful construction of the TGRET.\   

\vs

Finally, we amputate each external state $\FF_3^{} \!=\hsm\hS$ in the identity
\eqref{eq:GF-ID3} at tree level and derive the following:
%
\begin{align}
\mathcal{G} \big[{\FF}_3^{}(p),\cdots\hsmx\big]
& =
\epS^{\alpha\beta}
\mathcal{D}_{\alpha \beta}^{h~~\mu\nu}\hsm (p)\hs\mathcal{M}\big[h_{\mu\nu}^{}(p),\cdots\hsmx\big]
\nn\\
&= \frac{\,\ii\hs 2\hs\xi\,}{\,p^{2}\,}
\Big\{\hsm\epS^{\mn}\hsm\mathcal{M}\hsm\big[h_{\mu\nu}^{}(p),\cdots\hsmx\big]
\!\Big\}
= \frac{\,\ii\hs 2\hs\xi\,}{\,p^{2}\,}\mathcal{M}\hsm
\big[\hs{\FFB}_3^{}(p),\cdots\hsmx\big]\hs,
\end{align}
where the induced gauge-fixing function
$\,{\FFB}_3^{} \!=\! \epS^{\mn} h_{\mn}^{}\hsm\!=\!\hS\hs$.\
Thus, applying the LSZ reduction to the external states on the left-hand side 
of Eq.\eqref{eq:GF-ID3}, we derive the following amputated identity:
%
\begin{align}
\label{eq:BF3-ETID}
\MM \hsm\big[\hs{\hS}(p_1^{}),{\hS}(p_2^{}),\cdots\! ,{\hS}(p_N^{}), \Phi \big]
= 0\,.
\end{align}
%
The identity \eqref{eq:BF3-ETID} demonstrates that any scattering amplitude containing
at least one external state of scalar graviton $\hS \!=\hsm\epS^{\mn} h_{\mn}^{}$ 
must vanish.\

\vspace*{1.5mm}
\subsection{\hspace*{-2mm}Gravitational Power Counting Method and Energy Cancellations}
\label{sec:3.2}
\label{sec:3.3new}
\vspace*{1.5mm}

In this subsection, 
we develop a generalized gravitational power counting method
for the graviton (dilaton) scattering amplitudes 
in the WTMG theory
and demonstrate that the TGRET can provide a general mechanism
to guarantee the nontrivial large energy cancellations in the
$N$-point massive graviton scattering amplitudes 
(for $N\!\geqq\! 4$).\ 

\vs 

Consider the WTMG theory containing graviton and dilaton fields, as well as ghost fields 
(plus possible fermions if added).\ 
For a general $S$-matrix element $\hs\SS\hs$ with scattering energy $E$ in 3d spacetime,  
a relevant Feynman diagram contains $\EE$ number of external lines, 
$L$ number of loops, $I$ number of propagators,  and $\VV_j^{}$ number of vertices of type-$j\hs$,
where each type-$j$ vertex carries $d_j^{}$ derivatives and $n_j^{}$ legs.\ 
Thus, the scattering amplitude $\SS$ in 3d spacetime 
has a mass-dimension\,\cite{Hang:2021oso}:
\begin{equation}
\label{eq:DS}
D_{\SS}^{} \,=\, 3 - \fr{1}{2}\,\EE \,.
\end{equation}
We choose the Landau gauge ($\xi\!=\!\zeta\!=\!0\hs$) of the WTMG theory,  where 
the graviton propagator \eqref{eq:Dh-Landau} has the leading high-energy behavior of 
$\,m/p^{\omega}\hs$ 
with $\omega\!=\!3$ and $m$ the graviton mass.\ 
In general, a vertex of type-$j$ contains $\,d_j^{}\,$ derivatives,
$\,b_j^{}\,$ bosonic lines, and $\,f_j^{}\,$ fermionic lines, where the $\,b_j^{}\,$ bosonic lines
contains $\bar{b}_j^{}$ number of internal graviton lines and $\hat{b}_j^{}$ number of
all other bosonic fields (rather than internal graviton line) in this vertex-$j\hs$,
\beq 
b_j^{} = \bar{b}_j^{} + \hat{b}_j^{} \,.
\eeq 
Then, the energy-independent effective coupling constant in
the amplitude $\,\mathbb{S}\,$ has its total mass-dimension given by
\begin{equation}
\label{eq:DC}
D_{\rm{C}}^{} \,=\, \sum_j \hsm\VV_j^{}\!
\(\!3\!-\hsm d_j^{}\!-\! \fr{1}{\hs 2\hs} b_j^{}\!-\! f_j^{}\!+\!\fr{1}{\hs 2\hs}\bar{b}_j^{}\!\) \!.
\end{equation}
In this $D_C^{}$ formula the term $\fr{1}{\hs 2\hs}\bar{b}_j^{}$ is due to 
the leading high-energy behavior ($\hs m/p^3\hs$) of every internal graviton propagator 
\eqref{eq:Dh-Landau} having a graviton-mass factor $m$ splitted into the contributions 
to each effective coupling of its attached two vertices (one factor $\sqrt{m\,}$ for each vertex).\  
For each Feynman diagram contributing to the scattering amplitude $\SS$, we have the following 
general relations:
\begin{equation}
\label{eq:L-V-I}
L \!= \! 1+I-\VV\,, \quad~~
\sum_j \hsm\VV_j^{}n_j^{} = 2I+\EE\hs, 
\end{equation}
where $\,\VV=\sum_j\!\VV_j^{}\,$
denotes the total number of vertices in this Feynman diagram.
According to Weinberg\,\cite{Weinberg},
the leading high-energy-power dependence of a Feynman diagram is given by the difference 
$\,D_{\rm{E}}^{}\!=\! D_{\mathbb{S}}^{} - D_{\rm{C}}^{}\,$.\ 
Hence, using this formula and Eqs.\eqref{eq:DS}-\eqref{eq:L-V-I}, 
we derive the following:
\begin{equation}
\label{eq:DE} 
D_{\rm{E}}^{} \,=\, 2\hs (1\!-\!\VV)\hsm +
\sum_j\hsm\VV_j^{}\!\(\!d_j^{} \!-\!\fr{1}{\hs 2\hs}\bar{b}_j^{}
\!+\!\fr{1}{\hs 2\hs }f_j^{}\!\) \!+\hsm L
\,.
\end{equation}
In addition, we may consider a scattering amplitude containing $\EE_{\hP}^{}\!\!$ number of
external physical graviton states ($\hP$) with each polarization tensor $\epP^{\mn}$ scaling as
$E^2$.\ This contributes an extra energy-power factor $2\hs\EE_{\hP}^{}\!\!$ to the formula
\eqref{eq:DE}.\ Since the right-hand side of the TGRET identity \eqref{eq:TGET-ID} 
also includes amplitudes with external states like 
$\htd_v^{}\!=\!h_{\mn}^{}v^{\mn}\!=\!O(m/E)$, 
the  $\hs\EE_{v}^{}$ number of external $\htd_v^{}$ states 
will contribute an energy-power factor $-\hs\EE_{v}^{}\,$.\ 
Thus, for a scattering amplitude including $\EE_{\hP}^{}\!\!$ number of
external physical graviton states and $\EE_{v}^{}\!$ number of
external $\htd_v^{}$ states,  
we can extend the leading energy-power counting rule \eqref{eq:DE} as follows:
\begin{equation}
\label{eq:DE-hp-hv}
D_{\rm{E}}^{} \,=\hs \big(2\hs\EE_{\hP}^{}\!\!-\hsm\EE_{v}^{}\big)\!+\hsm
2\hs (1\!-\!\VV)\hsm +
\sum_j\hsm\VV_j^{}\!\(\!d_j^{} \!-\!\fr{1}{\hs 2\hs}\bar{b}_j^{}
\!+\!\fr{1}{\hs 2\hs }f_j^{}\!\) \!+\hsm L
\,.
\end{equation}

We note that the energy power counting rule \eqref{eq:DE} or \eqref{eq:DE-hp-hv} 
applies to the WTMG theory in the Landau gauge ($\xi\!=\!\zeta\!=\!0\hs$) 
where the graviton propagator 
\eqref{eq:Dh-Landau} has the leading high-energy behavior of 
$\,m/p^{3}\hs$ (rather than $1/p^2$) which adds a new contribution term
$\big(\hsm\!-\!\fr{1}{\hs 2\hs}\bar{b}_j^{}\big)$ on the right-hand side of Eq.\eqref{eq:DE}.\  
This differs from the previous energy power counting rule derived as Eq.(3.10) 
of Ref.\,\cite{Hang:2021oso}, 
which considers the conventional TMG theory (corresponding to
the unitary gauge $\zeta\!=\hsm\infty$ of the present WTMG theory) and has
its graviton propagator scale as $1/p^2\,$:
\begin{equation}
\label{eq:DE-uni} 
D_{\rm{E}}^{\rm{U}} \,=\, 
\big(2\hs\EE_{\hP}^{}\!\!-\hsm\EE_{v}^{}\big)\!+\hsm
2\hs (1\!-\!\VV)\hsm +
\sum_j\hsm\VV_j^{}\!\(\!d_j^{} 
\!+\!\fr{1}{\hs 2\hs }f_j^{}\!\) \!+\hsm L
\,.
\end{equation}
This energy power counting formula \eqref{eq:DE-uni} holds for the unitary gauge amplitudes 
and gives higher energy-dependence counting
by the amount of $\sum_j\!\(\!\fr{1}{\hs 2\hs}\bar{b}_j^{}\VV_j^{}\hsm\)\!=\!I_h^{}$,
which equals the number of internal graviton propagators.\ 
It means that in the conventional TMG theory the graviton scattering amplitudes have 
additional energy cancellations by the energy-power factor $E^{I_h}$.\ 

 
Alternatively, we can derive the same energy power counting rule \eqref{eq:DE}
by applying direct energy-power counting to the WTMG theory in the Landau gauge.\ 
Consider a general Feynman diagram containing $\EE$ number of external states, 
$I$ number of internal propagators, and $L$ number of loops, 
where $\EE\!=\hsm\EE_B^{}\hsm +\EE_F^{}$  with $\EE_B^{}$\,($\EE_F^{}$) number of external  
boson (fermion) states, and $I\!=\! I_B^{}\!+\!I_F^{}$ with 
$I_B$\,($I_F^{}$) number of internal boson (fermion) propagators.\ 
The $I_B^{}$ number of internal bosonic propagators
consists of $I_h^{}$ internal graviton propagators and $\hat{I}_B^{}$ internal
non-graviton bosonic propagators, namely,
$I_B^{}\!=\!I_h^{}\!+\!\hat{I}_B^{}\hs$.\ 
Thus, we directly count the leading energy-power dependence of this Feynman diagram,
\beq 
\label{eq:DE-2}
D_{\rm{E}}^{} = 3L \!-\! (\omega I_h^{}\!+\!2\hat{I}_B^{}\!+\!I_F^{})\!+\!
\sum_j\VV_j^{}d_j^{}\!+\!\frac{1}{\hs 2\hs}\EE_F^{}\,,
\eeq 
where the coefficient $\omega\!=\!3$ (or, $\omega\!=\!2$) for the internal graviton propagators
of the WTMG theory in Landau gauge (or, in unitary gauge, corresponding to the
conventional TMG theory).\ 
The total number $I$ of internal propagators includes $I_h^{}$ graviton propagators,
$\hat{I}_B^{}$ non-graviton bosonic propagators and $I_F^{}$ fermionic propagators.\ 
Thus, we have the following relations:
\beq 
\label{eq:I=Ih+Ibar-bjfj}
\begin{split}
& I = I_B^{}\!+\!I_F^{} = (I_h^{}\!+\hsm\hat{I}_B^{}) \!+\hsm I_F^{} 
= I_h^{}\!+\hsm\bar{I}\hs,
\hspace*{5mm}
\bar{I} \equiv \hat{I}_B^{}\!+\hsm I_F^{}\hs,
\\
& \sum_j\hsm\VV_j^{}\bar{b}_j^{} = 2I_h^{}\,,
\hspace*{6mm}
 \sum_j\hsm\VV_j^{}\bar{f}_j^{} = 2I_F^{}\,,
\end{split}
\eeq 
where $I_h^{}$ denotes the number of internal graviton propagators,
$\hat{I}_B$ the number of non-graviton bosonic propagators, 
$\bar{b}_j^{}$ the number of internal graviton lines in the vertex-$j$
and $\bar{f}_j^{}$ the number of internal fermionic lines in the vertex-$j\hs$.\ 
Using the relations \eqref{eq:L-V-I} and \eqref{eq:I=Ih+Ibar-bjfj},
we can further derive the energy power counting formula \eqref{eq:DE-2}
as follows:
\begin{equation}
\label{eq:DE-2F}
D_{\rm{E}}^{} \,=\, 2\hs (1\!-\!\VV)\hsm +
\sum_j\hsm\VV_j^{}\!\(\!d_j^{} \!-\!\fr{1}{\hs 2\hs}\bar{b}_j^{}
\!+\!\fr{1}{\hs 2\hs }f_j^{}\!\) \!+\hsm L
\,,
\end{equation}
where we have set $\hs\omega\!=\!3\hs$ for the leading energy behavior of the graviton propagator
\eqref{eq:Dh-Landau}  in the Landau gauge of the WTMG theory.\ 
We note that the above formula \eqref{eq:DE-2F} fully agrees with Eq.\eqref{eq:DE}
and thus serves as a nontrivial self-consistency check.\ 
For the unitary gauge of the WTMG theory (corresponding to the conventional
TMG theory\,\cite{Deser:1981wh}), we set $\omega\!=\!2$ and reproduce the energy
power counting rule of Eq.(3.10) of Ref.\,\cite{Hang:2021oso}
which differs from the above formula \eqref{eq:DE-2F} by the absence of the term 
$\big(\!\!-\!\fr{1}{\hs 2\hs}\bar{b}_j^{}\big)$.\ 

\vs 

Next, we apply the new power counting formula \eqref{eq:DE-hp-hv} to analyze
the leading energy-power dependence of the $N$-point physical graviton scattering amplitudes
and of the corresponding dilaton scattering amplitudes in the WTMG theory.\ 
For $\,\EE_{\hP}^{}\!\!\!=\!N\,$ number of external physical graviton states ($\hP$),
the leading-energy contributions are given by diagrams composed of 
cubic graviton vertices of the Chern-Simons term with 3 partial derivatives 
($d_j^{}\!=\!3$), where all internal lines are graviton propagators with 
$I\!=\!I_h^{}\!=\!\frac{1}{2}\hsm\sum_j\!\big(\VV_j^{}\bar{b}_j^{}\big)$.\ 
Thus, from the power counting formula \eqref{eq:DE-hp-hv}, we derive 
the leading energy-power dependence of the $N$-point physical graviton scattering amplitudes
in the Landau gauge of the WTMG theory:
\beq 
\label{eq:DE-Nhp-WTMG}
D_{\rm{E}}^{}[N\hP] = 2N\hsm\!+\hsm 2\!+\!(\VV\hsm\!-\!I_h^{}\!+\!L)
= 2N \! +3 \,,
\eeq
where in the last step we have used the general Euler relation 
$I\hsm\!=\!L\!+\!\VV\!-\!1$ as given by Eq.\eqref{eq:L-V-I}.\ 
If we consider the conventional TMG theory 
(corresponding to the unitary gauge of the WTMG theory),
we have the power counting formula \eqref{eq:DE-uni}, which gives the 
leading energy dependence of the $N$-point physical graviton scattering amplitude:
\beq 
\label{eq:DE-Nhp-TMG}
D_{\rm{E}}^{\rm{U}}\hsm [N\hP] 
= (2N\hsm\!+\!2)\!+\!\VV\!+\!L 
= (2N\hsm\!+\!3)\!+\!I_h^{}
\,,
\eeq
where in the second equality we have used the Euler relation of Eq.\eqref{eq:L-V-I}.\ 
At tree level, we have $L\!=\!0$ and the relation $N\hsm\!=\!\VV\!+\!2\hs$
for the $N$-point pure graviton scattering amplitudes 
composed of the cubic graviton vertices only.\ 
Thus, we further derive the leading energy dependence of the $N$-point 
physical graviton scattering amplitude in the unitary gauge:
\beq 
\label{eq:DE-Nhp-TMG}
D_{\rm{E}(0)}^{\rm{U}}\hsm [N\hP] = 3N \,,
\eeq
which agrees with Eq.(3.14) of Ref.\,\cite{Hang:2021oso}.\ 
Comparing the power counting formulas \eqref{eq:DE-Nhp-WTMG} and \eqref{eq:DE-Nhp-TMG},
we deduce a difference in the leading energy-power dependence at tree level:
\beq 
\label{eq:DE-DEU-NhP}
D_{\rm{E}(0)}^{\rm{U}}\hsm [N\hP] \!-\! D_{\rm{E}(0)}^{}\hsm [N\hP] = N\!-\hsm 3 \,.
\eeq 
This difference is due to the graviton propagator \eqref{eq:Dh-xi-zeta} 
has different high-energy behaviors in the Landau gauge [cf.\ Eq.\eqref{eq:DhDphi-Landau}]
versus unitary gauge [cf.\ Eq.\eqref{eq:Dh-unitary}].\ 
Thus, for the four-point physical graviton scattering amplitude ($N\hsm\!=\!4$), 
the individual leading energy-power dependence is given by  
$D_{\rm{E}}^{}[4\hP] \!=\! 11$
in the Landau gauge of the WTMG theory, and 
$D_{\rm{E}}^{\rm{U}}[4\hP] \!=\! 12\hs$
in the unitary gauge of the WTMG theory 
(corresponding to the conventional TMG theory).\ 
Since the $N$-point physical graviton scattering amplitudes must be gauge-invariant,
we conclude that the graviton scattering amplitude computed in the Landau gauge
formulation automatically encodes an energy power cancellation of $E^{N-3}$ 
relative to the same graviton amplitude computed in the unitary gauge.   

\vs

Then, we count the leading energy power dependence for the 
$N$-point dilaton scattering amplitudes at tree level.\ 
For the case of even number of external states $N\!\!=\!2n$ (with $n\!\!\geqq\!2\hs$),
we observe that the leading energy-power dependence is given by the Feynman diagram  
containing $n$ trilinear vertices of $\phi\hs\phi\hs h$ type 
(with two derivatives and with all $\phi$'s being external states) 
and $(n\!-\!2)$ cubic graviton vertices of
$h^3$ type (arising from the gravitational Chern-Simons term 
with three derivatives).\ 
Due to the conclusion below Eq.\eqref{eq:LphiHat},
we can ignore contributions from 
the pure dilaton self-interaction vertices 
for the power counting analysis.\ 
Using the formula \eqref{eq:DE} or \eqref{eq:DE-hp-hv}, we derive the 
leading energy-power dependence of the $N$-point dilaton scattering amplitude
as follows:
\beq 
\label{eq:DE-Nphi-N=2n}
D_{\rm{E}(0)}^{}\hsm [N\phi] = 3 \hsm -\hsm n = 3 \hsm-\!\Fr{1}{2}\hsm N \hs,
\hspace*{15mm}
(N\hsm\!=\!2n),
\eeq
at tree level for even number of $N\hsm\!=\!2\hs n$ external states.\ 
For odd number of external dilaton states ($N\hsm\!=\!2n +\hsm 1$), 
the leading energy-power dependence arises from the Feynman diagram 
having 1 trilinear vertex of $\phi\hs h\hs h$ type (with two derivatives 
and with $\phi$ being the external state), 
$n$ trilinear vertices of $\phi\hs\phi\hs h$ type 
(with two derivatives and with all $\phi$'s being external states) 
and $(n\!-\!2)$ cubic graviton vertices of
$h^3$ type (arising from the gravitational Chern-Simons term 
with three derivatives).\ 
Thus, we derive the leading energy-power dependence of the $N$-point  
dilaton scattering amplitude as follows:
\beq 
\label{eq:DE-Nhp-N=2n+1}
D_{\rm{E}(0)}^{}\hsm [N\phi] = 2 \hsm -\hsm n 
= \Fr{1}{\hs 2\hs}(5\!-\!N) \hs,
\hspace*{15mm}
(N\hsm\!=\!2\hs n\!+\!1), 
\eeq
at tree level for odd number of $N\hsm\!=\!2\hs n\!+\!1$ external states.\ 
According to the power counting rules \eqref{eq:DE-Nphi-N=2n} and
\eqref{eq:DE-Nhp-N=2n+1}, we deduce, for instance, 
that the four-point and five-point dilaton amplitudes
has leading energy-power dependence $E^1$ and $E^0$, respectively.\ 
Besides, in the above power counting analysis, 
there is no need to consider the contributions
of the pure dilaton self-interaction vertices of $\hs\phi^n(\partial\phi)^2\hs$
because they do not give any net contribution to the on-shell dilaton amplitudes
as we discussed below Eq.\eqref{eq:LphiHat}.\ 

\vs 

For the TGRET identity \eqref{eq:TGET-ID}, we consider the four-point
physical graviton scattering amplitude 
$\MM[h_{\rm{P}1}^{},h_{\rm{P}2}^{},h_{\rm{P}3}^{},h_{\rm{P}4}^{}]$
on the left-hand side of this identity, where we denote
$h_{\rm{P}j}^{}\!=\!h_{\rm{P}}^{}(p_j^{})$.\ 
Then, the right-hand side of the identity \eqref{eq:TGET-ID} contains 
the corresponding four-point leading dilaton scattering amplitude
$\MM[\phi_1^{},\phi_2^{},\phi_3^{},\phi_4^{}]$ plus the residual terms
including the $\htd_v^{}(\hsm =\!\vt^{\mn}h_{\mn}^{})$ suppressed amplitudes
such as $\MM[\htd_{v1}^{},\phi_2^{},\phi_3^{},\phi_4^{}]$ and so on,
where we denote $\phi_j^{}\!=\!\phi(p_j^{})$
and $\htd_{vj}^{}\!=\!\htd_v^{}(p_j^{})\hs$.\ 
By energy power counting, we deduce the following leading energy behaviors
in the Landau gauge for both sides of the TGRET identity \eqref{eq:TGET-ID}:
\\[-6mm]
\begin{align}
\MM[h_{\rm{P}1}^{},h_{\rm{P}2}^{},h_{\rm{P}3}^{},h_{\rm{P}4}^{}]
& = O(E^{11})\hs,
& \MM[\phi_1^{},\phi_2^{},\phi_3^{},\phi_4^{}] 
& = O(E^{1})\hs,
\nn\\
\MM[\htd_{v1}^{},\phi_2^{},\phi_3^{},\phi_4^{}]
& = O(E^0)\hs,
& 
\MM[\htd_{v1}^{},\htd_{v2}^{},\phi_3^{},\phi_4^{}]
& = O(E^0)\hs,
\label{eq:4hp-E-cancel}
\\
\MM[\htd_{v1}^{},\htd_{v2}^{},\htd_{v3}^{},\phi_4^{}]
& = O(E^{-1})\hs,
& 
\MM[\htd_{v1}^{},\htd_{v2}^{},\htd_{v3}^{},\htd_{v4}^{}]
& = O(E^{-1})\hs.
\nn 
\end{align}  
This shows that the amplitudes including one or more external lines of $\htd_v^{}$ 
are suppressed relative to the leading four-dilaton amplitude 
$\MM[\phi_1^{},\phi_2^{},\phi_3^{},\phi_4^{}]$ 
by at least a factor of $O(m/E)$.\ 
We see that for the four-point graviton scattering amplitude
$\MM[h_{\rm{P}1}^{},h_{\rm{P}2}^{},h_{\rm{P}3}^{},h_{\rm{P}4}^{}]$,
the TGRET ensures the large energy cancellations 
$E^{11}\!\ito E^1$ for the Landau gauge calculation of the four graviton amplitude,
and $E^{12}\!\ito E^1$ for the unitary gauge calculation of the four graviton amplitude.\ 
For the general $N$-point graviton scattering amplitudes ($N\!\geqq\!4$)  
at the tree level in comparison with the corresponding
dilaton scattering amplitudes, we use the power counting formulas \eqref{eq:DE-Nhp-WTMG} and
\eqref{eq:DE-Nphi-N=2n}-\eqref{eq:DE-Nhp-N=2n+1} to 
deduce the following large energy-power cancellations
in the $N$-point graviton scattering amplitudes 
as computed in the Landau gauge:
\beq 
\label{eq:DE-NhP-Nphi} 
D_{\rm{E}(0)}^{}\hsm [N\hP] \!-\! D_{\rm{E}(0)}^{}\hsm [N\phi] = 
\left\{
\begin{array}{ll} 
\hspace*{-1mm}
\Fr{\hs 5\hs}{2}\hsm N\hs, ~~&~~ (\rm{for}~N\!=\!\rm{even}),
\\[1.5mm]
\hspace*{-1mm}
\Fr{\hs 1\hs}{2}\hsm (5N\!+\!1)\hs, ~~&~~ (\rm{for}~N\!=\!{\rm{odd}}),
\end{array}
\right. 
\eeq 
and in the $N$-point graviton scattering amplitudes as computed in the unitary gauge:
\beq 
\label{eq:DE-NhP(U)-Nphi}
D_{\rm{E}(0)}^{\rm{U}}\hsm [N\hP] \!-\! D_{\rm{E}(0)}^{}\hsm [N\phi] = 
\left\{
\begin{array}{ll} 
\hspace*{-1mm}
\Fr{\hs 7\hs}{2}\hsm N\!-\! 3\hs, ~~&~~ (\rm{for}~N\!=\!\rm{even}),
\\[1.5mm]
\hspace*{-1mm}
\Fr{\hs 1\hs}{2}\hsm (7N\!-\!5)\hs, ~~&~~ (\rm{for}~N\!=\!{\rm{odd}}).
\end{array}
\right. 
\eeq 
These large energy cancellations are enforced by the TGRET \eqref{eq:TGRET}.

\vspace*{2mm}
\section{\hspace*{-2mm}Massive Graviton Scattering Amplitudes, TGRET and Double Copy}
\label{sec:4}
\label{sec:4new}
\vspace*{1mm}

The main purpose of this section is to explicitly verify {\it for the first time}
the newly formulated Topological Graviton Equivalence Theorem (TGRET)
in Eq.\eqref{eq:TGRET} as well as the large energy cancellations
in Eqs.\eqref{eq:DE-NhP-Nphi}-\eqref{eq:DE-NhP(U)-Nphi}
(predicted by the TGRET through the general energy power-counting analysis).\
For this purpose, we will compute both the scattering amplitudes of gravitons and 
of dilatons in the new WTMG theory in Section\,\ref{sec:4.1}.\ 
Our explicit calculations of the three-point and four-point graviton (dilaton) scattering
amplitudes are performed in the newly formulated WTMG theory {\it for the first time.}\ 
The dilaton scattering amplitudes exist only in the WTMG theory, but not the in the
conventional TMG theory (containing no dilaton field).\   
We compute the four-point graviton scattering amplitude in the Landau gauge of the WTMG theory 
{for the first time} where the Landau-gauge graviton propagator 
\eqref{eq:DhDphi-Landau} has better high-energy behavior 
(scaling as $m/p^3$) than the conventional unitary-gauge 
graviton propagtor \eqref{eq:Dh-unitary} (scaling as $1/p^2$) and 
additional new Feynman diagrams from dilaton exchange will
contribute to the graviton amplitude, causing different energy cancellations from 
those of the four-graviton amplitude computed for the conventional TMG theory
in the literature\,\cite{TMG-DCx} 
(corresponding to the unitary gauge of the WTMG theory).\  
With these, we will explicitly prove the equivalence between the graviton amplitudes computed 
both in the Landau gauge and in the unitary gauge.\ 
This gives nontrivial consistency checks on the gauge-invariance of the physical graviton
amplitudes based on our new BRST quantization of the WTMG theory 
(including the unphysical dilaton field).\ 
We further compute the four-point scattering amplitude of 
physical scalar particles in the WTMGS theory \eqref{eq:L-WTMGS},  
and explicitly prove its nontrivial equivalence 
to the four-point dilaton scattering amplitude
at the leading order of high energy expansion, 
in agreement with Eq.\eqref{eq:M0[Nphi]=M0[Npsi]}.

\vs 

In Section\,\ref{sec:4.2}, we will study the double copies for the 
graviton scattering amplitudes and the corresponding dilaton amplitudes,
where we pay special attentions to the energy structures 
of the scattering amplitudes both in the gauge sector 
(including the adjoint scalars) and in the gravity sector
(including the dilaton in the WTMG theory).\ 
This is because our main goal of this work is 
to understand the dynamics of the topological graviton mass generation through
the TGRET that controls the energy structures of the graviton (dilaton) scattering amplitudes,
in connection to the dynamics of the topological gauge-boson mass generation through
the TGAET that controls the energy structures of 
two types of gauge boson scattering amplitudes.\ 
In Section\,\ref{sec:4.2.1}, we will adopt the 
massive spinor-helicity formalism\,\cite{other2a-3d-CS}\cite{Arkani-Hamed:2017jhn} 
and show that the three-point graviton amplitude
can be obtained by the double copy of the corresponding three-point gauge boson amplitude,
whereas the three-point dilation-dilation-graviton amplitude can be newly obtained by the
double copy of the scalar-scalar-gauge-boson amplitude (where the scalar denotes the
colored adjoint scalar field).\ 

\vs 

Then, in Section\,\ref{sec:4.2.2},   
we first use the method of \cite{Arkani-Hamed:2017jhn} to 
show how the four-point gauge boson amplitudes can be constructed from the three-point
gauge boson amplitudes.\ With these, we further study the double copies of the four-point 
scattering amplitudes of gravitons and of dilatons, and analyze their relations in the
high energy limit.\ 
Despite that the double copy of the four-point graviton amplitude in the TMG theory
was derived previously\,\cite{TMG-DCx}\cite{Hang:2021oso},
Ref.\,\cite{TMG-DCx} did not study the energy-dependence structure of the graviton amplitude,
whereas Ref.\,\cite{Hang:2021oso} first explained the energy cancellations $E^{12}\!\ito E^4$
in the four graviton scattering amplitude by using the double copy construction, 
but found further cancellations $E^{4}\hsm\ito\hsm E^1$ that remain puzzling.\
This puzzle can be resolved by using the TGRET.\footnote{%
For this, we have explained in Section\,\ref{sec:3.3new} 
the series of striking energy cancellations $E^{12}\!\ito E^1$
in the four graviton amplitude by using the TGRET 
joined with the energy power counting   
on both sides of the TGRET where the four dilaton amplitude on 
its right-hand side has manifest energy dependence of $E^1$ and guarantees the
four graviton amplitude on its left-hand side to have 
energy cancellations $E^{12}\ito E^1$.}\ 
As we further verify explicitly in Section\,\ref{sec:4.1.2},  
the remaining nonzero four-graviton amplitude of $O(E^1)$ does equal quantitatively the leading-order four-dilaton amplitude 
on the right-hand side of the TGRET under high energy expansion.\   
Then, we construct the double copy of the four-point 
physical scalar amplitude in the WTMGS theory 
from corresponding amplitude of the colored adjoint scalars.\  
With these and using the equivalence between 
each $N$-point leading-order dilaton amplitude and 
the $N$-point leading-order physical scalar amplitude 
as shown in Eq.\eqref{eq:M0[Nphi]=M0[Npsi]},
we obtain the double-copied leading-order dilaton amplitude.\ 
We will show that the double-copied four-point leading-order 
dilaton amplitude has a manifest energy-dependence of 
$O(mE^1)$.\

\vspace*{0.6mm}

In Section\,\ref{sec:4.3}, we will take the massless limit 
and further discuss the relations between these amplitudes 
and those of the 3d massless theories.\
Finally, it is known that except for very special supersymmetric theories 
such as the 3d $\mathcal{N}\hsm\!=\!6$ ABJM theory\,\cite{ABJM},
there is no general proof of the BCFW recursions\,\cite{BCFW} 
for 3d non-supersymmetric field theories.\ 
Hence, the present explicit constructions 
of the four-point scattering amplitude of gauge bosons (adjoint scalars) and the four-point amplitudes of gravitons 
(of physical scalars or of dilatons)
via double copy in the WTMG (WTMGS) theory are nontrivial analyses,
whereas extensions to the $N$-point amplitudes 
($N\!\!\geqq\! 5$) are much harder and could be
discussed only case by case, which are beyond the current scope.\  
We will give a further comment on this at the end of Section\,\ref{sec:4.2}.\

\vspace*{1.5mm}
\subsection{\hspace*{-2mm}Graviton (Dilaton) Scattering Amplitudes and TGRET}
\label{sec:4.1}
\vspace*{1.5mm}

In this subsection, we compute the three-point and four-point scattering amplitudes
of gravitons and of dilatons in the WTMG theory.\ 
We pay special attentions to analyzing the energy-dependence structure  
of these scattering amplitudes.\ 
Then, we explicitly demonstrate how the TGRET \eqref{eq:TGRET} holds  
for their three-point and four-point scattering amplitudes, respectively.\
We will compute the four-point graviton scattering amplitude 
in the Landau gauge of the WTMG theory 
[including different graviton propagator \eqref{eq:DhDphi-Landau} 
and additional Feynman diagrams with dilaton exchanges]
and prove that it equals the four-point graviton amplitude 
computed in the unitary gauge.\ 
This serves as nontrivial consistency checks on the gauge-invariance of 
the graviton scattering amplitudes and on our new BRST quantization 
of the WTMG theory given in Section\,\ref{sec:2}.\  


\subsubsection{%
\hspace*{-2mm}%
Three-Point Graviton (Dilaton) Scattering Amplitudes and TGRET}
\label{sec:4.1.1}
\vspace*{1mm}

\begin{figure}[t]
\centering
\includegraphics[width=0.45\textwidth]{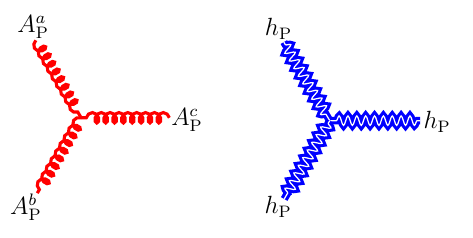}
\vspace*{-6mm}
\caption{\hspace*{-1mm}\small
Three-point gauge boson scattering in the TMYM theory (left diagram)
and the three-point graviton scattering in the TMG theory (right diagram).\
The amplitudes of these two scattering processes are connected by the double copy.}
\label{fig:2}
\end{figure}

For comparison, we first compute the three-point massive graviton scattering amplitudes 
in the WTMG theory,
corresponding to the right diagram of Fig.\,\ref{fig:2}.\
With the momentum conservation condition $p_1^{}\!+ p_2^{}\hsm + p_3^{}\!=\!0\hs$,
we express the three-point graviton scattering amplitude as follows:
\begin{equation}
\label{eq:Amp-3hP}
\MM[h^{}_{\PP 1},\hsm h^{}_{\PP 2},\hsm h^{}_{\PP 3}]
= -{\epsilon_{1\mu_1\nu_1}^{}\!\epsilon_{2\mu_2\nu_2}^{}\!\epsilon_{3\mu_3\nu_3}}^{}\!
\mathcal{V}_{3h}^{\mu_1\nu_1\hsm ,\hs\mu_2\nu_2,\hs\mu_3\nu_3}\hsm (p_1^{},p_2^{},p_3^{}) \,,
\end{equation}
where we denote  $h^{}_{\hsm \rm{P}j}\!\!=\!\hP (p_j^{})$ and
the trilinear graviton vertex function
$V_{3h}^{\mu_1\nu_1\hsm ,\hs\mu_2\nu_2,\hs\mu_3\nu_3}$ is derived in
Eq.\eqref{Beq:V3h} of Appendix\,\ref{app:B}.\
The polarization tensor $\ep_{j}^{\mn}(\equiv\!\epsilon_{\rm{P}j}^{\mu\nu})$
of each massive physical graviton in the WTMG (TMG) theory can be expressed as the product of
two polarization vectors of the corresponding physical massive gauge bosons in the TMYM theory, $\epsilon_{\rm{P}j}^{\mu\nu}
\!=\epsilon_{\rm{P}j}^\mu\epsilon_{\rm{P}j}^\nu$,
with $j\!=\!1,2,3$.\
With these, we derive the on-shell scattering amplitude \eqref{eq:Amp-3hP} as follows:
\begin{align}
\label{eq:Amp-3hp-EP}
\MM[h^{}_{\PP 1},\hsm h^{}_{\PP 2},\hsm h^{}_{\PP 3}]
& = \frac{\,- 16\hs\kappa\,}{m^4}
(p_1^{}\!\cdot\hsm\ep_{\PP 2}^{})^{2}(p_2^{}\!\cdot\hsm
\ep_{\PP 3}^{})^2(p_3^{}\!\cdot\hsm\ep_{\PP 1}^{})^2
\nn\\
& ={\frac{\kappa}{\,3\hs 2m^4\,}
\langle12\rangle^2 \langle23\rangle^2 \langle31\rangle^2} \,.
\end{align}
Although for the on-shell three-point amplitude \eqref{eq:Amp-3hp-EP}
the scattering energy is fixed by the mass $m$,
we may use it to explicitly test the validity of TGRET
by taking a mathematical limit, under which the masses of
particles\,1 and 2 are expanded to the leading order, and the mass of
particle\,3 is fixed without expansion.\
This is equivalent to taking the massless limit for particles\,1 and 2
and setting the mass of particle\,3 fixed
in the amplitude \eqref{eq:Amp-3hp-EP}.\
For this, we can derive the following three-point amplitude:
\beq
\label{eq:2hp0-hp}
\MM[h^{}_{\rm{P}1_0},\hsm h^{}_{\rm{P}2_0},\hsm h^{}_{\rm{P}3}]
= {\frac{\kappa}{\,32\hs m^2\,} \langle23\rangle^2 \langle31\rangle^2},
\eeq
where we have used shorthand notation $h^{}_{j\rm{P}0}\!=\!h_{\PP 0}^{}(p_j^{})$
and the subscript ``0'' denotes taking the massless limit
for the corresponding graviton state.\
In Eq.\eqref{eq:2hp0-hp},
we have taken the massless limit for the external states 1 and 2,
namely, $p_1^2\hsm =\hsm p_2^2\hsm =\hsm 0\hs$,
which leads to
\begin{equation}
\label{eq:m1m2=0-<12>}
\langle 12 \rangle^{2}= -(p_1^{}\!+\!p_2^{})^2 = -p_{3}^2 = m^2 \hs.
\end{equation}

Then, with the trilinear Feynman vertices 
$\psi\psi h$ and $hh\phi$
given in 
Eqs.\eqref{Beq:Vhhphi} and \eqref{Beq:WTMGS0-h-2psi},
we compute the following three-point scattering amplitudes involving dilatons:
\beqs
\label{eq:h2phi-hhphi}
\begin{align}
\MM [\psi_1^{},\psi_2^{},h_{\rm{P}3}^{}] & =
-\ep_{3\mu_3\nu_3}^{}\hsm \mathcal{V}_{\phi\phi h}^{\mu_3\nu_3}\hsm (p_1,p_2,p_3)
\nn\\
& = -\kappa (p_1^{}\!\cdot\hsm\ep_{\PP 3}^{})^2
=\frac{\kappa}{\,8\hs m^2\,} \langle23\rangle^2 \langle31\rangle^2 ,
\label{eq:2psi-hp} 
\\[1mm]
\MM [h_{\rm{P}1}^{},h_{\rm{P}2}^{},\phi_3^{}] & =
-\ep_{1\mu_1\nu_1}^{}\ep_{2\mu_2\nu_2}^{}\hsm
\mathcal{V}_{hh\phi}^{\mu_1\nu_1,\mu_2\nu_2}\hsm (p_1^{},p_2^{},p_3^{})
\nn\\
& =\frac{\,-3\hs\kappa\,}{\,2\hs m^2\,}(p_1^{}\!\cdot\hsm\ep_{\PP 2}^{})^2
(p_2^{}\!\cdot\hsm\ep_{\PP 1}^{}\hsm )^2
= \frac{\,-3\hs\kappa\,}{\,128\hs m^2\,}\!\left<12\right>^{\hsm 4}
, \hspace*{15mm}
\label{eq:hh-phi}
\end{align}
\eeqs
where the shorthand notation $\phi_j^{}\!=\!\phi(p_j^{})$ is used
and the dilaton states are massless.\
Then, from the WTMGS Lagrangian \eqref{eq:L-WTMGS} 
(with the scalar mass choice $m_s^{}\!=\!0\hs$) and 
Eq.\eqref{eq:L-phih/psih-h} (plus the discussions below it),
we see that the trilinear vertices $\phih\phih h_{\mn}^{}$ and 
$\psih\psih h_{\mn}^{}$ have exactly the same coupling
as shown in Eq.\eqref{Beq:WTMGS-V(h2psih)=V(h2psi)} 
and thus the two three-point on-shell amplitudes
are equal to each other,
$\MM[\phih_1^{},\phih_2^{},h^{}_{\rm{P}3}] \!=\! 
\MM[\psih_1^{},\psih_2^{},h^{}_{\rm{P}3}]\,$.\ 
Since on-shell scattering amplitudes are invariant under 
field-redefinitions, we deduce the following three-point 
dilaton-dilaton-graviton amplitudes at tree level,
\beq 
\label{eq:2phihp=2psihp}
\MM[\phi_1^{},\phi_2^{},h^{}_{\rm{P}3}] = 
\MM[\psi_1^{},\psi_2^{},h^{}_{\rm{P}3}] =
\frac{\kappa}{\,8\hs m^2\,} \langle23\rangle^2 \langle31\rangle^2 
\hs,
\eeq 
where both the dilaton $\phi$ and 
physical scalar $\psi$ ($\psih$) are massless.\ 

\vs

For the later usage, we compute two additional unphysical three-point graviton scattering amplitudes:
\\[-8mm]
\beqs
\label{eq:Amp-2hp-hall}
\begin{align}
\MM[h^{}_{1\PP},\hsm h^{}_{2\PP},\hsm h_3^{}]
& =
-\ep_{1\mu_1\nu_1}^{}\!\ep_{2\mu_2\nu_2}^{}\hsm\eta_{\mu_3\nu_3}^{}\!
\mathcal{V}_{3h}^{\mu_1\nu_1\hsm ,\hs\mu_2\nu_2,\hs\mu_3\nu_3}\hsm (p_1^{},p_2^{},p_3^{})
\nn\\[1.5mm]
& = \fr{\,3\hs\kappa\,}{\,2\hs m^2\,}
(p_1^{}\!\cdot\hsm\ep_{\PP 2}^{})^2 (p_2^{}\!\cdot\hsm\ep_{\PP 1}^{})^2
=\fr{\,3\hs\kappa\,}{\,128\hs m^2\,}
\langle 12\rangle^{\hsm 4} \,, \hspace*{10mm}
\label{eq:Amp-2hp-h}
\\[2mm]
\MM [h_{1\PP}^{},h_{2\PP}^{},h_{3\rm{S}}] & =
-\ep_{1\mu_1\nu_1}^{}\ep_{2\mu_2\nu_2}^{}\ep_{3\rm{S}\mu_3\nu_3}^{}\hsm
\mathcal{V}_{3h}^{\mu_1\nu_1,\mu_2\nu_2,\mu_3\nu_3}\hsm (p_1,p_2,p_3)
\nn\\
& =-\fr{1}{m^{2}}\ep_{1\mu_1\nu_1}\ep_{2\mu_2\nu_2}p_{3\mu}^{}p_{3\nu}^{}\hsm
\mathcal{V}_{3h}^{\mu_1\nu_1,\mu_2\nu_2,\mu_3\nu_3}\hsm (p_1,p_2,p_3)=0 \,,
\label{eq:Amp-2hp-hS}
\end{align}
\eeqs
where the shorthand notations $h_j^{}\!=\!h(p_j^{})$
and $h_{\hsm j\hs\rm{S}}^{}\!=\!\hS (p_j^{})$
are used.\

\vs

\begin{figure}[b]
\centering
\includegraphics[width=0.6\textwidth]{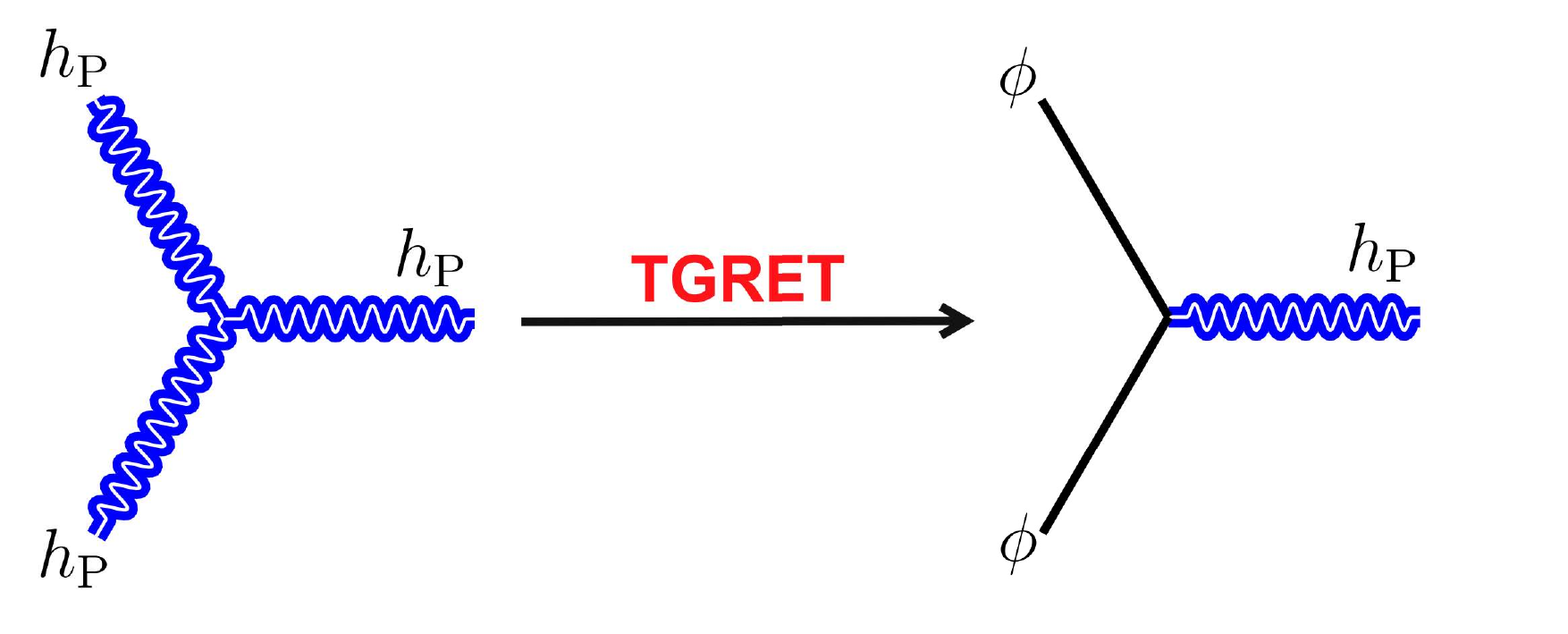}
\vspace*{-6mm}
\caption{\small\hspace*{-1mm}%
Feynman diagrams for the triple graviton scattering (left diagram)
and for the dilaton-dilaton-graviton ($\phi\hs\phi\hs\hP$)  
scattering (right diagram),
which are used for verifying the TGRET \eqref{eq:TGRET}.
}
\label{fig:3}
\end{figure}

Inspecting the graviton amplitude \eqref{eq:2hp0-hp} 
and the dilaton amplitude \eqref{eq:2phihp=2psihp},
we explicitly verify that the TGRET holds at the level of three-point amplitudes:
\begin{equation}
\label{eq:ET-3pt-h2phi}
\MM [h_{1\rm{P}_0}^{}, h_{2\rm{P}_0}^{},h_{3\rm{P}}^{}]=
\(\!\Fr{1}{\,2\,}\!\)^{\!2}\!{\MM}[\phi_1^{},\phi_2^{},h_{3\rm{P}}^{}] \hs.
\end{equation}
It agrees with the TGRET \eqref{eq:TGRET} for the case of $N\!\!=\!2$ 
with an additional physical state $h_{3\rm{P}}^{}\hs$.\ 
We illustrate this in Fig.\,\ref{fig:3}.\
The TGRET \eqref{eq:ET-3pt-h2phi} is expected since, 
according to the TGRET, two external graviton states 
$\hP(p_1^{})\hP(p_2^{})$ can be replaced
by the corresponding dilaton states $\phi(p_1^{})\phi(p_2^{})$ 
under the limit $m/E\!\ll\! 1$;
and the leading-order amplitude in this limit can be realized by either taking
the external graviton energy $E\ito \infty$ (with its mass $m$ fixed)
or taking the external graviton mass $m\ito 0\hs$ (with its energy $E$ fixed).\
In the above verification of the TGRET \eqref{eq:ET-3pt-h2phi},
we have chosen the massless limit $m\ito 0\hs$
for the external particles\,1 and 2 and set the mass of particle\,3 fixed.

\vs

\begin{figure}[h]
\centering
\includegraphics[width=0.65\textwidth]{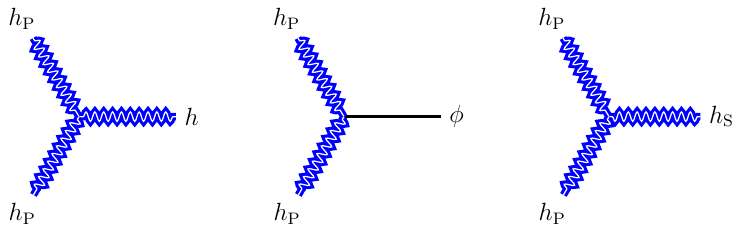}
\vspace*{-4mm}
\caption{\small\hspace*{-0.5mm}%
Relevant Feynman diagrams of the three-point graviton (dilaton) scattering
for verifying the Slavnov-Taylor-type identities
\eqref{eq:hhF2-TGETID} and \eqref{eq:hhF3-TGETID}.}
\label{fig:5}
\label{fig:4new}
\end{figure}

As the final examples of this subsection, we explicitly verify
the exact identities \eqref{eq:F2-MID} and \eqref{eq:BF3-ETID}
for the case of three-point amplitude
with two external physical graviton states and just one external
$\FFd_2^{}$ (or $\hS$) line:
\beqs
\begin{align}
\label{eq:hhF2-TGETID}
\MM \big[\hP(p_1^{}),\hsm\hP(p_2^{}),\hsm\FFd_2^{}(p_3^{})\big] &=
\MM\big[\hP(p_1^{}),\hsm\hP(p_2^{}),\hsm h(p_3^{})\big]\hsm +\hsm
\MM\big[\hP(p_1^{}),\hsm\hP(p_2^{}),\hsm\phi(p_3^{})\big]
\hspace*{8mm}
\nn\\
&= \(\!\!\fr{\,3\hs\kappa\,}{\,128\hs m^2\,}\langle 12 \rangle^{4}\!\)
   \!+\!\(\!\!\fr{\,-3\hs\kappa\,}{\,128\hs m^2\,}\langle 12 \rangle^{4}\!\)
	= 0 \,,
\\[1.5mm]
\label{eq:hhF3-TGETID}
\MM \big[\hP(p_1^{}),\hP(p_2^{}),\hS (p_3^{}) \big] &= 0 \,,
\end{align}
\eeqs
where $\,\FFd_2^{} \!=\hsm h +\phi$\,
as given by \eqref{eq:F2-under}
and $\hS\!=\!\epS^{\mn}h_{\mn}^{}$.\
In the above, we explicitly prove that the identities
\eqref{eq:hhF2-TGETID} and \eqref{eq:hhF3-TGETID}
hold at tree level.\
The $\hP\hP\phi$ vertex is given by Eq.\eqref{Beq:Vhhphi}
and its scattering amplitude is computed in Eq.\eqref{eq:hh-phi}.\
The scattering amplitude \eqref{eq:hhF3-TGETID} is given by Eq.\eqref{eq:Amp-2hp-hall}.\
The relevant Feynman diagrams for these three-point scattering amplitudes
are shown in Fig.\,\ref{fig:4new}.
%

\subsubsection{
\hspace*{-2mm}%
Four-Point Graviton (Dilaton) Scattering Amplitudes and TGRET}
\label{sec:4.1.2}
\vspace*{1mm}

Using the Feynman rules derived in Appendix\,\ref{app:Bnew},
we explicitly compute the four-point massive graviton scattering amplitude
in the Landau gauge of the WTMG theory.\
This calculation includes both the dilaton-exchange contributions and the graviton-exchange contributions 
in the $(s,t,u)$ channels and is thus highly nontrivial because the dilaton-exchange
diagrams are absent in the unitary gauge calculation (corresponding to the conventional
TMG theory\,\cite{Deser:1981wh} without dilation field) 
and the Landau-gauge graviton propapagtor \eqref{eq:Dh-Landau} 
has a different high-energy behavior from
the unitary-gauge graviton propagator \eqref{eq:Dh-unitary}.\ 
We will show that the four graviton amplitude as computed in the Landau gauge
equals the four graviton amplitude computed in the unitary gauge.\ 
This equivalence explicitly proves the gauge-invariance of the physical graviton amplitude
and serves as a nontrivial consistency check on our new BRST quantization of the WTMG theory
(including dilaton field).\ Hence, our Landau-gauge calculation of the four-point graviton
amplitude significantly differs from and goes beyond the previous calculation\,\cite{TMG-DCx}
of the four-point graviton amplitude in the conventional TMG theory 
(corresponding only to the unitary gauge of our WTMG theory).

\begin{figure}[t]
\centering
\includegraphics[width=0.8\textwidth]{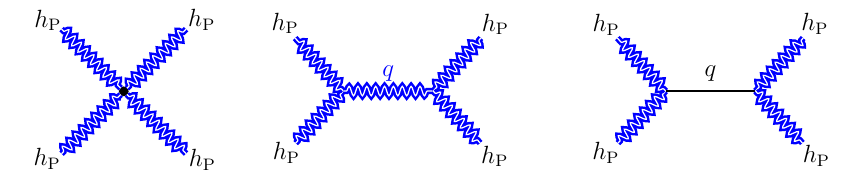}
\vspace*{-4mm}
\caption{\small\hspace*{-2mm}
Contributions to the four-point graviton amplitude at tree level 
and in Landau gauge ($\zeta\hsm\!=\!0$)
of the Weyl-transformed TMG (WTMG) theory.\ The first diagram is the four-point contact interaction,
the second diagram denotes the graviton-exchange channels, and the third diagram presents 
the dilaton-exchange channels (which is {\it absent} in the conventional TMG theory).\ 
}
\label{fig:5n}
\label{fig:new}
\end{figure}

We consider the $2\ito 2$ graviton scattering process
in the center-of-mass frame with external momenta:
\\[-6mm]
\begin{equation}
\begin{aligned}
p^\mu_1 & = E \big(1,0,\beta\big),
&~~~ p^\mu_3 & = E\big(1,\beta s_\theta,\beta c_\theta^{}\big),
\\
p^\mu_2 & = E\big(1,0,-\beta\big),
&~~~ p^\mu_4 & = E\big(1,-\beta s_\theta,-\beta c_\theta\big),
\end{aligned}
\end{equation}
where we have defined $\beta\!=\!\sqrt{1\!-\!m^2\!/E^2\,}$ and
$(s_\theta^{},\hs c_\theta^{})\!=\!(\sin\hsm\theta,\hs \cos\hsm\theta)$.\
The graviton's physical polarization tensor is obtained by ``squaring''
the corresponding polarization vector \eqref{eq.PVfromEOM} or \eqref{eq:polarization},
namely,
$\epsilon_{\text{P}}^{\mn} \!=\! \epsilon_{\text{P}}^{\mu} \epsilon_{\rm{P}}^{\nu}\,$,
as given by Eq.\eqref{eq:gravitonPol}.\
For the external states,
we derive the relevant physical polarization vectors as follows:
\begin{equation}
\begin{aligned}
\epsilon_{\rm{P}1}^{\mu} &=
\Fr{1}{\sqrt{2\,}\,}(\bar{E}\beta,\hs\ii ,\hs\bar{E}) \hs,
&~~ \epsilon_{\rm{P}3}^{\mu} &= \Fr{1}{\sqrt{2\,}\,}
(\bar{E}\beta,\hs \bar{E}s_\theta^{}\!+\!\ii\hs c_\theta^{},\hs
\bar{E}c_\theta^{}\!-\!\ii s_\theta^{})\hs,
\\
\epsilon_{\rm{P}2}^{\mu} &= \Fr{1}{\sqrt{2\,}\,}
		(\bar{E}\beta, -\ii ,-\bar{E})\hs,
&~~ \epsilon_{\rm{P}4}^{\mu} &= 
\Fr{1}{\sqrt{2\,}\,}
(\bar{E}\beta, -\bar{E}s_\theta^{}\!-\hsm \ii\hs c_\theta^{},
-\bar{E}c_\theta^{}\!+\hsm\ii\hs s_\theta^{}) \hs,
\end{aligned}
\end{equation}
where we have defined a dimensionless energy parameter $\bar{E}\!=\!E/m\hs$.\
Using the Feynman diagram approach, we compute the scattering amplitudes
for contributions from the contact diagram and the $(s,\hs t,\hs u)$-channel diagrams
including both graviton-exchanges and dilaton-exchanges 
in the Landau gauge ($\xi\hsm\!=\!\zeta\hsm\!=\!0$), 
as shown in Fig.\,\ref{fig:new}.\ 
This {\it nontrivially differs} from the unitary-gauge calculation (Appendix\,\ref{app:B3}) 
for the same scattering amplitude because our Landau-gauge calculation contains 
new contributions from the dilaton-exchange diagrams.\   
Since the graviton propagator \eqref{eq:DhDphi-Landau} in Landau gauge is much simpler than 
the propagator \eqref{eq:Dh-unitary} in unitary gauge 
(setting $\zeta\!=\!\infty$ and $\xi\!=\!0$), 
this explicit calculation becomes simpler.\ 
But it is still highly nontrivial and fairly lengthy due to the complicated 
cubic and quartic graviton vertices and 
physical polarization tensors of the external graviton states, 
as compared to the calculation of gauge boson amplitudes in the TMYM theory.\
For contributions of the contact diagram and the $(s,\hs t,\hs u)$-channel 
graviton-exchanges, we derive the following compact expressions:
\begin{subequations}
\label{eq:Amp-4hp-cstu-h}
\begin{align}
\MM_{c}^{h}[4\hP]
= & \frac{\,\kappa^2m^2\,}{\,131072\,}
\Big\{\!\!-\!(65 \bar{s}\hsm +\hsm 424) (\bar{s}\!-\!4)^4
\!+\! 12\hs (5\bar{s}^3\!+\!48\bar{s}^2\!-\!368\bar{s}\!-\!128) (\bar{s}\!-\!4)^2 c_{2\theta}
\nn\\
& +(5\hs\bar{s}^5\hsm\!+\!312\hs\bar{s}^4\hsm\!-\hsm 672\hs\bar{s}^3
\hsm\!-\!10752\hs\bar{s}^2
\!-\!768\hs\bar{s}\hsm +\hsm 2048) c_{4\theta}
\label{eq:Amp-4hp-c}
\\
& +\ii\hs 64\hs\bar{s}^{\frac{1}{2}}\hsm\!\left[6 (\bar{s}^2\!-\!3\bar{s}\!-\!12)
(\bar{s}\!-\!4)^2 s_{2\theta}^{}
\!+\!(\bar{s}^4\!+\!9\bar{s}^3
\!-\!84\bar{s}^2\!-\!144\bar{s}\!+\!64) s_{4\theta}\right] \!\!\bigg\},
\nn
\\[1.5mm]
\hspace*{-10mm}
\mathcal{M}_{s}^{h}[4\hP]
= &
\frac{\,\kappa^2m^2(\bar{s}\!-\!4)^2\,}
{\,16384(1\!-\!\bar{s})\bar{s}\,}
 \Big[(\bar{s}^5\!+\!59\hs\bar{s}^4\!-\!1448\hs\bar{s}^3\!-\!2740\hs \bar{s}^2\!-\!992\hs\bar{s}\!-\!64) c_{2\theta}^{}
\nn\\
& -\hsm\ii\hs\bar{s}^{\frac{1}{2}}(\bar{s}^5\!-\!37\hs\bar{s}^4\!+\!328\hs\bar{s}^3
\!+\!2588\hs\bar{s}^2\!+\! 1984\hs\bar{s}\!+\!320)s_{2\theta}^{}\Big],
\label{eq:Amp-4hp-s}
\\[1.5mm]
\MM_{t}^{h}[4\hP]=
& \frac{\kappa ^2 m^2 c_{\theta/2}^2
\big(\bar{s}^{\frac{1}{2}}\!+\!\ii\hs 2\tan\!\frac{\theta}{2}\big)^{\!4}}
{\,4194304\hs(\bar{s}\!-\!4)[(2\!-\!\bar{s})\!+\!(4\!-\!\bar{s}) c_{\theta}^{}]\,}\! \Big[\hsm\!-\!\ii\hs4\bar{s}^{\frac{1}{2}}
\hsm (\bar{s}^2\!+\!26\hs\bar{s}\!+\!24)(\bar{s}\!-\!4)^3 s_{5\theta}^{}
\!+\hsm \ii\hs 16 \bar{s}^{\frac{1}{2}}
\nn\\
& \times\!(19\hs\bar{s}^2\!-\!266\hs\bar{s}\!+\!168)(\bar{s}\!-\!4)^2 s_{4\theta}^{}
\!+\!\ii\hs 4\hs\bar{s}^{\frac{1}{2}}(21\bar{s}^4
\!\!+\!482\hs\bar{s}^3\!\!-\!4120\hs\bar{s}^2\!\!-\hsm\!14240\hs\bar{s}\!+\!94848)
\nn\\
& \times\!(\bar{s}\!-\!4) s_{3\theta}^{}\!+\!\ii\hs 32\hs \bar{s}^{\frac{1}{2}}(8\bar{s}^5\!\!+\!69\bar{s}^4\!\!+\!306\hs\bar{s}^3
\hsm\!-\!22472\hs\bar{s}^2\!\!+\!140704\hs\bar{s}\!-\!249216) s_{2\theta}^{}
\nn\\
& +\!\ii\hs8\bar{s}^{\frac{1}{2}}(35\hs\bar{s}^4\hsm\!+\!282\hs\bar{s}^3\hsm\!+\!8704\hs\bar{s}^2
\hsm\!-\hsm\!128352\hs\bar{s}\!+\!360960)(\bar{s}\!-\!4) s_{\theta}^{}
\nn\\
&-\!(33\hs\bar{s}^2\hsm\!+\!152\hs\bar{s}\hsm +\!16) (\bar{s}\!-\!4)^3 c_{5\theta}^{}
\!+\! 2 (43\hs\bar{s}^3\hsm\!-\!366\hs\bar{s}^2\hsm\!-\!2144\hs\bar{s}\hsm +\!2208) 
(\bar{s}\!-\!4)^2 c_{4\theta}^{}
\nn\\
&+\! (691\hs\bar{s}^4\!+\!400\hs\bar{s}^3\!-\!60896\hs\bar{s}^2\!+\!149760\hs\bar{s}
\!+\!238336)(\bar{s}\!-\!4) c_{3\theta}^{}
\nn\\
&+\!8(97\hs\bar{s}^5\!+\!934\hs\bar{s}^4\!-\!16224\hs\bar{s}^3\!-\!9952\hs\bar{s}^2
\!+\!522752\hs\bar{s}\!-\!1235968) c_{2\theta}^{}
\nn\\
&-\! 2 (457\hs\bar{s}^4\!-\!9344\hs\bar{s}^3\!-\!112544\hs\bar{s}^2\!+\!1507840\hs\bar{s}
\!-\!3860224)(\bar{s}\!-\!4)c_{\theta}^{}
\nn\\
&-\! 2(559\hs\bar{s}^3\!-\!2614\hs\bar{s}^2\!-\!137312\hs\bar{s}\!+\!689056) 
(\bar{s}\!-\!4)^2\Big],
\label{eq:Amp-4hp-t}
%
\\
\MM_{u}^{h}[4\hP]=
&\frac{\kappa ^2 m^2 s_{\theta/2}^2
	\big(\bar{s}^{\frac{1}{2}}\!-\!\ii\hs 2\cot\!\frac{\theta}{2}\big)^{\!4}}
{\,4194304\hs(\bar{s}\!-\!4)[(2\!-\!\bar{s})\!-\!(4\!-\!\bar{s}) c_{\theta}^{}]\,}\! \Big[\ii\hs4\bar{s}^{\frac{1}{2}}
\hsm (\bar{s}^2\!+\!26\hs\bar{s}\!+\!24)(\bar{s}\!-\!4)^3 s_{5\theta}^{}\!+\hsm \ii\hs16 \bar{s}^{\frac{1}{2}}
\nn\\
& \times\!(19\hs\bar{s}^2\!-\!266\hs\bar{s}\!+\!168)(\bar{s}\!-\!4)^2 s_{4\theta}^{}
\!-\!\ii\hs 4\hs\bar{s}^{\frac{1}{2}}\big(21\bar{s}^4\!\!+\!482\hs\bar{s}^3\!\!-\!4120\hs\bar{s}^2
\hsm\!-\!14240\hs\bar{s}\!+\!94848\big)
\nn\\
& \times\!(\bar{s}\!-\!4) s_{3\theta}^{}\!+\!\ii\hs32\hs \bar{s}^{\frac{1}{2}}(8\bar{s}^5\!+\!69\bar{s}^4\!+\!306\hs\bar{s}^3\hsm\!-\!22472\hs\bar{s}^2\!+\!140704\hs\bar{s}\!-\!249216) s_{2\theta}^{}
\nn\\
& -\!\ii\hs8\bar{s}^{\frac{1}{2}}(35\hs\bar{s}^4\hsm\!+\!282\hs\bar{s}^3\hsm\!+\!8704\hs\bar{s}^2\hsm\!-\hsm\!128352\hs\bar{s}\!+\!360960)(\bar{s}\!-\!4) s_{\theta}^{}
\nn\\
& +\!(33\hs\bar{s}^2\hsm\!+\hsm\!152\hs\bar{s}\!+\hsm\!16) (\bar{s}\!-\!4)^3 c_{5\theta}^{}
\!+\! 2 (43\hs\bar{s}^3\!\!-\!366\hs\bar{s}^2\!\!-\!2144\hs\bar{s}\!+\!2208) 
(\bar{s}\!-\!4)^2 c_{4\theta}^{}
\nn\\
& -\! (691\hs\bar{s}^4\hsm\!+\!400\hs\bar{s}^3\hsm\!-\!60896\hs\bar{s}^2\hsm\!+\!149760\hs\bar{s}
\!+\!238336)(\bar{s}\!-\!4) c_{3\theta}^{}
\nn\\
& +\!8(97\hs\bar{s}^5\hsm\!+\!934\hs\bar{s}^4\hsm\!-\!16224\hs\bar{s}^3\hsm\!-\!9952\hs\bar{s}^2
\hsm\!+\!522752\hs\bar{s}\!-\!1235968) c_{2\theta}^{}
\nn\\
& +\! 2 (457\hs\bar{s}^4\hsm\!-\!9344\hs\bar{s}^3\hsm\!-\!112544\hs\bar{s}^2\hsm\!+\!1507840\hs\bar{s}
\!-\!3860224)(\bar{s}\!-\!4)c_{\theta}^{}
\nn\\
&-\! 2(559\hs\bar{s}^3\hsm\!-\!2614\hs\bar{s}^2\hsm\!-\!137312\hs\bar{s}\hsm +\hsm 689056) 
(\bar{s}\!-\!4)^2\Big],
\label{eq:Amp-4hp-u}
\end{align}
\end{subequations}
%
where we define $\bar{s}\!=\! s/m^2\hs$.\
For contributions of the $(s,\hs t,\hs u)$-channel 
dilaton-exchanges, we derive the following:
\\[-7mm]
\beqs 
\label{eq:Amp-4hp-stu-phi}
\begin{align}
\MM_{s}^{\phi}[4\hP]
= & \frac{\kappa^2 m^2}{\,16384\hs\bar{s}\,}(\bar{s}\!+\!2)^{2}(\bar{s}\!-\!4)^{4} \hs,
\\[1.5mm]
\MM_{t}^{\phi}[4\hP]
= & \frac{\,\kappa^2 m^2 c_{\theta/2}^{6} \big(\bar{s}^{\frac{1}{2}}\!+\!\ii\hs 2\tan\!\frac{\theta}{2}\big)^{\!8}\,}{65536(4\!-\!\bar{s})}[(8\!-\!\bar{s})
\!+\!(4\!-\!\bar{s})c_{\theta}^{}]^2 \hs, 
\\[1.5mm]
\MM_{u}^{\phi}[4\hP]
= & \frac{\,\kappa^2 m^2 s_{\theta/2}^{6} \big(\bar{s}^{\frac{1}{2}}\!-\!\ii\hs 2\cot\!\frac{\theta}{2}\big)^{\!8}\,}{65536(4\!-\!\bar{s})}
[(8\!-\!\bar{s})\!-\!(4\!-\!\bar{s})c_{\theta}^{}]^2 \hs. 
\end{align}
\eeqs 
The above four-point graviton sub-amplitudes are computed in Landau gauge
with the graviton propagator \eqref{eq:Dh-Landau}.\
From the above formulas, we see that the individual amplitudes
$\MM_c$ and $(\MM_s^{h,\phi}, \MM_t^{h,\phi}, \MM_u^{h,\phi})$
have the leading energy-power terms of $E^{10}$ and $E^{11}$, respectively.\
We have also computed these sub-amplitudes by using the general graviton propagator
\eqref{eq:Dh-xi-zeta} [including the Feynman gauge propagator \eqref{eq:Dh-xi=zeta=1}
and unitary gauge propagator \eqref{eq:Dh-unitary} as special cases],
where their leading energy-dependence of the individual terms scales like $E^{12}$.\
But we find that the $O(E^{12})$ contributions vanish
after imposing the on-shell condition for external graviton states.\
Thus, the remaining individual leading-energy contributions scale as
$E^{11}$ at most.\

\vs 

Then, we sum up the individual contributions of all the sub-amplitudes 
\eqref{eq:Amp-4hp-cstu-h}-\eqref{eq:Amp-4hp-stu-phi} 
and derive the complete four-point graviton scattering amplitude in Landau gauge:
\begin{align}
\label{eq:app-Amp-4hp}  
\hspace*{-4mm}
\MM_{\rm{L}}^{}\hsm [4h_{\mathrm{P}}] =
\frac{~\kappa^2 m^2\csc^2\!\theta
\big(\mathbb{Y}_{0}\!+\!\mathbb{Y}_{2}c_{2\theta}^{}
\!+\!\mathbb{Y}_{4}c_{4\theta}^{}
\!+\!\mathbb{Y}_{6}c_{6\theta}^{}
\!+\!{\mathbb{Y}}_{2}'s_{2\theta}^{}
\!+\!{\mathbb{Y}}_{4}'s_{4\theta}^{}
\!+\!{\mathbb{Y}}_{6}'s_{6\theta}^{}\big)~}
{\,4096\hs (\bar{s}\!-\!1)\hs\bar{s}\hs
[(2\!-\!\bar{s})^2\!-\!(4\!-\!\bar{s})^2 c_{\theta}^{2}\hs]\,},
\end{align}
with the quantities $(\mathbb{Y}_j^{},{\mathbb{Y}}'_j)$ defined as
\begin{align}
\mathbb{Y}_{0}^{} &=
4(692\hs\bar{s}^4\!-\hsm 6305\hs\bar{s}^3
\!+\!17220\hs\bar{s}^2\!-\!12272\hs\bar{s}\hsm -\!64)\hs,
\nn\\
\mathbb{Y}_{2}^{} &=
-505\hs\bar{s}^4\!+\!19008\hs\bar{s}^3\!-\hsm 65568\hs\bar{s}^2
\!+\!45568\hs\bar{s}\hsm +\!768\hs,
\nn\\
\mathbb{Y}_{4}^{} &=
-4(58\hs\bar{s}^4\!+\hsm 635\hs\bar{s}^3\!+\hsm 20\hs\bar{s}^2
\!-\!176\hs\bar{s} \hsm +\!192)\hs,
\nn\\
\mathbb{Y}_{6}^{} &=
17\hs\bar{s}^4 \!+\hsm 560\hs\bar{s}^3
\!+\hsm 2912\hs\bar{s}^2\!+\!2816\hs\bar{s}\hsm +\hsm 256\hs,
\label{eq:app-Amp-4hp-Y}
\\
{\mathbb{Y}}_{2}' &=
-\ii\hs\bar{s}^{\frac{1}{2}}(475\hs\bar{s}^4
\!-\!8960\hs\bar{s}^3\!+\!21312\hs\bar{s}^2
\!-\!256\hs\bar{s}\!+\!1280) \hs,
\nn\\
{\mathbb{Y}}_{4}' &=
-\ii\hs 4\hs\bar{s}^{\frac{1}{2}}(5\hs\bar{s}^4\!+\!272\hs\bar{s}^3\!+\!676\hs\bar{s}^2
\!-\!544\hs\bar{s}\!+\!320) \hs,
\nn\\
{\mathbb{Y}}_{6}' &=
\ii\hs\bar{s}^{\frac{1}{2}} (\bar{s}^4\!+\!128\hs\bar{s}^3\!+\!1568\hs\bar{s}^2
\!+\!3584\hs\bar{s}\!+\!1280) \hs.
\nn
\end{align}
The above Landau-gauge calculation within the WTMG theory is 
nontrivial and has important difference from the unitary-gauge
calculation [using the graviton propagator \eqref{eq:Dh-unitary}
with $\zeta\!=\!\infty$].\ 
This is because the unitary-gauge amplitude 
$\MM_{\rm{U}}^{}\hsm [4\hP]$
does not contain any dilaton-exchange contributions, as shown in 
Eqs.\eqref{eq:Amp-4hp-cstu-unitary}-\eqref{eq:app-Amp-4hp-U}  
of Appendix\,\ref{app:B}.\   
We have demonstrated that the four-point graviton scattering amplitude \eqref{eq:app-Amp-4hp-U}
as computed in the unitary gauge 
equals the above Landau-gauge result \eqref{eq:app-Amp-4hp},
namely, 
$\MM_{\rm{U}}^{}\hsm [4h_{\mathrm{P}}]
\!=\!\MM_{\rm{L}}^{}\hsm [4h_{\mathrm{P}}]$.\
We note that Ref.\,\cite{TMG-DCx}
used the unconventional Breit coordinate system 
to explicitly calculate
four-graviton amplitude of the TMG (without dilaton field) 
with rather different and lengthy expressions 
in its eqs.(C1)-(C2) which cannot be simply compared
with our unitary-gauge formulas in Appendix\,\ref{app:B3}.\ 
Hence, the full agreement between our current independent calculations 
in both the Landau gauge (Section\,\ref{sec:4.1.2}) and unitary gauge 
(Appendix\,\ref{app:B3}) is important, which gives nontrivial consistency checks
on the gauge-invariance of the four-point physical graviton amplitude
and on our new BRST quantization of the WTMG theory.\

\vs

Inspecting the above full amplitude \eqref{eq:app-Amp-4hp}, we see that the individual
leading energy-dependence is only $\bar{s}^{\frac{1}{2}}\!\propto\!E^1$.\
In comparison with the leading energy dependence of each sub-amplitude
in Eqs.\eqref{eq:Amp-4hp-c}-\eqref{eq:Amp-4hp-u},
we further find that there are striking energy-power cancellations of
$E^{11}\!\ito\! E^{1}$ in the four-point graviton scattering amplitude
\eqref{eq:app-Amp-4hp}.\
This agrees with the energy cancellation in Eq.\eqref{eq:DE-NhP-Nphi} 
(for $N\hsm\!=\!4$) as predicted by our general power counting analysis joined with the TGRET.\ 
Taking the high energy expansion for the full scattering amplitude \eqref{eq:app-Amp-4hp},
we can derive its leading-energy contribution of
$O(\bar{s}^{\frac{1}{2}})$
and its sub-leading-energy contribution of $O(\bar{s}^{\,0})$ as follows:
\begin{equation}
\label{eq:Amp-4hp-LO+NLO}
\begin{aligned}
\MM_{\rm{L}}^{}\hsm [4h_{\mathrm{P}}]=
& -\!\frac{~\ii\hs\kappa^2 m^2\,}{2048} \hsm\csc^3\!\theta
\big(494\hs c_{\theta}^{}\!+\! 19\hs c_{3\theta}^{}\!-\!c_{5\theta}^{}\big)
\bar{s}^{\frac{1}{2}}
\\
& +\!\frac{~\kappa^2 m^2\,}{4096} \hsm\csc^4\!\theta
\big(2768\!-\!505\hs c_{2\theta}^{}\!-\!232\hs c_{4\theta}^{}\!+\!17\hs c_{6\theta}^{}\big)
\bar{s}^{\hs 0}
\!+\hsm {O}(\bar{s}^{-\frac{1}{2}}) \hs.
\end{aligned}
\end{equation}
We note that the above leading-energy contribution of $O(\bar{s}^{\frac{1}{2}})$
is important for our explicit demonstration of the TGRET \eqref{eq:TGRET}
in the following analysis.\

\vs

Then, we compute the four-point dilaton scattering amplitude
$\MMT [4\phi]$,
which includes contributions from the pure dilaton self-interactions 
and from the exchanges of massive gravitons.\
We perform this calculation for general gauge-fixing parameters $(\xi,\hs\zeta)$.\ 
We first compute the contributions from pure dilaton self-interactions 
to the four-point dilaton amplitude $\MM [4\phi]$ and
find that their sum vanishes:
\begin{subequations}
\begin{align}
\MMT_{4c}^{\phi}[4\phi] & =0 \,,  \hspace*{3.55cm}
\MMT_{4s}^{\phi}[4\phi] = -\Fr{1}{\,4\,}\hs a_0^{}\hs\kappa^{2}m^{2}\bar{s} \,,
\\
\MMT_{4t}^{\phi}[4\phi] & =
\Fr{1}{\,8\,}\hs a_0^{}\hs\kappa^{2}m^{2}(1\!+\!c_{\theta}^{})\hs\bar{s} \,,
\hspace*{6mm}
\MMT_{4u}^{\phi}[4\phi] =
\Fr{1}{\,8\,}\hs a_0^{}\hs\kappa^{2}m^{2}(1\!-\!c_{\theta}^{})\hs\bar{s} \,,
\\[1mm]
\Longrightarrow\hspace*{7mm} & 
\MMT_{4}^{\phi}
= \MMT_{4c}^{\phi} + \MMT_{4s}^{\phi} + \MMT_{4t}^{\phi} +\MMT_{4u}^{\phi} = 0 \,,
\label{appeq:4phi-phi-sum-p0}
\end{align}
\end{subequations}
where the coefficient $\hs a_0^{}\!=\!(\xi\!+\!\zeta\!-\!4)^{-1}$,
the dilaton $\phi$ is massless, and each external line obeys the on-shell condition $p_j^2\!=\!0\,$.\
Eq.\eqref{appeq:4phi-phi-sum-p0} shows that the sum of
pure dilaton contributions vanishes and this fact agrees with 
our conclusion given below Eq.\eqref{eq:LphiHat}.\ 
Then, we compute the contributions to the four-point dilaton amplitude $\MM [4\phi]$
from exchanging massive gravitons via $(s,\hs t,\hs u)$ channels:
\begin{subequations}
\label{eq:Amp-4phi-stu(h)}
\begin{align}
\label{eq:Amp-4phi-s(h)}
\MMT_{4s}^{h}
& =\!\frac{\kappa ^2 m^2\hs\bar{s}}{~16 (\bar{s}\!-\!1)~}
		\big[\ii\hs\bar{s}^{\frac{1}{2}} s_{2\theta}+c_{2\theta}+(4a_0\!+\!1)(\bar{s}\!-\!1)\big] \hs,
\\
\label{eq:Amp-4phi-t(h)}
\MMT_{4t}^{h} & =
\!\frac{\kappa ^2 m^2\hs\bar{s}}
		{\,256\hs c_{\theta/2}^2\big[2\!+\!(1\!\!+\! c_{\theta}^{})\bar{s}\hs\big]\,}
		\Big\{\!\hsm-\!\ii\hs16\hs\bar{s}^{\frac{1}{2}}\hsm (s_{2\theta}^{}\!-\!6\hs s_{\theta}^{})
		\!-\!16\big[a_0 c_{2\theta}^{}\!+\!4(a_0\!+\!2)c_{\theta}^{}\!+\!(3a_0\!-\!8)\big]
\nn\\
&\hspace{10.9em}-\!(4a_0\!+\!1)(c_{3\theta}^{}\!+\!6\hs c_{2\theta}^{}\!+\!15\hs c_{\theta}^{}\!+\!10)\bar{s}\Big\} \hs,
\\
\label{eq:Amp-4phi-u(h)}
\MMT_{4u}^{h} &=
\!\frac{\kappa ^2 m^2\hs\bar{s}}
{\,256\hs s_{\theta/2}^2\big[2\!+\!(1\!\!-\! c_{\theta}^{})\bar{s}\hs\big]\,}
\Big\{\!\hsm-\!\ii\hs16\hs\bar{s}^{\frac{1}{2}}\hsm (s_{2\theta}^{}\!+\!6\hs s_{\theta}^{})
\!-\!16\big[a_0 c_{2\theta}^{}\!-\!4(a_0\!+\!2)c_{\theta}^{}\!+\!(3a_0\!-\!8)\big]
\nn\\
&\hspace{10.9em}+\!(4a_0\!+\!1)(c_{3\theta}^{}\!-\!6\hs c_{2\theta}^{}\!+\!15\hs c_{\theta}^{}\!-\!10)\bar{s}\Big\} \hs.
\end{align}
\end{subequations}
We see that the sub-amplitude in each channel contains  
the $O(\sB^1)$ terms proportional to the $(4a_0^{}\!+\!1)$ factor  
and the $O(\sB^{\frac{1}{2}})$ terms independent of $a_0^{}\hs$.\ 
But the coefficient $(4a_0^{}\!+\!1)$ of the $O(\sB^1)$ term in each sub-amplitude  
vanishes for Landau gauge ($\xi\!=\!\zeta\!=\!0$) where $a_0^{}\!=\!-\frac{1}{4}\hs$.\ 
Hence, in the Landau gauge, each sub-amplitude of Eq.\eqref{eq:Amp-4phi-stu(h)} has manifest
leading-order energy dependence of $O(\sB^{\frac{1}{2}})$ 
which agrees with our direct energy-power counting 
on the four-dilaton scattering amplitude 
as shown in the first line of Eq.\eqref{eq:4hp-E-cancel}.\ 
Then, we sum up the above sub-amplitudes
\eqref{eq:Amp-4phi-s(h)}-\eqref{eq:Amp-4phi-u(h)}
of the graviton-exchanges from all three channels,
\\[-7mm]
\begin{align}
\MMT_{4}^{h} & = \MMT_{4s}^{h} + \MMT_{4t}^{h} +\MMT_{4u}^{h}
\nn\\
& = \frac{\bar{s}\hs\kappa^{2}m^{2}\hsm\csc^{2}\!\theta}
	{~256\hs (1\!-\!\bar{s})[(2\!+\!\bar{s})^2\!-\!\bar{s}^{2}c_{\theta}^{2}]~}\!
	\left\{\ii\hs\bar{s}^{\frac{1}{2}}\hsm
	\Big[\hsm (475\hs\bar{s}^2\hsm\!-\!544)s_{2\theta}^{}\!+\!
	(20\hs\bar{s}^2\!+\!16)s_{4\theta}^{}\!-\!\bar{s}^2 s_{6\theta}^{}\Big]  \right.
\nn\\
& \left.\hspace*{5mm}
-\Big[(759\hs\bar{s}^{2}\hsm\!-\!240\hs\bar{s}\!\!-\!\!448)c_{2\theta}^{}
\!-\!(20\hs\bar{s}\!+\!16)c_{4\theta}^{}\!+\!\bar{s}^2 c_{6\theta}^{}\Big]
\!-\!(1288\bar{s}^2\!+\!260\bar{s}\!-\!1584)\!\right\} \hsm.
\label{appeq:4phi-h-sum-p0}
\end{align}
We find that all $a_0^{}$-dependent terms exactly cancel and the summed amplitude
\eqref{appeq:4phi-h-sum-p0} does not depend on gauge parameters.\ 

\vs 

Combining Eqs.\eqref{appeq:4phi-phi-sum-p0} and
\eqref{appeq:4phi-h-sum-p0}, 
we obtain the full four-point dilaton scattering amplitude
and further derive its leading order amplitude 
under high energy expansion:
\beqs
\label{appeq:Amp-4phi-full+LO}
\begin{align}
\label{appeq:Amp-4phi-full}
\MM_{\rm{L}}^{}\hsm [4\phi] 
& = \MMT_{4}^{\phi} + \MMT_{4}^{h} = \MMT_{4}^{h}
\\[1mm]
& = -\frac{\,\ii\hs\kappa^2 m\,}{128}\csc^3\!\theta\hs
(494\hs c_{\theta}^{}\!+\! 19\hs c_{3\theta}^{}\!-\!c_{5\theta}^{})
\hs {s}^{\frac{1}{2}}
+ {O}({s}^{0}) \hs.
\label{appeq:Amp-4phi-LO}
\end{align}
\eeqs
It shows that the leading-order dilaton scattering amplitude \eqref{appeq:Amp-4phi-LO}
is of $O(s^{\frac{1}{2}})$ and agrees with
the leading-order term of the graviton scattering amplitude
\eqref{eq:Amp-4hp-LO+NLO} up to an overall factor $(1/2)^4$.\
Moreover, comparing the leading-energy contributions in the
full amplitude \eqref{appeq:4phi-h-sum-p0}-\eqref{appeq:Amp-4phi-full}
and the individual amplitudes
\eqref{eq:Amp-4phi-s(h)}-\eqref{eq:Amp-4phi-u(h)} by choosing the Landau gauge,
we see that they both have the same leading energy behavior of
$O(s^{\frac{1}{2}})$ and hence there is no any 
large energy-power cancellation 
when summing up the individual dilaton amplitudes
\eqref{eq:Amp-4phi-s(h)}-\eqref{eq:Amp-4phi-u(h)} (for Landau gauge)
into the full dilaton amplitude
\eqref{appeq:4phi-h-sum-p0}-\eqref{appeq:Amp-4phi-full+LO}.\

\vs

Note that we could compute the four-point dilation scattering amplitude by setting
each external momentum to obey $p_j^2\!=\!-m^2$ as required by the formulation
of the TGRET \eqref{eq:TGRET}.\
We will perform this calculation in Landau gauge ($\xi\!=\!\zeta\!=\!0$) for simplicity.\ 
Thus, imposing $p_j^2\!=\!-m^2$ for each external momentum,
we recompute the pure dilaton-interaction-induced four-point dilaton scattering amplitude
\eqref{appeq:4phi-phi-sum-p0} as follows:
%
\begin{equation}
\label{appeq:4phi-phi-p=m}
\MMT_4^{\phi} = -
\frac{\,\kappa^2 m^2\csc^2\!\theta\,}{8\hs\bar{s}(\bar{s}\!-\!4)}
\hsm\!\left[(2\hs\bar{s}^2\hsm\!-\!7\hs\bar{s}\!-\!4)c_{2\theta}^{}
\!-\!(2\hs\bar{s}^2\hsm\!-\!15\hs\bar{s}\!-\!4)\right] \!.
\end{equation}
%
Then, with the condition $p_j^2\!=\!\!-m^2$,
we recompute the contributions to the four-point dilaton amplitude \eqref{appeq:4phi-h-sum-p0}
from exchanging massive gravitons:
\begin{align}
\hspace*{-5mm}
\MMT_4^h =& \frac{-\ii\hs\kappa^{2}m^{2}\csc^{2}\!\theta}
	{\,256\hs\bar{s}(\bar{s}\!-\!1)(\bar{s}\!-\!4)
		[(\bar{s}\!-\!2)^2\!-\!(\bar{s}\!-\!4)^{2}c_{\theta}^{2}]\,}
	\Big\{\bar{s}^{\fr{1}{2}}
	\big[(\bar{s}\!-\!4)(475\hs\bar{s}^4\!-\!2736\hs\bar{s}^3\!+\!4736\hs\bar{s}^2\!-\!2560\hs\bar{s}
    \nn\\
    &+\!256)s_{2\theta}^{}+\!4(\bar{s}\!-\!4)^3(5\hs\bar{s}^{2}\hsm\!-\!8\hs\bar{s}\!+\!4)
	s_{4\theta}^{}\!-\!(\bar{s}\!-\!4)^5 s_{6\theta}^{}\big]
	\!+\!\ii\hs(\bar{s}\!-\!4)(759\hs\bar{s}^{4}\hsm\!-\!3552\hs\bar{s}^{3}
	\hsm\!+\!4448\hs\bar{s}^{2}
    \nn\\
    &-\!1920\hs\bar{s}\!+\!256)c_{2\theta}^{}\!+\hsm\ii\hs 4(\bar{s}\!-\!4)^3 (3\hs\bar{s}\!-\!4)c_{4\theta}^{}
	\!+\hsm\ii\hs(\bar{s}\!-\!4)^5 c_{6\theta}^{}\!+\!\ii\hs4(322\hs\bar{s}^5\hsm\!-\!2447\hs\bar{s}^4\hsm\!+\!6088\hs\bar{s}^3
    \nn\\
    &-\!5344\hs\bar{s}^2\hsm\!+\!1152\hs\bar{s}\!+\!256)\hsm\Big\} .
	\label{appeq:Amp-4phi-h-p=m}
\end{align}
Using the above and making high energy expansion,
we find that the amplitude $\MMT_4^{\phi}$ scales like $\hs\bar{s}^{\,0}$
and the amplitude $\MMT_4^h$ has its leading contribution behave as $\bar{s}^{\frac{1}{2}}\,$:
\begin{subequations}
\begin{align}
\label{appeq:Amp-4phi-phi-LO-p=m}
\MMT_4^{\phi} &
=\frac{1}{\hs 2\hs}{\,\kappa^2 m^2\,} \hsm +\hsm {O}(\bar{s}^{-1}) \hs ,
\\
\MMT_4^h &
= -\frac{\,\ii\hs\kappa^2 m^2\hs}{128} \hsm\csc^3\!\theta
\big(494\hs c_{\theta}^{}\hsm +\hsm 19\hs c_{3\theta}^{}
  \!-\hsm c_{5\theta}\big)\hs\bar{s}^{\frac{1}{2}}
+\hsm {O}(\bar{s}^{0})\hs .
\label{appeq:Amp-4phi-h-LO-p=m}
\end{align}
\end{subequations}
Summing up both contributions \eqref{appeq:4phi-phi-p=m} and \eqref{appeq:Amp-4phi-h-p=m}
with $p_j^2\!=\!-m^2$,
we derive the full four-point dilaton scattering amplitude
in the Landau gauge,
\begin{align}
\hspace*{-5mm}
\MM_{\rm{L}}^{}\hsm [4\phi] =\, &
\frac{~\kappa^2 m^2\csc^2\!\theta\big(\YT_{\hsm 0}^{}\!+\!\YT_{\hsm 2}^{}c_{2\theta}^{}
\!+\!\YT_{\hsm 4}^{}c_{4\theta}^{}
\!+\!\YT_{\hsm 6}^{}c_{6\theta}^{}\!+\!\YT'_{\hsm 2}s_{2\theta}^{}\!+\!\YT'_{\hsm 4}s_{4\theta}^{}
\!+\!\YT'_{\hsm 6}s_{6\theta}^{}\big)~}
{\,256\hs\bar{s}\hs (\bar{s}\!-\!1)(\bar{s}\!-\!4)
[(\bar{s}\!-\!2)^2\!-\!(\bar{s}\!-\!4)^{2}c_{\theta}^2\hs ]\,},
\label{appeq:Amp-4phi-sum-p=m}
\end{align}
with the quantities $(\YT_j^{}, \YT'_j)$ defined as
\begin{align}
\YT_{\hsm 0}^{} &=
4(334\hs\bar{s}^5-2565\hs\bar{s}^4\!+\hsm 6282\hs\bar{s}^3
\!-\hsm 5112\hs\bar{s}^2\!+\hsm 832\bar{s}\hsm +\hsm 256)\hs,
\nn\\
\YT_{\hsm 2}^{} &=
(\bar{s}\!-\!4)(695\hs\bar{s}^4\!-\hsm 3136\hs\bar{s}^3
\!+\!3456\hs\bar{s}^2\!-\hsm 1408\hs\bar{s}\hsm+\hsm 384)\hs,
\nn\\
\YT_{\hsm 4}^{} &=
4(\bar{s}\hsm-\hsm 4)^3(4\hs\bar{s}^2\!+\!\bar{s}\hsm-\!6)\hs,
\nn\\
\YT_{\hsm 6}^{} &=
(\bar{s}\hsm-\hsm 4)^5\hs,
\label{appeq:Amp-4phi-Yj-p=m}
\\
\YT_{\hsm 2}' &=
-\ii\hs\bar{s}^{\frac{1}{2}}(\bar{s}\hsm-\hsm 4)(475\hs\bar{s}^4\!-\hsm 2736\hs\bar{s}^3 \!+\hsm 4736\hs\bar{s}^2\!-\hsm 2560\hs\bar{s}\hsm +\hsm 256) \hs,
\nn\\
\YT_{\hsm 4}' &=
-\ii\hs 4\hs\bar{s}^{\frac{1}{2}}(\bar{s}\hsm-\hsm 4)^3(5\hs\bar{s}^2\!-\hsm 8\hs\bar{s}\hsm+\hsm 4) \hs,
\nn\\
\YT_{\hsm 6}' &=
\ii\hs\bar{s}^{\frac{1}{2}}(\bar{s}\hsm -\hsm 4)^5 \hs.
\nn
\end{align}
Thus, making a high energy expansion of the dilaton amplitude 
\eqref{appeq:Amp-4phi-sum-p=m}-\eqref{appeq:Amp-4phi-Yj-p=m},
we derive its leading contribution as follows:
\begin{align}
\MM_{\rm{L}}^{}\hsm [4\phi] 
= -\frac{\,\ii\hs\kappa^2 m^2\,}{128}\hsm 
\csc^3\!\theta \big(494\hs c_{\theta}^{}\hsm +\! 19\hs c_{3\theta}^{}\!-\hsm c_{5\theta}\big)
\bar{s}^{\frac{1}{2}} +\hsm {O}(\bar{s}^{0})\,.
\label{appeq:Amp-4phi-LO-p=m}
\end{align}
This agrees with the leading-order amplitude of graviton-exchange 
in Eq.\eqref{appeq:Amp-4phi-h-LO-p=m}
as expected because the $\MMT_4^{\phi}$ scales as $O(\bar{s}^0)$ 
in the high energy limit.\
It also agrees with the leading-order four-dilaton scattering amplitude
\eqref{appeq:Amp-4phi-LO} with each external momentum $p_j^2\!=\hsm 0\hs$.\
This shows that the leading-order dilaton scattering amplitude does not depend on the
choice of external momentum-squared being zero or not.\

\vs

\begin{figure}[t]
\centering
\includegraphics[width=0.7\textwidth]{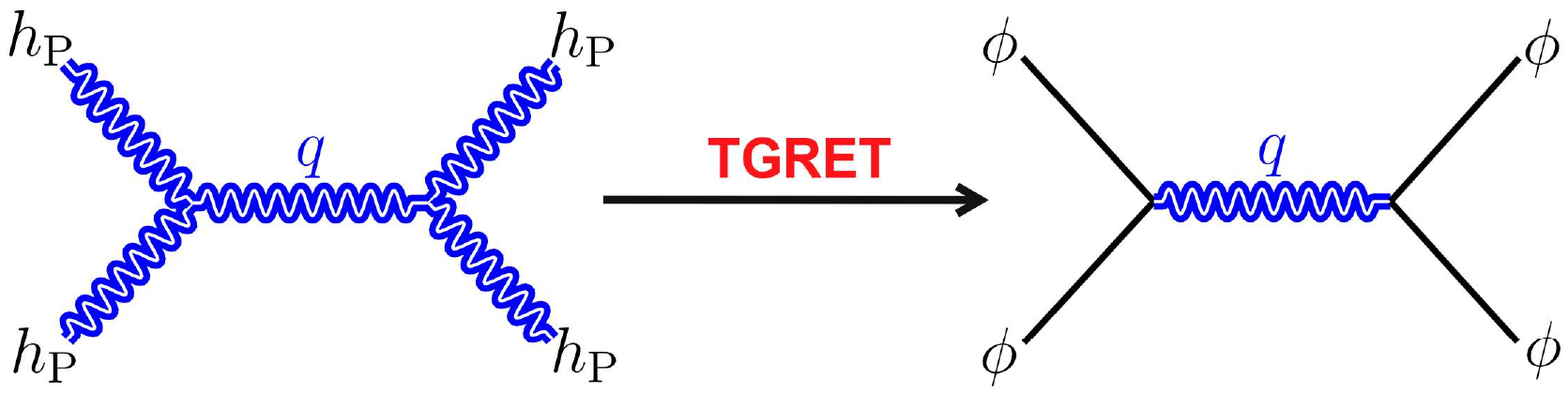}
\vspace*{-27mm}
\caption{\small\hspace*{-1mm}
Four-point massive graviton scattering (left diagram) versus the corresponding
four dilaton scattering (right diagram), and the demonstration of the
topological gravitational equivalence theorem (TGRET) in the high energy limit.\
The $4\hP$ diagram represents the dressed four-point scattering amplitudes (which have absorbed the
contributions from the $4\hs\hP$ contact diagram), 
whereas the leading-order four-dilaton amplitude
is solely given by the graviton-exchange contributions 
(right diagram).}
\label{fig:6}
\label{fig:5new}
\end{figure}

Finally, we compare the four-dilation amplitude \eqref{appeq:Amp-4phi-LO}
or \eqref{appeq:Amp-4phi-LO-p=m} with the four-graviton amplitude
\eqref{eq:Amp-4hp-LO+NLO} at the leading order of high energy expansion.\
From this comparison, we explicitly demonstrate that
for the four-point graviton (dilaton) scattering amplitudes
the TGRET holds as follows:
\begin{equation}
\label{eq:TGRET-4pt}
\MM_{\rm{L}}^{}\hsm [4\hP]
=\(\!\Fr{1}{2}\!\)^{\!4}\!
\MM_{\rm{L}}^{}\hsm [4\phi]+ {O}(\bar{s}^{0})\,, 
\end{equation}
in agreement with our general TGRET formula \eqref{eq:TGRET}.\ 
This is illustrated in Fig.\,\ref{fig:6}.\
From the above TGRET, we can understand
the striking energy-power cancellations of
$E^{11}\hsm\ito E^{1}$ in the four-point graviton scattering amplitude $\MM_{\rm{L}}^{}\hsm [4\hP]$
as computed in Landau gauge of the WTMG theory.\ 
This can be understood because on the right-hand side of Eq.\eqref{eq:TGRET-4pt}
the dilaton scattering amplitude $\MMT_{\rm{L}}^{}\hsm [4\phi]$
[as given by Eqs.\eqref{appeq:4phi-h-sum-p0}\eqref{appeq:Amp-4phi-full} or
Eq.\eqref{appeq:Amp-4phi-sum-p=m}]
has the apparent high-energy scaling behavior of $E^1$
and it has a much simpler structure of energy dependence
due to its external scalar dilaton states having no polarization
and the $\phi\hs\phi\hs h$ vertex containing only two (rather than three) partial derivatives.\
Thus, the leading-order energy-dependence of $s^{\frac{1}{2}}$ 
in the dilaton scattering amplitude 
$\MM_{\rm{L}}^{}\hsm [4\phi]$ is manifest,
without any extra large cancellations among the individual terms.\
Hence, we conclude that the right-hand side of the TGRET \eqref{eq:TGRET-4pt}
provides a proof to guarantee the striking energy cancellations of
$E^{11}\!\ito E^{1}$ in the graviton amplitude 
$\MM_{\rm{L}}^{}\hsm [4\hP]$ of Eq.\eqref{eq:app-Amp-4hp}  
as computed in the Landau gauge
and of $E^{12}\!\ito E^{1}$ in the graviton amplitude 
$\MM_{\rm{U}}^{}\hsm [4\hP]$
of Eq.\eqref{eq:app-Amp-4hp-U}  
as computed in the unitary gauge (cf.\ Appendix\,\ref{app:B3}).

\vs 

For comparison, we have also explicitly computed in Appendix\,\ref{app:B2} 
the $4\hTT$ scattering amplitudes \eqref{eq:Amp-4hT} and 
the $4\hs h$ scattering amplitudes \eqref{eq:Amp-4h}.\  
We find that none of their leading-order amplitudes \eqref{eq:Amp-4hT-exp} and 
\eqref{eq:Amp-4h-exp} equal the leading-order 
$4\hP$ physical graviton amplitude \eqref{eq:Amp-4hp-LO+NLO}  
in the high energy limit.\  
In contrast,  Eq.\eqref{eq:TGRET-4pt} proves that 
the $4\hs\phi$ dilaton scattering amplitude equals 
the $4\hP$ amplitude in the high energy limit.\  
This demonstrates that our formulation of the WTMG theory
(including the dilaton field) is essential for the successful construction of the TGRET.\ 

\vs 


\vspace*{1mm}
\subsubsection{\hspace*{-2mm}%
Four-Point Scalar Scattering Amplitudes in WTMGS Theory}
\label{sec:4.1.3}
\vspace*{1mm}

In this subsection, we consider a physical massive real
scalar field $\psi$ (with mass $m_s^{}$) coupled to the WTMG theory
\eqref{eq:LTMG-phi}, which we defined in Eq.\eqref{eq:L-WTMGS} and 
called the WTMGS theory.\ 
We will compute the four-point physical scalar scattering amplitude
in both the Landau gauge and unitary gauge, with which we explicitly prove 
the gauge invariance of the physical scalar amplitude.\ 
Then, we explicitly establish the equivalence 
between the physical scalar amplitudes
and the four-point dilaton amplitudes 
at tree level and at the leading order of high energy expansion.\ 
Finally, we will explicitly prove that the four-point the physical scalar amplitude is invariant under the field redefinition
\eqref{eq:transf-phi-psi}, showing the consistency of our analysis.

\vs

From the WTMGS Lagrangian \eqref{eq:L-WTMGS}, we derive the trilinear vertex 
($\phi\psi\psi$) as follows:
\begin{equation}
\label{eq:V-phi-psi2}
\ii\hs \mathcal{V}_{\phi\psi\psi}(p_1^{},p_2^{},p_3^{})
=\fr{\,\ii\hs\kappa\,}{2}
\big(3m_s^2\!-\!p_2^{}\!\cdot\hsm  p_3^{}\big) .
\end{equation}
Then, we choose the Landau gauge 
($\hs\zeta\hsm\!=\!\xi\hsm\!=\!0$)
of the WTMGS theory and 
compute the four-point scalar scattering amplitude 
$\MM_{\rm{L}}^{}[\psi\psi\ito\psi\psi]\!\equiv\!
\MM_{\rm{L}}^{}[4\psi]\hs$, 
which includes the contributions 
from both graviton-exchanges and dilaton-exchanges 
via the $(s,t,u)$ channels.\  
In the unitary gauge ($\zeta\!=\!\infty$), the WTMGS theory 
reduces to the TMG-scalar (TMGS) theory, 
where the scattering amplitude 
$\MM_{\rm{U}}^{}[\psi\psi\ito\psi\psi]\!\equiv\!
 \MM_{\rm{U}}^{}[4\psi]\hs$ 
contains only the graviton-exchange
contribution via $(s,t,u)$ channels.\ 
Since the physical amplitude of $\psi\psi\ito\psi\psi\hs$
is gauge-invariant, both the Landau-gauge and unitary-gauge 
calculations should give the same result, which will be shown
below.\ We further note that taking the massless limit for
graviton ($m\!\to\!0$), the Landau-gauge graviton propagator
\eqref{eq:DhDphi-Landau} vanishes 
as shown in Eq.\eqref{eq:Dh-m=0-Landau}.\ 
Hence, in the massless limit ($m\!\to\!0$), the Landau-gauge
amplitude $\MM_{\rm{L}}^{}[4\psi]|_{m=0}^{}$ contains only 
the contribution of dilaton-exchanges and it should equal 
the unitary-gauge amplitude  
$\MM_{\rm{U}}^{}[4\psi]|_{m=0}^{}$
that contains the contribution of graviton-exchanges alone.\  
(Note that the unitary gauge in the WTMGS theory is 
the same as the TMGS theory.) 
We will demonstrate these nontrivial properties 
by the following explicit calculations. 

\vs 

In the Landau gauge ($\zeta\!=\hsm\xi\!=\!0$) of the WTMGS theory, 
the four-point physical scalar scattering amplitude contains
contributions from the graviton-exchanges and dilaton exchanges
as well as a contribution of the contact vertex of $\psi^4$,
\beq 
\label{eq:M[4psi]-sum-Landau}
\MM_{\rm{L}}^{}[4\psi]= 
\MM_{\rm{L}}^{h}[4\psi]+ \MM_{\rm{L}}^{\phi}[4\psi] +\MM_{\rm{L}}^{c}[4\psi]\,,
\eeq 
where $\MM_{\rm{L}}^{h}[4\psi]$ and $\MM_{\rm{L}}^{\phi}[4\psi]$
receive contributions from $(s,t,u)$ channels respectively,
as shown in Appendix\,\ref{app:Cnew}.\ 
We sum up the graviton-exchange contributions of 
the $(s,t,u)$ channels as follows: 
\beq 
\label{eq:M[4psi]h-sum-Landau}
\MM_{\rm{L}}^{h}[4\psi] = \frac{\kappa^2 m\big(\mathbb{Z}_{0}^{}\!+\!\mathbb{Z}_{2}^{}c_{2\theta}^{}
\!+\!\mathbb{Z}_{4}^{}c_{4\theta}^{}
\!+\!\mathbb{Z}_{6}^{}c_{6\theta}^{}\!+\!\mathbb{Z}'_{2}s_{2\theta}^{}\!+\!\mathbb{Z}'_{4}s_{4\theta}^{}
\!+\!\mathbb{Z}'_{6}s_{6\theta}^{}\big)}
{~1024\hs s(s\!-\!m^2)(s\!-\!4m_s^2)s_{\theta}^2\hs 
\big[(s\!-\!4m_s^2)s_{\theta/2}^2\!+\!m^2\big]\!
\big[(s\!-\!4m_s^2)c_{\theta/2}^2\!+\!m^2\big]~}\,, 
\eeq
where the numerator coefficients $(\mathbb{Z}_j^{}, \mathbb{Z}'_j)$ 
are given by \eqref{Aeq:Z-Z'-M[4psi]h} of Appendix\,\ref{app:Cnew}.\ 
It is clear that the above graviton-exchange amplitude 
$\MM_{\rm{L}}^{h}[4\psi]$ in Landau gauge 
vanishes in the massless limit 
$m\ito 0\hs$ as expected, because the Landau-gauge graviton
propagator vanishes in the massless limit 
as shown in Eq.\eqref{eq:Dh-m=0-Landau}.\ 
Then, under high energy expansion, we derive 
the graviton-exchange amplitude \eqref{eq:M[4psi]h-sum-Landau}
as follows:
\begin{align}
\label{eq:M[4psi]L-h-expand}
\MM_{L}^{h}[4\psi]=&
-\!\frac{\,\ii\hs\kappa^2\,}{\,128\,}\hs m\hs s^{\frac{1}{2}}
 \big(494\hs c_{\theta}^{}\hsm +\! 19\hs c_{3\theta}^{}\!-\hsm c_{5\theta}\big)\!\csc^3\!\theta
\nn\\
&+\!\frac{\,\kappa^2\,}{\,256\,}\hs m^2 s^0
\big(1288\!+\!759\hs c_{2\theta}\!+\hsm c_{6\theta}\big)
\!\csc^4\!\theta \hsm +\hsm {O}\big(s^{-\frac{1}{2}}\big)\hs,
\end{align}
where the leading-order contribution in the first line coincides with the
leading-order dilaton amplitude \eqref{appeq:Amp-4phi-LO-p=m}.\  
Next, we compute the dilaton-exchange contributions 
to the four-scalar amplitude via the $(s,t,u)$ channels,  
\begin{align}
\MM_{\rm{L}}^{\phi}[4\psi]
=-\frac{\,\kappa^2 m_s^2
\big[(7s^2\!-\!24m_s^2 s\!-\!16m_s^4)c_{2\theta}^{}
\!-\!(7s^2\!-\!56m_s^2 s\!-\!16m_s^4)\big]\,}
{8\hs s(s\!-\!4m_s^2)s_{\theta}^{2}}  \,,
\label{eq:M[4psi]phi-sum-Landau}
\end{align}
which is independent of graviton mass $m\hs$.\ 
Under high energy expansion, we find that the leading-order
contribution of the amplitude 
$\MM_{\rm{L}}^{\phi}[4\psi]$ is energy-independent:
\begin{equation}
\MM_{\rm{L}}^{\phi}[4\psi] 
=\frac{\hs 7\hs}{4}\kappa^2 m_s^2+{O}\big(s^{-1}\big)\hs. 
\end{equation}
With the sub-amplitudes \eqref{eq:M[4psi]h-sum-Landau} and
\eqref{eq:M[4psi]phi-sum-Landau}, we derive the complete 
four-scalar scattering amplitude \eqref{eq:M[4psi]-sum-Landau}
as follows: 
\begin{align}
\MM_{\rm{L}}^{}[4\psi] &= 
\MM_{\rm{L}}^{h}[4\psi]+ \MM_{\rm{L}}^{\phi}[4\psi] +\MM_{\rm{L}}^{c}[4\psi]
\nn\\
&=\frac{~\kappa^2\big(\widetilde{\mathbb{Z}}_{0}^{}\!+\!\widetilde{\mathbb{Z}}_{2}^{}c_{2\theta}^{}
\!+\!\widetilde{\mathbb{Z}}_{4}^{}c_{4\theta}^{}
\!+\!\widetilde{\mathbb{Z}}_{6}^{}c_{6\theta}^{}\!+\!\widetilde{\mathbb{Z}}'_{2}s_{2\theta}^{}\!+\!\widetilde{\mathbb{Z}}'_{4}s_{4\theta}^{}
\!+\!\widetilde{\mathbb{Z}}'_{6}s_{6\theta}^{}\big)~}
{256s(s\!-\!m^2)[(s\!-\!4m_s^2\!+\!2m^2)^2\!-\!(s\!-\!4m_s^2)^2c_{\theta}^{2}]s_{\theta}^2} -\kappa^2 m_s^2\, ,
\label{eq:M[4psi]-sum1-Landau}
\end{align}
where the numerator coefficients $(\widetilde{\mathbb{Z}}_j^{}, \widetilde{\mathbb{Z}}'_j)$ 
are given by \eqref{Aeq:Zt-Zt'-M[4psi]L-full} 
of Appendix\,\ref{app:Cnew}.\ 
In Eq.\eqref{eq:M[4psi]-sum1-Landau}, 
we can readily deduce the contribution of contact diagram, 
$\MM_{\rm{L}}^{c}[4\psi]\!=\!-\kappa^2 m_s^2$.\ 
Under high energy expansion, we derive expanded form of the four-point physical scalar   
scattering amplitude \eqref{eq:M[4psi]-sum1-Landau} as follows:
\begin{equation}
\label{eq:M[4psi]L-expand}
\begin{aligned}
\hspace*{-5mm}
\MM_{\rm{L}}[4\psi]=
&-\!\frac{\,\ii\hs\kappa^2\,}{~128~}\hs
m\hs s^{\frac{1}{2}}
\big(494\hs c_{\theta}^{}\hsm +\! 19\hs c_{3\theta}^{}\!-\hsm c_{5\theta}\big)\! \csc^3\!\theta
\!+\!\frac{\kappa^2}{\,256\,}\hsm
\big[m^2 c_{6\theta}\hsm +\hsm 24\hs m_s^2 c_{4\theta}^{}
\\
&+\!(759\hs m^2\!-\!96\hs m_s^2)c_{2\theta}
\!+\!(1288\hs m^2\!+\!72\hs m_s^2\hs)\big]
\!\csc^4\!\theta\hsm 
+\hsm {O}\big(s^{-\frac{1}{2}}\big). 
\end{aligned}
\end{equation}
We see that its leading-order amplitude is of 
$O(m\hs s^{\frac{1}{2}})$ and has no explicit dependence 
on the scalar mass $m_s^{}\hs$.\ 
This leading-order amplitude is given by the leading-order contribution 
of graviton-exchanges as shown in Eq.\eqref{eq:M[4psi]L-h-expand} 
and equals the leading-order dilaton amplitude 
of $O(m\hs s^{\frac{1}{2}})$ as shown in 
Eq.\eqref{appeq:Amp-4phi-LO} or Eq.\eqref{appeq:Amp-4phi-LO-p=m}.\ 

\vs 

Then, we note that in the massless limit of graviton ($m\!\ito 0$),
the graviton-exchange contribution vanishes in the Landau gauge,
$\MM_{\rm{L}}^{h}[4\psi]|_{m=0}^{}\!=\!0\hs$.\ 
Hence, for the graviton mass $m\ito 0\hs$,
the four-scalar amplitude in the Landau gauge 
contains the dilaton-exchange contribution 
$\MM_{\rm{L}}^{\phi}[4\psi]$ 
[given by Eq.\eqref{eq:M[4psi]phi-sum-Landau}] 
and the four-scalar contact contribution  
$\MM_{\rm{L}}^{c}[4\psi]\!=\!\kappa^2m_s^2$
(which remains the same in the unitary gauge).\ 
Thus, we have
\beq
\label{eq:Amp4S-Landau:m=0}
\MM_{\rm{L}}^{}[4\psi]\!\left|_{m=0}^{}= 
\MM_{\rm{L}}^{\phi}[4\psi]+\MM_{\rm{L}}^{c}[4\psi]\,. 
\right. 
\eeq 

Next, we consider the unitary gauge ($\zeta\!=\!\infty$) of
the WTMGS theory, which corresponds to TMGS the Lagrangian \eqref{eq:L-TMGS-UG}.\ 
We further choose $\xi\!=\!0$ to simplify the graviton propagator
\eqref{eq:Dh-unitary} in the unitary gauge.\ 
With these, we can recompute the 
($4\psi$) scattering amplitude using the TMGS Lagrangian
\eqref{eq:L-TMGS-UG} (equivalent to the unitary gauge of 
the WTMGS theory), which contains the contribution of
graviton-exchanges plus the four-scalar contact contribution,
as given by \eqref{Aeq:M[4psi]U-cstu} of Appendix\,\ref{app:Cnew}.\ 
(The four-scalar contact contribution remains the same in the Landau gauge.) 
Then, we find that the four-scalar amplitude 
$\MM_{\rm{U}}^{}[4\psi]$
in the unitary gauge precisely equals our Landau-gauge amplitude 
\eqref{eq:M[4psi]-sum1-Landau}
(including {\it contributions from both the graviton-exchanges 
and dilaton-exchanges} in addition to the four-scalar contact contribution):
\beq
\label{eq:M[4psi]U=L} 
\MM_{\rm{U}}^{}[4\psi]= \MM_{\rm{L}}^{}[4\psi]\,. 
\eeq 
This proves the gauge-invariance of the physical four-scalar 
amplitude $\MM [4\psi]$ and gives a consistency check 
of our WTMGS formulation whose unitary gauge 
($\zeta\!=\!\infty$) corresponds to the conventional 
TMGS theory.\ 

\vs 

We further take the massless limit of graviton ($m\ito 0$)
for the unitary-gauge amplitude $\MM_{\rm{U}}^{}[4\psi]$
[which is also the ($4\psi$) amplitude given by the TMGS
Lagrangian \eqref{eq:L-TMGS-UG}],
where only the graviton-exchange diagrams 
and the four scalar contact-diagram contribute.\ 
The contact-diagram contribution is 
$\MM_{\rm{L}}^{c}[4\psi]\!=\!\MM_{\rm{U}}^{c}[4\psi]
\!=\!-\kappa^2m^2_s$,
which does not depend on graviton mass and remains the same
in both the Landau gauge and unitary gauge.\ 
So $\MM_{\rm{L}}^{c}[4\psi]$ and $\MM_{\rm{U}}^{c}[4\psi]$
cancels out on the two sides
of the equality \eqref{eq:M[4psi]U=L}.\ 
Then, taking the massless graviton limit and 
using Eq.\eqref{eq:Amp4S-Landau:m=0}, 
we deduce the following relation from our explicit calculations 
on both sides of Eq.\eqref{eq:M[4psi]U=L}:
\beq
\label{eq:Amp4S|m=0|U=L(phi)}
\MM_{\rm{U}}^{h}[4\psi]\!\left|_{m=0}^{}
= \MM_{\rm{L}}^{\phi}[4\psi]\!\left|_{m=0}^{}
= \MM_{\rm{L}}^{\phi}[4\psi]
= \rm{Eq.}\eqref{eq:M[4psi]phi-sum-Landau}\,,
\right.\right. 
\eeq 
where $\MM_{\rm{U}}^{h}[4\psi]$ 
denotes the graviton-exchange amplitude in the unitary gauge,
and the second equality holds 
because the dilaton-exchange amplitude 
$\MM_{\rm{L}}^{\phi}[4\psi]$
in Eq.\eqref{eq:M[4psi]phi-sum-Landau} does not depend on 
the graviton mass $m\hs$ at tree level.\ 
Eq.\eqref{eq:Amp4S|m=0|U=L(phi)} shows that 
in the massless graviton limit  
($m\!\ito 0$), the survival ($4\psi$)-amplitude in the unitary gauge
(contributed solely by the graviton-exchanges)
just equals the corresponding Landau-gauge ($4\psi$)-amplitude
from the dilaton-exchanges (which is independent of the 
graviton mass $m$).\  
Hence, the unitary-gauge survival amplitude $\MM_{\rm{U}}^{h}[4\psi]|_{m=0}^{}$ 
(from graviton-exchanges) has its origin from
the massless-dilaton-exchange amplitude
$\MM_{\rm{L}}^{\phi}[4\psi]$ in the Landau gauge.\

\vs 

We note that Ref.\,\cite{other2a-3d-CS} 
showed a surviving term as a hint of dilaton-like contribution   
of the four-point scalar amplitude in the massless graviton limit 
for the conventional TMGS theory.\ 
But it did not introduce any dilaton field at Lagrangian level
and it remains unclear whether this surviving term is the 
real contribution of a dilaton field or not.
In contrast with Ref.\,\cite{other2a-3d-CS}, 
from our above Landau-gauge analysis within 
the WTMGS theory \eqref{eq:L-WTMGS} 
and in comparison with the unitary-gauge analysis 
[corresponding to the TMGS theory \eqref{eq:L-TMGS-UG}],  
we have explicitly proved in Eq.\eqref{eq:Amp4S|m=0|U=L(phi)} 
that such a surviving four-scalar amplitude 
(with graviton-exchanges) in the massless graviton limit
of the TMGS theory indeed has its origin from 
the dilaton-exchange contribution 
\eqref{eq:M[4psi]phi-sum-Landau} in the Landau gauge
of our WTMGS theory.\

\vs

Finally, for the later comparison with our double-copy analysis
in Sec.\,\ref{sec:4.2.2}, 
we consider a special choice of the scalar mass $m_s^{}\!=\!m$
in the WTMGS Lagrangian \eqref{eq:L-WTMGS}, namely, the
scalar mass equals the graviton mass.\ 
Thus, we can reduce the four-scalar amplitude 
\eqref{eq:M[4psi]-sum1-Landau} to the following compact form:  
\begin{equation}
\label{eq:Amp-4psi|L|ms=m}
\hspace*{-7mm}
\MM_{\rm{L}}[4\psi]\!\left|_{m_s^{}=m}^{}\right.
=\frac{~\kappa^2 m^2\big(\widehat{\mathbb{Z}}_{0}^{}
\!+\!\widehat{\mathbb{Z}}_{2}^{}c_{2\theta}^{}
\!+\!\widehat{\mathbb{Z}}_{4}^{}c_{4\theta}^{}
\!+\!\widehat{\mathbb{Z}}_{6}^{}c_{6\theta}^{}
\!+\!\widehat{\mathbb{Z}}'_{2}s_{2\theta}^{}
\!+\!\widehat{\mathbb{Z}}'_{4}s_{4\theta}^{}
\!+\!\widehat{\mathbb{Z}}'_{6}s_{6\theta}^{}\big)~}
{256\hs \sB\hs (\sB\!-\!1)
\big[(\sB\!-\!2)^2\!-\!(\sB\!-\!4)^2c_{\theta}^{2}\big]
s_{\theta}^2}\,,
\end{equation}
with the numerator coefficients 
$(\widehat{\mathbb{Z}}_j^{}, \widehat{\mathbb{Z}}'_j)$ 
given by
\begin{align}
\Zh_{\hsm 0}^{} &= 4\big(340\hs\bar{s}^4\!-\hsm 1329\hs\bar{s}^3\!+\hsm 1028\hs\bar{s}^2
\!+\hsm 16\hs\bar{s}\hsm -\hsm 64) \hs,
\nn\\
\Zh_{\hsm 2}^{} &= 663\hs\bar{s}^4\!-\hsm 2688\hs\bar{s}^3\!+\hsm 1760\hs\bar{s}^2
\!-\hsm 512\hs\bar{s} \hsm +\hsm 768 \hs,
\nn\\
\Zh_{\hsm 4}^{} &= 4\hs (\bar{s}\!-\hsm 4)^2
\big(6\hs\bar{s}^2\!+\hsm 5\hs\bar{s}\!-\hsm 12\big) 
\hs,
\nn\\
\Zh_{\hsm 6} &= \big(\bar{s}\hsm -\hsm 4\big)^4 ,
\label{eq:Zhat-4psi|L|ms=m}
\\
\Zh'_{\hsm 2} &=
-\ii\hs\bar{s}^{\frac{1}{2}}
\big(475\hs\bar{s}^4\!-\hsm 2736\hs\bar{s}^3
\!+\hsm 4736\hs\bar{s}^2\!-\hsm 2560\hs\bar{s}
\hsm +\hsm 256) \hs,
\nn\\
\Zh'_{\hsm 4} &=
-\ii\hs4\hs\bar{s}^{\frac{1}{2}}
(\bar{s}-4)^2
(5\hs\bar{s}^2\!-\hsm 8\hs\bar{s}\hsm +\hsm 4) \hs,
\nn\\
\Zh'_{\hsm 6} &=
\ii\hs\bar{s}^{\frac{1}{2}} (\bar{s}\hsm -\hsm 4)^4,
\nn
\end{align}
where $\sB\!=\!s/m^2$.\ Under high energy expansion, we derive 
the leading-order amplitude of Eq.\eqref{eq:Amp-4psi|L|ms=m} as follows:
\begin{equation}
\label{eq:M[4psi]L-LO}
\begin{aligned}
\hspace*{-5mm}
\MM_{\rm{L}}[4\psi]\!\left|_{m_s^{}=m}^{}\right.
= -\frac{\,\ii\hs\kappa^2\,}{~128~}\hs
m\hs s^{\frac{1}{2}}
\big(494\hs c_{\theta}^{}\hsm +\! 19\hs c_{3\theta}^{}\!-\hsm c_{5\theta}\big)\! \csc^3\!\theta
\!+\!\hsm {O}\big(s^{0}\big),
\end{aligned}
\end{equation}
which equals the leading-order amplitude of Eq.\eqref{eq:M[4psi]L-expand}.\
This is because Eq.\eqref{eq:M[4psi]L-expand} shows that
its leading-order amplitude is contributed by
the graviton-exchange [cf.\ Eq.\eqref{eq:M[4psi]L-h-expand}]
and is thus independent of the scalar mass $m_s^{}\hs$.\ 
We further note that the leading-order physical scalar amplitude \eqref{eq:M[4psi]L-LO}
just equals the leading-order four-dilaton amplitude \eqref{appeq:Amp-4phi-LO}
(derived by the Feynman-diagram method).\ This is because both 
the ($4\psi$)-amplitude and the ($4\phi$)-amplitude have their leading-order contributions
given by the graviton-exchange constributions, and the trilinear vertices
$\phi\phi h_{\mn}^{}$ and $\psi\psi h_{\mn}^{}$ of the WTMGS Lagrangian 
\eqref{eq:L-WTMGS} have the same coupling.\   
Moreover, the equivalence between the four-point 
leading-order amplitudes
\eqref{eq:M[4psi]L-LO} and \eqref{appeq:Amp-4phi-LO}
also agrees with the general statement 
in Eq.\eqref{eq:M0[Nphi]=M0[Npsi]} for $N\!=\!4\hs$.

\vs 

Finally, as a consistency check our analysis, 
we have further explicitly proved that
the four-point physical scalar amplitude 
\eqref{eq:M[4psi]-sum1-Landau}
is invariant under the field redefinitions
\eqref{eq:transf-phi-psi}, namely, 
$\MM_{\rm{L}}^{}[4\psi]\!=\!\MM_{\rm{L}}^{}[4\psih]$.\ 
This is shown at the end of Appendix\,\ref{app:Cnew}.\ 

\vspace*{2mm}
\subsection{\hspace*{-2mm}%
Graviton (Dilaton) Scattering Amplitudes from Massive Double Copy}
\label{sec:4.2}
\vspace*{1.5mm}

In this subsection, we use the extended massive double-copy method to construct the
three-point and four-point graviton (dilaton) scattering amplitudes
in the WTMG theory from
the corresponding gauge boson (adjoint scalar) scattering amplitudes
in the TMYM (TMYM-Scalar) theory.\
Different from the previous double-copy analysis of the four graviton amplitude\,\cite{TMG-DCx},
our focus is to understand the energy-dependence structures 
of the double-copied graviton amplitude and dilaton amplitude 
in the WTMG theory,
in connection to the energy-dependence structures 
of the gauge amplitude 
and adjoint-scalar amplitude in the TMYM (TMYM-Scalar) theory.\ 
With the extended double-copy approach, 
we construct the three-point 
scattering amplitudes of gravitons and of dilatons 
as well as the four-point graviton amplitudes at tree level.\  
But the four-point dilaton scattering amplitudes can be
constructed only at the leading order of high energy expansion.\
This is because the dilaton is an unphysical field 
in the WTMG theory and 
there is no guarantee for the double copy of 
such unphysical dilaton amplitudes.\ 
But our TGRET \eqref{eq:TGRET} ensures that  
each leading-order dilaton amplitude equals the corresponding 
physical graviton amplitude at the leading order,
hence each leading-order dilaton amplitude 
is a gauge-invariant physical amplitude.\ 
Moreover, we proved in Eq.\eqref{eq:M0[Nphi]=M0[Npsi]} that 
each $N$-point leading-order dilaton amplitude equals 
the $N$-point leading-order physical scalar amplitude
under high energy expansion.\
The quantitative construction of the leading-order
dilaton amplitudes is highly nontrivial and will be 
explicitly demonstrated for the four-point scattering 
in Section\,\ref{sec:4.2.2}, 
which was not studied in Refs.\,\cite{TMG-DCx}\cite{Hang:2021oso}.\ 
We note that there is no general proof of the BCFW recursions 
for 3d non-supersymmetric field theories, 
so the extension of the present study to the $N$-point amplitudes 
(with $N\!\!\geqq\! 5$) is much harder and 
could be discussed only case by case, 
as we will comment further at the end of this subsection.

\vspace*{1.5mm}
\subsubsection{
\hspace*{-2mm}%
Three-Point Graviton (Dilaton) Amplitudes from Massive Double Copy}
\label{sec:4.2.1}
\vspace*{1.5mm}

In this subsection, we study the extended massive double-copy construction of the three-point
scattering amplitudes of gravitons and of dilatons in the WTMG theory from the 
corresponding gauge boson scattering amplitudes in the TMYM (TMYM-Scalar) theory.\
For this, we first consider the gauge boson amplitudes of the TMYM theory.\
The three-point scattering of on-shell massive gauge bosons $(\AP\AP\AP)$
is shown by the left diagram of Fig.\,\ref{fig:2}.\
Thus, we derive the three-point amplitude of $(\AP\AP\AP)$ as follows:
\begin{equation}
\hspace*{-3mm}
\TT[1,2,3]
= \ii\hs g\hs f^{abc} \!\left[ \eta_{\mn}^{}(p_1^{}\!-\hsm p_2^{})_{\rho}^{} \!+\hsm
\eta_{\mu\rho}^{}(p_3^{}\!-\hsm p_1^{})_{\nu}^{}\! +\hsm \eta_{\nu\rho}^{}(p_2\!-\hsm p_3^{})_{\mu}^{}
\!+\hsm \ii\hs m\hs\varepsilon_{\mu\nu\rho}^{} \right]\!
\epsilon_{\text{P}1}^{\mu}\epsilon_{\text{P}2}^{\nu}\epsilon_{\text{P}3}^{\rho}\,,
\end{equation}
where the amplitude $\TT[1,2,3]\!\equiv\!\TT[A_{\rm{P}1}^-,\!A_{\rm{P}2}^-,\!A_{\rm{P}3}^-]$
and $A_{\rm{P}j}^-\!=\!\AP^-(p_j^{})$ with $j\!=\!1,2,3\hs$.\
The superscript of $\AP^{-}$ denotes
have helicity $-1$ for each external gauge boson.\
(Since in the 3d TMYM theory the spin angular momentum is a pseudoscalar and
each gauge boson has only one physical degree of freedom 
with helicity either $-1$ or $+1$
\cite{Deser-CS1982PRL}\cite{Jackiw:1991},
we choose the helicity $-1$ throughout our analysis
without losing generality.\
For simplicity, we will suppress the superscript ``$^-$''
in our notation for the helicity state of each gauge boson, unless specified otherwise.)\
With the 3d gauge polarization vector \eqref{eq:polarization},
we further derive this scattering amplitude as follows:
\begin{align}
\hspace*{-4mm}
\TT [1,2,3] &\!=\!\frac{\ii\hs g\hs f^{abc}}{\,16\sqrt{2\,}m^{3}\,}\!\hsm
\(\hsm\ii\hs 4\hs m\langle 12 \rangle\hsm \langle 23 \rangle\hsm \langle 31 \rangle
\hsm\!-\! 4\langle 3| p_{1} |3 \rangle\hsm \langle 12 \rangle^{\!2}
\hsm\!+\! 4\langle 2| p_{1} |2 \rangle\hsm\langle 13 \rangle^{\!2}
\hsm\!-\! 4\langle 1| p_{2} |1 \rangle \hsm\langle 32 \rangle^{\!2}\hsm\)
\nn\\
&\!=\! \frac{~gf^{abc}\,}{~2\sqrt{2\,}\hs m^{2}~}
\langle 12 \rangle \langle 23 \rangle \langle 31 \rangle,
\label{eq:threeTMYM}
\end{align}
where in the first line we have computed the last three terms as follows:
\beqs
\begin{align}
\label{eq:3p13}
\langle 3| p_{1} |3 \rangle\hsm \langle 12 \rangle^2
& = \langle 12 \rangle\hsm \langle 31 \rangle
\big(\!\!-\!\langle 2| p_{1}^{} |3 \rangle + \ii\hs m \langle 23 \rangle \hsm\big)
= \ii\hs m \langle 12 \rangle \langle 23 \rangle \langle 31 \rangle ,
\\
\label{eq:2p12}
\langle 2| p_{1} |2 \rangle\hsm \langle 13 \rangle^2
&= \langle 12 \rangle\hsm\langle 31 \rangle \big(\!-\!\langle 2| p_{1} |3 \rangle \!-\hsm\ii\hs m \langle 23 \rangle \hsm\big)
= -\ii\hs m\langle 12 \rangle \langle 23 \rangle\hsm \langle 31 \rangle ,
\\
\label{eq:1p21}
\langle 1| p_{2} |1 \rangle\hsm \langle 32 \rangle^2
&= \langle 12 \rangle \langle 23 \rangle
\big(\!-\!  \langle 3| p_{2} |1 \rangle \!+\hsm \ii\hs m \langle 31 \rangle \hsm\big)
= \ii\hs m\langle 12 \rangle \langle 23 \rangle\hsm\langle 31 \rangle .
\end{align}
\eeqs
In the above derivations we have made use of the Dirac equations \eqref{eq:DiracEq2} to prove
$\langle 2| p_1^{} |3 \rangle\!=\!0\hs$ and
$\langle 3| p_2^{} |1 \rangle\!=\!0\hs$.\

Next, we reexpress the three-point gauge boson scattering amplitude \eqref{eq:threeTMYM}
in the following form:
\\[-8mm]
\beqs
\label{eq:3pt-Amp-Gauge}
\begin{align}
\TT[A_{\rm{P}1},\!A_{\rm{P}2},\!A_{\rm{P}3}] & = gf^{abc}\NN_{\!A}^{} [1,2,3]\,,
\\
\NN_{\!A}^{} [1,2,3] & = \frac{~1~}{~2\sqrt{2\,}\hs m^{2}~}
\langle 12 \rangle \langle 23 \rangle \langle 31 \rangle \hs.
\end{align}
\eeqs
For the color-kinematics duality, we have the following correspondence to realize the double copy for
the three-point gauge boson amplitudes versus graviton amplitudes:
\beq
\label{eq:3pt-CK-duality}
g\to \frac{\,\kappa\,}{4}\hs,
\hspace*{5mm}
f^{abc} \!\to \NN_{\!A}^{}[1,2,3]\hs.
\eeq

Then, we consider the three-point scattering amplitude of massive gravitons in the TMG theory,
as shown by the right diagram of Fig.\,\ref{fig:2}.\
Using the three-point gauge boson amplitude \eqref{eq:3pt-Amp-Gauge}
and the color-kinematics duality correspondence \eqref{eq:3pt-CK-duality},
we can derive the three-point massive graviton scattering amplitude via the double copy:
\begin{equation}
\label{eq:DC-3hp-Amp}
\MM_{\rm{DC}}[h^{}_{1\PP},h^{}_{2\PP},h^{}_{3\PP}]
= \frac{\,\kappa~}{~32\hs m^4~}\langle 12 \rangle^{\!2} \langle 23 \rangle^{\!2} \langle 31 \rangle^{\!2} ,
\end{equation}
where we have used the shorthand notation
$h^{}_{\rm{P}j}\!\equiv\!h_{\rm{P}}^{-}(p_j^{})$
with $j=1,2,3$ and the superscript of $h_{\rm{P}}^{-}(p_j^{})$
denotes the graviton helicity $-2\hs$ in 3d spacetime.\
Since in the 3d TMG theory, the spin angular momentum is a pseudoscalar and
each massive graviton has only one physical degree of freedom with helicity either $-2$ or $+2$
\cite{Deser:1981wh},
we will choose the helicity $-2$ in our analysis without losing generality.\
In the above and hereafter, for simplicity we suppress the superscript ``$^-$''
of the polarization tensor $\epP^{\mn -}$
and of each helicity state of the physical graviton $h_{\PP}^-$
unless specified otherwise.\
We note that Eq.\eqref{eq:DC-3hp-Amp} agrees with
the direct calculation of the three-point graviton scattering amplitude
in the TMG theory as given in Eq.\eqref{eq:Amp-3hp-EP}.\
Then, taking the massless limit for the external states 1 and 2,
we reduce the amplitude \eqref{eq:DC-3hp-Amp} to the following form:
\begin{equation}
\label{eq:DC-3hp-Amp-zero12}
{\MM}_{\rm{DC}}[h^{}_{1\PP_0},h^{}_{2\PP_0},h^{}_{3\PP}]
= \frac{\,\kappa~}{~32\hs m^2~}\langle 23 \rangle^{\!2} \langle 31 \rangle^{\!2} ,
\end{equation}
which agrees with Eq.\eqref{eq:2hp0-hp}.\

\vs

Next, we use the extended double-copy method to construct
the three-point dilaton-dilaton-graviton scattering amplitude
from the corresponding scalar-scalar-gluon amplitude
in the TMYM theory (coupled to adjoint scalar field).\
For this, we include a massive scalar field $\Phi^a$
in the adjoint representation of the TMYM gauge group:
\begin{equation}
\label{eq:TMYM-PhiAdj}
\mathcal{L}_{\rm{TMYMS}}^{} = -\frac{1}{2}{\tr}\!\hsm
\left[\mathbf{F}_{\mn}^{2}\!+\!(\mathbf{D}_{\mu}\hsm\PhiB )^{2}\!+\!m_s^2\PhiB^2 \hsm\right]
\!+\hsm \tilde{m}\hs\varepsilon^{\mn\rho}
{\tr}\!\hsm\(\!\!{\mathbf{A}_{\mu}\partial_{\nu}^{}\mathbf{A}_{\rho}\!-\!\frac{\,\ii\hs 2\hs g\,}{3}
\mathbf{A}_{\mu}\mathbf{A}_{\nu}\mathbf{A}_{\rho}}\!\)\!,
\end{equation}
which we denote as the TMYM-Scalar (TMYMS) theory.\
In the above, we have used notations
$\mathbf{F}_{\mn}\!=\!F_{\mu \nu}^{a}T^{a}$,
$\mathbf{A}_{\mu}\!=\!A_{\mu}^{a}T^{a}$,
$\PhiB\!=\!\Phi^{a}T^{a}$, and the covariant derivative is given by
$\,\mathbf{D}_{\mu}^{}\PhiB\hsm =\hsm\partial_{\mu}^{}\PhiB\hsm -\hsm\ii\hs g[\mathbf{A}_{\mu},\PhiB]$.\
For the three-point double-copy analysis in this subsection, 
we will choose the adjoint-scalar mass 
$m_s^{}\hsm\!=\!0\hs$.\  
%
Thus, we derive the following gluon-scalar-scalar vertex:
\begin{equation}
\label{eq:scalar-scalar-gluon}
\ii\hs\mathcal{V}^{\mu}_{A\Phi\Phi} = gf^{abc}(p_{1}^{}\!-\hsm p_{2}^{})^{\mu} \hs.
\end{equation}
With this, we compute the three-point scattering amplitude of the scalar-scalar-gluon ($\Phi^a\Phi^b\hsm\AP^c$) as follows:
\\[-8mm]
\beqs
\label{eq:Amp-2phi-Ap}
\begin{align}
\TT [\Phi_1^{a}\hsm,\Phi_2^{b}\hsm,\hsm A^{c}_{\rm{P}3}]
&\,= g f^{abc}\NN_{\Phi}^{} [1,2,3]\hs,
\\
\NN_{\Phi}^{} [1,2,3]
&\,= \frac{\ii}{\sqrt{2\,}\hs m\,}\langle 3| p_{1} |3 \rangle \hs.
\end{align}
\eeqs
This scattering amplitude corresponds to the left diagram of Fig.\,\ref{fig:3}.\

\vs

\begin{figure}[]
\centering
\includegraphics[width=0.6\textwidth]{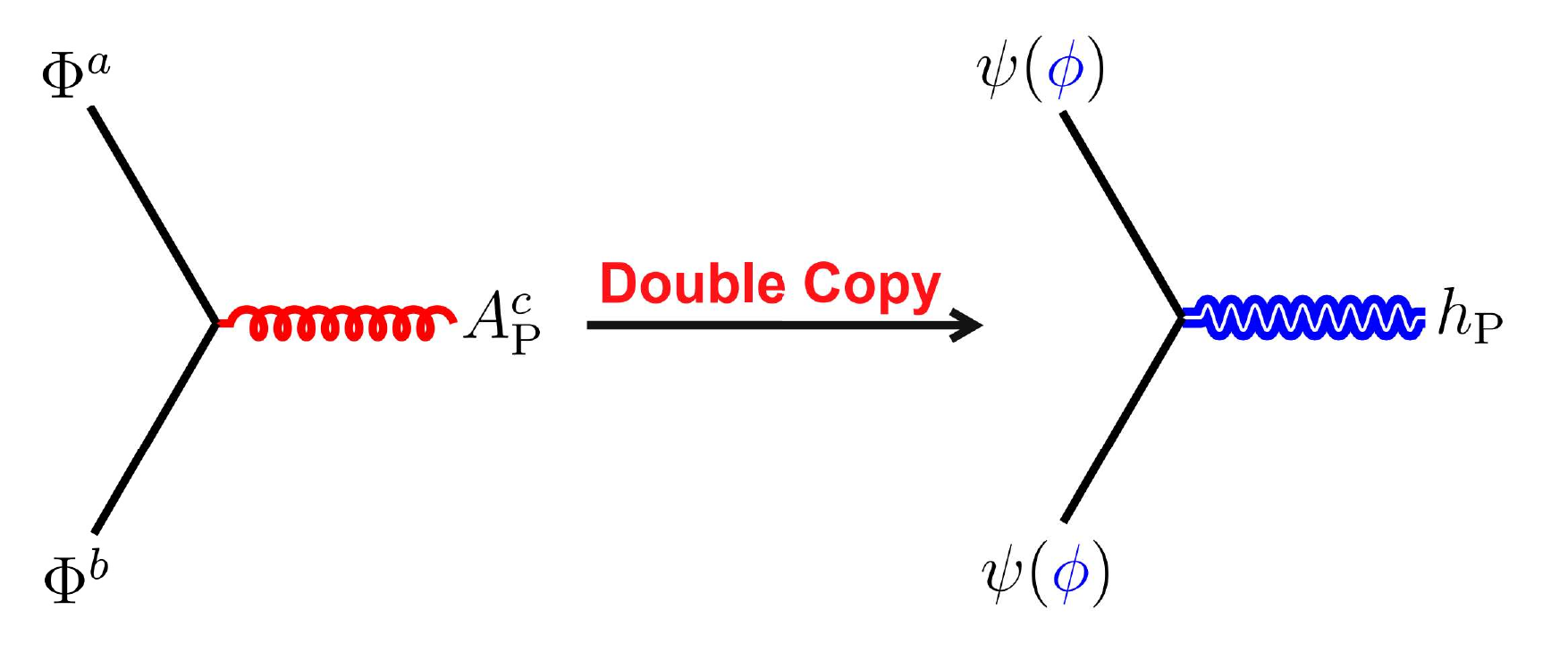}
\vspace*{-3mm}
\caption{\small\hspace*{-1mm}%
Three-point adjoint-scalar scattering $\Phi^a\Phi^b\AP^c$
of Eq.\eqref{eq:Amp-2phi-Ap} in the TMYMS theory (left diagram) 
and three-point physical-scalar scattering
$\psi\hs\psi\hs\hP$ of Eq.\eqref{eq:DC-MT-hpsipsi}
[dilaton scattering $\phi\hs\phi\hs\hP$ of 
Eq.\eqref{eq:DC-MT-hphiphi}] 
in the WTMG theory (right diagram).\
Their scattering amplitudes can be connected through
the extended double-copy construction as discussed in the text.}
\label{fig:4}
\label{fig:6new}
\end{figure}

For the color-kinematics duality,
we have the following correspondence to realize the double copy construction of 
the three-point $\Phi^a\hs\Phi^b\hsm\AP^c$ amplitude
from $\phi\hs\phi\hs\hP$ amplitude:
\beq
\label{eq:3pt-CK-duality}
g\to \frac{\,\kappa\,}{4}\hs,
\hspace*{5mm}
f^{abc} \!\to \NN_{\Phi}^{} [1,2,3]\hs.
\eeq
With these, we construct the double-copy of the three-point 
$(\psi\hs\psi\hs\hP)$ amplitude
from the corresponding gluon-scalar-scalar 
($\Phi^a\Phi^b\!\AP^c$) amplitude:
\begin{align}
& \MM_{\rm{DC}}^{} [\psi_1^{}\hsm,\psi_2^{}\hsm,\hsm h^{}_{\rm{P}3}]
 = \frac{\,\kappa\,}{4} \big(\NN_{\Phi}^{}[1,2,3]\big)^{\!2}
\nn\\
& = \frac{~-\!\kappa~}{\,8\hs m^2\,} \langle 3| p_1 |3 \rangle^{\hsm 2}
= \hsm\frac{\kappa}{~8\hs m^2~} 
{\langle 23 \rangle^{\!2}\langle 31 \rangle^{\!2}}
\hs,
\label{eq:DC-MT-hpsipsi}
\end{align}
where $\psi$ denotes a massless physical scalar coupled to the
WTMG as defined in Eq.\eqref{eq:L-WTMGS} by choosing the scalar
mass $m_s^{}\!=\!0\hs$.\  
This agrees with the direct calculation of the three-point scattering amplitude $\MM[\psi_1^{},\psi_2^{},h^{}_{\rm{P}3}]$ 
for $m_s^{}\!=\!0$ 
as given by Eqs.\eqref{eq:2psi-hp}.\ 
In Eq.\eqref{eq:DC-MT-hphiphi}, we can use Eq.\eqref{eq:3p13} 
to deduce
$\langle 3| p_1^{} |3 \rangle
\!=\!\ii\hs m \langle 23 \rangle\langle 31 \rangle\hsm
/\hsm\langle 12 \rangle$,
and we further have 
$\langle 12 \rangle^2\hsm\!=\!m^2$ 
for particles\,1 and 2 being massless 
according to Eq.\eqref{eq:m1m2=0-<12>}.\
According to our predicted equality \eqref{eq:2phihp=2psihp},
we can thus obtain the corresponding double-copied three-point
on-shell dilation-dilaton-graviton amplitude:
\beq 
\label{eq:DC-MT-hphiphi}
\MM_{\rm{DC}}^{} [\phi_1^{}\hsm,\phi_2^{}\hsm,\hsm h^{}_{\rm{P}3}]
=\MM_{\rm{DC}}^{} [\psi_1^{}\hsm,\psi_2^{}\hsm,\hsm h^{}_{\rm{P}3}]
= \hsm\frac{\kappa}{~8\hs m^2~} 
{\langle 23 \rangle^{\!2}\langle 31 \rangle^{\!2}}
\hs. 
\eeq 
The double-copied scattering amplitude \eqref{eq:DC-MT-hphiphi}
corresponds to the right diagram of Fig.\,\ref{fig:4}.\  
We see that as expected, 
the double-copied three-point graviton amplitude \eqref{eq:DC-3hp-Amp-zero12} and
dilaton amplitude \eqref{eq:DC-MT-hphiphi} just obey the TRGET \eqref{eq:ET-3pt-h2phi} 
at the level of three-point amplitudes.

\vspace*{2mm}
\subsubsection{
\hspace*{-2mm}%
Four-Point Graviton (Dilaton) Amplitudes from Massive Double Copy}
\label{sec:4.2.2}
\vspace*{1.5mm}

\vs
In this subsection, we first briefly review the basic steps of the 
double-copy construction of the four-point graviton amplitude from
the corresponding gauge boson amplitude\,\cite{TMG-DCx}\cite{Hang:2021oso},
which will be used to compare with our double-copy construction 
of the four-point dilaton amplitude.\ 
But different from Refs.\,\cite{TMG-DCx}\cite{Hang:2021oso}, 
we will use the method of Ref.\,\cite{Arkani-Hamed:2017jhn} to 
construct the four-point gauge boson amplitude
from the sum of products of two three-point gauge boson amplitudes
connected by the relevant pole terms and with the spinor-helicity formulation.\ 
Then, we study the double copy of the 
four-point dilaton scattering amplitude 
at the leading order of high energy expansion,  
through the double-copy construction of the four-point physical
scalar amplitude [which equals the four-point dilaton amplitude
at the leading order of high energy expansion
as shown in Eq.\eqref{eq:M0[Nphi]=M0[Npsi]}].\ 
These were not studied in Refs.\,\cite{TMG-DCx}\cite{Hang:2021oso}.\
Moreover, different from Ref.\,\cite{TMG-DCx}, 
our focus is to understand the energy structures
of the double-copied graviton amplitudes and 
dilaton amplitudes (physical scalar amplitudes).\

\vs 

For the 3d TMYM gauge theory, we can write the tree-level
four-point physical gauge boson scattering amplitude
as the sum of dressed $(s,t,u)$-channel contributions
(including the contact diagram contribution):
\begin{equation}
\label{eq:T-Gauge-4pt}
\TT[4\AP] = g^2\!
\(\!\frac{\CC_{s}\hs\NN_{s}}{\,s\!-\!m^2\,} + \frac{\CC_{t}\hs\NN_{t}}{\,t\!-\!m^2\,}
+ \frac{\CC_{u}\hs\NN_{u}}{\,u\!-\!m^2\,}\!\)\!,
\end{equation}
where the color factors are given by
$(\CC_s,\hs \CC_t,\hs \CC_u)\!=\!(f^{abe}f^{cde},f^{ade}f^{bce},f^{ace}f^{dbe})$.\ 
%
%
For this study, we define the $S$-matrix element
$\mathcal{S}\hsm\!=\! 1\hsm -\ii\hs \TT$
as our convention, where $\TT$ represents a general
scattering amplitude under consideration and
its overall sign does not affect physics.\
For the present analysis, we treat all momenta as incoming
(unless specified otherwise)
and define the Mandelstam variables,
%
$s \!=\! -(p_{1}^{}\!+\hsm p_{2}^{})^{2}$,
$t \!=\! -(p_{1}^{}\!+\hsm p_{4}^{})^{2}$, and 
$u \!=\! -(p_{1}^{}\!+\hsm p_{3}^{})^{2}$.\
%
For non-Abelian gauge theories, the color Jacobi identity holds,
%
$\CC_s \!+ \CC_t \!+ \CC_u \!=\! 0\hs$,
%
and the gauge symmetry should further hold 
the kinematic Jacobi identity 
for the massive numerators\,\cite{TMG-DCx}\cite{Hang:2021oso},
%
$\NN_{s}\!+\NN_{t}\!+\NN_{u}\!=\!0\hs$.\
%
The gauge-boson amplitude \eqref{eq:T-Gauge-4pt} 
is found to be invariant under the generalized massive gauge 
transformation\,\cite{TMG-DCx}\cite{Hang:2021oso}, 
%
$\NN_j^{} \to \NN_j' \!=\! 
 \NN_j^{} + \Delta\hs (s_j^{}\!-\!m^2)\hs$,
%
where $j\!\in\!(s,t,u)$ and $\Delta$ is an arbitrary coefficient.\
By summing up both sides of this transformation and
for $m\!\neq\! 0\hs$, the coefficient
$\Delta$ is determined by the difference of numerator sums,
%
$\Delta \!=\!
\sum_j\!\hsm\big(\NN_j' \!-\! \NN_j^{}\big)/m^2$.\ 
%

\vs

For the massless gauge and gravity theories, 
the color-kinematics duality\,\cite{BCJ}-\cite{BCJ-rev}
exchanges the color factor and kinematic factor,  
$\CC_{j}\hsm\ito \NN_{j}$, which allows the
double-copy construction of massless graviton amplitudes 
from massless gauge boson amplitudes 
\`{a} la Bern-Carrasco-Johansson (BCJ)\,\cite{BCJ}-\cite{BCJ-rev}.\
For the 3d amplitudes of massive gauge bosons 
in the TMYM gauge theory and of massive gravitons in the 
TMG theory, the massive kinematic Jacobi identity 
can hold only after making a generalized gauge transformation
as mentioned above.\
Thus, extending the massless BCJ double-copy formulation and 
using the color-kinematics duality $\CC_{j}\ito \NN_{j}$,
we construct the corresponding four-point 
massive graviton amplitude as follows:
\begin{equation}
\label{eq:doublecopy}
\MM [4\hP] = \frac{~\kappa^2\hs}{16}\!
\(\!\!\frac{\NN_{s}^2}{\,s\!-\!m^2\,}+\frac{\NN_{t}^2}{\,t\!-\!m^2\,}
+\frac{\NN_{u}^2}{\,u\!-\!m^2\,}\!\hsm\)\!,
\end{equation}
where the gauge/gravity coupling correspondence 
($g\ito \kappa/4\hs$) is used.\


\vs

\begin{figure}[t]
\centering
\includegraphics[width=0.4\textwidth]{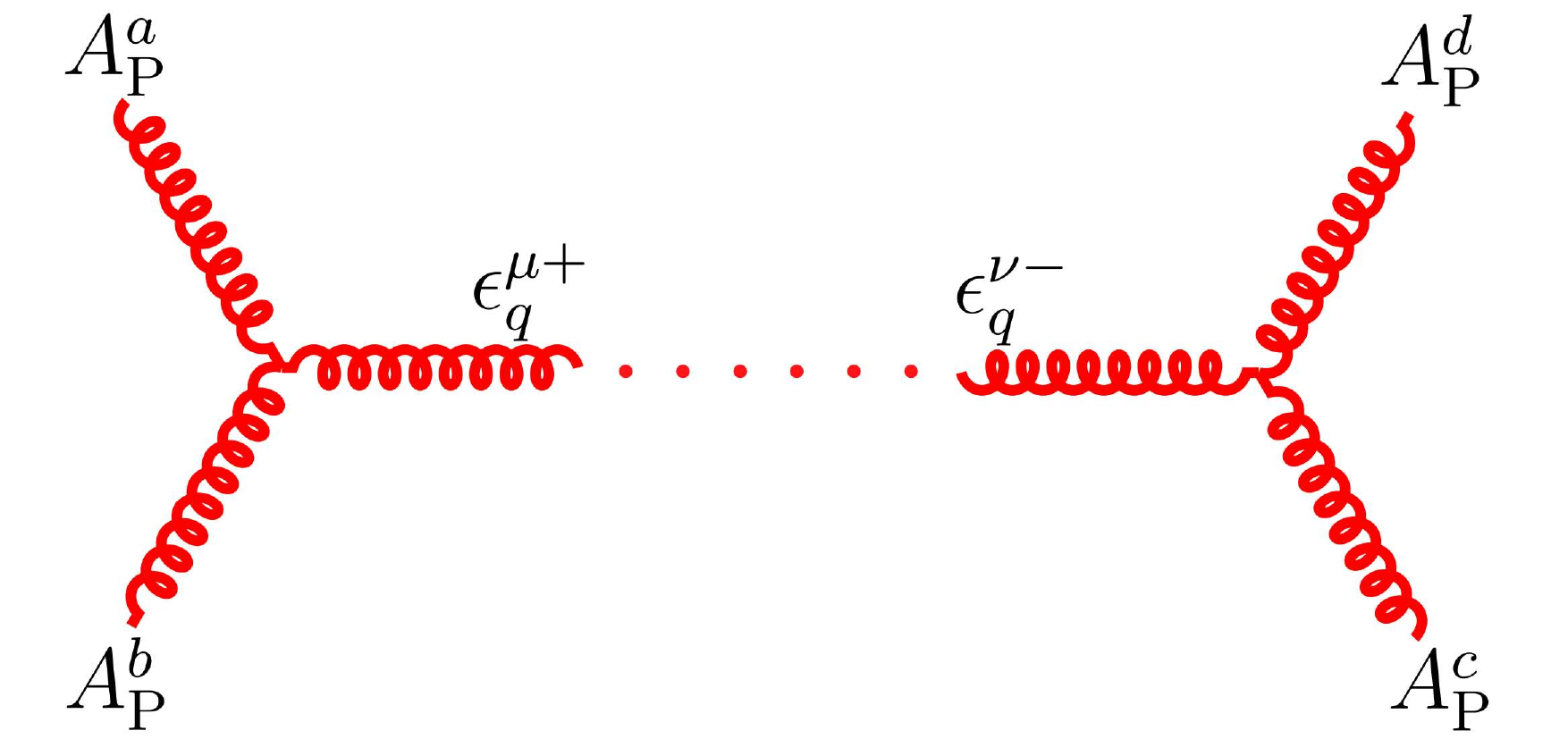}
\vspace*{-4mm}
\caption{\small\hspace*{-2mm}
Four-point gauge boson scattering amplitude in the $s$-channel of the TMYM theory,
which can be constructed from combining two three-point gauge boson amplitudes.}
\label{fig:7}
\label{fig:8new}
\end{figure}

Using the method of Ref.\,\cite{Arkani-Hamed:2017jhn}, 
we construct the four-point gauge boson amplitude
from the sum of products of two three-point gauge boson amplitudes
connected by the relevant pole terms.\
For the $s$-channel of the four-point gauge boson scattering,
we first obtain the two three-point gauge boson amplitudes
from Eq.\eqref{eq:threeTMYM},
{\small
\beq
\label{eq:T3L-T3R}
\begin{split}
\hspace*{-2mm}
\TT_{3,L}^{} [1,2,q_s^{+}]
&= \frac{\,\ii\hs gf^{a_1^{}a_2^{}b}\,}{16\hs m^{3}} \!
\Big(\ii\hs 4m\langle 12 \rangle \langle 2\bar{q} \rangle \langle \bar{q}1 \rangle
\!-\! 4\langle \bar{q}| p_1^{} |\bar{q} \rangle \langle 12 \rangle^{\!2}
\!+\! 4\langle 2| p_{1} |2 \rangle \langle 1\bar{q} \rangle^{\!2}
\!-\! 4\langle 1| p_{2} |1 \rangle \langle \bar{q}2 \rangle^{\!2}\Big),
\\
\hspace*{-2mm}
\TT_{3,R}^{} [q_s^{-}\!,3,4]
&= \frac{\,\ii\hs gf^{ba_{3}a_{4}}\,}{16\hs m^{3}} \!
\Big(\ii\hs 4m\langle q3 \rangle \langle 34 \rangle \langle 4q \rangle
\!-\! 4\langle q| p_3^{} |q \rangle \langle 34 \rangle^{\!2}
\!+\! 4\langle 4| p_{3} |4 \rangle \langle q3 \rangle^2
\!-\! 4\langle 3| p_4^{} |3 \rangle \langle 4q \rangle^{\!2} \Big),
\end{split}
\eeq
}
\hspace*{-2.5mm}
where $q_s^{}\hsm =\hsm p_{1}^{}\!+\!p_{2}^{}\!=\!-(p_{3}^{}\!+\!p_{4}^{})$.\
These two three-point amplitudes are illustrated in Fig.\,\ref{fig:7}.\
Thus, we can construct the four-point scattering amplitude 
by combining the above three-point amplitudes 
for the (s,\hs t,\hs u) channels respectively, 
\\[-8mm]
\beqs
\label{eq:T4-stu}
\begin{align}
\label{eq:T4-s}
\TT_{4}^{(s)} &= \frac{~\ii\hs\TT_{3,L}^{}[1,2,q_s^{+}]\hs\TT_{3,R}^{}[q_s^{-}\!,3,4]~} {~s - m^2~}\,,
\\
\label{eq:T4-t}
\TT_{4}^{(t)} &= \frac{~\ii\hs\TT_{3,L}^{}[1,4,q_t^{+}]\hs\TT_{3,R}^{}[q_t^{-}\!,2,3]~} {~t - m^2~},
\\
\label{eq:T4-u}
\TT_{4}^{(u)} &= \frac{~\ii\hs\TT_{3,L}^{}[1,3,q_u^{+}]\hs\TT_{3,R}^{}[q_u^{-}\!,2,4]~} {~u - m^2~},
\end{align}
\eeqs
where $q_t^{}\hsm =\hsm (p_1^{}\!+\!p_4^{})\!=\!-(p_3^{}\!+\!p_{2}^{})$
and $q_u^{}\hsm =\hsm (p_1^{}\!+\!p_3^{})\!=\!-(p_2^{}\!+\!p_4^{})$.\

\vs

Then, we derive the contact-term contribution 
to the four-point gauge boson amplitude:
\\[-7mm]
\begin{align}
\label{eq:Amp-Contact4}
\TT_{4,C}^{} =\,&
\frac{- g^2}{\,16\hs m^4\,}\bigg[\hsm
f^{abe}f^{cde}
\Big( \!\langle 13 \rangle^{\!2} \langle 24 \rangle^{\!2} \!-\! \langle 14 \rangle^{\!2} \langle 23 \rangle^{\!2} \Big)
\!+\!f^{ace}f^{bde}\hsm\Big(\!\langle 12 \rangle^{\!2} \langle 34 \rangle^{\!2}\!-\!
\langle 14\rangle^{\!2} \langle 23\rangle^{\!2} \Big)
\nn\\
& \hspace*{12mm}
+f^{ade}f^{bce} \Big(\!\langle 12 \rangle^2 \langle 34 \rangle^2  \!-\! \langle 13 \rangle^2 \langle 24 \rangle^2 \Big)
\!\bigg] .
\end{align}
Including the contact contribution \eqref{eq:Amp-Contact4} and combining \eqref{eq:propagator}  and \eqref{eq:threeTMYM},
we construct the four-point amplitude from the product of two three-point gauge-boson amplitudes
through Eq.\eqref{eq:T4-stu}.\
For instance, considering the $s$-channel contribution (shown in Fig.\,\ref{fig:8new})
with the contact diagram contribution included,
we express the kinematic numerator $\NN_s^{}$
of the four-point gauge-boson amplitude as follows:
{\small
\begin{align}
\hspace*{-5mm}
\NN_{s} =\,& \frac{\ii}{\,16\hs m^2\,}\!\hsm\left\{\!
\langle 12 \rangle^{\!2} \langle 34 \rangle^{\!2}
\Big[ (\pB_1^{}\!-\!\pB_2^{})(\pB_3^{}\!-\!\pB_4^{}) -
\frac{\,\ii\hs\varepsilon^{\mu\nu\rho}\,}{\,\qB^2\hs}
(\pB_1^{}\!-\!\pB_2^{})_{\mu}^{}
(\pB_3^{}\!-\!\pB_4^{})_{\nu}^{}\qB_{\rho}^{}\Big]\right.
\notag
\\
& -\!4\langle 12 \rangle^2 \langle 3| \pB_4^{} |3 \rangle
\Big[ \langle 4| \pB_1^{}\!-\!\pB_2^{} |4 \rangle \!-\!
\frac{\,\ii\hs \varepsilon^{\mu\nu\rho}\,}{\qB^2}
\qB_{\mu}^{} (\pB_1^{}\!-\!\pB_2^{})_{\nu}^{} \langle 4| \gamma_{\rho} |4 \rangle \Big] \notag
\\
&
+ \!4\langle 1| \pB_2^{} |1 \rangle \langle 4| \pB_3^{} |4 \rangle
\Big[\langle 23 \rangle^2 \!+\! \frac{\,\ii\,}{\,\qB^2\hs}
\langle 2| q |3 \rangle \langle 23 \rangle \Big]
- \!4 \langle 1| \pB_2^{} |1 \rangle \langle 3| \pB_4^{} |3 \rangle
\hsm\Big[ \langle 24 \rangle^2 \!+\! \frac{\,\ii\,}{\,\qB^2\hs}
\langle 2| \qB |4 \rangle \langle 24 \rangle \Big]
\notag
\\
&- \!2\langle 12 \rangle^2 \langle 34 \rangle\!
\Big[ \ii\langle 3| \pB_1^{}\!-\!\pB_2^{} |4 \rangle
\!+\! \frac{\,\varepsilon^{\mu\nu\rho}\hs}{\,\qB^2\hs}
\qB_{\mu}^{} (\pB_1^{}\!-\!\pB_2^{})_{\nu}^{} \langle 3| \gamma_{\rho} |4 \rangle \Big] 
\!+\!\big( \qB^2 \!+\! 1\big) \!\hsm
\(\! \langle 13 \rangle^2 \langle 24 \rangle^2 \!-\! \langle 14 \rangle^2 \langle 23 \rangle^2 \)
\notag
\\
&- \!{4}\langle 1| \pB_2^{} |1 \rangle \langle 34 \rangle
\Big[ \ii\hs 2\langle 23 \rangle \langle 24 \rangle \!-\!\frac{1}{\,m\hs\qB^2\,}\!\hsm
\(\!\langle 23 \rangle^2 \langle \bar{3}4 \rangle \!+\! \langle 24 \rangle^2 \langle \bar{4}3 \rangle\!\)\!\hsm \Big]
\notag
\\
& \left. 
+\!\langle 12 \rangle \langle 34 \rangle\!
\Big[ \langle 13 \rangle \langle 24 \rangle \!+\! \langle 23 \rangle \langle 14 \rangle
\!-\!\frac{\ii}{\,m\hs \qB^2\hs} \!\hsm\left(\langle 23 \rangle \langle 1\bar{2} \rangle \langle 24 \rangle
\!+\! \langle 14 \rangle \langle 2\bar{1} \rangle \langle 13 \rangle\right)\!\hsm \Big] \!\right\}\!,
%
\label{eq:kinematicfactor}
\end{align}
}
\hspace*{-3mm}
where $q\!=\!q_s^{}\!=\!(p_1^{}\!\!+\!p_2^{})\!=\!-(p_3^{}\hsm\!+\!p_4^{})$,
and we have defined $\bar{p}_j^{}\!\!=\!p_j^{}/m$ and $\bar{q}\!=\!q/m\hs$.\
The above formula of the numerator $\NN_s$ agrees with the result
that we obtained by direct calculation of the 4-point Feynman diagrams.\  

\vs

Thus, using Eq.\eqref{eq:p1-4-CMF}, we further derive the 
numerator $\NN_s$ from Eq.\eqref{eq:kinematicfactor}
and the numerators $\NN_t$ and $\NN_u$ in a similar way,
{\small 
\beqs
\label{eq:N-stu}
\begin{align}
\label{eq:N-s}
\hspace*{-2mm}
\NN_{s} &=  -\frac{\,\ii\hs (s\!-\!4m^2)\,}{~16\hs m^3s^{\frac{1}{2}}\,}\hsm\! \left[
4ms^{\frac{1}{2}}\hsm\big(4\hs s\!+\!5m^2\big)c_{\theta}^{} \!+\!
\ii\big(3\hs s^2 \hsm\!+\!29\hs m^2\hsm s\!+\!4m^4\big)s_{\theta}^{}\right]\!,
\\
\label{eq:N-t}
\hspace*{-2mm}
\NN_t^{} &= \frac{\ii\hs\cht}{\,16\hs m^3\,}\!
\(\! s^{\frac{1}{2}} \!+\! \ii 2\hs m \tan\!\fr{\theta}{2} \)^{\!\!2}\!
\!\LB 4 m  [(s\!-\!8m^2)\ct\!+\! (3\hs s\!-\! 13 m^2) ]\cht
\hsm +\hsm \ii\hs s^{\frac{1}{2}}[(3\hs s\!-\!20\hs m^2 )\ct \!+\! (3\hs s \!-\!22\hs m^2)] \sht \RB \!, \hspace*{7mm}
\\[-3mm]
\label{eq:N-u}
\hspace*{-2mm}
\NN_u^{} &= \frac{\ii\hs\sht}{\,16\hs m^3\,}\!
\(\!s^{\frac{1}{2}}\!-\!\ii\hs 2\hs m\cot\!\fr{\theta}{2} \hsm\)^{\!\!2}\!\!
\LB 4 m  [ (s \!-\! 8\hs m^2 )\ct  \!-\! (3\hs s \!-\!13 m^2) ]\sht
\!+\hsm \ii \hs s^{\hf}  [ (3\hs s \!-\!20\hs m^2 )\ct\!-\! (3\hs s \!-\!22\hs m^2)] \cht \RB \!,   \hspace*{7mm}
\\[-10mm]
~&~ \nn 
\end{align}
\eeqs
}
\hspace*{-3.5mm}
where 
$(s_{\theta}^{},\hs c_{\theta}^{})\!=\!(\sin\hsm\theta,\hs \cos\hsm\theta)$ and
$(\sht,\hs \cht)\hsm\!=\!\!\(\sin\hsm\frac{\hs\theta\hs}{2},\hs \cos\hsm\frac{\hs\theta\hs}{2}\hsm\)$.\
The $t$ and $u$ channels have a crossing symmetry which imposes   
a kinematic relation, 
$\NN_t^{}(\theta)\!=\!-\NN_u^{}(\pi \hsm +\hsm\theta)$.\
In each numerator above, we have dropped an irrelevant 
overall phase factor,
\beq
\label{eq:phase-factor}
\frac{\,E\cos\hsm\theta \!-\! \ii\hs m\sin\hsm\theta\,}{\,E\cos\hsm\theta \!+\! \ii m\sin\hsm\theta\,}
= e^{-\ii\hs 2\omega},
\eeq
with $\dis\omega \hsm =\hsm \arctan\!\Big(\frac{m}{E\,}\hsm\tan\hsm\theta\hsm\Big)$.\
Different from Refs.\,\cite{TMG-DCx}\cite{Hang:2021oso}
which directly computed the four-point Feynman diagrams, 
we construct the four-point gauge-boson amplitude 
from the products of two three-point amplitudes 
via Eqs.\eqref{eq:T4-s}-\eqref{eq:T4-u} in each channel
together with Eq.\eqref{eq:Amp-Contact4},
where the three-point amplitudes are given by Eq.\eqref{eq:T3L-T3R} 
using the 3d spinor-helicity formalism.\ 
We find that above results of Eq.\eqref{eq:N-stu} coincide with 
Ref.\,\cite{Hang:2021oso} except the overall phase factor
\eqref{eq:phase-factor} which does not affect physics. 
From Eqs.\eqref{eq:N-s}-\eqref{eq:N-u}, 
the sum of the three numerators is given by
{\small 
\begin{equation}
\begin{split}
\Omega \equiv 
\NN_s^{}+\NN_t^{}+\NN_u^{} 
= \frac{\,\big(3s^2 \!+\!24\hs m^2s \!-\!16\hs m^4\big)c_{2\theta}^{} 
\!-\!\big(3s^2\!-\!8\hs m^2s \!-\!16\hs m^4 \big)
\!+\hsm\ii\hs 16\hs m s^{\frac{3}{2}} s_{2\theta}^{}\,}
{32\hs m\hs s^{\frac{1}{2}}\sin\hsm\theta}\hs,
\end{split}
\end{equation}
}
\hspace*{-3.4mm}
where 
$(s_{2\theta}^{},\hs c_{2\theta}^{})
\!=\!(\sin\hsm 2\theta,\hs \cos\hsm 2\theta)$.\
We find that the sum of numerators does not obey 
the kinematic Jacobi identity.\
Then, making the following 
generalized gauge transformation\,\cite{TMG-DCx}\cite{Hang:2021oso},
${\NN}_j' = \NN_j^{} \!-\! (\Omega/m^2)(s_j^{}\!-\hsm m^2)\hs$,
%
with $j\!=\!s,t,u\hs$, we can verify the kinematic Jacobi identity 
for the sum of the new numerators,
%
${\NN}_s'\hsm+\hsm {\NN}_t' \hsm+\hsm {\NN}_u'
= \sum_j\!{\NN}_j^{} \!-\hsm \Omega \!=\! 0 \hs$.\
%
Under the generalized gauge transformation, 
the four-point gauge-boson amplitude \eqref{eq:T-Gauge-4pt} is invariant and its structure remains unchanged:
\begin{equation}
\label{eq:T4-Nhat-Gauge}
\TT[4\AP]  = g^2\!
\(\!\frac{\CC_{s}\hs\NN_{s}'}{~s\!-\!m^2~} + \frac{\CC_{t}\hs\NN_t'}{~t\!-\!m^2~}
+ \frac{\CC_{u}\hs\NN_u'}{~u\!-\!m^2~}\!\)\!.\ 
\end{equation}
\begin{figure}[t]
\centering
\includegraphics[width=0.7\textwidth]{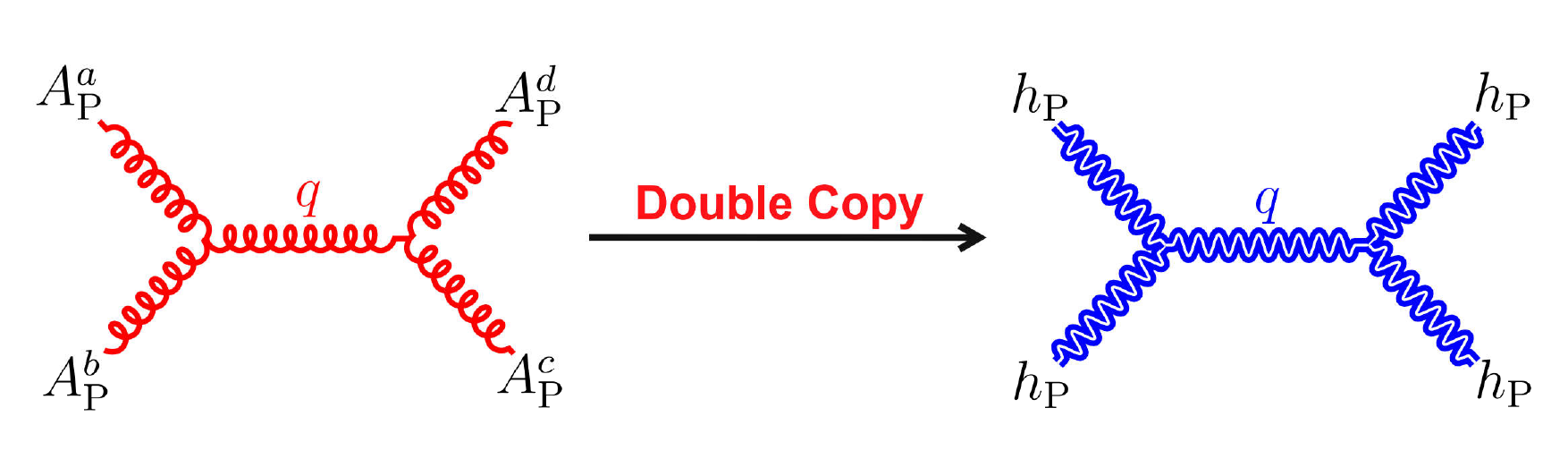}
\vspace*{-8mm}
\caption{\small\hspace*{-1mm}
Four-point gauge boson scattering in the TMYM theory (left) and the four-point graviton scattering in the TMG theory (right),
where each diagram with massive gauge boson (graviton) exchange has absorbed the contribution from the $4\AP$ ($4\hP$)
contact diagram and the internal momentum $q\!=\!p_i^{}\!+\!p_j^{}$ denotes the kinematic channels of $(s,t,u)$.\
As indicated by the arrow, the 3d four-point graviton scattering amplitude can be constructed from the corresponding
gauge boson scattering amplitudes via massive double copy.}
\label{fig:8}
\label{fig:9new}
\end{figure}

\vs

Based upon the massive color-kinematics duality, we have the following correspondence to realize the
extended massive double copy:
\beq
\label{eq:4pt-CK-duality}
g\to \frac{\,\kappa\,}{4}\hs,
\hspace*{5mm}
\CC_j^{} \to \NN_j' \hs.
\eeq
Following Ref.\,\cite{Hang:2021oso} and using Eq.\eqref{eq:T4-Nhat-Gauge} together with the 
correspondence \eqref{eq:4pt-CK-duality},
we construct the four-point graviton amplitude 
through the extended massive double copy,
\begin{equation}
\label{eq:M4-DC}
\MM_{\rm{DC}}^{}[4\hP ] = \frac{\,\kappa^2\,}{16}\!
\(\!\hsm\frac{{\NN_s'}^2}{~s\!-\!m^2~} + \frac{{\NN_t'}^2}{~t\!-\!m^2~} + \frac{{\NN_u'}^2}{~u\!-\!m^2~} \hsm\!\)\!.
\end{equation}
This double-copy operation is illustrated in Fig.\,\ref{fig:8}.\
Expanding the numerators of Eq.\eqref{eq:M4-DC}, 
we have verified that the double-copied 
four graviton amplitude \eqref{eq:M4-DC} 
equals the direct calculation result 
of Eqs.\eqref{eq:app-Amp-4hp}-\eqref{eq:app-Amp-4hp-Y}
which we derived in the Landau gauge of the WTMG theory.\
It also agrees with the literature\,\cite{Hang:2021oso}\cite{TMG-DCx}.\ 
Unlike Refs.\,\cite{TMG-DCx}\cite{Hang:2021oso}, 
we have used the method of 
Ref.\,\cite{Arkani-Hamed:2017jhn} to construct the four-point
massive gauge-boson amplitude and to obtain the 
double-copy of the four-graviton amplitude.\  
This is mainly for the comparison with our following 
new double-copy construction of the dilaton amplitude.\

\vs

Then, we make high energy expansion for Eq.\eqref{eq:M4-DC} and
derive the following leading-order (LO) contribution
of the four-point graviton scattering amplitude:
\begin{equation}
\label{eq:M0-4hp-DC-LO}
\mathcal{M}_{0}^{\text{DC}}[4\hP] =
-\frac{~\ii\hs\kappa^{2}m~}{\,{2048}\,} \csc^3\!\theta
\big(494\hs c_{\theta}^{}\hsm +\hsm 19\hs c_{3\theta}^{}\!-\hsm c_{5\theta}\big) s^{\frac{1}{2}}  \,.
\end{equation}
The above double-copied four-graviton amplitude
has the LO contribution of 
$O(m\hs s^{1/2})$  and equals the LO four-graviton
amplitude as given in Eqs.\eqref{appeq:Amp-4phi-LO-p=m} and Eq.\eqref{eq:M[4psi]L-expand}.\ 
It also agrees with the LO double-copy formula 
of Ref.\,\cite{Hang:2021oso} and 
will be used to compare with our following double-copy 
construction of the four-point dilaton amplitude.

\vs   

Next, we construct the four-point dilaton ($\phi$)
scattering amplitude
from the corresponding massive colored adjoint scalar 
($\Phi^a$) scattering amplitude via double copy.\
For this, we consider the TMYM theory coupled 
to massive adjoint scalar fields 
[which is called the TMYM-scalar (TMYMS) theory] 
and its Lagrangian is given by Eq.\eqref{eq:TMYM-PhiAdj}.\
We choose $m_s^{}\hsm\!=\!m\hs$ for the double-copy analysis
and will explain later the reason of this choice around 
Eq.\eqref{eq:DC-4Phi|ms-m}.\ 
It is expected that the double-copy of the adjoint-scalar amplitudes
in the TMYMS theory of Eq.\eqref{eq:TMYM-PhiAdj} should give the corresponding amplitudes of physical scalars ($\psi$) 
in the WTMGS theory of Eq.\eqref{eq:L-WTMGS}.\ 
As shown in Eq.\eqref{eq:M0[Nphi]=M0[Npsi]}, we have proved that 
each leading-order dilaton amplitude equals 
the corresponding leading-order physical scalar amplitude,
which are all contributed by the graviton-exchanges 
(and jointly by possible graviton self-interactions).\
Hence, the second equality of Eq.\eqref{eq:M0[Nphi]=M0[Npsi]} 
allows us to construct each leading-order dilaton amplitude from the 
leading-order of the double-copied physical scalar amplitude 
in the WTMGS theory of Eq.\eqref{eq:L-WTMGS},  
which will be constructed by the double copy of 
the corresponding adjoint-scalar amplitude in the TMYMS theory
of Eq.\eqref{eq:TMYM-PhiAdj}.\ 
In addition, we also note that the TGRET connects the leading-order 
dilation amplitudes to the corresponding physical graviton amplitudes 
at the leading order.\ 
Hence, the leading-order dilation amplitudes are actually gauge-invariant physical amplitudes,  
which are generated by the graviton-dilaton interactions.\

\begin{figure}[t]
\centering
\includegraphics[width=0.7\textwidth]{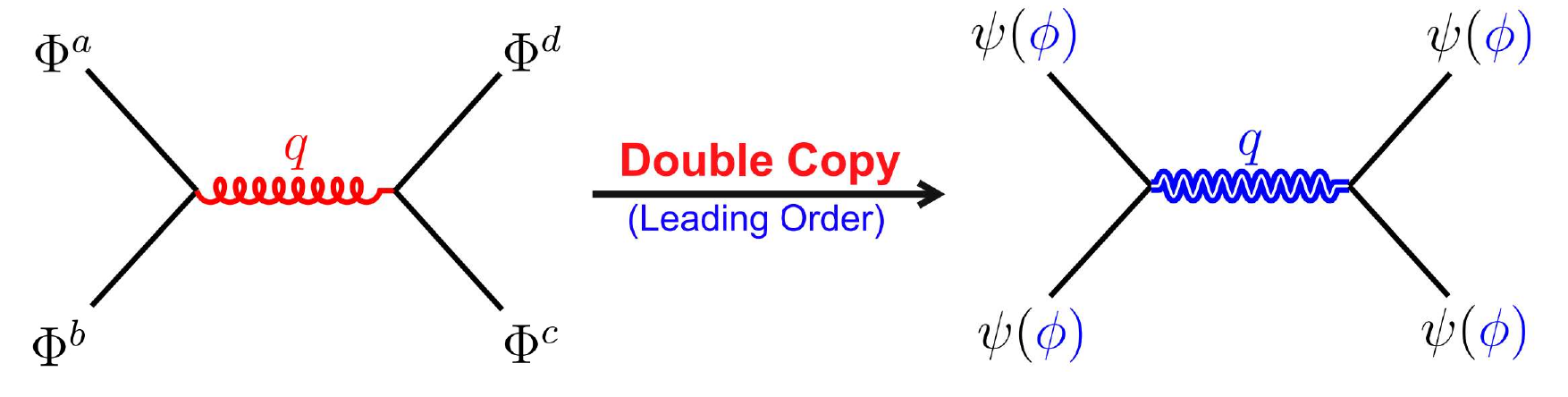}
\vspace*{-5mm}
\caption{\small\hspace*{-2.6mm}
Four-point scattering of the colored adjoint scalar bosons 
(left diagram) and the corresponding 
physical scalar (dilaton) scattering 
in the WTMG theory (right diagram).\
As indicated by the arrow, the double-copy construction gives 
the physical scalar scattering amplitude 
[shown in Eqs.\eqref{eq:Amp-4phi-DC3}-\eqref{eq:Amp-4phi-DC2-RD-Yt}] 
and the leading-order dilaton scattering amplitude
[shown in Eq.\eqref{eq:Amp-4phi-DC-LO}].
}
\label{fig:9}
\label{fig:10new}
\end{figure}

In passing, we note that Ref.\,\cite{2112.08401} discussed double copy
of the four scalar amplitude in the TMGS theory for a different 
purpose and perspective
by taking the eikonal limit and connecting the eikonal 
amplitudes to the classical shockwave solutions
where $s\!\gg\!m^2$ and $s\!\gg\!|t|$ with $t$ fixed 
for small angle scattering.\ 
But, our study on the physical scalar amplitudes in the WTMGS theory
is very different because our goal is to analyze the energy structure
of the physical scalar amplitudes under high energy expansion 
(for general scattering angle) 
and its connection to the dilaton amplitudes via 
Eq.\eqref{eq:M0[Nphi]=M0[Npsi]} at the leading order 
of high energy expansion.\ Our main goal here is to construct
the double copy for the corresponding leading-order dilaton amplitudes.

\vs 

The four-point massive adjoint-scalar scattering contains gluon-exchange contributions via
$(s,t,u)$ channels as illustrated by the left diagram of Fig.\,\ref{fig:10new}.\    
Thus, we can compute the four-point massive adjoint-scalar scattering amplitude as follows:
\begin{equation}
\label{eq:T-4Phi}
\TT[4\Phi] = g^2\!
\(\!\frac{\CC_{s}\hs\NN_s^{\Phi}}{~s\!-\!m^2~} + \frac{\CC_{t}\hs\NN_t^{\Phi}}{~t\!-\!m^2~}
+ \frac{\CC_{u}\hs\NN_u^{\Phi}}{~u\!-\!m^2~}\!\)\!,
\end{equation}
where the kinematic numerators are given by
\beqs
\label{eq:4Phi-Nstu0}
\begin{align}
\NN_{s}^{\Phi} &= {\ii\hs}(p_1^{}\!-\!p_{2}^{})\!\cdot\!(p_{3}^{}\!-\!p_{4}^{})
\hsm +\! \frac{\,m\,}{q^2} \varepsilon^{\mu \nu \rho}(p_{1}^{}\!-\!p_{2}^{})_{\mu}^{}
(p_{3}^{}\!-\!p_{4}^{})_{\nu}^{}q_{\rho}^{}\,,
\\
\NN_{t}^{\Phi} &= {\ii\hs}(p_{1}^{}\!-\!p_{4}^{})\!\cdot\!(p_{2}^{}\!-\!p_{3}^{})
\hsm +\! \frac{\,m\,}{q^2} \varepsilon^{\mu \nu \rho}(p_{1}^{}\!-\!p_{4}^{})_{\mu}^{}
(p_{2}^{}\!-\!p_{3}^{})_{\nu}^{}q_{\rho}^{}\,,
\\
\NN_{u}^{\Phi} &= \ii\hs (p_{1}^{}\!-\!p_{3}^{})\!\cdot\!(p_{4}^{}\!-\!p_{2}^{})
\hsm +\! \frac{\,m\,}{q^2} \varepsilon^{\mu \nu \rho}(p_{1}^{}\!-\!p_{3}^{})_{\mu}^{}
(p_{4}^{}\!-\!p_{2}^{})_{\nu}^{}q_{\rho}^{}\,.
\end{align}
\eeqs
These numerators can be further evaluated in the center-of-mass frame
by using the momentum formulas of Eq.\eqref{eq:p1-4-CMF},
\beqs
\label{eq:4Phi-Nstu0-2}
\begin{align}
\NN_{s}^{\Phi} &= -{\ii\hs}m^{2}\!\hsm\left[\hsm
(\bar{s}\!-\!4)\ct \hsm +\hsm
\ii\hsm\(\!\!\sqrt{\bar{s}\,}\!-\!\fr{4}{\sqrt{\bar{s}\,}\,}\hsm\!\)
\!\hsm\st\right]\!,
\\
\NN_{t}^{\Phi} &=\fr{\,{\ii\hs}m^2\,}{2}\!\hsm\left[-(3\hs\bar{s}\!-\!4)
\hsm +\hsm (\bar{s}\!-\!4)\ct
\hsm +\hsm\ii\hs 4\sqrt{\bar{s}\,}\hsm\tan\hsm{\fr{\theta}{2}}\hs\right]\!,
\\
\NN_{u}^{\Phi} &=\fr{\,{\ii\hs}m^2\,}{2}\!\hsm
\left[(3\hs\bar{s}\!-\!4)\hsm +\hsm (\bar{s}\!-\!4)\ct \hsm +\hsm
\ii\hs 4\sqrt{\bar{s}\,}\cot\hsm{\fr{\theta}{2}}\hs\right]\!,
\end{align}
\eeqs
where $\sB\!=\!s/m^2$.\
Thus, we derive the following sum of the three kinematic numerators:
\begin{equation}
\label{eq:Nj-Phi-Sum}
\Omega^{}_{\Phi} \equiv
\NN_{s}^{\Phi}\!+\NN_{t}^{\Phi}\!+\NN_{u}^{\Phi}
= - \frac{\,m^2\hs}{\sqrt{\sB\,}~}
\big[4\hs\bar{s}\sec\hsm\theta \!-\!(\bar{s}\!-\!4)\st \big],
\end{equation}
which is non-vanishing and does not obey the kinematic Jacobi identity.\

\vs

Then, we make the following generalized gauge transformation:
\begin{equation}
\label{eq:GGT-Phi}
\NN_j^{\Phi} ~\longrightarrow~ 
\hat{\NN}_j^{\Phi}\! = \NN_j^{\Phi} \!-\!
\frac{\,\Omega^{}_{\Phi}\,}{\,m^2\hs}(s_j^{}\!-\hsm m^2)\hs,
\end{equation}
with index $j\!=\!s,t,u\hs$.\
Thus, we can readily verify that the new sum of numerators vanishes:
\begin{equation}
\label{eq:KJacobi-N'(Phi)-stu}
\hat{\NN}_s^{\Phi}\hsm+\hsm \hat{\NN}_t^{\Phi} \hsm+\hsm \hat{\NN}_u^{\Phi}
= 0 \,,
\end{equation}
which is a scalar-type kinematic Jacobi identity.\
From Eqs.\eqref{eq:GGT-Phi} and \eqref{eq:Nj-Phi-Sum},
we explicitly compute the gauge-transformed numerators $\{\hat{\NN}_j^{\Phi}\}$ as follows:
\beqs
\label{eq:4Phi-hatNstu0}
\begin{align}
\hat{\NN}_s^{\Phi} &=
\frac{~{\ii\hs}g^{2}m^{2}\,}{\,\sqrt{\bar{s}\,}\,}\!
\Big[\sqrt{\sB\,}(4\!-\!\sB)\ct \!+\hsm\ii (\sB\!-\!4)(\sB\!-\!2)\st\! -\hsm\ii\hs 4\hs\sB (\sB\!-\!1)\hsm\csc\hsm\theta \Big] ,
\\
\hat{\NN}_t^{\Phi} &=
\frac{~{\ii\hs}g^2m^2\,}{\,4\sqrt{\sB\,}\,}\!\!
\left\{\hsm 2\sqrt{\sB\,}\big[(\sB\!-\!4)\ct \!-\!(3\sB\!-\!4)\big]
\hsm\!+\hsm \ii\hs 8\hs\sB\bigg[2\hsm\tan\hsm{\frac{\theta}{2}}
\!+\!(\sB\!-\!3)\hsm\cot\hsm{\frac{\theta}{2}}\bigg]
\right.
\nn\\
& \hspace*{17mm}
-\ii (\sB\!-\!4)\!\left[(\bar{s}\!-\!4)\stt \!+\! 2(\sB\!-\!2)\st \right]\hsm\!\Big\} \hs,
\\
\hat{\NN}_u^{\Phi} &=
\frac{~{\ii\hs}g^2m^2\,}{\,4\sqrt{\sB\,}\,}\!\!
\left\{\hsm 2\sqrt{\sB\,}\big[(\sB\!-\!4)\ct \!+\!(3\sB\!-\!4)\big]
\hsm\!+\hsm \ii\hs 8\hs\sB\bigg[2\hsm\cot\hsm{\frac{\theta}{2}} \!+\!(\sB\!-\!3) \hsm\tan\hsm{\frac{\theta}{2}}\bigg]
\right.
\nn\\
& \hspace*{17mm}
+\ii (\sB\!-\!4)\!\left[(\bar{s}\!-\!4)\stt \!-\! 2(\sB\!-\!2)\st \right]\hsm\!\Big\} \hs.
\end{align}
\eeqs

Under the generalized gauge transformation \eqref{eq:GGT-Phi},
the four-point adjoint scalar scattering amplitude \eqref{eq:T-4Phi} is invariant:
\begin{equation}
\label{eq:T-4Phi-GGT}
\TT[4\Phi] = g^2\!
\(\!\frac{\CC_{s}\hs\hat\NN_s^{\Phi}}{~s\!-\!m^2~} + \frac{\CC_{t}\hs\hat\NN_t^{\Phi}}{~t\!-\!m^2~}
	+ \frac{\CC_{u}\hs\hat\NN_u^{\Phi}}{~u\!-\!m^2~}\!\)\!,
\end{equation}
where the kinematic numerators $\{\hat\NN_j^{\Phi}\}$
are given by Eq.\eqref{eq:GGT-Phi} and Eqs.\eqref{eq:4Phi-Nstu0-2}-\eqref{eq:Nj-Phi-Sum}.\
With these, we construct the four-point physical scalar scattering amplitude from the corresponding
adjoint scalar amplitude via the extended massive double copy:
\begin{equation}
\label{eq:Amp-4phi-DC}
\MM_{\rm{DC}}^{}[4\psi]= \(\!\frac{\,\kappa\,}{4\,}\!\)^{\!\!2} \!\hsm
\left[\hsm \frac{(\hat{\NN}_{s}^{\Phi})^2}{\,s\!-\!m^2\,}
\hsm+ \frac{(\hat{\NN}_{t}^{\Phi})^2}{\,t\!-\!m^2\,}
\!+\! \frac{(\hat{\NN}_{u}^{\Phi})^2}{\,u\!-\!m^2\,} \hsm\right]\!.
\end{equation}
Using the above double-copy formula \eqref{eq:Amp-4phi-DC},
we explicitly compute the four-point physical scalar 
scattering amplitude as follows:
\begin{align}
\label{eq:Amp-4phi-DC2}
\hspace*{-8mm}
\MM_{\rm{DC}}^{}[4\psi] =& -\frac{\,{\ii\hs}\kappa^2m^2\,}{8}
\!\left\{\!\left[ 8\sqrt{\sB\,}(\sB\!-\!2)\!\cot\hsm\theta
\!-\!\frac{\,(\sB\!-\!4)^2\stt\,}{\,2\sqrt{\sB\,}(\sB\!-\!1)\,}
\!-\!\frac{\,4\sqrt{\sB\,}(2\sB\!-\!3)(\sB\!-\!4)^2\stt\,}
          {(\sB^2\!-\!8)\!-\!(\sB\!-\!4)^2\ctt\,}
\right]\right.
\nn\\
&\left.
+\ii\!\left[\! 16\hs\sB\csc^2\!\theta
-\frac{\,4(8\sB^3\!-\!40\sB^2\!+\!57\sB\!-\!18)\,}{\,(\sB^2\!-\!8)\!-\!(\sB\!-\!4)^2\ctt\,}
+\frac{\,(\sB\!-\!4)^2\ctt\,}{\,2\sB (\sB\!-\!1)\,} \!+\!7 \!+\!\frac{8}{\,\sB\,}
\right]\! \right\}\!.
\end{align}
We can further express this amplitude in the following compact form:
\begin{align}
\label{eq:Amp-4phi-DC3}
\MM_{\rm{DC}}^{}[4\psi]=&
\frac{~\kappa^{2}m^{2}(\YT_{\hsm 0}\!+\!\YT_{\hsm 2}c_{2\theta}^{}\!+\!\YT_{\hsm 4}c_{4\theta}^{}\!+\!\YT_{\hsm 6}c_{6\theta}^{}
\!+\!\YT'_{\hsm 2}s_{2\theta}^{}\!+\!\YT'_{\hsm 4}s_{4\theta}^{}\!+\!
\YT'_{\hsm 6}s_{6\theta}^{})~}
{\,256\hs\bar{s}\hs (\bar{s}\!-\!1)[(\bar{s}\!-\!2)^2\!-\!
(\bar{s}\!-\!4)^{2}c_{\theta}^2]s_{\theta}^2\,},
\end{align}
with the numerator coefficients $(\YT_j^{}, \YT'_j)$ 
given by
\begin{align}
\YT_{\hsm 0} &= 4(340\hs\bar{s}^4\!-\hsm 1329\hs\bar{s}^3\!+\hsm 1028\hs\bar{s}^2
\!+\hsm 16\hs\bar{s} \hsm -\hsm 64) \hs,
\nn\\
\YT_{\hsm 2} &= 663\hs\bar{s}^4\!-\hsm 2688\hs\bar{s}^3\!+\hsm 1760\hs\bar{s}^2\!-\hsm 512\hs\bar{s} \hsm +\hsm 768  \,,
\nn\\
\YT_{\hsm 4} &= 4\big(\bar{s} \hsm -\hsm 4\big)^2
(6\hs\bar{s}^2\!+\hsm 5\hs\bar{s}-12)\hs,
\nn\\
\YT_{\hsm 6} &= (\bar{s}\hsm -\hsm 4)^4,
\label{eq:Amp-4phi-DC2-RD-Yt}
\\
\YT'_{\hsm 2} &=
-\ii\hs\bar{s}^{\frac{1}{2}}(475\hs\bar{s}^4
\!-\hsm 2736\hs\bar{s}^3
\!+\hsm 4736\hs\bar{s}^2\!-\hsm 2560\hs\bar{s} 
\hsm +\hsm 256),
\nn\\
\YT'_{\hsm 4} &=
-\ii\hs 4\hs\bar{s}^{\frac{1}{2}}(5\hs\bar{s}^2
\!-\hsm 8\hs\bar{s} \hsm +\hsm 4)(\bar{s}\hsm -\hsm 4)^2,
\nn\\
\YT'_{\hsm 6} &=
\ii\hs\bar{s}^{\frac{1}{2}} (\bar{s}\hsm -\hsm 4)^4,
\nn
\end{align}
where $\sB\!=\!s/m^2$.\
Under high energy expansion, we derive the double copy amplitude
\eqref{eq:Amp-4phi-DC3} at its leading order,
\begin{equation}
\label{eq:Amp-4phi-DC-LO-0}  
{\MM}_{0}^{\rm{DC}}[4\psi]  
= -\frac{~\ii\hs\kappa^2m\,}{128}
\csc^3\!\theta\hsm\big(494\hs c_{\theta}^{}\hsmx +\hsmx 19\hs c_{3\theta}^{}\!-\hsm c_{5\theta}\big)
s^{\fr{1}{2}}  .
\end{equation}
We compare the above double-copied amplitude 
\eqref{eq:Amp-4phi-DC3}-\eqref{eq:Amp-4phi-DC2-RD-Yt}
with our Feynman-diagram-calculation of the physical scalar amplitude
\eqref{eq:Amp-4psi|L|ms=m}-\eqref{eq:Zhat-4psi|L|ms=m}
and find that they are precisely equal to each other.\
This is expected because the double copy of the four-point amplitude 
of physical adjoint-scalars (having the same mass
$m$ as the gauge bosons) in the YMYMS theory should
correspond to the four-point amplitude of physical scalars 
(having the same mass as the gravitons) in the WTMGS theory.\  
Moreover, we note that the above double-copied full amplitude \eqref{eq:Amp-4phi-DC3} 
differs from the original gravitational dilaton scattering amplitude
\eqref{appeq:Amp-4phi-sum-p=m}.\
This is expected since the dilaton $\phi$ is an unphysical field  
in our WTMG formulation, and there is no guarantee a priori 
to ensure an exact double copy for the unphysical dilaton amplitude 
despite that its kinematic Jacobi identity \eqref{eq:KJacobi-N'(Phi)-stu} holds.\ 
We further observe that the above leading-order amplitude \eqref{eq:Amp-4phi-DC-LO-0}   
agrees with the leading-order four-dilation amplitude \eqref{appeq:Amp-4phi-LO} (computed by
using the Feynman-diagram approach) and also coincides with the double-copied leading-order 
four-graviton amplitude \eqref{eq:M0-4hp-DC-LO} (up to an overall factor $1/16$)  
consistent with the prediction of the TMGRET \eqref{eq:TGRET-4pt}.\ 
Because the TGRET \eqref{eq:TGRET-4pt} requires 
that the leading-order dilaton amplitude equals the
corresponding leading-order physical graviton amplitude (up to a known overall factor)
under high energy expansion,
the leading-order dilaton amplitude is actually a gauge-invariant physical amplitude
and it is expected to be given by a proper double-copied amplitude 
at its leading order (just like what happens to the corresponding 
graviton amplitude at its leading order).\
In fact, explicit calculation shows that 
the leading-order dilaton amplitude equals 
the leading-order physical scalar amplitude \eqref{eq:M[4psi]L-LO} 
[as explained in the text below Eq.\eqref{eq:M[4psi]L-LO}].\  
Hence the leading-order dilaton amplitude can be derived from the leading-order result \eqref{eq:Amp-4phi-DC-LO-0}
of the double-copied amplitude \eqref{eq:Amp-4phi-DC3} as follows:
\begin{equation}
\label{eq:Amp-4phi-DC-LO}  
\MM_0^{}[4\phi] 
= \left.\MM_{\rm{L},0}^{}[4\psi]\right|_{m_s=m}^{}\!\! 
= {\MM}_{0}^{\rm{DC}}[4\psi] 
= -\frac{~\ii\hs\kappa^2m\,}{128}
s^{\fr{1}{2}} \hsm\big(494\hs c_{\theta}^{}\hsm +\hsm 19\hs c_{3\theta}^{}\!-\hsm c_{5\theta}\big)\!\csc^3\! \theta \,.
\end{equation}
This is illustrated in Fig.\,\ref{fig:10new}.\
The first equality of Eq.\eqref{eq:Amp-4phi-DC-LO} holds because the leading-order $(4\phi)$-amplitude
and the leading-order $(4\psi)$-amplitude are both given by the graviton-exchange diagrams and
their trilinear vertices $\phi\phi h_{\mn}^{}$ and $\psi\psi h_{\mn}^{}$ have the same coupling.\
A general proof of this equality is given by Eq.\eqref{eq:M0[Nphi]=M0[Npsi]}.\ 
The second equality of Eq.\eqref{eq:Amp-4phi-DC-LO}
holds because the full double-copied amplitude 
\eqref{eq:Amp-4phi-DC3}-\eqref{eq:Amp-4phi-DC2-RD-Yt} 
is shown to coincide with the our Feynman-diagram-calculation 
of the physical scalar amplitude
\eqref{eq:Amp-4psi|L|ms=m}-\eqref{eq:Zhat-4psi|L|ms=m} 
and this massive double-copy construction 
obeys the color-kinematics duality.\

\vs 

Finally, we examine the general case with the ajoint-scalar
mass $m_s^{}$ being arbitrary (including   
$m_s^{}\!\neq\! m\hs$).\  	 
We will explain why it is crucial to set up
$\,m_s^{}\!=\! m\,$ 
for realizing double copy of the leading-order dilaton amplitude.\
For the adjoint scalar having a general mass $m_s^{}$, 
we can rederive the numerators $\mathcal{N}_{j}^{\prime\Phi}$ in Eq.\eqref{eq:4Phi-Nstu0-2} of the four-point dilaton amplitude
as follows:
\\[-5mm]
\beqs
\begin{align}
\NN_{s}^{\prime\Phi} &= -{\ii\hs}\!\hsm\left[\hsm
(s\!-\!4\hs m_s^2)\ct \hsm +\hsm
\ii\hs m\hsm\(\!\!\sqrt{s\,}
\!-\!\fr{4\hs m_s^2}{\sqrt{s\,}\,}\hsm\)
\!\hsm\st\right]\!,
\\
\NN_{t}^{\prime\Phi} &=\frac{\,\ii\,}{\,2\,}
\!\hsm\left[-(3\hs s\!-\!4\hs m_s^2)
\hsm +\hsm (s\!-\!4\hs m_s^2)\ct
\hsm +\hsm\ii\hs 4\hs m\sqrt{s\,}\hsm
\tan\hsm{\fr{\theta}{2}}\hs\right]\!,
\\
\NN_{u}^{\prime\Phi} &=\frac{\,\ii\,}{\,2\,}\!\hsm
\left[(3\hs s\!-\!4m_s^2)\hsm 
+\hsm (s\!-\!4\hs m_s^2)\ct \hsm +\hsm
\ii\hs 4\hs m\sqrt{s\,}\hsm\cot\hsm{\fr{\theta}{2}}\hs\right]\!,
\end{align}
\eeqs
which have the following nonzero sum:
\begin{equation}
\Omega^{\prime}_{\Phi}\equiv
\NN_{s}^{\prime\Phi}\!+\hsm\NN_{s}^{\prime\Phi}
\!+\hsm\NN_{s}^{\prime\Phi}
=-\frac{m}{\sqrt{s\,}~}\big[4\hs s\csc\hsm\theta \!-\!(s\!-\!4m_s^2)\st \big]. 
\end{equation}
Then, different from Eq.\eqref{eq:GGT-Phi}, we derive
the generalized gauge transformation as follows:
\begin{equation}
\NN_j^{\prime\Phi} ~\longrightarrow~ \hat{\NN}_j^{\prime\Phi} = \NN_j^{\prime\Phi} \!-\!
\frac{\,\Omega^{\prime}_{\Phi}\,}{\,(4m_s^2\!-\!3m^2)\hs}(s_j^{}\!-\hsm m^2)\hs,
\end{equation}
where the gauge-transformed numerators are given by
{\small 
\beqs
\label{eq:hat{N}j'-4Phi}
\begin{align}
\hspace*{-8mm}
\hat{\NN}_{s}^{\prime\Phi} =& -\!\frac{\ii}{(4m_s^2\!-\!3m^2)\sqrt{s\,}}
\Big[\hsm\big(4m_s^2\!-\!3m^2\big)\sqrt{s\,}
\big(s\!-\!4m_s^2\big)c_{\theta}^{}
\!+\!\ii\hs4\hs m\hs s\big(s\!-\!m^2\big)\!\csc\hsm\theta
\\
&\hspace{8.5em}
\!-\!\ii\hs m\big(s\!-\!4m_s^2)(s\!-\!4m_s^2\!+\!2m^2\big)
\hsm\Big],
\nn\\ 
\hspace*{-8mm}
\hat{\NN}_{t}^{\prime\Phi} =&
\fr{\ii}{\,2(4m_s^2\!-\!3m^2)\sqrt{s\,}\,}
\Big\{\!\!-\!(4m_s^2\!-\!3m^2)\sqrt{s\,}(3s\!-\!4m_s^2)
\!+\!(4m_s^2\!-\!3m^2)\sqrt{s\,}(s\!-\!4m_s^2)c_{\theta} 
\\
&\left.
+\ii\hs m\Big[8s(2m_s^2\!-\!m^2)\hsm\tan\hsm\fr{\theta}{2}
\!+\!4\hs s\hs (s\!-\!4m_s^2\!+\!m^2)\hsm\cot\hsm\fr{\theta}{2}
\!-\!2(s\!-\!4m_s^2)\hsm\big(m^2\!+\!(s\!-\!4m_s^2)c_{\theta/2}^2\big)
s_{\theta}^{}\Big]\hsmx\right\} \!,
\nn\\[0mm]
\hspace*{-8mm}
\hat{\NN}_{u}^{\prime\Phi} =&
\fr{\ii}{\,2(4m_s^2\!-\!3m^2)\sqrt{s\,}\,}
\Big\{\!(4m_s^2\!-\!3m^2)\sqrt{s\,}(3s\!-\!4m_s^2)\!+\!(4m_s^2\!-\!3m^2)\sqrt{s\,}(s\!-\!4m_s^2)c_{\theta}
\\
&+\!\ii\hs  m\Big[8s(2m_s^2\!-\!m^2)\cot\hsm\fr{\theta}{2}\!+\!4s(s\!-\!4m_s^2\!+\!m^2)\tan\hsm\fr{\theta}{2}\!-\!2(s\!-\!4m_s^2)(m^2\!+\!(s\!-\!4m_s^2)s_{\theta/2}^2)s_{\theta}\Big]\hsmx\bigg\} . 
\nn 
\end{align}
\eeqs
}
\hspace*{-3mm}
Then, applying the double-copy formula \eqref{eq:Amp-4phi-DC}, 
we derive the new double-copied four-point dilaton amplitude
as follows:
\\[-2mm]
\begin{equation}
\label{eq:Amp'-4phi-DC-ms}
\MM_{\rm{DC}}^{\hs\prime}[4\psi]= 
\(\!\frac{\,\kappa\,}{4\,}\!\)^{\!\!2} \!\hsm
\left[\hsm \frac{(\hat{\NN}_{s}^{\prime\hs\Phi})^2}{\,s\!-\!m^2\,}
\hsm+ \frac{(\hat{\NN}_{t}^{\prime\hs\Phi})^2}{\,t\!-\!m^2\,}
\!+\! \frac{(\hat{\NN}_{u}^{\prime\hs\Phi})^2}{\,u\!-\!m^2\,} 
\hsm\right]\!.
\end{equation}
With Eq.\eqref{eq:hat{N}j'-4Phi} and 
using the numerator expressions of 
Eq.\eqref{eq:hat{N}j'-4Phi},
we explicitly derive the following double-copied amplitude
(with general scalar mass parameter $m_s^{}\hs$):
\begin{align}
{\MM}_{\rm{DC}}^{\hs\prime}[4\psi]
=\frac{\kappa^2\big(\widehat{\mathbb{Y}}_{0}^{}\!+\!\widehat{\mathbb{Y}}_{2}^{}c_{2\theta}^{}
\!+\!\widehat{\mathbb{Y}}_{4}^{}c_{4\theta}^{}
\!+\!\widehat{\mathbb{Y}}_{6}^{}c_{6\theta}^{}\!+\!\widehat{\mathbb{Y}}'_{2}s_{2\theta}^{}\!+\!\widehat{\mathbb{Y}}'_{4}s_{4\theta}^{}
\!+\!\widehat{\mathbb{Y}}'_{6}s_{6\theta}^{}\big)}
{~1024(4m_s^2\!-\!3m^2)s(s\!-\!m^2)[(s\!-\!4m_s^2\!+\!2m^2)^2\!-\!(s\!-\!4m_s^2)^2c_{\theta}^{2}]s_{\theta}^2~}\, ,
\label{eq.psi.DC}
\end{align}
with the numerator coefficients 
$(\widehat{\mathbb{Y}}_j^{}, \widehat{\mathbb{Y}}'_j)$ 
given by 
{\small 
\begin{align}
\hspace*{-4mm}
\widehat{\mathbb{Y}}_{0}^{} =&\,
4\big[4m^8(99s^2\!-\!88m_s^2 s\!+\!48m_s^4)\!-\!m^6(195s^3\!+\!1776m_s^2 s^2\!-\!1552m_s^4 s\!+\!1152m_s^6)\!-\!m^4(259s^4
\nn\\
&\!-\!1890m_s^2 s^3\!-\!904m_s^4 s^2\!+\!1440m_s^6 s\!-\!2176m_s^8)\!+\!m^2(85s^5\!-\!m_s^2 s^4\!-\!1808m_s^4 s^3\!+\!864m_s^6 s^2
\nn\\
&\!-\!1280m_s^{10})\!-\!m_s^2 s(s\!-\!4m_s^2)(85s^3\!-\!260m_s^2 s^2\!+\!176m_s^4 s\!+\!64m_s^6)\big]\hs,
\nn\\
\hspace*{-4mm}
\widehat{\mathbb{Y}}_{2}^{} =&\, 
64\hs m^8(7s^2\!+\!24m_s^2 s\!-\!16m_s^4)\!+\!16m^6(45s^3\!-\!288m_s^2s^2\!-\!496m_s^4 s\!+\!384m_s^6)\!-\!m^4(1095s^4
\nn\\
&\!-\!1680m_s^2 s^3\!-\!8160m_s^4 s^2\!-\!13568m_s^6 s\!+\!12032m_s^8)\!-\!m^2(286s^5\!-\!3646m_s^2 s^4\!+\!8480m_s^4 s^3
\nn\\
&\!+\!704m_s^6 s^2\!+\!7168m_s^8 s\!-\!7680m_s^{10})\!+\!2m_s^2 s(143s^4\!-\!944m_s^2 s^3\!+\!1696m_s^4 s^2\!-\!768m_s^6 s
\!-\!256m_s^8)\hs,
\nn\\
\hspace*{-4mm}
\widehat{\mathbb{Y}}_{4}^{} =&\, 
4\big(s\!-\!4m_s^2\big)^2[4m^8\!+\!3m^6(5s\!-\!8m_s^2)\!+\!7m^4(3s^2\!-\!14m_s^2 s\!+\!8m_s^4)\!-\!m^2(13s^3\!-\!m_s^2 s^2\!-\!104m_s^4 s
\nn\\
&\!+\!48m_s^6)\!+\!m_s^2 s(13s^2\!-\!16m_s^2 s\!-\!16m_s^4)]\hs,
\nn\\
\hspace*{-4mm}
\widehat{\mathbb{Y}}_{6}^{} =&
-\!(s\!-\!4\hs m_s^2)^4
\big[m^4\!+\!2m^2(s\!-\!m_s^2)\!-\!2m_s^2 s\big]\hs,
\\
\hspace*{-4mm}
\widehat{\mathbb{Y}}_{2}' =&\, 
\ii\hs m \big(4m_s^2\!-\!3m^2\big)
s^{\fr{1}{2}}\big[32(m^4\!-\!4m^2 m_s^2)(17s^2\!-\!40\hs m_s^2 s\!+\!16\hs m_s^4)\!-\!(475\hs s^4\!-\!2736\hs m_s^2 s^3
\nn\\
&\!+\!3104\hs m_s^4 s^2
\!+\!1280\hs m_s^6 s\!-\!1280\hs m_s^8)\big]\hs,
\nn\\
\hspace*{-4mm}
\widehat{\mathbb{Y}}_{4}' =&\, 
\!-\!\ii\hs 4\hs m \big(4m_s^2\!-\!3m^2\big)
s^{\fr{1}{2}}(s\!-\!4m_s^2)^2
\big[5s^2\!-\!8m_s^2 s\!+\!4(m^2\!-\!2m_s^2)^2\big]\hs,
\nn\\
\hspace*{-4mm}
\widehat{\mathbb{Y}}_{6}' =&\, 
\ii\hs m(4m_s^2\!-\!3m^2) s^{\fr{1}{2}}(s\!-\!4m_s^2)^4\hs. 
\nn
\end{align}
}
\hspace*{-3mm}
Under high energy expansion, we derive the double-copy 
formula \eqref{eq.psi.DC} as follows:
{\small 
\beq
\label{eq:DC-4Phi|ms-m}
\hspace*{-6mm}
{\MM}_{\rm{DC}}^{\hs\prime}[4\psi]
=-\frac{\,\hs\kappa^2(m_s^2\!-\!m^2)\hs s^1\,}{\,16(4m_s^2\!-\!3m^2)\,}
\hsm\frac{\,(7\!+\!c_{2\theta}^{})^2\,}{\sin^2\!\theta} 
\!-\!\frac{\hs\,\ii\hs\kappa^2 m\hs s^{\frac{1}{2}}\,}{128}\hsm 
\frac{\,(494\hs c_{\theta}^{}\hsm +\! 19\hs c_{3\theta}^{}\!-\hsm c_{5\theta}^{})\,}
{\sin^3\!\theta }
\!+\hsm {O}(s^{0})\hs. 
\eeq
}
\hspace*{-3mm}
Note that the above leading energy-power term of $O(s^1)$
is proportional to $(m_s^2\!-\!m^2)$ and vanishes for 
$m_s^{}\!=\!m\hs$.\ 
Hence, for the choice $m_s^{}\hsm\!=\!m\hs$, the remaining 
leading-order amplitude is of the order of 
$s^{\frac{1}{2}}$ and precisely coincides with 
our double-copy result \eqref{eq:Amp-4phi-DC-LO}
(by choosing $m_s^{}\!=\!m\hs$ from the beginning)
for the leading-order dilaton amplitude;  
it also fully coincides with the leading-order
dilaton amplitude \eqref{appeq:Amp-4phi-LO-p=m} 
(which is derived by the conventional Feynman diagram method),
namely, 
\beq 
\left. {\MM}^{\hs\prime}_{\rm{DC},0}
[4\psi]\right|_{m_s^{}=m}^{} 
= {\MM}_{\rm{L},0}^{}[4\phi]
\hs,
\eeq 
where the subscript ``0'' denotes the leading-order amplitude.\ 
This shows that the equality $m_s^{}\!=\!m\hs$ is a necessary
condition for realizing the double copy of 
the leading-order dilaton amplitude.

\vs 

As the final remark of this subsection, we note that due to the special kinematics 
of the 3d spacetime the massive double copy for constructing the $N$-point graviton amplitudes 
in the TMG (WTMG) theory from the $N$-point gauge boson amplitudes in the TMYM theory 
is not yet fully established at tree level for $N\!\!\geqq\!5\hs$.\ 
There are special BCJ relations for the color-ordered 
gauge amplitudes in the 3d TMYM theory.\ 
For five-point amplitudes, the determinant of massive KLT kernal is proportional to 
the Gram determinant of arbitrary four out of five external momenta, 
which vanishes because any four momenta cannot be independent of each other in 3d spacetime.\     
So this implies unique BCJ relations for the 3d theory.\ 
Ref.\,\cite{TMG-DCx} showed that there is an extra BCJ relation that 
can be satisfied by the five-point amplitudes in the 3d TMYM theory, 
so the double-copy works for the five-point graviton amplitudes in the TMG theory.\ 
It is expected that such unique BCJ relations may also exsit for higher-point amplitudes 
(with $N\!\!>\!5$).\ But, since there is no general proof of the BCFW recursions for 
3d non-supersymmetric field theories, the present explicit constructions 
of the four-point scattering amplitude of gauge bosons (adjoint scalars) and of  
the gravitons (dilatons) via double copy are already nontrivial,
and further extensions to the $N$-point ($N\!\!\geqq\!5$) amplitudes are much harder 
and could be studied only case by case, which are beyond the current scope
and worth of future studies.\  

\vs 

Finally, we stress that the main focus of this work is to establish the TGRET for understanding the
topological graviton mass-generation and the energy-dependence structures of the  
graviton (dilaton) scattering amplitudes.\ For this purpose, we have already presented 
the general energy-power counting analysis for any $N$-point ($N\!\!\geqq\! 4$) scattering amplitudes 
of gravitons and of dilatons with the aid of the TGRET 
in Section\,\ref{sec:3.3new}.\ 
The main goal of Section\,\ref{sec:4.1} is to explicitly verify 
{\it for the first time} the TGRET 
for the three-point and four-point graviton (dilaton) amplitudes and prove the gauge-invariance
of four-point physical graviton amplitude as 
the nontrivial consistency check of our new BRST 
quantization of the WTMG theory (including the dilaton field).\ 
We further presented the double-copy analysis 
in Section\,\ref{sec:4.2} 
in order to explicitly understand how the graviton ampplitudes 
and dilaton amplitides (lying on the two sides of the TGRET relation) 
are constructed from the double copy.\

\vspace*{2mm}
\subsection{\hspace*{-2mm}Relations to 3d Massless Scattering Amplitudes and Double Copy}
\label{sec:4.3}
\vspace*{1.5mm}

In this subsection, we consider the 3d massless YM theory 
and the 3d massless gravity theory coupled to the massless dilaton [introduced through
the Weyl transformation \eqref{eq:CT-gmunu}] 
which we denote as GRD$_3^{}$ theory.\ The 3d massless YM Lagrangian
\begin{equation}
\label{eq:YM}
\mathcal{L}_{\text{YM}} = -\frac{1}{2}\text{tr}({\mathbf{F}_{\mu \nu}^2})\,,
\end{equation}
can be considered as the massless limit of the TMYM theory 
(by removing the Chern-Simons term), whereas the GRD$_3^{}$
can be regarded as the massless limit of the WTMG theory (including the massless dilaton)
as shown in Eq.\eqref{eq:limitTMG}.\

\vs

\begin{figure}[t]
\centering
\includegraphics[width=0.7\textwidth]{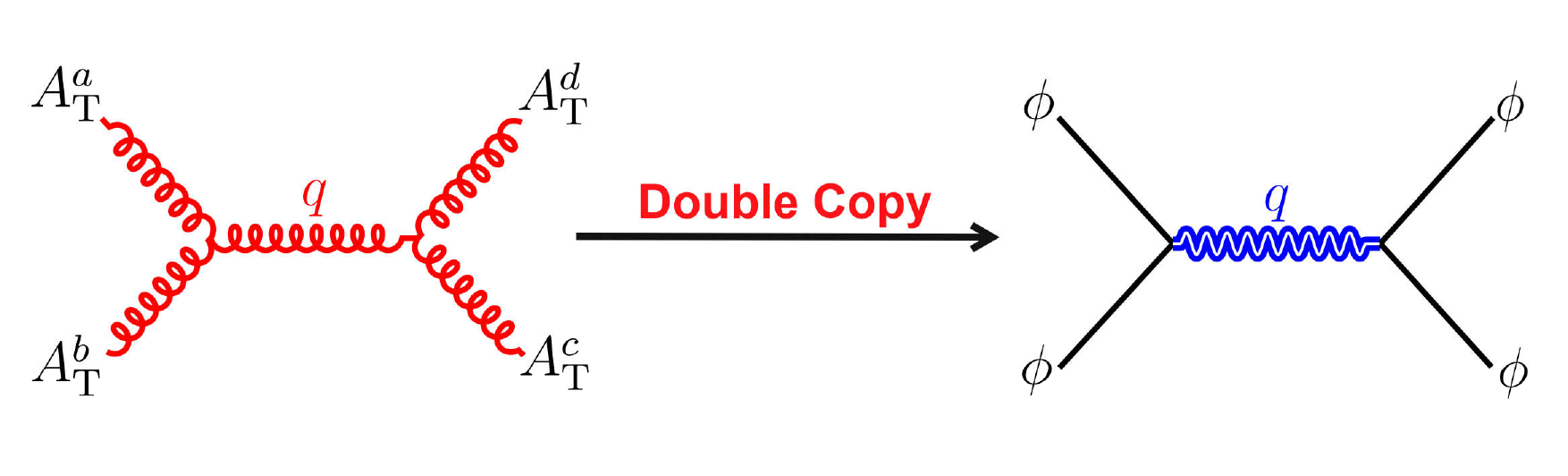}
\vspace*{-8.5mm}
\caption{\small\hspace*{-1mm}
Four-point gauge boson scattering in the 3d massless YM gauge theory (left diagram)
and the corresponding four-point dilaton scattering in the GRD$_3^{}$ theory (right diagram).\
These two diagrams represent the dressed four-point amplitudes which have absorbed the
contributions from the $4\AT$ and $4\phi$ contact diagrams.\
The dilaton scattering amplitude can be constructed from the gauge boson amplitude
through the double copy, as indicated by the right arrow.
}
\label{fig:10}
\label{fig:11new}
\end{figure}

We can establish a double-copy relation between the four-point gauge boson scattering amplitude in the 3d massless YM gauge theory
and the corresponding four-point dilaton scattering amplitude in the GRD$_3^{}$ theory.\
This is illustrated by Fig.\,\ref{fig:10}.\
Using the gauge boson polarization vector $\epT^{\mu}$ in Eq.\eqref{eq:epL-epT} or \eqref{eq:epT-spinor},
we can derive the 3d four-point massless gauge boson scattering amplitude:
\begin{equation}
\label{eq:T-4AT}
\TT[\AT^a\AT^b\AT^c\AT^d] = g^2\!
\(\!\frac{~\CC_{s}\hs\NN_{\rm{T}}^{s}\,}{~s~} + \frac{~\CC_{t}\hs\NN_{\rm{T}}^{t}\,}{~t~}
	+ \frac{~\CC_{u}\hs\NN_{\rm{T}}^{u}\,}{~u~}\!\)\!,
\end{equation}
with the following kinematic numerators,
%
\begin{align}
\NN_{\rm{T}}^{s} & = {\ii\hs}(p_{1_{0}}^{}\!\!-\hsm p_{2_{0}}^{})\!\cdot\!(p_{3_{0}}^{}\!\!-\hsm p_{4_{0}}^{})\,,
\nn\\
\NN_{\rm{T}}^{t} & = {\ii\hs}(p_{1_{0}}^{}\!\!-\hsm p_{4_{0}}^{})\!\cdot\!(p_{2_{0}}^{}\!\!-\hsm p_{3_{0}}^{})\,,
\label{eq:T-4AT-Nj0}
\\
\NN_{\rm{T}}^{u} & = {\ii\hs}(p_{1_{0}}^{}\!\!-\hsm p_{3_{0}}^{})\!\cdot\!(p_{4_{0}}^{}\!\!-\hsm p_{2_{0}}^{})\,,
\nn
\end{align}
%
where the momentum $p_{j_{0}}^{}$ can be obtained by taking the massless limit of $p_{j}^{}$.\
Then, it is straightforward to verify that the massless kinematic Jacobi identity holds:
\beq
\NN_{\rm{T}}^{s} + \NN_{\rm{T}}^{t} + \NN_{\rm{T}}^{u} = 0\hs.\
\eeq
Using the expressions \eqref{eq:T-4AT-Nj0}, we explicitly compute the four-point massless gauge boson scattering amplitude
\eqref{eq:T-4AT} as follows:
\begin{equation}
\label{eq:4AT=4Phi-1}
\TT[\AT^a\AT^b\AT^c\AT^d] =
g^{2}\! \left[\mathcal{C}_{s}(-{\ii\hs}\ct)
\!+ \mathcal{C}_{t}\bigg(\!{\ii\hs}\frac{~3\!-\!\ct\,}{\,1\!+\!\ct\,}\!\bigg)
\!+ \mathcal{C}_{u}\bigg(\!{-\ii\hs}\frac{\,3\!+\!\ct\,}{\,1\!-\!\ct\,}\!\bigg)
\!\right]\!.
\end{equation}

With these, we construct the four-point massless dilaton scattering amplitude from the corresponding
four-point gauge boson amplitude through the double copy:
\begin{equation}
\label{eq:MT-4phi-DC}
\begin{split}
\MM_{\rm{T}}^{\rm{DC}}[4\phi] &=
\(\hsmx\!\frac{\,\kappa\,}{\,4\,}\!\hsm\)^{\!\!2}\!
\left[\!\frac{\,(\NN_{\rm{T}}^{s})^2\,}{s} \!+\!
\frac{\,(\NN_{\rm{T}}^{t})^2\,}{t} \!+\! \frac{\,(\NN_{\rm{T}}^{u})^2\,}{u} \hsm\right]
\\
&= \frac{\,\kappa^2\hs s~}{64}\big(7\hsm +\hsm\cos\hsm 2\theta)^2\!\csc^2\!\theta \,.
\end{split}
\end{equation}
This double copy construction is illustrated in Fig.\,\ref{fig:10}.\
The above massless dilaton amplitude has energy-dependence $s^{1}\!\propto\!E^2$,
which differs from the energy-dependence $s^{\frac{1}{2}}\!\propto\!E^1$ 
of the leading-order dilaton amplitude
\eqref{appeq:Amp-4phi-LO-p=m} or \eqref{eq:Amp-4phi-DC-LO}  
in the WTMG theory.\
Instead, this double-copy result \eqref{eq:MT-4phi-DC}
should correspond to the scattering amplitude of dilatons ($\phi$) in the GRD$_3^{}$ theory.\
This is because the gauge field $A^{a}_{\mu}$ in the massless YM theory has only one physical degree of freedom
(as given by $\AT^a\!=\!\ep^\mu_{\rm{T}}A^a_\mu$) and the massless gravity theory of GRD$_3^{}$
also has only one physical degree of freedom (as given by the dilaton $\phi\hs$).\
Hence the double copy of the scattering amplitude of physical gauge bosons $\AT^a$'s should correspond to the
scattering amplitude of physical dilatons $\phi$ in the GRD$_3^{}$ theory.\footnote{%
Note that in the 3d pure gravity theory (GR$_3^{}$), the 3d massless graviton field $h_{\mn}^{}$ alone
has no physical degree of freedom, so their scattering amplitudes are unphysical and cannot correspond to
any double-copied physical scattering amplitudes of $\AT^a$.}\
This double-copy procedure is presented in Eqs.\eqref{eq:AxA=hmunu+phi-0} and \eqref{eq:m0-AxA=0.hmunu+1.phi}.\

\vs 

In the massless limit, the WTMG theory reduces to the GRD$_3$ in Eq.\eqref{eq:limitTMG}.\ 
Using its dilaton-dilaton-graviton vertex \eqref{Beq:scalar-scalar-graviton},
we explicitly compute the four-point dilaton amplitude as follows:
\begin{align}
\label{eq:4phi-Amp-GRD3}
\MM[4\phi] &=
\frac{\,\kappa^2 s\,}{4}
\!\hsm\(\hsm\frac{1}{4}\sin^2\!\theta \hsm +\hsm\cot^2\!\frac{\theta}{2} \hsm +\hsm\tan^2\!\frac{\theta}{2}\)
\nn\\
&= \frac{\,\kappa^2 s\,}{64}(7\hsm +\hsm\cos\hsm 2\theta)^2\!\csc^2 \!\theta \,,
\end{align}
which is of $O(s^1)$ and precisely equals the four-point massless dilaton scattering amplitude 
\eqref{eq:MT-4phi-DC} from double copy construction as expected.\
Note that as shown in Section\,\ref{sec:4.2} 
the double copy of the scattering amplitude of the massive physical gauge bosons ($\AP^a$) 
in the TMYM theory gives the corresponding scattering amplitude of the massive physical gravitons 
($\hP$) in the TMG theory.\ 
In the massless limit, the TMYM theory reduces to the YM theory.\  
Thus, the double copy of the scattering amplitude of the massless physical gauge bosons ($\AT^a$) 
in the YM theory gives the corresponding scattering amplitude of the massless dilatons ($\phi$) 
in the GRD$_3$ theory.\ This indicates that the massless dilaton $\phi$ 
in the GRD$_3$ theory does carry a physical degree of freedom, so its scattering amplitude
can be constructed via double-copy method from the scattering amplitude of 
the massless physical gauge bosons ($\AT^a$) in the YM theory.\ 
It also implies that the massless limit of the WTMG theory just corresponds to 
the $\rm{GRD}_{3}$ theory, where the physical degree of freedom originating from 
the massive graviton $\hP$ decouples in the massless limit and is converted to 
the massless dilaton $\phi$ of the $\rm{GRD}_{3}$.\  

\vs 

\begin{figure}[t]
\centering
\includegraphics[width=0.7\textwidth]{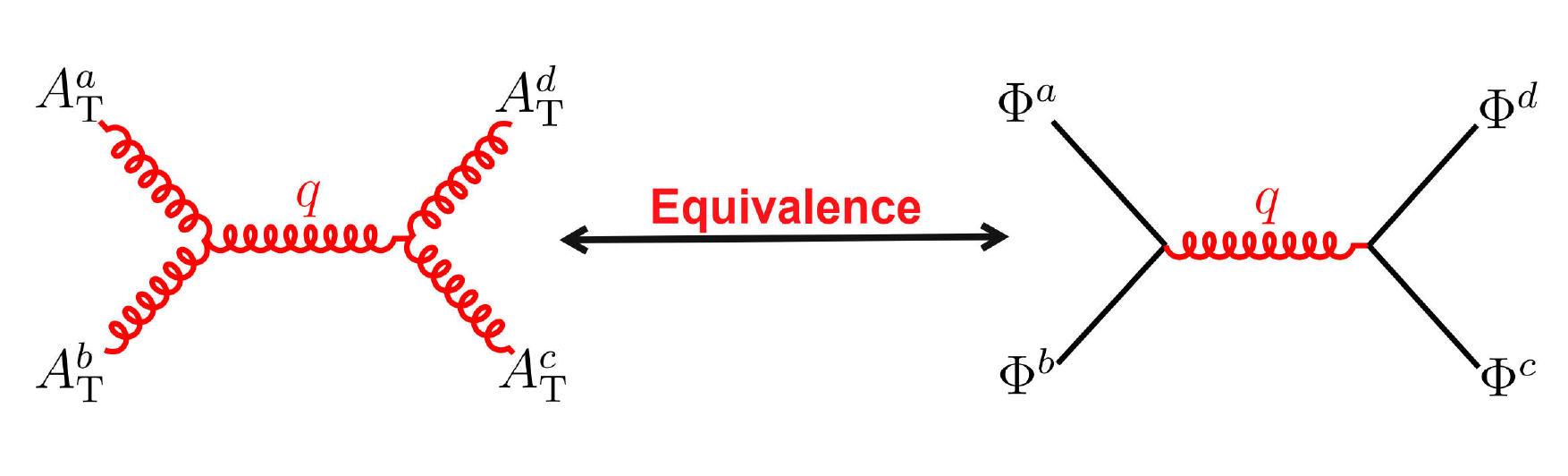}
\vspace*{-7mm}
\caption{\small%
Four-point gauge boson scattering in the 3d massless YM gauge theory 
(left diagram) and the corresponding scattering of the adjoint scalars 
in the 3d massless YM-Scalar theory (right diagram).\
The two types of scattering amplitudes are equal to each other, 
as indicated.\ Here, the left diagram denotes the dressed 
scattering amplitude of gauge bosons which has absorbed 
the contribution of the $4\AT^a$ contact diagram.}
\label{fig:11}
\label{fig:12}
\end{figure}

\vs

Moreover, we find that in the massless limit, the scattering amplitude of
the adjoint scalars ($\Phi^a$) in Eqs.\eqref{eq:4Phi-hatNstu0}-\eqref{eq:T-4Phi-GGT}
reduces to the scattering amplitude \eqref{eq:T-4AT}-\eqref{eq:T-4AT-Nj0} of gauge bosons ($\AT^a$)
in the 3d massless YM theory,
\beq
\label{eq:Amp-4AT=4Phi-ML}
\TT_{\rm{ML}}^{}[4\AT^a] =\TT_{\rm{ML}}^{}[4\Phi^a] \,,
\eeq
where the subscript ``ML'' stands for massless gauge bosons or adjoint scalars.\
This is illustrated in Fig.\,\ref{fig:11}.\
This is expected since each 3d massless gauge boson has only one physical degree of freedom ($\AT^a$)
and the spin angular momentum is a 
pseudoscalar in 3d spacetime\,\cite{Deser-CS1982PRL}\cite{Jackiw:1991}.\
The equality between the four-point $\AT^a$-amplitude and the corresponding $\Phi^a$-amplitude
can be traced back to the equality of their kinematic numerators.\
This is clear by comparing the numerators \eqref{eq:T-4AT-Nj0} of the $\AT^a$-amplitude
with the numerators of the $\Phi^a$-amplitude
that are obtained by taking the massless limit of Eq.\eqref{eq:4Phi-Nstu0}, 
namely, $\NN^j_{\rm{T}} \!=\! \NN_j^{\Phi}$,
in the massless case.\ 
For explicit calculation, we consider 
the Lagrangian of the massless YM gauge fields coupled to
the massless adjoint scalar fields $\Phi^a$ as follows: 
\begin{equation}
	\label{eq:YM-adjointPhi-m0}
	\mathcal{L}_{\rm{YM}\Phi}^{} = -\frac{1}{2}\text{tr}{\(\mathbf{F}_{\mu \nu}^{\,2}
		\!+\hsm \mathbf{D}_{\mu}^{}\hsm\Phi\hs \mathbf{D}^{\mu}\Phi\)} ,
\end{equation}
where the adjoint scalar field is $\Phi\!=\!\phi^{a}T^{a}$ and
the covariant derivative is defined as
$\mathbf{D}_{\mu}^{}\Phi\hsm =\partial_{\mu}^{}\Phi - \ii\hs g[\mathbf{A}_{\mu},\Phi]$.\
We may call this Lagrangian the YM-Scalar theory,
which can be obtained from Eq.\eqref{eq:TMYM-PhiAdj}  
by taking the massless limit.\
Using Eqs.\eqref{eq:T-4AT}-\eqref{eq:T-4AT-Nj0} and 
the fact of equal numerators 
$\NN^j_{\rm{T}} \hsm =\hsm \NN_j^{\Phi}$, 
we explicitly derive the $\AT^a$-amplitude and $\Phi^a$-amplitude 
as follows:
%
\begin{equation}
\label{eq:4AT=4Phi-2}
\TT_{\rm{ML}}^{}[4\AT^a]=\TT_{\rm{ML}}^{}[4\Phi^a] =
\ii\hs g^{2}\! \left[\mathcal{C}_{s}(-\ct)
\!+\hsm \mathcal{C}_{t}\bigg(\!\frac{~3\!-\!\ct\,}{\,1\!+\!\ct\,}\!\bigg)
\!+\!\mathcal{C}_{u}\bigg(\!\frac{\,-3\!-\!\ct\,}{\,1\!-\!\ct\,}\!\bigg)
\!\right]\!.
\end{equation}

\vs

In addition, we note that for the TMYM theory, the four-point scattering amplitude of massive physical gauge bosons ($\AP^a$)
is connected to the corresponding scattering amplitude of the transverse gauge bosons ($\AT^a$) via the
topological gauge-boson equivalence theorem (TGAET) at the leading order of high energy expansion\,\cite{Hang:2021oso}:
\beq
\label{eq:TMG-T0(4AP)=T0(4AT)}
\TT_0^{}[4\AP^a]= \frac{1}{4}\tT_0^{}[4\AT^a] =
\frac{~\ii\hs g^2\hs}{4}\hsm\!\left[\mathcal{C}_{s}(-\ct)
\!+\! \mathcal{C}_{t}\bigg(\!\!\frac{~3\!-\!\ct\,}{\,1\!+\!\ct\,}\!\bigg)
\!+\!\mathcal{C}_{u}\bigg(\!\!\frac{\,-3\!-\!\ct\,}{\,1\!-\!\ct\,}\!\bigg)
\!\right]\!,
  \eeq
which are mass-independent.\
Comparing the leading-order scattering amplitudes \eqref{eq:TMG-T0(4AP)=T0(4AT)} 
with the scattering amplitudes
\eqref{eq:4AT=4Phi-2} of the massless gauge fields $\AT^a$ 
(or the massless adjoint scalar fields $\Phi^a$),
we find that they have identical structure up to an overall factor of $1/4\hs$.\ 
Thus, we deduce the following relations:
\beq
\label{eq:4AP=4AT=4Phi}
\TT_0^{}[4\AP^a]=\frac{1}{4}\tT_0^{}[4\AT^a] = 
\frac{1}{4}\TT_{\rm{ML}}^{}[4\AT^a]= \frac{1}{4}\TT_{\rm{ML}}^{}[4\Phi^a] \,.
\eeq

\vspace*{2mm}
\section{\hspace*{-2mm}Conclusions}
\label{sec:5}

Gravitons naturally acquire topological masses in 3d spacetime 
through the topologically massive gravity (TMG) theory
that includes the gravitational Chern-Simons term.\
It is important to understand the topological mass-generation for gravitons 
both at the Lagrangian level and
at the level of $S$-matrix scattering amplitudes.\
In this work, we studied the structure of the massive graviton 
scattering amplitudes in the 3d TMG theory.\  
We analyzed the topological mass-generation for gravitons
and presented a new formulation\,---\,the Weyl-transformed TMG 
(WTMG) theory that includes the dilaton field $\phi$ through 
Weyl transformation.\ 
With these, we performed the BRST quantization of the WTMG theory 
and constructed the Topological Graviton Equivalence Theorem (TGRET) 
that formulates the topological mass-generation of gravitons 
by connecting the massive graviton scattering amplitudes to 
the corresponding scalar dilaton scattering amplitudes 
in the high energy limit.\ 
This provides a general mechanism to guarantee the 
large energy cancellations in the $N$-point 
massive graviton scattering amplitudes ($N\!\!\geqq\!4$).\
We further studied the 3d massive double-copy construction of 
the gauge-gravity duality connection, 
(Gravity)\,=\,(Gauge\,Theory)$^2$, 
for both massive graviton scattering amplitudes 
and dilaton scattering amplitudes 
(up to three- and four-point amplitudes) in the WTMG theory 
by using the corresponding gauge boson (adjoint scalar) scattering amplitudes.\

\vspace*{0.5mm}

In Section\,\ref{sec:2.1},
we presented the WTMG theory in Eq.\eqref{eq:LTMG-phi} 
by introducing an unphysical dilaton field
through the Weyl transformation \eqref{eq:CT-gmunu}.\ 
Then, we performed the BRST quantization of this WTMG theory
by constructing Lorentz-covariant gauge-fixing terms \eqref{eq:gaugefix-1}-\eqref{eq:gaugefix-2} 
and the corresponding ghost terms.\
We showed that the Landau-gauge graviton propagator 
\eqref{eq:Dh-Landau} (with $\xi\!=\!\zeta\hsm\!=\!0$) 
has good high-energy behavior $m/p^3$, 
which is better than that of the Feynman-gauge graviton propagator 
\eqref{eq:Dh-xi=zeta=1} 
(with $\xi\!=\!\zeta\!=\!1$ and scaling as $1/p^2\hs$)
and that of the unitary-gauge graviton propagator \eqref{eq:Dh-unitary} 
(with $\zeta\!=\!\infty$ and scaling as $1/p^2\hs$).\ 
We demonstrated that the WTMG theory conserves the physical degrees of freedom (DoF) in the massless limit, under which the physical massive 
graviton becomes an unphysical massless graviton
and its physical DoF is converted to the massless dilaton.\
In Section\,\ref{sec:2.2new}, we analyzed the pure dilaton 
(scalar) self-interactions under field redefinition 
in the WTMG (WTMGS) theory and proved that in the WTMG (WTMGS) theory 
the pure dilaton (scalar) self-interactions do not contribute to 
the on-shell leading-order dilaton (scalar) amplitudes
as shown in Eq.\eqref{eq:LphiHat} and Eq.\eqref{eq:L-phih/psih-kin}.\
We further proved in Eq.\eqref{eq:M0[Nphi]=M0[Npsi]} that for the WTMGS theory, the $N$-point dilaton amplitudes and physical scalar amplitudes 
are equal at the leading order of high energy expansion. 

\vspace*{0.5mm}

In Section\,\ref{sec:3.1new}, we derived relations for the 3d massive polarization 
vectors and tensors that are used for the proof of the TGRET in Section\,\ref{sec:3.2new}
and for the scattering amplitude analysis of Section\,\ref{sec:4.1}.\
In Section\,\ref{sec:3.2new},
we newly constructed the formulation for the TGRET, as given by Eq.\eqref{eq:TGRET},
which connects the scattering amplitudes of physical gravitons to the corresponding scattering amplitudes
of the dilatons in the high energy limit.\
The TGRET is derived from the 3d BRST identity \eqref{eq:TGET-ID}.\ 
Then, in Section\,\ref{sec:3.3new}, we developed a generalized gravitational power counting method  
to extract the leading energy-power dependence of the $N$-point graviton (dilaton) scattering
amplitudes as given by Eq.\eqref{eq:DE-hp-hv} (Landau gauge) and Eq.\eqref{eq:DE-uni} (unitary gauge)
of the WTMG theory.\ 
With these, we established the TGRET as a general mechanism to ensure the striking energy cancellations
in the massive graviton scattering amplitudes.\ 
Applying our power counting rules, we proved that the $N$-point massive graviton amplitudes 
($N\!\!\geqq\!4$) have large energy cancellations 
by powers proportional to $\frac{5}{2}N$ ($\frac{7}{2}N$) in the Landau (unitary) gauge.\   
This explains the large energy cancellations of 
$E^{11}\hsm\!\!\to\hsm\!E^1$ (Landau gauge) and 
$E^{12}\!\!\to\hsm\!E^1$ (unitary gauge) 
for the four-point graviton scattering amplitudes.\ 

\vspace*{0.5mm}

In Section\,\ref{sec:4.1}, we presented explicit calculations of the three-point and four-point
graviton and dilaton scattering amplitudes 
in the Landau gauge of the WTMG theory.\ 
We first computed the basic three-point graviton (dilaton) 
scattering amplitudes 
as in Eqs.\eqref{eq:Amp-3hp-EP}\eqref{eq:2hp0-hp} and Eq.\eqref{eq:h2phi-hhphi}.\   
With these, we verified the TGRET \eqref{eq:ET-3pt-h2phi}
for the three-point graviton (dilaton) amplitudes 
at the leading order of $m/E$
expansion for two of the external states.\ 
We demonstrated that the Landau-gauge and unitary-gauge calculations 
of the four-point physical graviton scattering amplitudes 
contain different energy cancellations.\ 
But they both give precisely the same total amplitude
as in Eq.\eqref{eq:app-Amp-4hp} and Eq.\eqref{eq:app-Amp-4hp-U},
proving the gauge-independence as in Eq.\eqref{Aeq:M[4hp]U=L}.\  
We further computed the four-point dilaton scattering amplitude 
\eqref{appeq:4phi-h-sum-p0}-\eqref{appeq:Amp-4phi-full}
in Landau gauge, and found that the leading-order dilaton scattering amplitude \eqref{appeq:Amp-4phi-LO}     
equals the leading-order physical graviton scattering amplitude of Eq.\eqref{eq:Amp-4hp-LO+NLO}.\  
Hence, we explicitly proved the TGRET \eqref{eq:TGRET-4pt} for the four-point scattering amplitudes of
gravitons and of dilatons in the high energy limit, 
in agreement with our general TGRET formula \eqref{eq:TGRET}.\ 
Moreover, we proved that the four-point leading-order
dilaton amplitude \eqref{appeq:Amp-4phi-LO} equals the 
corresponding leading-order physical scalar amplitude 
\eqref{eq:M[4psi]L-LO}, which verifies the general relation
of Eq.\eqref{eq:M0[Nphi]=M0[Npsi]} for $N\!=\!4$.\

\vspace*{0.5mm}

Then, in Section\,\ref{sec:4.2},
we extend the double-copy method to systematically construct the three-point and four-point massive graviton (dilaton) scattering amplitudes in the TMG theory from the corresponding gauge-boson (adjoint scalar) scattering amplitudes in the TMYM (TMYMS) theory, 
in agreement with the explicit calculations
of these scattering amplitudes.\
For instance, we presented the double-copied three-point 
graviton amplitude ($3\hP$) in Eq.\eqref{eq:DC-3hp-Amp} 
and three-point dilaton amplitude ($\phi\hs\phi\hs\hP$) 
in Eq.\eqref{eq:DC-MT-hphiphi}, 
which agree with our direct calculation results 
in Eqs.\eqref{eq:Amp-3hp-EP} and 
\eqref{eq:2phihp=2psihp} respectively.\ 
We further constructed the double-copied four-point graviton amplitude \eqref{eq:M4-DC}
and the double-copied four-point physical scalar amplitude 
\eqref{eq:Amp-4phi-DC3}-\eqref{eq:Amp-4phi-DC2-RD-Yt}.\ 
We found that the expanded form of the double-copied four graviton ($4\hP$) amplitude \eqref{eq:M4-DC} precisely equals 
the direct calculation result
given in Eqs.\eqref{eq:app-Amp-4hp}-\eqref{eq:app-Amp-4hp-Y}.\ 
The double-copied four-point physical scalar amplitude \eqref{eq:Amp-4phi-DC3}-\eqref{eq:Amp-4phi-DC2-RD-Yt}
differs from the original four dilaton amplitude \eqref{appeq:Amp-4phi-sum-p=m} as expected.\    
But we derived the leading-order result of the double-copied 
four physical scalar amplitude in Eq.\eqref{eq:Amp-4phi-DC-LO-0} 
and demonstrated that it just equals the leading-order result
\eqref{appeq:Amp-4phi-LO} 
of the original dilaton amplitude.\
In fact, we proved a general relation in Eq.\eqref{eq:M0[Nphi]=M0[Npsi]}
stating that the leading-order $N$-point dilaton amplitude equals the leading-order $N$-point physical scalar amplitude.\ 
Hence, we can derive the leading-order dilaton amplitude  
from the double-copied physical scalar amplitude 
at the leading order.\

\vspace*{0.5mm}

Finally, in Section\,\ref{sec:4.3} we analyzed the massless limit of 
the graviton scattering amplitudes in the TMG theory and their relations 
to the corresponding amplitudes in the 3d massless theories 
as well as their double-copy constructions.\
We find that, in the massless limit the TMYM theory reduces to the massless YM theory, 
where the double copy of the four-point scattering amplitude 
of massless physical gauge bosons ($\AT^a$) shown in Eqs.\eqref{eq:T-4AT}\eqref{eq:4AT=4Phi-1} 
gives a gravitational scalar amplitude \eqref{eq:MT-4phi-DC}, which just equals  
the scattering amplitude \eqref{eq:4phi-Amp-GRD3} 
of massless dilatons ($\phi$) in the $\rm{GRD}_3$ theory.\ 
This shows that the massless limit 
of the WTMG theory corresponds to the $\rm{GRD}_3$ theory: 
the physical degree of freedom carried by the massive graviton $\hP$ 
in the TMG theory 
decouples under $m\ito 0$ and is converted to the massless dilaton of the $\rm{GRD}_3$ theory.\ 
Furthermore, we demonstrate that in the massless limit of the TMYM-Scalar theory, the
four-point $\AT^a$ amplitude and the corresponding $\Phi^a$ amplitude are equal, 
as in Eq.\eqref{eq:Amp-4AT=4Phi-ML}.\ 
Accordingly, for the massless YM-Scalar theory, 
the four-point scattering amplitude of gauge bosons $\AT^a$ 
equals the corresponding $\Phi^a$ amplitude at leading order 
of high energy expansion, as in Eq.\eqref{eq:4AT=4Phi-2}.\
For the TMYM theory the four-point scattering amplitude of $\AP^a$ 
coincides with the corresponding $\AT^a$ amplitude (up to an overall factor) 
at leading order of high energy expansion, as in Eq.\eqref{eq:TMG-T0(4AP)=T0(4AT)}.\ 
These equality relations are summarized in Eq.\eqref{eq:4AP=4AT=4Phi}.\ 
As a final remark, we note that there is no general proof of the BCFW recursions for 
3d non-supersymmetric field theories, the current explicit constructions 
of the four-point scattering amplitude of gauge bosons (adjoint scalars) and of  
the gravitons (dilatons) via double copy in Sec.\,\ref{sec:4} are already nontrivial,
and further extensions to the $N$-point ($N\!\!\geqq\!5$) amplitudes are much harder 
and could be studied only case by case, which are beyond the current scope
and merit future studies.\  

\vspace*{7mm}
\noindent
{\bf\Large Acknowledgements}
\\[1mm]
This research was supported in part by the National Natural Science Foundation of China (NSFC)
under Grants No.\,12435005 and No.\,12175136,
by Shenzhen Science and Technology Program 
(Grant No.\,JCYJ2024 0813150911015),
by the State Key Laboratory of Dark Matter Physics,
by the Key Laboratory for Particle Astrophysics and Cosmology (MOE), 
and by the Shanghai Key Laboratory for Particle Physics and Cosmology.\

\vspace*{8.5mm}
\appendix

\noindent
{\Large\bf Appendix}
\vspace*{-2.5mm}

\section{\hspace*{-2mm}3d Massive Spinor-Helicity Basis and Polarizations}
\label{app:A}
\label{sec:appendix1}

In this Appendix, for the usage in the calculations of Section\,\ref{sec:4}, 
we introduce our conventions and notations,  
and briefly review the massive spinor-helicity variables in 3d spacetime based on Ref.\,\cite{other2a-3d-CS} 
and also the early literature\,\cite{Chiou:2005jn}-\cite{TorresDelCastillo2003}.\ 
Then, we derive the relevant formulas for the 3d polarization vectors and tensors.

\vspace*{0.5mm}

In the present work, we choose the 3d Minkowski metric with signatures
$\,\eta_{\mu\nu}^{}\!=\!\eta^{\mu\nu}\!=\!\text{diag}$ $(-1,+1,+1)$,
and the rank-3 antisymmetric Levi-Civita tensor obeys the convention:
%
$\varepsilon^{012} \!=\hsm -\varepsilon_{012}^{} \hsm =\! +1$\,.\
%
Thus, the momentum on-shell condition is given by
$p^2\!=\!-m^2$.\ Then, we define three-dimensional real symmetric and antisymmetric
$2\!\times\!2$ matrix bases of the $\sigma$ and $\epsilon$ as follows:
\begin{equation}
\label{eq:sigma-ep}
\hspace*{-5mm}
[\sigma^{0}]^{\alpha \beta} \!=\!
\begin{pmatrix} \!-1 & 0\! \\ \!0 & -1\hsm \end{pmatrix}\!,~~
[\sigma^{1}]^{\alpha\beta} \!=\! \begin{pmatrix} 1 & 0\! \\ 0 & -1\hsm \end{pmatrix}\!,~~
[\sigma^{2}]^{\alpha\beta} \!=\! \begin{pmatrix} \!0 & -1 \\ \!-1 & 0 \end{pmatrix}\!,~~
\varepsilon^{\alpha\beta} \!=\! \begin{pmatrix} 0 & -1\hsm \\ 1 & 0\! \end{pmatrix}\!.
\end{equation}
The conjugate base of the reduced exponent is defined as
$\sigma^{\mu}_{\alpha \beta}\!=\varepsilon_{\alpha\gamma}^{}\varepsilon_{\beta \delta}^{}
[\sigma^{\mu}]^{\gamma \delta}$,
with
$\varepsilon_{\alpha\beta}^{}\!=\varepsilon_{\alpha \gamma}^{}
 \varepsilon_{\beta \delta}^{}\varepsilon^{\gamma \delta}$.\
With these, we express the conjugate base of $\sigma$ and $\varepsilon$ matrices as follows:
\begin{equation}
\label{eq:sigma-ep-conj}
\hspace*{-5mm}
[\sigma^{0}]^{}_{\alpha \beta} \!=\!
\begin{pmatrix} \!-1 & 0\! \\ \!0 & -1\hsm \end{pmatrix}\!,~~
[\sigma^{1}]^{}_{\alpha\beta} \!=\!
\begin{pmatrix} \hsm -1~ & 0 \\ \!0 & 1
\end{pmatrix}\!,~~
[\sigma^{2}]^{}_{\alpha\beta} \!=\!
\begin{pmatrix} \!0~ & 1 \\ 1~ & 0 \end{pmatrix}\!,~~
{\varepsilon_{\alpha\beta}} \!=\! \begin{pmatrix} \!0 & 1\\ \hsm -1 & 0\! \end{pmatrix}\!.
\end{equation}
With these matrices, the 3d momentum of a massive particle is represented as a bispinor:
%
\begin{equation}
p_{\alpha\beta}^{} = p_{\mu}^{}\sigma^{\mu}_{\alpha \beta}
= \begin{pmatrix}
		-p_{0}^{}\!-\hsm p_{1}^{} & p_{2}^{} \\
		p_{2}^{} & -p_{0}\!+\hsm p_{1}
\end{pmatrix}\hsm\!,
\hspace*{7mm}
p_{\mu}^{} = -\Fr{1}{2} \sigma_{\mu}^{\alpha \beta} p^{}_{\alpha \beta} \,,
\end{equation}
and the determinent
$\det (p_{\alpha\beta}^{})\!=\!-p^2\!=\!m^2$
is equivalent to the on-shell condition $p^2\!=\!-m^2$.\ 

\vs 

For massive particles in 3d spacetime, the little group is SO(2),  
and one works with the angle spinor brackets $|i\rangle$ only.\ 
The spinor contractions are given by 
$\langle ij\rangle\!=\! \epsilon^{\alpha \beta} \lambda_{i\beta}^{}\lambda_{j\alpha}^{}$.\ 
A convenient explicit realization of the spinors $\lambda_{\alpha}^{}$ and 
$\bar{\lambda}_{\alpha}^{}$ takes the form:
\begin{equation}
	\lambda_{\alpha}^{} = \frac{1}{\sqrt{p_0^{}\!-\!p_1^{}\,}\,}\!\!
	\begin{pmatrix} p_2^{}\!-\hsm \ii\hs m \\ p_1^{}\!-\hsm p_0^{} \end{pmatrix}\!,
	\hspace*{7mm}
	\bar{\lambda}_{\alpha} = -\frac{1}{\sqrt{p_{0}^{}\!-\hsm p_{1}^{}\,}\,}\!\!
	\begin{pmatrix} p_2^{}\!+\hsm\ii\hs m \\ p_{1}^{}\!-\hsm p_{0}^{} \end{pmatrix}\!.
\end{equation}
If $p_{\mu}^{}$ is real, 
then $p_{\alpha \beta}^{}$ is also real and admits the following symmetric decomposition:
\begin{equation}
p_{\alpha \beta} = \Fr{1}{\,2\,}\!\(\lambda_{\alpha} \bar{\lambda}_{\beta}+\bar{\lambda}_{\alpha} {\lambda}_{\beta} \)\!.
\end{equation}
Thus, we have the following relation:
\begin{equation}
\label{eq:lala=p-im}
\lambda_{\alpha}^{} \bar{\lambda}_{\beta}^{} 
= p_{\alpha \beta}-\ii\hs m\hs\epsilon_{\alpha \beta}\hs,
\hspace*{7mm}
\bar{\lambda}_{\alpha}^{} {\lambda}_{\beta}^{} 
= p_{\alpha \beta}^{}+\ii\hs m\hs\epsilon_{\alpha \beta} \hs.
\end{equation}
One can introduce the antisymmetric scalar product of spinors, 
which may be called the twistor bracket:
\begin{equation}
\langle \lambda \bar{\lambda} \rangle = \lambda^{\alpha} \bar{\lambda}_{\alpha}
= \epsilon^{\alpha \beta} \lambda_{\beta} \bar{\lambda}_{\alpha}
= -\ii\hs 2\hs m\hs,
\hspace*{6mm}
\langle \bar{\lambda} \lambda \rangle = \ii\hs 2\hs m \hs.
\end{equation}

The $\gamma$ matrices can defined from the contraction,
$[\gamma^{\mu}]_{\al}^{~\be}=\sigma^{\mu}_{\alpha\gamma}\varepsilon^{\gamma \beta}$,
which takes the following explicit form:
\begin{equation}
\hspace*{-10mm}
[\gamma^{0}]_{\alpha}^{\ \beta} \!=\! \ii\hs\sigma^{2}
\!=\! \begin{pmatrix} \!0 & 1 \\ \!-1\, & 0 \end{pmatrix}\hsm\!,~~~~
[\gamma^{1}]_{\alpha}^{\ \beta} \!=\! \sigma^{1}
\!=\! \begin{pmatrix} 0~ & 1 \\  1~ & 0 \end{pmatrix}\!,~~~~
[\gamma^{2}]_{\alpha}^{\ \beta} \!=\! \ii\hs\sigma^{3}
\!=\! \begin{pmatrix} 1 & 0\! \\ 0 & -1\hsm
\end{pmatrix}\!.
\end{equation}
In the above, the $\sigma^i$ matrices denote the standard Pauli matrices, and the $\gamma^\mu$ matrices
obey algebraic relation:
\begin{equation}
\gamma^{\mu}\gamma^{\nu} =\hs \eta^{\mu \nu}\!+\varepsilon^{\mu \nu \rho}\gamma_{\rho}^{}\,.
\end{equation}
Thus, we can further derive the algebraic relation for the product 
$\gamma^{\mu}\gamma^{\nu}\gamma^{\rho}$ and compute the traces of $\gamma$ matrices:
\\[-9mm]
%
%
%
\begin{align}
\label{eq:gamma-algebraic-relation}
& \gamma^{\mu}\gamma^{\nu}\gamma^{\rho}
= \varepsilon^{\mu\nu\rho}\!-\!\eta^{\mu\rho}\gamma^{\nu}\!\!+\hsm\eta^{\rho \nu}\gamma^{\mu}\!+\hsm\eta^{\mu \nu}\gamma^{\rho},
\nn\\
& {\tr}(\gamma^{\mu}\gamma^{\nu}) = 2\hs\eta^{\mu \nu},
\hspace*{5mm}
{\tr}(\gamma^{\mu}\gamma^{\nu}\gamma^{\rho}) = 2\hs\varepsilon^{\mu \nu \rho}\,,
\\
&
{\tr}(\gamma^{\mu}\gamma^{\nu}\gamma^{\rho}\gamma^{\sigma})
= 2(\eta^{\mu \nu}\eta^{\rho \sigma}\!-\hsm\eta^{\mu \rho}\eta^{\nu \sigma}
\!+\hsm\eta^{\mu \sigma}\eta^{\rho\nu}) \hs.
\nn 
\end{align}
%
We note that the $\gamma$ matrices obey the completeness relation,
\begin{equation}
\label{eq:gamma.gamma}
[\gamma^{\mu}]_{\alpha}^{\ \beta}[\gamma_{\mu}]_{\rho}^{\ \sigma}
= -\!\(\hsm\varepsilon_{\alpha \rho}\varepsilon^{\beta \sigma}
\!+\delta_{\alpha}^{\sigma}\delta_{\rho}^{\beta}\hs\) \!,
\end{equation}
and the Levi-Civita identity,
\begin{equation}
\varepsilon^{\rho \mu \nu} q_{\rho}^{}\langle i| \gamma_{\mu}^{} |j \rangle \langle k| \gamma_{\nu}
|\ell \rangle = 2\langle k | q | j \rangle \langle i\ell \rangle
\!-\! \langle i | q | j \rangle \langle k\ell \rangle
\!-\! \langle k | q | \ell \rangle \langle ij \rangle \hs .
\end{equation}
With Eq.\eqref{eq:lala=p-im}, we may write the following Dirac equations:
\begin{equation}
p_{\alpha \beta}^{} \lambda^{\beta} \!= -\ii\hs m\lambda_{\alpha}^{} \hs,
\hspace*{7mm}
p_{\alpha \beta}^{} \bar{\lambda}^{\beta} \!= \ii\hs m\bar{\lambda}_{\alpha}^{} \hs.
\end{equation}
Using the twistor brackets, we express the above equations as follows:
\begin{equation}
\label{eq:DiracEq2}
p_{i}^{} |i\rangle \!=\hsm -\ii\hs m_{i}^{} |i\rangle\hs,
\quad
p_{i} |\bar{i}\rangle \!=\hsm \ii\hs m_{i}^{} |\bar{i}\rangle\hs,
\quad
\langle i| p_{i}^{} \!=\hsm \ii\hs m_{i} \langle i| \hs,
\quad
\langle \bar{i}| p_{i}^{} \!=\hsm -\ii\hs m_{i}^{} \langle \bar{i}|\hs.
\end{equation}
Making contraction with the $\sigma^\mu$ matrices, we obtain the identity:
\beq
\lambda_{\alpha} \bar{\lambda}_{\beta}[\sigma^{\mu}]^{\alpha \beta} =
p_{\alpha \beta}^{}[\sigma^{\mu}]^{\alpha \beta} \!+\hsm \ii\hs m\hs {\tr}(\gamma^{\mu})\hs,
\eeq
from which we derive the following relations,
\begin{equation}
\label{eq:mass-p}
\langle i \left\lvert \gamma^{\mu}\right\rvert \bar{i}\hs\rangle
= \langle \bar{i} \left\lvert \gamma^{\mu} \right\rvert i \hs\rangle = -2\hs p^{\mu} \hs.
\end{equation}
For the massless case, they take the form,
%
$\langle i \left\lvert \gamma^{\mu}\right\rvert {i}\hs\rangle
\!=\! \langle \bar{i} \left\lvert \gamma^{\mu} \right\rvert \bar{i} \hs\rangle 
\!=\! -2 \pu^{\mu}, 
$
%
with the momentum $\pu^{\mu}\!\equiv\hsm p^{\mu}|_{m=0}^{}\,$.\
Thus, making high-energy expansion, we derive the relation:
\begin{equation}
\label{Aeq:high-energyexpansion}
\lambda_{\alpha}^{}\bar{\lambda}_{\beta}^{} = \bar{\lambda}_{\alpha}\lambda_{\beta}
= -\lambda_{\alpha}^{}\lambda_{\beta}^{} = -\bar{\lambda}_{\alpha}\bar{\lambda}_{\beta}
= \pu_{\hs\alpha\beta}^{} \!+ {O}({m}) ,
\end{equation}
where the momentum $\pu_{\hs\alpha\beta}^{}\!\equiv\hsm p_{\hs\alpha\beta}^{}|_{m=0}^{}\,$.\
We also obtain the formulas in the massless limit,
\beq
\label{Aeq:pp=lala}
|\hs \under{p}\hs\rangle\langle \under{p}\hs| = -\lambda_{\alpha}^{}\lambda_{\beta}^{}\,,
\hspace*{8mm}
|\hs\under{\bar{p}}\rangle\langle \under{\bar{p}}\hs| = -\lambda_{\alpha}^{}\lambda_{\beta}^{}  \,.
\eeq
Using the completeness relation \eqref{eq:gamma.gamma}, we deduce the following identity:
\begin{equation}
\langle i| \gamma^{\mu} |j \rangle \langle k| \gamma_{\mu} |\ell \rangle
= -\!\( \langle ik \rangle \langle j\ell \rangle \!+\! \langle i\ell \rangle \langle jk \rangle \) .
\end{equation}
Moreover, for a momentum $k$ with mass $m_k^{}$, 
we have the identities:
\begin{equation}
\begin{split}
\langle i\bar{k}  \rangle \langle k\ell \rangle 
= -\langle i |k| \ell \rangle - \ii\hs m_{k}\langle i\ell \rangle \hs ,
\\
\langle ik  \rangle \langle \bar{k}\ell \rangle 
= -\langle i |k| \ell \rangle + \ii\hs m_{k}\langle i\ell \rangle \hs .
\end{split}
\end{equation}
%


With the above 3d spinor-helicity formalism, we proceed to discuss the polarization
vectors and tensors using the spinor-helicity basis.\ 
We present an explicit 3d spinor-helicity realization of the polarization vectors and tensors  
and summarize the resulting relations.\ 
The polarization vector of the gauge field $A^{\mu a}$ 
in the TMYM theory can be expressed as follows:
\begin{equation}
\label{eq:polarization}
\epP^{\mu -} =
\frac{\,\langle p| \gamma^{\mu} |p \rangle\,}{2\sqrt{2\,}\hs |p|},
\hspace*{6mm}
\epP^{\mu +} =
\frac{\,\langle \bar{p}| \gamma^{\mu} |\bar{p} \rangle\,}{2\sqrt{2\,}\hs |p|} ,
\end{equation}
where $|p|\!\equiv\!\sqrt{|p_\mu^2|\,}\!=\!m\hs$ and
we have imposed the on-shell condition $p^{2}\!=\!-m^2$.\
In the above, the polarization vectors are normalized as
$\epP^{\mu-}\epsilon_{\rm{P}\mu}^+\!=\!1\hs$.\
The polarization vector $\epsilon_{\text{P}}^{\mu-}$ or $\epsilon_{\text{P}}^{\mu+}$
satisfies the equations of motion of $A^{\mu a}$ for the TMYM theory \eqref{eq:TMYM-L}:
\begin{equation}
\label{eq.EOM}
\(\!\eta_{\mu \rho}\partial^{\nu}\! + 
\frac{\hs\mt\hs}{2}\hs\varepsilon_{\mu \nu \rho} \!\)\hsm\! F^{\nu \rho} 
= 0 \,.
\end{equation}
The 3d polarization vectors obey the relation,  $\epsilon_{\text{P}}^{\mu+}\hsm\!=\!(\epsilon_{\text{P}}^{\mu-})^{*}$.\ 
In 3d spacetime, there is only one independent physical polarization vector,
which reflects the fact that the gauge boson of the
3d YM theory has only one physical degree of freedom.\
In the following analysis, we will choose the polarization vector
$\epP^{\mu-}\hsm\!\equiv\hsm\epP^{\mu}$
without losing generality.

\vs

We can explicitly compute the physical polarization vector $\epP^{\mu}$
of Eq.\eqref{eq:polarization} in the spinor formulation
and compare it with the expression \eqref{eq.PVfromEOM}.\
We find that the formula \eqref{eq:polarization} differs from that of Eq.\eqref{eq.PVfromEOM}
only by an overall phase factor,
\beq
\mathbb{P} = \frac{~\bar{E}(\beta \!+\! s_{\theta}^{})\hsm +\hsm \ii\hs c_{\theta}^{}~}
{~\bar{E}(1 \!+\!\beta\hs s_{\theta}^{})~}\,,
\eeq
with modulus  $|\mathbb{P} |=1$.\
This only causes an overall phase factor to the scattering amplitude and does not affect physics.\
Hence the formulas of the polarization vector  $\epP^{\mu}$
in Eqs.\eqref{eq:polarization} and \eqref{eq.PVfromEOM} are physically equivalent.\
A similar overall-phase difference was found in the literature\,\cite{Emond:2025nxa}
(cf.\ its Appendix\,B).\
From the definitions in Eq.\eqref{eq:epL-epT} and below Eq.\eqref{eq:epP=epL+epT}, 
we note that the transverse polarization vector
$\epT^{\mu}$ or $\bepT^{\,\mu}$
is independent of the momentum and mass of the gauge boson,
so we may reexpress it in the spinor formulation
by using massless momenta:
\begin{equation}
\label{eq:epT-spinor}
\epT^{\mu}
= \frac{\,\langle p_0^{}| \gamma^{\mu} |q_0^{} \rangle\,}{\langle p_0^{}\hs q_0^{} \rangle} \,,
\end{equation}
where $p_0^{\mu}\!=\!p^{\mu}|_{m=0}^{}$ with $p_0^2\! =\! 0\,$, and $q_0^{}$ is a reference momentum obeying
$q_0^{2}\!=\!0\hs$.\
Without losing generality,
we can explicitly express massless momenta $(p_0^\mu,\hs q_0^\mu)$ as
$\hs p_0^\mu\!=\!E(1,\hs s_{\theta}^{},\hs c_{\theta}^{})\hs$
and $\hs q_0^\mu\!=\!E(1, -s_{\theta}^{}, -c_{\theta}^{})\hs$.\
The transverse polarization vector $\epT^{\mu}$ satisfies the normalization condition
$\,\epsilon_T^{\mu}\epsilon_{T\mu}^{*}\hsm\!=\!1\hs$.\
Then, using the spinor formula \eqref{eq:epT-spinor}, we compute the transverse polarization tensor
as follows:
\beq
\label{eq:epT-munu-sp}
\epT^{\mn} = \epT^{\mu}\epT^{\nu}
=
\frac{\,\langle p_0^{}| \gamma^{\mu}|q_0^{} \rangle \langle p_0^{}| \gamma^{\nu}|q_0^{} \rangle\,}
{\langle p_0^{}\hs q_0^{} \rangle^2}
=  \eta^{\mn} \!+\hsm
\frac{~p_0^\mu q_0^\nu \!+\hsm p_0^\nu q_0^\mu\,}{p_0^{\al}p_0^{\be}\delta_{\al\be}^{}}\,.
\eeq
Furthermore, using Eq.\eqref{eq:polarization},
we can define the physical graviton
$\hP\!\!=\!\epP^{\mn}h_{\mn}^{}$
with its polarization tensor $\epP^{\mn}$ expressed via the spinor formulation:
\beq
\label{eq:epP-munu-sp}
\epP^{\mn} = \epP^{\mu-}\!\epP^{\nu-} =
\frac{\,\langle p| \gamma^{\mu}|p \rangle\hsm
\langle p| \gamma^{\nu}|p\rangle\,}{8\hs p^2} \,,
\eeq
where we impose the on-shell condition $p^2\!=\!p_\mu^2\!=\!-m^2$.\
Since in the 3d TMG theory, the spin angular momentum is a pseudoscalar and
each massive graviton has only one physical degree of freedom with helicity either $-2$ or $+2$
\cite{Deser:1981wh},
we will choose the helicity $-2$ in our analysis without losing generality.\
In the above and hereafter, for simplicity, we will suppress the superscript ``$^-$''
of the polarization tensor $\epP^{\mn -}$
and of each helicity state of the physical graviton $h_{\PP}^-$
unless specified otherwise.\

\vs

Next, we verify that using the polarization vectors \eqref{eq:polarization}
can give the numerator of the gauge boson propagator in the Landau gauge
of the TMYM theory:
\begin{align}
\Delta^{\mu \nu} &= \epsilon_{\text{P}}^{\mu +}\epsilon_{\text{P}}^{\nu -}
= \frac{\,\langle \bar{p}| \gamma^{\mu} |\bar{p} \rangle \langle p| \gamma^{\nu} |p \rangle~}{8\hs p^{2}}
\nn\\
&= \frac{-p^{2}\eta^{\mu \nu}\!+\!2p^{\mu}p^{\nu} \!+\! \ii\hs m\hs \varepsilon^{\mu \nu \rho}p_{\rho}
	\!-\! \ii\hs m\big( \ii\hs m\eta^{\mu \nu}\!-\! \varepsilon^{\mu \nu \rho}p_{\rho}^{} \big)}{4\hs p^{2}}
\nn\\
&= -\frac{1}{\hs 2\hs}\!\(\! \eta^{\mu \nu} \!-\!\frac{\,p^{\mu}p^{\nu}\,}{p^{2}} \!-\!
\frac{\,\ii\hs m\hs\varepsilon^{\mu \nu \rho}p_{\rho}^{}\,}{p^{2}}\!\)\!,
\label{eq:propagator}
\end{align}
where we have used Eqs.\eqref{eq:gamma-algebraic-relation} and \eqref{eq:mass-p}.\
To derive the correct propagator \eqref{eq:propagator},
we impose the on-shell condition $p^2\!=\!-m^2$
for the numerator, but not for the denominator.\

\vspace*{2mm}
\section{\hspace*{-3.5mm} Feynman Rules for WTMG Theory and Scattering Amplitudes}
\label{app:B}
\label{app:Bnew}
\label{sec:appendix3}
\vspace*{1.5mm}

In this Appendix, 
we derive Feynman vertices for the cubic and quartic graviton
interactions and dilaton interactions.\
Then, we present some four-point graviton scattering amplitudes with unphysical polarizations
which are needed for the discussion in the main text.\
Finally, we explicitly compute the four-point physical graviton scattering amplitudes in
the unitary gauge of the WTMG theory in comparison with our Landau-gauge calculation
of the four graviton scattering amplitudes presented in the main text.

\vspace*{1.5mm}
\subsection{\hspace*{-2mm}Feynman Rules for the WTMG Theory}
\label{app:B1}
\vspace*{1.5mm}

In this part, we present Feynman vertices for the cubic and quartic graviton
interactions and dilaton interactions.\
These will be used to compute the three-point and four-point
graviton (dilaton) scattering amplitudes explicitly in Section\,\ref{sec:4.1}, 
and to verify the TGRET and double-copy results presented in Sections\,\ref{sec:4.1}-\ref{sec:4.2}
of the main text.\

\vs

We have derived the Feynman rules of the above WTMG Lagrangian \eqref{eq:LTMG-phi}
using the standard Feynman diagram approach.\
To verify the TGRET, we need the $h\phi\phi$ vertex to compute
three-point and four-point dilaton scattering amplitudes.\
After the Weyl transformation, we derive the relevant Lagrangian as follows:
\begin{equation}
\mathcal{L}_{h\phi\phi} \,\supset\,
\frac{1}{2}h_{\mu\nu}\partial^{\mu}\phi\partial^{\nu}\phi-\frac{1}{4}h\partial_{\mu}\phi\partial^{\mu}\phi
-\frac{1}{8}\partial_{\mu}\partial_{\nu}h^{\mu\nu}\phi^2+\frac{1}{8}\partial^2 h\hs\phi^2 .
\end{equation}
From the above, we derive the trilinear dilaton-dilaton-graviton vertex ($\phi\hs\phi\hs h$):
\begin{equation}
\label{Beq:scalar-scalar-graviton}
\ii\hs \mathcal{V}^{\mn}_{\phi\phi h} = \fr{\,\ii\hs\kappa\,}{4}\hsm
\bigg[ p_{3}^{\mu} p_{3}^{\nu}\!-\hsm p_{3}^2\hs\eta^{\mu\nu} \hsm\! -\hsm
2\big(p_1^{\mu} p_{2}^{\nu}\!+\!p_{1}^{\nu}p_{2}^{\mu}\big)
\!+\hsm 2(p_1^{}\!\cdot\hsm p_{2}^{})\eta^{\mu\nu}\bigg] .
\end{equation}
Throughout our analysis, we treat all momenta as incoming,
unless specified otherwise.\
Then, we present the relevant Feynman vertices containing
$\phi$-interactions:
\begin{subequations}
\begin{align}
\ii\hs \mathcal{V}_{3\phi}(p_1,p_2,p_3)
&=-\fr{\,\ii\hs\kappa\,}{2}
(p_1^{} \!\cdot\hsm  p_2^{} + p_1^{} \!\cdot\hsm p_3+p_2 \!\cdot\hsm p_3^{}\big) \hs,
\\
\ii\hs \mathcal{V}_{4\phi}^{}(p_1^{},p_2^{},p_3^{},p_4^{})
&=\frac{\,\ii\hs\kappa^2\hs}{4}(p_1^{}\!\cdot\hsm p_2^{}+p_1^{} \!\cdot\! p_3^{}
\!+\hsm p_1^{} \!\cdot\hsm p_4^{}\!+\hsm p_2^{} \!\cdot\hsm p_3^{}\!+\hsm p_2^{} \!\cdot\hsm p_4^{}
\!+\hsm p_3^{}\! \cdot\hsm p_4) \hs.
\end{align}
\end{subequations}

Using the symmetrization and permutations of the Lorentz indices of the graviton field,
we derive the following three-point graviton vertex,
{\small 
\begin{align}
\hspace{-10mm}
& \ii\hs \mathcal{V}^{\mu_1\nu_1,\mu_2\nu_2,\mu_3\nu_3}_{3h} (p_1^{},p_2^{},p_3^{})
\nn\\
&=\fr{\hs\ii\hs\kappa\hs}{2}\hs\rm{Sym}\Bigg\{\!\mathrm{P}
\Big\{\!\hsm +\!\Fr{1}{4}(p_1^{}\!\cdot\hsm p_2^{})\hs \eta^{\mu_1\nu_1}\eta^{\mu_2\nu_2}\eta^{\mu_3\nu_3}
\!-\!\Fr{1}{4}(p_1^{}\!\cdot\hsm p_2^{})\hs \eta^{\mu_1\mu_2}\eta^{\nu_1\nu_2}\eta^{\mu_3\nu_3}
\!+\hsm (p_1^{}\!\cdot\hsm p_2^{})\hs\eta^{\mu_1^{}\nu_1^{}}\eta^{\mu_2^{}\mu_3^{}}\eta^{\nu_2^{}\nu_3^{}}
\nn\\[-2mm]
& \hspace*{4mm}
-\!(p_1^{}\!\cdot\hsm p_2^{})\hs \eta^{\mu_1\nu_2}\eta^{\mu_2\nu_3}\eta^{\mu_3\nu_1}
\!-\!\Fr{1}{2}\hs p_1^{\mu_1} p_2^{\nu_1}\eta^{\mu_2\nu_2}\eta^{\mu_3\nu_3}
\!+\!\Fr{1}{2}\hs p_2^{\mu_1} p_1^{\mu_2}\eta^{\nu_1\nu_2}\eta^{\mu_3\nu_3}
\!-\hsm p_2^{\mu_1} p_1^{\mu_2}\eta^{\nu_1 \nu_3}\eta^{\mu_3\nu_2}
\nn\\[1mm]
& \hspace*{4mm}
-\!2\,p_3^{\mu _1} p_1^{\mu_2}\eta^{\nu_1 \nu_3}\eta^{\mu_3 \nu_2}
\!+\hsm p_2^{\mu_1} p_1^{\mu_3}\eta^{\nu_1\nu_3}\eta^{\mu_2\nu_2}
\!+\hsm p_2^{\mu_2} p_1^{\mu_3}\eta^{\nu_2\nu_3}\eta^{\mu_1\nu_1}
\!-\!\Fr{1}{2}\hs p_1^{\mu _3} p_2^{\nu _3} \eta^{\mu _1 \nu _1} \eta^{\mu _2 \nu _2}
\nn\\[1mm]
& \hspace*{4mm}
+\!\Fr{1}{2}\hs p_1^{\mu_3} p_2^{\nu_3}\eta^{\mu_1\mu_2}\eta^{\nu_1\nu_2^{}}
\!+\hsm p_1^{\mu _3} p_3^{\nu_3}\eta^{\mu_2\nu_1} \eta^{\mu_1\nu_2}
\!+\!\Fr{\ii}{\,2m\,}\!\Big[{\varepsilon}^{\mu_1\mu_2 p_1^{}}p_1^{\mu_3} p_2^{\nu_3}\eta^{\nu_1\nu_2}
\!+\hsm {\varepsilon }^{\mu _1 \mu _2 p_2}p_1^{\mu _3} p_2^{\nu _1} \eta^{\nu _2 \nu _3}
\nn\\[1mm]
& \hspace*{4mm}
-\!{\varepsilon }^{\mu_1\mu_2 p_1}p_1^{\mu_3}p_2^{\nu_1}\eta^{\nu_2\nu_3}\!+\hsm {\varepsilon}^{\mu_1\mu_2 p_1}
(p_1 \!\cdot\hsm p_2)\eta^{\nu_2\nu_3}\eta^{\mu_3\nu_1}
\!+\hsm {\varepsilon}^{\mu_1 p_2 p_3} p_1^{\nu_2}\eta^{\mu_3\nu_1} \eta^{\mu_2\nu _3}
\nn\\[-2mm]
& \hspace*{4mm}
+{\varepsilon}^{\mu_1\mu_2\mu_3}(p_1 \!\cdot\hsm p_2) p_3^{\nu_1}\eta^{\nu_2\nu_3}
\!-\hsm \Fr{1}{3}\hs{\varepsilon }^{\mu _1 \mu _2 \mu _3}p_3^{\nu _1} p_1^{\nu _2} p_2^{\nu_3} \hsm\Big]\hsm\Big\}\!\Bigg\} \hs ,
\label{Beq:V3h}
\end{align}
}
and the four-point graviton vertex,
%
{\small 
\begin{align}
& \ii\hs \mathcal{V}^{\mu_1\nu_1,\mu_2\nu_2,\mu_3\nu_3,\mu_4\nu_4}_{4h} (p_1,p_2,p_3,p_4)
\nn\\[-1mm]
& =  \fr{\,\ii\hs\kappa^2\hs}{2}\hs \mathrm{Sym}\Bigg\{\!\mathrm{P}\Big\{\!\hsm
+\!(p_3 \!\cdot\hsm p_4)\hs \eta^{\mu_1\mu_2}\eta^{\nu_2\mu_4}\eta^{\nu_1\nu_4}\eta^{\mu_3\nu_3}
\!-\!\Fr{1}{2}(p_3 \!\cdot\hsm p_4)\hs \eta^{\mu_1\nu_1}\eta^{\mu_2\mu_4}\eta^{\nu_2\nu_4}\eta^{\mu_3\nu_3}
\nn\\[-1mm]
& \hspace*{4mm}
-\!\Fr{1}{8}(p_3 \!\cdot\hsm p_4)\hs \eta^{\mu_1\mu_2}\eta^{\nu_1\nu_2}\eta^{\mu_3\nu_4}\eta^{\mu_4\nu_4}
\!+\!\Fr{1}{16}(p_3 \!\cdot\hsm p_4)\hs \eta^{\mu_1\nu_1}\eta^{\mu_2\nu_2}\eta^{\mu_3\nu_3}\eta^{\mu_4\nu_4}
\nn\\[1mm]
& \hspace*{4mm}
+\! \Fr{1}{2}(p_{3} \!\cdot\hsm p_{4})\hs \eta^{\mu_1\mu_4}\eta^{\nu_1\nu_4}\eta^{\mu_2\mu_3}\eta^{\nu_2\nu_3}
\!-\!\Fr{1}{2}(p_3 \!\cdot\hsm p_4)\hs \eta^{\mu_1\mu_4}\eta^{\nu_1\mu_3}\eta^{\mu_2\nu_4}\eta^{\nu_2\nu_3}
\nn\\[1mm]
& \hspace*{4mm}
-\!(p_{3} \!\cdot\hsm p_{4})\hs \eta^{\mu_1\mu_2}\eta^{\nu_2\mu_4}\eta^{\nu_1\mu_3}\eta^{\nu_3\nu_4}
\!+\!\Fr{1}{2}(p_{3} \!\cdot\hsm p_{4})\hs \eta^{\mu_1\nu_1}\eta^{\mu_2\mu_4}\eta^{\nu_2\mu_3}\eta^{\nu_3\nu_4}
\nn\\[1mm]
& \hspace*{4mm}
+\!\Fr{1}{8}(p_{3} \!\cdot\hsm p_{4})\hs \eta^{\mu_1\mu_2}\eta^{\nu_1\nu_2}\eta^{\mu_3\mu_4}\eta^{\nu_3\nu_4}
\!-\!\Fr{1}{16}(p_{3} \!\cdot\hsm p_{4})\hs \eta^{\mu_1\nu_1}\eta^{\mu_2\nu_2}\eta^{\mu_3\mu_4}\eta^{\nu_3\nu_4}
\nn\\[1mm]
& \hspace*{4mm}
+ p_{3}^{\mu_1}p_{4}^{\mu_2}\eta^{\nu_1\mu_4} \eta^{\nu_2\mu_3}\eta^{\nu_3\nu_4}
\!-\!\Fr{1}{2}\hs p_{3}^{\mu_2}p_{4}^{\mu_1} \eta^{\nu_1\nu_2}\eta^{\mu_3\mu_4}\eta^{\nu_3\nu_4}
\!+\!\Fr{1}{4}\hs p_{3}^{\mu_2}p_{4}^{\nu_2} \eta^{\mu_1\nu_1}\eta^{\mu_3\mu_4}\eta^{\nu_3\nu_4}
\nn\\[1mm]
& \hspace*{4mm}
+\!\Fr{1}{2}\hs p_{3}^{\mu_2}p_{4}^{\mu_1} \eta^{\nu_1\nu_2}\eta^{\mu_3\nu_3}\eta^{\mu_4\nu_4}
\!-\!\Fr{1}{4}\hs p_{3}^{\mu_2}p_{4}^{\nu_2} \eta^{\mu_1\nu_1}\eta^{\mu_3\nu_3}\eta^{\mu_4\nu_4}
\!-\!p_{3}^{\mu_1}p_{4}^{\mu_4} \eta^{\nu_1\mu_2}\eta^{\nu_2\nu_4}\eta^{\mu_3\nu_3}
\nn\\[1mm]
& \hspace*{4mm}
+\!\Fr{1}{2}\hs p_{3}^{\mu_2}p_{4}^{\mu_4} \eta^{\mu_1\nu_1}\eta^{\nu_2\nu_4}\eta^{\mu_3\nu_3}
\!-\!p_{3}^{\mu_2}p_{4}^{\nu_2} \eta^{\mu_1\mu_3}\eta^{\nu_1\mu_4}\eta^{\nu_3\nu_4}
\!-\!p_{3}^{\mu_1}p_{4}^{\mu_2} \eta^{\nu_1\mu_3}\eta^{\nu_2\nu_3}\eta^{\mu_4\nu_4}
\nn\\[1mm]
& \hspace*{4mm}
+ p_{3}^{\mu_2}p_{4}^{\nu_2} \eta^{\mu_1\mu_3}\eta^{\nu_1\nu_3}\eta^{\mu_4\nu_4}
\!-\hsm p_{3}^{\mu_1}p_{4}^{\mu_3} \eta^{\nu_1\mu_2}\eta^{\nu_2\nu_3}\eta^{\mu_4\nu_4}
\!+\!\Fr{1}{2}\hs p_{3}^{\mu_2}p_{4}^{\mu_3} \eta^{\mu_1\nu_1}\eta^{\nu_2\nu_3}\eta^{\mu_4\nu_4}
\nn\\[1mm]
& \hspace*{4mm}
- p_{3}^{\mu_2}p_{4}^{\mu_4} \eta^{\mu_1\mu_3}\eta^{\nu_1\nu_3}\eta^{\nu_2\nu_4}
\!-\!p_{3}^{\mu_4}p_{4}^{\nu_4} \eta^{\mu_1\mu_2}\eta^{\nu_2\mu_3}\eta^{\nu_1\nu_3}
\!+\!\Fr{1}{2}\hs p_{3}^{\mu_4}p_{4}^{\nu_4} \eta^{\mu_1\nu_1}\eta^{\mu_2\mu_3}\eta^{\nu_2\nu_3}
\nn\\[1mm]
& \hspace*{4mm}
+\!\Fr{1}{4}\hs p_{3}^{\mu_4}p_{4}^{\nu_4} \eta^{\mu_1\mu_2}\eta^{\nu_1\nu_2}\eta^{\mu_3\nu_3}
\!-\!\Fr{1}{8}\hs p_{3}^{\mu_4}p_{4}^{\nu_4} \eta^{\mu_1\nu_1}\eta^{\mu_2\nu_2}\eta^{\mu_3\nu_3}
\!-\! p_{3}^{\mu_2}p_{4}^{\mu_3} \eta^{\mu_1\mu_4}\eta^{\nu_1\nu_4}\eta^{\nu_2\nu_3}
\nn\\[1mm]
& \hspace*{4mm}
+\hsm 2\,p_{3}^{\mu_2}p_{4}^{\mu_3} \eta^{\mu_1\mu_4}\eta^{\nu_1\nu_3}\eta^{\nu_2\nu_4}
\!+\!2\hs p_{3}^{\mu_1}p_{4}^{\mu_3}\eta^{\nu_1\mu_2}\eta^{\nu_2\mu_4}\eta^{\nu_3\nu_4}
\!-\!p_{3}^{\mu_2}p_{4}^{\mu_3}\eta^{\mu_1\nu_1}\eta^{\nu_2\mu_4}\eta^{\nu_3\nu_4}
\nn\\[1mm]
& \hspace*{4mm}
+p_{3}^{\mu_4}p_{4}^{\mu_3}\eta^{\mu_1\mu_2}\eta^{\nu_2\nu_4}\eta^{\nu_1\nu_3}
\!-\!\Fr{1}{2}\,p_{3}^{\mu_4}p_{4}^{\mu_3}\eta^{\mu_1\nu_1}\eta^{\mu_2\nu_4}\eta^{\nu_2\nu_3}
\!-\!\Fr{1}{4}\hs p_{3}^{\mu_4}p_{4}^{\mu_3}\eta^{\mu_1\mu_2}\eta^{\nu_1\nu_2}\eta^{\nu_3\nu_4}
\nn\\[1mm]
& \hspace*{4mm}
+\!\Fr{1}{8}\hs p_{3}^{\mu_4}p_{4}^{\mu_3}\eta^{\mu_1\nu_1}\eta^{\mu_2\nu_2}\eta^{\nu_3\nu_4}
\!+\!\Fr{\ii}{\,2\hs m\,}\hsm \Big[\varepsilon^{\mu_2\mu_3\mu_4}(p_3 \!\cdot\hsm p_4)\hs p_{2}^{\mu_1}\eta^{\nu_1\nu_4}\eta^{\nu_2\nu_3}
\!+\hsm\varepsilon^{\mu_2\mu_4\mu_3}(p_3 \!\cdot\hsm p_4)p_{2}^{\nu_3}\eta^{\mu_1\nu_2}\eta^{\nu_1\nu_4}
\nn\\[1mm]
& \hspace*{4mm}
+\hsm \varepsilon^{\mu_3 p_2 p_4}p_{3}^{\mu_4}\eta^{\mu_1\mu_2}\eta^{\nu_1\nu_4}\eta^{\nu_2\nu_3}
\!-\hsm \varepsilon^{\mu_2 p_3 p_4}p_{2}^{\mu_1}\eta^{\nu_1\mu_3}\eta^{\nu_3\mu_4}\eta^{\nu_2\nu_4}
\!-\hsm \varepsilon^{\mu_3 p_4 p_2}p_{3}^{\mu_4}\eta^{\mu_1\mu_2}\eta^{\nu_1\nu_3}\eta^{\nu_2\nu_4}
\nn\\[1mm]
& \hspace*{4mm}
-\hsm \varepsilon^{\mu_2\mu_4\mu_3}p_{2}^{\mu_1}p_{3}^{\nu_1}p_{4}^{\nu_3}\eta^{\nu_2\nu_4}
\!+\hsm \varepsilon^{\mu_2\mu_3\mu_4}p_{2}^{\mu_1}p_{3}^{\nu_4}p_{4}^{\nu_2}\eta^{\nu_1\nu_3}
\!-\hsm \varepsilon^{\mu_3\mu_4 p_4}(p_3 \!\cdot\hsm p_4)\hs \eta^{\mu_1\mu_2}\eta^{\nu_2\nu_3}\eta^{\nu_1\nu_4}
\nn\\[1mm]
& \hspace*{4mm}
+\hsm \varepsilon^{\mu_3\mu_4 p_4}p_{3}^{\mu_2}p_{4}^{\mu_1}\eta^{\nu_1\nu_3}\eta^{\nu_2\nu_4}
\!-\hsm \varepsilon^{\mu_3\mu_4 p_4}p_{3}^{\mu_2}p_{4}^{\mu_1}\eta^{\nu_1\nu_2}\eta^{\nu_3\nu_4}
\!+\hsm \varepsilon^{\mu_3\mu_4 p_4}p_{3}^{\nu_4}p_{4}^{\mu_1}\eta^{\nu_1\mu_2}\eta^{\nu_2\nu_3}
\nn\\[-2mm]
& \hspace*{4mm}
-\hsm \varepsilon^{\mu_3\mu_4 p_4}p_{3}^{\mu_2}p_{4}^{\nu_2}\eta^{\mu_1\nu_3}\eta^{\nu_1\nu_4}
\!+\hsm \varepsilon^{\mu_3\mu_4 p_4}p_{3}^{\mu_2}p_{4}^{\nu_3}\eta^{\mu_1\nu_2}\eta^{\nu_1\nu_4}\Big]\!\Big\}\!\Bigg\},
\label{Beq:V4h}
\end{align}
}
\hspace*{-3.5mm}
where ``P'' represents a summation over all permutations of graviton indices and ``Sym'' denotes a symmetrization
on each pair of Lorentz indices $\mu_i^{}\nu_i^{}$.\
In the above, we have used the {\sf Schoonschip} notation for simplicity, i.e.,
$\varepsilon^{\alpha\beta p}\hsm\equiv \varepsilon^{\alpha\beta\gamma}p_{\gamma}$.\

\vs

Making the Weyl transformation \eqref{eq:CT-gmunu} and expanding the Lagrangian
for the trilinear terms
of $hh\phi\hs$,
we derive the following graviton-graviton-dilaton interaction Lagrangian,
{\small 
\begin{align}
\mathcal{L}_{hh\phi} \,=\, &+\!\frac{3}{8}\partial_{\lambda}h_{\mu\nu}\partial^{\lambda}h^{\mu\nu}\phi-\!\frac{1}{4}\partial_{\mu}h_{\lambda\nu}\partial^{\nu}h^{\lambda\mu}\phi+\!\frac{1}{2}\partial_{\mu}h\hs\partial_{\nu}h^{\mu\nu}\phi-\!\frac{1}{2}\partial_{\lambda}h^{\lambda\mu}\partial^{\nu}h_{\mu\nu}\phi-\!\frac{1}{8}\partial_{\lambda}h\hs\partial^{\lambda}h\hs\phi
\nn\\
&-\!\frac{1}{4}h\hs\partial^2 h\hs\phi+\!\frac{1}{4}h\hs\partial_{\mu}\partial_{\nu}h^{\mu\nu}\phi+\!\frac{1}{2} h^{\mu\nu}\partial^2 h_{\mu\nu}\phi-\! h^{\mu\nu}\partial_{\mu}\partial^{\lambda}h_{\lambda\nu}\phi+\! \frac{1}{2}h^{\mu\nu}\partial_{\mu}\partial_{\nu}h\hs\phi
\end{align}
}
\hspace*{-3mm}
and the corresponding $hh\phi\hs$ Feynman vertex in the momentum space,
{\small 
\begin{align}
& \ii\hs \mathcal{V}_{hh\phi}^{\mu_1\nu_1,\mu_2\nu_2}(p_1,p_2,p_3)
\nn\\
&=-\fr{\,\ii\hs\kappa\hs}{2}\,\mathrm{Sym}_{2}^{}\bigg\{\mathrm{P}_{2}^{}\hsm\Big[\hsm\!+\!\Fr{3}{4}
(p_1\!\cdot\hsm p_2)\hs\eta^{\mu_1\mu_2}\eta^{\nu_1\nu_2}
\!-\!\Fr{1}{2}\hs p_1^{\nu_2}p_2^{\nu_1}\eta^{\mu_1 \mu_2}
\!+\hsm p_1^{\mu_2}p_2^{\nu_2}\eta^{\mu_1 \nu_1}
\!-\hsm p_1^{\mu_1}p_2^{\mu_2}\eta^{\nu_1\nu_2}
\nn\\
& \hspace*{4mm}
-\!\Fr{1}{4}(p_1 \!\cdot\hsm p_2)\hs \eta^{\mu_1\nu_1}\eta^{\mu_2\nu_2}
\!-\!\Fr{1}{2}\hs p_1^2\eta^{\mu_1\nu_1}\eta^{\mu_2\nu_2}
\!+\!\Fr{1}{2}p_1^{\mu_1}p_1^{\nu_1}\eta^{\mu_2\nu_2} \!+\hsm p_1^2\eta^{\mu_1\mu_2}\eta^{\nu_1\nu_2}
\nn\\
& \hspace*{4mm}
-\!2\hs p_1^{\nu_1}p_1^{\nu_2}\eta^{\mu_1\mu_2} \!+\hsm p_1^{\mu_2}p_1^{\nu_2}\eta^{\mu_1\nu_1}\hsm\Big]\!\bigg\} \hs ,
\label{Beq:Vhhphi}
\end{align}
%
}
\hspace*{-3mm}
where the symbols Sym$_2$ stand for the symmetrization of Lorentz indices of the two gravitons in the vertex, and P$_2$ denotes the summation of permutations
of Lorentz indices of the two graviton fields.\

\vs 

In addition, from the TMGS Lagrangian \eqref{eq:L-TMGS-UG}
or the WTMGS Lagrangian \eqref{eq:L-WTMGS},
we extract the following $\psi\psi h\hs$ trilinear interaction Lagrangian, 
\begin{equation}
\label{Beq:h-psi-psi}
\mathcal{L}_{\psi\psi h} \,=\,
-\frac{1}{\hs 2\hs}h_{\mu\nu}\partial^{\mu}\psi\partial^{\nu}\psi
\hsm +\!\frac{1}{\hs 4\hs}h\partial_{\mu}\psi\partial^{\mu}\psi 
\hsm +\!\frac{1}{\hs 4\hs}m_s^2 h\psi^2 \hs,
\end{equation}
and derive the corresponding $\psi\psi h\hs$ trilinear coupling 
of as follows:
\beq
\begin{aligned}
\label{Beq:WTMGS-h-2psi} 
\ii\hs\mathcal{V}^{\mn}_{\psi\psi h} &=
\frac{\,\ii\hs\kappa\,}{2}
\!\hsm\(\hsm p_{1}^{\mu}p_{2}^{\nu}+p_{1}^{\nu}p_{2}^{\mu}
\hsm -\hsm\eta^{\mu\nu}p_{1}\!\cdot\hsm p_{2}\hsm +\hsm\eta^{\mu \nu}m_s^2\hsm\)\hsm . 
\end{aligned}
\eeq
For the physical scalar $\psi$ being massless, 
the above $\psi\psi h\hs$ vertex reduces to the following forms:
\beq
\label{Beq:WTMGS0-h-2psi}
\ii\hs\mathcal{V}^{\mn (0)}_{\psi\psi h} = 
\frac{\,\ii\hs\kappa\,}{2}\hsm 
\!\hs\(\hs p_{1}^{\mu}p_{2}^{\nu}+p_{1}^{\nu}p_{2}^{\mu}
\!-\!\eta^{\mu \nu}p_1^{}\!\cdot\hsm p_2^{}\) \hsm .
\eeq
Then, making the field redefinitions \eqref{eq:transf-phi-psi},
we change from the field basis $(\phi,\psi)\ito (\phih,\psih)$
and find that $\psih\psih h$ trilinear interaction Lagrangian
takes the same form as Eq.\eqref{Beq:h-psi-psi},
\begin{equation}
\label{Beq:h-psih-psih}
\mathcal{L}_{\psih\psih h} \,=\,
-\frac{1}{\hs 2\hs}h_{\mu\nu}\partial^{\mu}\psih\hs\partial^{\nu}\psih
\hsm +\!\frac{1}{\hs 4\hs}h\partial_{\mu}\psih\hs\partial^{\mu}\psih 
\hsm +\!\frac{1}{\hs 4\hs}m_s^2 h\psih^2 \hs.
\end{equation}
Hence, it gives the same graviton-scalar-scalar
trilinear coupling ($\psih\psih h$) as in Eq.\eqref{Beq:WTMGS-h-2psi}, 
\beq 
\label{Beq:WTMGS-V(h2psih)=V(h2psi)}
\mathcal{V}^{\mn}_{\psih\psih h} \,=\,
\mathcal{V}^{\mn}_{\psi\psi h} \,.
\eeq 

Finally, for the four-body elastic scattering amplitudes 
(with all external states having the same mass $m\hs$)
in the center-of-mass frame of the particles\,1 and 2,
we define momenta of the external states as follows:
\\[-7mm]
\beq
\label{eq:p1-4-CMF}
\begin{split}
	p_{1}^{\mu} &= E(1,0,\beta), \hspace*{11.1mm}
	p_{3}^{\mu} = -E(1,\beta \st,\beta\ct),
	\\
	p_{2}^{\mu} &= E(1,0,-\beta),
	\hspace*{8mm}
	p_{4}^{\mu} = -E(1,-\beta \st,-\beta \ct ),
\end{split}
\eeq
where $\beta\!=\!\sqrt{1\!-\!m^2\!/E^2\,}$ and
$(s_\theta^{},\hs c_\theta^{})\!=\!(\sin\hsm\theta,\hs \cos\hsm\theta)$.\

\vspace*{2mm}
\subsection{\hspace*{-2mm}Some Unphysical Graviton Amplitudes in the TMG Theory}
\label{app:B2}
\vspace*{1.5mm}

In this Appendix, we present some four-point unphysical scattering amplitudes 
by choosing the gauge parameters $(\xi,\zeta)\!=\!(0,\infty)$ in the WTMG theory,
where $\zeta\!=\!\infty$ corresponds to the unitary gauge and the theory reduces 
to the conventional TMG theory (without dilaton field)\,\cite{Deser:1981wh}.\
This is to show that without introducing the dilaton field, the scattering amplitudes
of the unphysical components of gravitons do not quantitatively mimic the high-energy
behaviors of the physical graviton scattering amplitudes, and thus could not realize
the TGRET.\  Hence, it is essential to use the WTMG theory (including
the dilaton field) for the formulation of the TGRET as in Section\,\ref{sec:3.1}.\  
 
\vs 

We first compute the four-point scattering amplitude of the transverse gravitons
($\hTT\!=\!\epT^{\mn}h_{\mn}^{}$) 
at the tree level in the TMG theory, where the graviton's transverse polarization tensor
is given by $\epT^{\mu\nu}\!\!=\hsm\epT^{\mu}\epT^{\nu}$
as defined in Eq.\eqref{eq:epS-epT-munu}.\  
By taking lengthy calculations of the contact diagram and graviton exchanges in the
$(s,t,u)$-channels, we derive the following compact expressions:
{\small 
\begin{subequations}
\begin{align}
\hspace*{-2mm}
\MM_{c}[4h_{\rm{T}}]= &
-\!\frac{\,\kappa^2 m^2 \bar{s}^{\frac{1}{2}}\,}{32}\!\hsm 
\left[\ii\hs\bar{s}\hs(14\hs s_{2\theta}^{}\!+\!s_{4\theta}^{})\!-\!\bar{s}^{\frac{1}{2}}(2\hs c_{2\theta}^{}\!+\!c_{4\theta}^{}\!-\!3)\!-\!\ii\hs 32\hs s_{2\theta}^{}\right]\!, 
\\
\hspace*{-2mm}
\MM_{s}[4h_{\rm{T}}]= &\,
\frac{\kappa^2 m^2}{\,16(\bar{s}\!-\!1)\,}\!
\left[\ii\hs\bar{s}^{\frac{1}{2}}(4\hs\bar{s}^{2}\!\!-\!11\hs\bar{s}\!+\!8)s_{2\theta}^{}\!-\!(3\hs\bar{s}\!-\!4)c_{2\theta}^{}\!+\!\bar{s}\hs(\bar{s}\!-\!1)\right]\!, 
\\
\hspace*{-2mm}
\MM_{t}[4h_{\rm{T}}] =&\, 
\frac{\ii\hs\kappa^2 m^2}{\,1024(\bar{s}\!-\!4)[(2\!-\!\bar{s})\!+\!(4\!-\!\bar{s})c_{\theta}^{}]\,}
\Big[\hsm\!-\!\bar{s}^{\frac{1}{2}}(\bar{s}\!-\!4)^3 s_{6\theta}^{}\!-\!2\hs\bar{s}^{\frac{1}{2}}(\bar{s}\!-\!4)^2(5\bar{s}\!-\!2)s_{5\theta}^{}
\!-\!4\hs\bar{s}^{\frac{1}{2}}
\nn\\
& \times\hsm\!(11\hs\bar{s}^3\!-\!75\hs\bar{s}^2\!+\!172\hs\bar{s}\!-\!192)s_{4\theta}^{}
\!-\!2\hs\bar{s}^{\frac{1}{2}}(55\hs\bar{s}^3\!-\!514\hs\bar{s}^2\!+\!1616\hs\bar{s}\!-\!1120)
s_{3\theta}^{}\!-\!3\hs\bar{s}^{\frac{1}{2}}
\nn\\
&
\times\hsm\!(55\hs\bar{s}^3\!-\!620\hs\bar{s}^2\!+\!1968\hs\bar{s}\!-\!2496)s_{2\theta}^{}
\!-\!12\hs\bar{s}^{\frac{1}{2}}(11\hs\bar{s}^3\!-\!124\hs\bar{s}^2\!+\!520\hs\bar{s}\!-\!608)
s_{\theta}^{}
\nn\\
&
-\!\ii\hs 8\hs\bar{s}\hs(\bar{s}\!-\!4)^2 c_{5\theta}^{}\!-\!\ii\hs\bar{s}\hs(33\hs\bar{s}^2
\!-\!32\hs\bar{s}\!-\!400) c_{4\theta}^{}\!-\!\ii\hs 16\hs\bar{s}\hs(2\hs\bar{s}^2\!-\!11\hs\bar{s}\!+\!52) c_{3\theta}^{}\!+\!\ii\hs 4\hs\bar{s}\hs (\bar{s}^2
\nn\\
& 
+\!88\hs\bar{s}\!+\!16) c_{2\theta}^{}
\!-\!\ii\hs 8\hs\bar{s}\hs(3\hs\bar{s}^2\!-\!34\hs\bar{s}
\!+\!264)c_{\theta}^{}\!-\!\ii\hs(35\hs\bar{s}^3\!-\!128\hs\bar{s}^2\!-\!48\hs\bar{s}\!-\!2048)\Big]
\hsm ,
\\
\hspace*{-2mm}
\MM_{u}[4h_{\rm{T}}]=&\,
\frac{\ii\hs\kappa^2 m^2}{\,1024(\bar{s}\!-\!4)[(2\!-\!\bar{s})\!-\!(4\!-\!\bar{s})c_{\theta}^{}]\,}
\Big[\hsm\!-\!\bar{s}^{\frac{1}{2}}(\bar{s}\!-\!4)^3 s_{6\theta}^{}\!+\!2\hs\bar{s}^{\frac{1}{2}}(\bar{s}\!-\!4)^2(5\bar{s}\!-\!2)s_{5\theta}^{}
\!-\!4\hs\bar{s}^{\frac{1}{2}}
\nn\\
&
\times\hsm\!(11\hs\bar{s}^3\!-\!75\hs\bar{s}^2\!+\!172\hs\bar{s}\!-\!192)s_{4\theta}^{}\!+\!2\hs\bar{s}^{\frac{1}{2}}(55\hs\bar{s}^3\!-\!514\hs\bar{s}^2\!+\!1616\hs\bar{s}\!-\!1120)s_{3\theta}^{}
\!-\!3\hs\bar{s}^{\frac{1}{2}}
\nn\\
&
\times\hsm\!(55\hs\bar{s}^3\!-\!620\hs\bar{s}^2
\!+\!1968\hs\bar{s}\!-\!2496)s_{2\theta}^{}\!+\!12\hs\bar{s}^{\frac{1}{2}}(11\hs\bar{s}^3\!-\!124\hs\bar{s}^2\!+\!520\hs\bar{s}\!-\!608)s_{\theta}^{}
\nn\\
&
+\!\ii\hs 8\hs\bar{s}\hs(\bar{s}\!-\!4)^2 c_{5\theta}^{}\!-\!\ii\hs\bar{s}\hs(33\hs\bar{s}^2
\!-\!32\hs\bar{s}\!-\!400) c_{4\theta}^{}\!+\!\ii\hs 16\hs\bar{s}\hs(2\hs\bar{s}^2\!-\!11\hs\bar{s}\!+\!52) c_{3\theta}^{}\!+\!\ii\hs 4\hs\bar{s}\hs(\bar{s}^2
\nn\\
&
+\!88\hs\bar{s}\!+\!16) c_{2\theta}^{}\!+\!\ii\hs 8\hs\bar{s}\hs(3\hs\bar{s}^2\!-\!34\hs\bar{s}
\!+\!264)c_{\theta}^{}\!-\!\ii\hs(35\hs\bar{s}^3\!-\!128\hs\bar{s}^2\!-\!48\hs\bar{s}\!-\!2048)\Big]
\hsm .
\end{align}
\end{subequations}
}
Summing up the above sub-amplitudes, we derive the full scattering amplitude of $4\hTT$
as follows:
{\small 
\begin{align}
\MM[4h_{\rm{T}}] =&
\frac{\kappa^2 m^2}
{\,512(\bar{s}\!-\!1)(\bar{s}\!-\!4)[(2\!-\!\bar{s})^2\!-\!(4\!-\!\bar{s})^2 c_{\theta}^{2}]\,}
\!\Big[\ii\hs 8\hs\bar{s}^{\frac{1}{2}}(10\hs\bar{s}^4\!-\!82\hs\bar{s}^3\!+\!159\hs\bar{s}^2\!+\!68\hs\bar{s}
\!-\!128)s_{4\theta}^{}
\nn\\
&
-\!\ii\hs 16\hs\bar{s}^{\frac{3}{2}}(10\hs\bar{s}^3\!-\!2\hs\bar{s}^2\!-\!97\hs\bar{s}\!+\!68)s_{2\theta}^{}\!-
\!(13\hs\bar{s}^5\!+\!\bar{s}^4\!-\!566\hs\bar{s}^3\!+\!1264\hs\bar{s}^2\!+\!1120\hs\bar{s}\!-\!2048)
c_{4\theta}^{}
\nn\\
&
+\!4\hs(13\hs\bar{s}^5\!-\!51\hs\bar{s}^4\!+\!538\hs\bar{s}^3\!-\!1856\hs\bar{s}^2\!+\!928\hs\bar{s}
\!+\!512)c_{2\theta}^{}\!-\!(39\hs\bar{s}^5\!-\!205\hs\bar{s}^4\!-\!610\hs\bar{s}^3
\nn\\
&
+\!1552\hs\bar{s}^2\!+\!5152\hs\bar{s}\!-\!6144)\Big]\hsm .
\label{eq:Amp-4hT}
\end{align}
}
Under high energy expansion, we derive the following expression: 
\beq
\label{eq:Amp-4hT-exp}
\begin{aligned}
\MM[4h_{\rm{T}}] =
-\frac{\,13\hs\kappa^2 s\,}{64} s_{\theta}^2 
\!-\!\frac{\,\ii\hs 5\hs\kappa^2 m\hs{s}^{\frac{1}{2}}\,}{8} s_{2\theta}^{}
\!+\!\frac{\,\kappa^2 m^2\,}{64}(85\hs c_{2\theta}^{}\!+\!19) + {O}({s}^{-\frac{1}{2}})\hs. 
\end{aligned}
\eeq
We see that the above $4\hTT$ scattering amplitude has the leading contribution
of $O({s}^1)\hs$, which differs from the leading-order contribution of $O(s^{\frac{1}{2}})$
in the four physical graviton ($4\hP$) scattering amplitude 
\eqref{eq:Amp-4hp-LO+NLO}.\  

\vs 

Next, we compute the four-point scattering amplitude of $h\,(=\!\eta^{\mn}h^{}_{\mn})$ 
at the tree level in the TMG theory: 
\beqs 
\label{eq:Amp-4h-cstu}
\begin{align}
\MM_{c}[4h]=&-\!\frac{15}{2}\hs\kappa^2 m^2\hs, 
\\
\MM_{s}[4h]=&\frac{\,\kappa^2 m^2\,}{4\hs\bar{s}}(\bar{s}\!+\!4)^2 \hs ,
\\
\MM_{t}[4h]=&
-\!\frac{\kappa^2 m^2}{8(\bar{s}\!-\!4)(1\!+\!c_{\theta}^{})}
[(\bar{s}\!-\!4)c_{\theta}^{}\!+\!(\bar{s}\!-\!12)]^2\hs ,
\\
\MM_{u}[4h]=&
-\!\frac{\kappa^2 m^2}{8(\bar{s}\!-\!4)(1\!-\!c_{\theta}^{})}
[(\bar{s}\!-\!4)c_{\theta}^{}\!-\!(\bar{s}\!-\!12)]^2\hs . 
\end{align}
\eeqs 
With these sub-amplitudes, we derive the full $4h$ scattering amplitude as follows:
\beq
\label{eq:Amp-4h}
\MM[4h]=\frac{\,\kappa^2 m^2\,}{\,4\hs\bar{s}\hs(\bar{s}\!-\!4)\,}\!\hsm 
\left[(\bar{s}^2\!-\!12\hs\bar{s}\!+\!32)c_{2\theta}^{}\!-\!(\bar{s}^2\!+\!52\hs\bar{s}\!+\!32)
\right]\!\csc^{2}\!\theta \,.
\eeq
Under high-energy expansion, we derive the following expression:
\beq
\label{eq:Amp-4h-exp}
\MM[4h]=\!
-\frac{1}{2}\hs\kappa^2 m^2\!-\!2\hs\kappa^2 m^2\bar{s}^{-1}\hsm 
(c_{2\theta}^{}\!+\!7)\hsm\csc^2\!\theta \hsm +\hsm {O}(\bar{s}^{-2}) \,,
\eeq
which contains the leading-order contribution of $O(\bar{s}^{\,0})\hs$.\ 
The above leading-order amplitude of $4h$ scattering differs from
the leading-order contribution of $O(s^{\frac{1}{2}})$ 
in the four-physical-graviton ($4\hP$) scattering amplitude 
\eqref{eq:Amp-4hp-LO+NLO}.\

\vspace*{2mm}
\subsection{\hspace*{-2mm}Four Graviton Scattering Amplitude in Unitary Gauge}
\label{app:B3}
\vspace*{1.5mm}

In Section\,\ref{sec:4.1.2} of the main text, we have computed the four-point physical graviton
scattering amplitude \eqref{eq:app-Amp-4hp}-\eqref{eq:app-Amp-4hp-Y} in Landau gauge
($\xi\!=\!\zeta\!=\!0$) of the WTMG theory, 
which uses the Landau-gauge graviton propagator \eqref{eq:DhDphi-Landau}
for the graviton-exchange diagrams and includes additional dilaton-exchange diagrams 
in the $(s,t,u)$-channels.\ 
As a nontrivial consistency check, we compute in this subsection the four-point
physical graviton scattering amplitude ($4\hP$) in unitary gauge 
($\zeta\!=\!\infty$)
including no dilaton-exchange diagrams and using the unitary-gauge graviton propagator
\eqref{eq:Dh-unitary} (with also $\xi\!=\!0$) [which contains extra complicated 
terms beyond the Landau-gauge propagator \eqref{eq:DhDphi-Landau}].\ 
We demonstrate that the physical graviton scattering amplitude takes the same form  
in both Landau and unitary gauges, and is thus gauge-invariant.\  
Through lengthy calculations, we finally derive the following 
sub-amplitudes in the unitary gauge, 
including the contact contribution and the graviton-exchange 
contributions via $(s,t,u)$ channels,
{\small 
\begin{subequations}
\label{eq:Amp-4hp-cstu-unitary}
\begin{align}
			\hspace*{-17mm}
			\MM_{c}[4\hP]
			= & \frac{\,\kappa^2m^2\,}{\,131072\,}
			\Big\{\!\!-\!(65 \bar{s}\hsm +\hsm 424) (\bar{s}\!-\!4)^4
			\!+\! 12\hs (5\bar{s}^3\!+\!48\bar{s}^2\!-\!368\bar{s}\!-\!128) (\bar{s}\!-\!4)^2 c_{2\theta}
			\nn\\
			& +(5\hs\bar{s}^5\hsm\!+\!312\hs\bar{s}^4\hsm\!-\hsm 672\hs\bar{s}^3
			\hsm\!-\!10752\hs\bar{s}^2
			\!-\!768\hs\bar{s}\hsm +\hsm 2048) c_{4\theta}
			\label{eq:Amp-4hp-c-unitary}
			\\
			& +\ii\hs 64\hs\bar{s}^{\frac{1}{2}} \big[6 (\bar{s}^2\!-\!3\bar{s}\!-\!12)
			(\bar{s}\!-\!4)^2 s_{2\theta}^{}
			\!+\!(\bar{s}^4\!+\!9\bar{s}^3
			\!-\!84\bar{s}^2\!-\!144\bar{s}\!+\!64) s_{4\theta}\big] \!\Big\},
			\nn
			\\[1.5mm]
			\hspace*{-10mm}
			\mathcal{M}_{s}[4\hP]
			= &
			\frac{\,\kappa^2m^2(\bar{s}\!-\!4)^2\,}
			{\,16384(\bar{s}\!-\!1)\bar{s}\,}
			\Big[\hsm (\bar{s}\!-\!4)^2 (\bar{s}\!-\!1) (\bar{s}\!+\!2)^2
			\!-\!(\bar{s}^5\!+\!59\hs\bar{s}^4\!-\!1448\hs\bar{s}^3\!-\!2740\hs \bar{s}^2
			\nn\\
			& -\!992\hs\bar{s}\!-\!64) c_{2\theta}^{}
			\!+\hsm\ii\hs\bar{s}^{\frac{1}{2}}(\bar{s}^5\!-\!37\hs\bar{s}^4\!+\!328\hs\bar{s}^3
			\!+\!2588\hs\bar{s}^2\!+\! 1984\hs\bar{s}\!+\!320)s_{2\theta}^{}\Big],
			\label{eq:Amp-4hp-s-unitary}
			\\[1.5mm]
			\MM_{t}[4\hP]=
			&\frac{\kappa ^2 m^2 c_{\theta/2}^2
				\big(\bar{s}^{\frac{1}{2}}\!+\!\ii\hs 2\tan\!\frac{\theta}{2}\big)^{\!4}}
			{\,1048576\hs [(2\!-\!\bar{s})\!+\!(4\!-\!\bar{s}) c_{\theta}^{}]\,}\! \Big[\hsm\!-\!\ii\hs\bar{s}^{\frac{1}{2}}
			\hsm (\bar{s}^2\!+\!24\hs\bar{s}\!+\!16)(\bar{s}\!-\!4)^2 s_{5\theta}^{}\!+\hsm \ii\hs4 \bar{s}^{\frac{1}{2}}\hsm
			(23\hs\bar{s}^2\!-\!280\hs\bar{s}\!+\!112)
			\nn\\
			&\times\!(\bar{s}\!-\!4) s_{4\theta}^{}
			\!+\!\ii\hs\bar{s}^{\frac{1}{2}}(21\bar{s}^4\!+\!536\hs\bar{s}^3\!-\!4672\hs\bar{s}^2\!-\!13184\hs\bar{s}\!+\!96000) s_{3\theta}^{}\!+\!\ii\hs8\hs \bar{s}^{\frac{1}{2}}(8\bar{s}^4\!+\!113\bar{s}^3
			\nn\\
			&+\!564\hs\bar{s}^2\hsm\!-\!19184\hs\bar{s}\!+\!62016) s_{2\theta}^{}
			\!+\!\ii\hs2\bar{s}^{\frac{1}{2}}(35\hs\bar{s}^4\hsm\!+\!324\hs\bar{s}^3
			\hsm\!+\!8144\hs\bar{s}^2\hsm\!-\hsm\!126272\hs\bar{s}\!+\!359424) s_{\theta}^{}
			\nn\\
			&-\! 8\hs\bar{s}\hs(\bar{s}\!+\!4) (\bar{s}\!-\!4)^2 c_{5\theta}^{}
			\!+\! 8 (3\hs\bar{s}^3\!-\!20\hs\bar{s}^2\!-\!160\hs\bar{s}\!+\!128) (\bar{s}\!-\!4) c_{4\theta}^{}\!+\! 8(23\hs\bar{s}^4\!+\!8\hs\bar{s}^3
			\nn\\
			&-\!1984\hs\bar{s}^2\!+\!4992\hs\bar{s}\!+\!7424) c_{3\theta}^{}\!+\! 32(7\hs\bar{s}^4\!+\!74\hs\bar{s}^3\!-\!680\hs\bar{s}^2\!-\!3280\hs\bar{s}\!+\!19264) c_{2\theta}^{}
			\nn\\
			&-\! 16 (11\hs\bar{s}^4\!-\!250\hs\bar{s}^3\!-\!3672\hs\bar{s}^2\!+\!47264\hs\bar{s}\!-\!120704)c_{\theta}^{}\!-\! 8(31\hs\bar{s}^3\!-\!124\hs\bar{s}^2\!-\!8672\hs\bar{s}
			\nn\\
			&+\!43072) (\bar{s}\!-\!4)\Big],
			\label{eq:Amp-4hp-t-unitary}
		%
		\\
		\MM_{u}[4\hP]=
		&\frac{\kappa^2 m^2 s_{\theta/2}^2
			\big(\bar{s}^{\frac{1}{2}}\!-\!\ii\hs 2\cot\!\frac{\theta}{2}\big)^{\!4}}
		{\,1048576\hs [(2\!-\!\bar{s})\!-\!(4\!-\!\bar{s})c_{\theta}^{}]\,}\!
		\Big[\ii\hs\bar{s}^{\frac{1}{2}}
		\hsm (\bar{s}^2\!+\!24\hs\bar{s}\!+\!16)(\bar{s}\!-\!4)^2 s_{5\theta}^{}
		\!+\hsm\ii\hs 4 \bar{s}^{\frac{1}{2}}\hsm
		(23\hs\bar{s}^2\!-\!280\hs\bar{s}\!+\!112)
		\nn\\
		& \times\!(\bar{s}\!-\!4) s_{4\theta}^{}
		\!-\!\ii\hs\bar{s}^{\frac{1}{2}}(21 \bar{s}^4\!+\!536\hs\bar{s}^3\!-\!4672\hs\bar{s}^2
		\!-\!13184\hs\bar{s}\!+\!96000) s_{3\theta}^{}\!+\!\ii\hs
		8\hs\bar{s}^{\frac{1}{2}}(8\bar{s}^4\!+\!113\bar{s}^3
		\nn\\
		&+\!564\hs\bar{s}^2\hsm\!-\!19184\hs\bar{s}\!+\!62016) s_{2\theta}^{}
		\!-\!\ii\hs2 \bar{s}^{\frac{1}{2}}(35\hs\bar{s}^4\hsm\!+\!324\hs\bar{s}^3\hsm\!+\!8144\hs\bar{s}^2
		\!\!-\!126272\hs\bar{s}\!+\!359424) s_{\theta}
		\nn\\
		& +\! 8\hs\bar{s}\hs
		(\bar{s}\!+\!4) (\bar{s}\!-\!4)^2 c_{5\theta}^{}
		\!+\! 8 (3\hs\bar{s}^3\!-\!20\hs\bar{s}^2\!-\!160\hs\bar{s}\!+\!128)
		(\bar{s}\!-\!4) c_{4\theta}^{}\!-\! 8(23\hs\bar{s}^4\!+\!8\bar{s}^3
		\nn\\
		& -\!1984\hs\bar{s}^2
		\!+\!4992\hs\bar{s}\!+\!7424) c_{3\theta}^{}
		\!+\! 32 (7\hs\bar{s}^4\!+\!74\hs\bar{s}^3\!-\!680\hs\bar{s}^2
		\!-\!3280\hs\bar{s}\!+\!19264)
		c_{2\theta}^{}
		\nn\\
		&+\! 16 (11\bar{s}^4\hsm\!-\!250\hs\bar{s}^3
		\!\!-\!3672\bar{s}^2\!\!+\!47264\hs\bar{s}\!-\!120704) c_{\theta}^{}
		\!-\! 8 (31\bar{s}^3\!\!-\hsm\!124\hs\bar{s}^2\!\!-\!8672\hs\bar{s}
		\nn\\
		& +\!43072)(\bar{s}\!-\!4)\Big].
\label{eq:Amp-4hp-u-unitary}
\end{align}
\end{subequations}
}
Summing up the above sub-amplitudes, we derive the full amplitude of four-graviton scattering in the unitary gauge as follows:
\begin{align}
\label{eq:app-Amp-4hp-U}  
\hspace*{-4mm}
\MM_{\rm{U}}^{}\hsm [4h_{\mathrm{P}}] =
\frac{~\kappa^2 m^2\csc^2\!\theta
\big(\mathbb{Y}_{0}\!+\!\mathbb{Y}_{2}c_{2\theta}^{}
\!+\!\mathbb{Y}_{4}c_{4\theta}^{}
\!+\!\mathbb{Y}_{6}c_{6\theta}^{}
\!+\!{\mathbb{Y}}'_{2}s_{2\theta}^{}
\!+\!{\mathbb{Y}}'_{4}s_{4\theta}^{}
\!+\!{\mathbb{Y}}'_{6}s_{6\theta}^{}\big)~}
{\,4096\hs (\bar{s}\!-\!1)\hs\bar{s}\hs
[(2\!-\!\bar{s})^2\!-\!(4\!-\!\bar{s})^2 c_{\theta}^{2}\hs]\,},
\end{align}
where the quantities $(\YY_{\!j}^{},\hs \YY_{\!j}')$ 
are same as those given by Eq.\eqref{eq:app-Amp-4hp-Y}.\ 
Hence, we find that the four-graviton scattering amplitude 
\eqref{eq:app-Amp-4hp-U} as computed in the unitary gauge 
precisely equals our Landau-gauge result 
in Eqs.\eqref{eq:app-Amp-4hp}-\eqref{eq:app-Amp-4hp-Y},
\beq 
\label{Aeq:M[4hp]U=L}
\MM_{\rm{U}}^{}[4\hP] = \MM_{\rm{L}}^{}[4\hP]\,.
\eeq 
This explicitly demonstrates the gauge-invariance of the
four physical graviton amplitudes, serving as nontrivial consistency checks
on our calculations and on our BRST quantization of the WTMG theory.\   

\vs 

We also note that Ref.\,\cite{TMG-DCx} used the unconventional
Breit coordinate system\,\cite{TMG-DCx} to explicitly calculate four-graviton amplitude
of the conventional TMG (without dilaton field) 
with rather different and lengthy expressions in its eqs.(C1)-(C2)
which cannot be simply compared to our above formulas
\eqref{eq:app-Amp-4hp-U} [and \eqref{eq:app-Amp-4hp-Y}].\
(The explicit calculation\,\cite{TMG-DCx} of the graviton amplitude of in the TMG theory also agrees with their
double-copy result.)
Thus, our current independent calculations 
in both the Landau gauge (Section\,\ref{sec:4.1.2}) 
and unitary gauge 
(Appendix\,\ref{app:B3}) are highly nontrivial, which agree with each other
and pass all the consistency checks including the verification of gauge-invariance,  
the demonstration of various striking energy cancellations, 
and the agreements with the TGRET \eqref{eq:TGRET} and the double-copy
(Section\,\ref{sec:4.2.2}).\
Moreover, different from \cite{TMG-DCx},
our main purpose is to analyze the energy structure of the graviton amplitude
\eqref{eq:app-Amp-4hp}-\eqref{eq:app-Amp-4hp-Y} and compare it with our calculation of
the corresponding dilation amplitude, so that we can explicitly demonstrate
the validity of the TGRET \eqref{eq:TGRET} in the high energy limit.\


\section{\hspace*{-2mm}Four-Point Scalar Amplitudes 
in the WTMGS and TMGS}
\label{app:Cnew}

In the Landau gauge ($\zeta\!=\hsm\xi\!=\!0$) of the WTMGS theory, 
the four-point physical scalar scattering amplitude contains
contributions from the graviton-exchanges and dilaton-exchanges
as well as the $4\psi$ contact term,
\beq 
\MM_{\rm{L}}^{}[4\psi]= 
\MM_{\rm{L}}^{h}[4\psi]+ \MM_{\rm{L}}^{\phi}[4\psi]+\MM_{\rm{L}}^{c}[4\psi]\,,
\eeq 
We explicitly compute these individual contributions from the $(s,t,u)$ channels and
the contact term.\ We present their analytical formulas as follows:
{\small 
\begin{subequations}
\begin{align}
\label{Aeq:M[4psi]-Contact}
\MM_{c}[4\psi] =& -\kappa^2 m_s^2 \,,
\\
\label{Aeq:M[4psi]h-s}
\MM_{s}^{h}[4\psi]  =& \frac{\,\kappa^2 m (s\!-\!4m_s^2)^2\,}{\,16\hs s\hs (s\!-\!m^2)\,}
(\ii s^{\frac{1}{2}}s_{2\theta}^{}\!+\!m c_{2\theta}^{})\hs ,
\\
\label{Aeq:M[4psi]h-t}
\MM_{t}^{h}[4\psi]  =& \frac{\kappa^2 m}
{\,128c_{\theta/2}^2(s\!-\!4m_s^2)[(s\!-\!4m_s^2)c_{\theta/2}^2\hsm\!+\!m^2]\,}
\Big[\!-\ii\hs 4\hs s^{\frac{1}{2}}(s\!-\!4m_s^2)^2 s_{2\theta}^{}
\nn\\
& 
+\!\ii\hs 8\hs s^{\frac{1}{2}} (3s^2\!-\!16m_s^2 s
+\!16m_s^4)s_{\theta}^{}\!+\!m(s\!-\!4m_s^2)^2 c_{2\theta}^{}\!-\!4m(7s^2\!-\!32m_s^2 s\!+\!16m_s^4)c_{\theta}^{}
\nn\\
& +\!m(35s^2\!-\!120m_s^2 s \!+\!48m_s^4)\hsm\Big] \hs,
\\
\label{Aeq:M[4psi]h-u}
\MM_{u}^{h}[4\psi]  =& \frac{\kappa^2 m}
{\,128s_{\theta/2}^2(s\!-\!4m_s^2)[(s\!-\!4m_s^2)s_{\theta/2}^2\hsm\!+\!m^2]\,}
\Big[\!-\ii\hs 4 s^{\frac{1}{2}}(s\!-\!4m_s^2)^2 s_{2\theta}^{}
\nn\\
& -\!\ii\hs 8 s^{\frac{1}{2}}(3s^2\!-\!16m_s^2 s
\!+\!16m_s^4)s_{\theta}^{}\!+\!m(s\!-\!4m_s^2)^2 c_{2\theta}^{}\!+\!4m(7s^2\!-\!32m_s^2 s\!+\!16m_s^4)c_{\theta}^{}
\nn\\
& \!+\!m(35s^2\!-\!120m_s^2 s +\!48m_s^4)\Big] \hs,
\\
\label{Aeq:M[4psi]phi-s}
\MM_{s}^{\phi}[4\psi] =& \frac{\kappa^2}{\,16\hs s\,}(s\!+\!4m_s^2)^2 \,,
\\
\label{Aeq:M[4psi]phi-t}
\MM_{t}^{\phi}[4\psi] =& -\!\frac{\kappa^2}{\,64\hs c_{\theta/2}^2(s\!-\!4m_s^2)\,}
\Big[\hsm (s\!-\!4m_s^2)c_{\theta}^{}\!+\!(s\!-\!12m_s^2)\hsm\Big]^2 ,
\\
\label{Aeq:M[4psi]phi-u}
\MM_{u}^{\phi}[4\psi] =& -\!\frac{\kappa^2}{\,64\hs s_{\theta/2}^2(s\!-\!4m_s^2)\,}
\Big[\hsm (s\!-\!4m_s^2)c_{\theta}^{}\!-\!(s\!-\!12m_s^2)\hsm\Big]^2 ,
\end{align}
\end{subequations}
}
\hspace*{-3mm}
Thus, we sum up the individual contributions 
\eqref{Aeq:M[4psi]h-s}-\eqref{Aeq:M[4psi]h-u}
from graviton-exchanges in the $(s,t,u)$ channels,
\beq
\label{Aeq:M[4psi]-h} 
\MM_{\rm{L}}^{h}[4\psi] = \frac{\kappa^2 m\big(\mathbb{Z}_{0}^{}\!+\!\mathbb{Z}_{2}^{}c_{2\theta}^{}
\!+\!\mathbb{Z}_{4}^{}c_{4\theta}^{}
\!+\!\mathbb{Z}_{6}^{}c_{6\theta}^{}\!+\!\mathbb{Z}'_{2}s_{2\theta}^{}\!+\!\mathbb{Z}'_{4}s_{4\theta}^{}
\!+\!\mathbb{Z}'_{6}s_{6\theta}^{}\big)}{1024 s(s\!-\!m^2)(s\!-\!4m_s^2)s_{\theta}^2[m^2\!+\!(s\!-\!4m_s^2)s_{\theta/2}^2][m^2\!+\!(s\!-\!4m_s^2)c_{\theta/2}^2]}
\eeq 
where the numerator coefficients $(\mathbb{Z}_j^{}, \mathbb{Z}'_j)$ are given by
{\small 
\begin{align}
\mathbb{Z}_{0}^{} 
=&\, 4m\big[322s^5\!+\!s^4(65m^2\!-\!2512m_s^2)\!-\!4s^3(99m^4\!-\!281m^2 m_s^2\!-\!1340m_s^4)
\nn\\
& +\!16m_s^2 s^2(95m^4\!-\!329m^2 m_s^2\!-\!100m_s^4)\!-\!64m_s^4 s(13m^4\!-\!51m^2 m_s^2\!+\!20m_s^4)\big]\hs,
\nn\\
\mathbb{Z}_{2}^{} =&\,
m(s\!-\!4m_s^2)\big[(759s^4\!-\!3312m_s^2 s^3\!+\!2464m_s^4 s^2\!-\!1792m_s^6 s+1792m_s^8)
\nn\\
& -\!16m^2(15s^3\!-\!152m_s^2 s^2\!+\!16m_s^4 s\!+\!128m_s^6)\!-\!64m^4(7s^2\!-\!2m_s^2 s\!-\!8m_s^4)\big]\hs,
\nn\\
\mathbb{Z}_{4}^{} =&\,
4m(s\!-\!4m_s^2)^3
\big[s(8m_s^2\!-\!5m^2)\!-\!4(2m_s^2\!-\!m^2)^2\big]\hs,
\nn\\
\mathbb{Z}_{6}^{} =&\, m(s\!-\!4m_s^2)^5\hs,
\label{Aeq:Z-Z'-M[4psi]h}
\\
{\mathbb{Z}}_{2}' =&\hs -\!\ii\hs  s^{\frac{1}{2}}(s\!-\!4m_s^2)\big[475s^4\!-\!2736m_s^2 s^3\!-\!32s^2(17m^4\!-\!68m^2 m_s^2\!-\!97m_s^4)
\nn\\
& +\!1280m_s^2 s(m^4\!-\!4m^2 m_s^2\!+\!m_s^4)\!-\!256m_s^4(2m^4\!-\!8m^2 m_s^2
\!+\!5m_s^4)\big]\hs,
\nn\\
{\mathbb{Z}}_{4}' =&
-\!\ii\hs 4\hs s^{\frac{1}{2}}(s\!-\!4m_s^2)^3
\big[5s^2\!-\!8m_s^2 s\!+\!4(2m_s^2\!-\!m^2)^2\big]\hs,
\nn\\
{\mathbb{Z}}_{6}' =&\,\ii\hs s^{\frac{1}{2}}(s\!-\!4m_s^2)^5\hs. 
\nn
\end{align}
}
\hspace*{-3mm}
Then, we sum up the individual contributions 
\eqref{Aeq:M[4psi]phi-s}-\eqref{Aeq:M[4psi]phi-u} 
from dilaton-exchanges in the $(s,t,u)$ channels,
\begin{align}
\label{Aeq:M[4psi]-phi} 
\MM_{\rm{L}}^{\phi}[4\psi]
=-\frac{\,\kappa^2 m_s^2
\big[(7s^2\!-\!24m_s^2 s\!-\!16m_s^4)c_{2\theta}^{}
\!-\!(7s^2\!-\!56m_s^2 s\!-\!16m_s^4)\big]\,}
{8\hs s(s\!-\!4m_s^2)s_{\theta}^{2}}  \hs. 
\end{align}
Using the amplitudes \eqref{Aeq:M[4psi]-Contact}, \eqref{Aeq:M[4psi]-h} and \eqref{Aeq:M[4psi]-phi}
of sub-channels, we derive the full four-point physical scalar 
amplitude as follows:
\begin{align}
\MM_{\rm{L}}^{}[4\psi] &= 
\MM_{\rm{L}}^{h}[4\psi]+ \MM_{\rm{L}}^{\phi}[4\psi] +\MM_{\rm{L}}^{c}[4\psi]
\nn\\
&=\frac{\,\kappa^2\big(\widetilde{\mathbb{Z}}_{0}^{}\!+\!\widetilde{\mathbb{Z}}_{2}^{}c_{2\theta}^{}
\!+\!\widetilde{\mathbb{Z}}_{4}^{}c_{4\theta}^{}
\!+\!\widetilde{\mathbb{Z}}_{6}^{}c_{6\theta}^{}
\!+\!\widetilde{\mathbb{Z}}'_{2}s_{2\theta}^{}
\!+\!\widetilde{\mathbb{Z}}'_{4}s_{4\theta}^{}
\!+\!\widetilde{\mathbb{Z}}'_{6}s_{6\theta}^{}\big)\,}
{256\hs s\hs (s\!-\!m^2)\big[(s\!-\!4m_s^2\!+\!2m^2)^2\!-\!(s\!-\!4m_s^2)^2c_{\theta}^{2}\big]
s_{\theta}^2} -\kappa^2 m_s^2\, ,
\label{Aeq:M[4psi]L-full}
\end{align}
where the numerator coefficients $(\widetilde{\mathbb{Z}}_j^{}, \widetilde{\mathbb{Z}}'_j)$ 
are given by 
{\small 
\begin{align}
\hspace*{-14mm}
\widetilde{\mathbb{Z}}_{0}^{} =&
-\!48\hs m^6(33\hs s^2\!+\!24\hs m_s^2 s\!+\!16\hs m_s^4)
\!+\!4\hs m^4(65\hs s^3\!+\!1384\hs m_s^2 s^2\!+\!1168\hs m_s^4 s\!+\!768\hs m_s^6)
\nn\\
& +\!8\hs m^2(161s^4\!-\!521m_s^2 s^3\!-\!444\hs m_s^4 s^2\!-\!624\hs m_s^6 s\!-\!320\hs m_s^8)
\nn\\
& +\!8\hs m_s^2 s\hs (s\!-\!4\hs m_s^2)(21s^2\!-\!136\hs m_s^2 s\!-\!48\hs m_s^4)\hs,
\nn\\
\hspace*{-14mm}
\widetilde{\mathbb{Z}}_{2}^{} =&
-\!64\hs m^6\big(7s^2\!-\!16\hs m_s^2 s\!-\!16\hs m_s^4\big)
\!-\!16m^4(15\hs s^3\!-\!152\hs m_s^2 s^2\!+\!240\hs m_s^4 s\!+\!156\hs m_s^6)
\nn\\
& +\!m^2(759\hs s^4\!-\!3984\hs m_s^2 s^3\!+\!3360\hs m_s^4 s^2\!+\!4864\hs m_s^6 s\!+\!3840\hs m_s^8)
\nn\\
& -\!32\hs m_s^2 s \hs\big(7s^3\!-\!68m_s^2 s^2\!+\!144m_s^4 s\!+\!64m_s^6\big) \hs,
\label{Aeq:Zt-Zt'-M[4psi]L-full}
\\
\hspace*{-14mm}
\widetilde{\mathbb{Z}}_{4}^{} =&
-\!4(s\!-\!4m_s^2)^2\big[4m^6\!+\!m^4(5s\!-\!16m_s^2)\!+\!6m^2 m_s^2\big(s\!+\!4m_s^2\big)
\!-\!2m_s^2 s\big(7s\!+\!4m_s^2\big)\big]\hs,
\nn\\
\hspace*{-14mm}
\widetilde{\mathbb{Z}}_{6}^{} =&\,
m^2\big(s\!-\!4m_s^2\big)^4\hs,
\nn\\
\hspace*{-14mm}
\widetilde{\mathbb{Z}}_{2}' =&
-\!\ii\hs m\hs s^{\fr{1}{2}}[475s^4\!-\!2736m_s^2 s^3\!-\!32(17m^4\!-\!68m^2 m_s^2\!-\!97m_s^4)s^2
\nn\\
& +\!1280\hs m_s^2(m^4\!-\!4m^2 m_s^2\!+\!m_s^4)s\!-\!256m_s^4(2m^4\!-\!8m^2 m_s^2\!+\!5m_s^4)]\hs,
\nn\\
\hspace*{-14mm}
\widetilde{\mathbb{Z}}_{4}' =&
-\ii\hs 4\hs m\hs s^{\fr{1}{2}}(s\!-\!4m_s^2)^2
\big[5s^2\!-\!8m_s^2 s\!+\!4(m^2\!-\!2m_s^2)^2\big]\hs,
\nn\\
\hspace*{-14mm}
\widetilde{\mathbb{Z}}_{6}' =&
\,\ii\hs m\hs s^{\fr{1}{2}}(s\!-\!4m_s^2)^4\hs.
\nn
\end{align}
}

Next, we compute the four-point physical scalar amplitude (including the case of $m_s^{}\!\neq\! m$)
in the unitary gauge of the WTMGS theory
by choosing $(\zeta,\hs\xi)\!=\!(\infty,\hs 0)$.\ 
This is equivalent to doing the calculation by using the TMGS Lagrangian \eqref{eq:L-TMGS-UG}.\ 
We derive the individual contributions from each channel as follows:
{\small 
\begin{subequations}
\label{Aeq:M[4psi]U-cstu}
\begin{align}
\hspace*{-5mm}
\MM_{c}[4\psi] =& -\kappa^2 m_s^2\,,
\label{Aeq:M[4psi]U-c}
\\
\hspace*{-5mm}
\MM_{s}[4\psi]  =& \frac{\kappa^2}{16s(s\!-\!m^2)}[\ii m s^{1/2}(s\!-\!4m_s^2)^2 s_{2\theta}^{}\!+\!m^2(s\!-\!4m_s^2)^2 c_{2\theta}^{}\!+\!(s\!+\!4m_s^2)^2(s\!-\!m^2)]\,,
\label{Aeq:M[4psi]U-s}
\\
\hspace*{-5mm}
\MM_{t}[4\psi]  =&\frac{\kappa^2}{512c_{\theta/2}^2[m^2\!+\!(s\!-\!4m_s^2)c_{\theta/2}^2]}\{-\!16\ii m s^{1/2}[(s\!-\!4m_s^2)s_{2\theta}^{}\!-\!2(3s\!-\!4m_s^2)s_{\theta}]
\nn\\
&-\!(s\!-\!5m_s^2)^2 c_{3\theta}^{}\!-\!2(s\!-\!4m_s^2)(3s\!-\!28m_s^2)c_{2\theta}^{}\!-\![(5s\!-\!36m_s^2)(3s-28m_s^2)
\nn\\
&+\!128m^2(s\!-\!2m_s^2)]c_{\theta}^{}
\!-\!2[(5s^2\!-\!88m_s^2 s\!+\!400m_s^4)\!-\!64m^2(s\!+\!2m_s^2)]\} \,,
\label{Aeq:M[4psi]U-t}
\\
\hspace*{-5mm}
\MM_{u}[4\psi]  =&\frac{\kappa^2}{512s_{\theta/2}^2[m^2\!+\!(s\!-\!4m_s^2)s_{\theta/2}^2]}\{-\!16\ii m s^{1/2}[(s\!-\!4m_s^2)s_{2\theta}^{}\!+\!2(3s\!-\!4m_s^2)s_{\theta}]
\nn\\
&+\!(s\!-\!4m_s^2)^2 c_{3\theta}^{}\!-\!2(s\!-\!4m_s^2)(3s\!-\!28m_s^2)c_{2\theta}^{}\!+\![(5s\!-\!36m_s^2)(3s-28m_s^2)
\nn\\
&+\!128m^2(s\!-\!2m_s^2)]c_{\theta}^{}\!-\!2[(5s^2\!-\!88m_s^2 s\!+\!400m_s^4)
\!-\!64m^2(s\!+\!2m_s^2)]\} \,,
\label{Aeq:M[4psi]U-u}
\end{align}
\end{subequations}
}
\hspace*{-3mm}
Summing up the individual contributions \eqref{Aeq:M[4psi]U-c}-\eqref{Aeq:M[4psi]U-u},
we derive the four-scalar amplitude $\MM_{\rm{U}}^{}[4\psi]$
in the unitary gauge and find that it precisely equals the corresponding Landau-gauge 
scattering amplitude \eqref{eq:M[4psi]-sum1-Landau},
\beq
\label{Aeq:M[4psi]U=L} 
\MM_{\rm{U}}^{}[4\psi]= \MM_{\rm{L}}^{}[4\psi]\,. 
\eeq 
This explicitly demonstrates the gauge-invariance of the physical scalar amplitude
and confirms the consistency of our WTMGS (WTMG) formulation, as discussed in the main text.\ 

\vs

As the last part of this Appendix, we can explicitly prove that
the four-point physical scalar amplitude 
\eqref{eq:M[4psi]-sum1-Landau}
is invariant under the field redefinitions
\eqref{eq:transf-phi-psi}, namely, 
$\MM_{\rm{L}}^{}[4\psi]\!=\!\MM_{\rm{L}}^{}[4\psih]$, 
which serves as a consistency check our analysis.\ 
Expanding Eq.\eqref{eq:transf-phi-psi}, we resolve the fields
$(\phi,\psi)$ in terms of the redefined basis $(\phih,\psih )$ 
as follows:
\begin{equation}
	\begin{aligned}
		\phi &\hs =\, 
		\phih\hsm +\!\frac{\hs\kappa\hs}{8}\big(\phih^2\!-\!\psih^2\big)
		\!+\!\frac{\kappa^2}{\hs48\hs}\big(\phih^3\!-\!3\psih^2\phih\big)
		\!+\!{O}(\kappa^3) \hs,
		\\
		\psi &\hs =\, 
		\psih\hsm +\!\frac{\hs\kappa\hs}{4}\psih\phih
		\!+\!\frac{\,\kappa^2\hs}{48}\big(3\psih\phih^2\!-\!\psih^3\big)
		\!+\!{O}(\kappa^3) \hs.
	\end{aligned}
\end{equation}
Thus, we can express the WTMGS Lagrangian \eqref{eq:L-WTMGS}
in terms of the field basis $(\phih,\psih )$.\   
From Eqs.\eqref{Beq:WTMGS-h-2psi} and 
\eqref{Beq:WTMGS-V(h2psih)=V(h2psi)}, 
we see that the graviton-scalar-scalar trilinear coupling 
does not change under the field redefinitions.\ 
In addition to the $(\phih,\psih )$ derivative terms given  
in Eq.\eqref{eq:L-phih/psih-h}, we derive
the following scalar potential terms for $(\phih,\psih )$
up to quartic interactions:
\beq
-V(\phih,\psih ) 
=\sqrt{-g\,}\!\left[\hsm -\frac{\hs m_s^2\hs}{2}\psih^2
\!+\!\frac{\hs\kappa\hs m_s^2\hs}{2}\phih\hs\psih^2 
\!-\!\frac{\,\kappa^2m_s^2\hs}{96}
\big(11\psih^4\!+\!18\phih^2\psih^2\big)
\!+\hsm O(\kappa^3)
\hsm\right]\!.
\eeq 
Thus, we derive the couplings of the trilinear vertex $\phih\psih\psih$
and the quartic vertex $\psih^4$ as follows:
%
\begin{align}
	\label{eq:V(phih-psih2)-V(4psi)}
	\ii\hs \mathcal{V}_{\phih\psih\psih}^{}
	=\ii\hs\kappa\hs m_s^2\,,
	\hspace*{8mm}
	\ii\hs \mathcal{V}_{\!4\psih}^{}
	=-\frac{\hs\ii\hs 11\hs}{4}\kappa^2 m_s^2 \,,
\end{align}
%
where the $\phih\psih\psih$ trilinear coupling
$\mathcal{V}_{\phih\psih\psih}^{}$ has no momentum-dependence
and differs from the $\phi\psi\psi$ coupling
$\mathcal{V}_{\phi\psi\psi}^{}$ 
in Eq.\eqref{eq:V-phi-psi2}, 
as expected.\ 
The above $\psih^4$ quartic coupling 
$\mathcal{V}_{4\psih}^{}$ also differs from 
the original $\psi^4$ quartic coupling 
$\mathcal{V}_{4\psi}^{}\!\!=\hsm\!-\kappa^2m_s^2\hs$ 
in the WTMGS Lagrangian \eqref{eq:L-WTMGS}.\ 
Hence, the four-point physical scalar scattering amplitude
$\MM_{\rm{L}}^{}[4\psih]$ receives contributions from the
graviton-exchanges, the dilaton-exchanges, and the $\psih^4$ 
contact term:
\beq 
\label{eq:ML4psih=h+phi+CT}
\MM_{\rm{L}}^{}[4\psih] = 
\MM_{\rm{L}}^{h}[4\psih]+ \MM_{\rm{L}}^{\phih}[4\psih] +\MM_{\rm{L}}^{c}[4\psih]\,.
\eeq 
Comparing Eq.\eqref{eq:ML4psih=h+phi+CT} with 
Eq.\eqref{eq:M[4psi]-sum1-Landau} and making use of 
Eq.\eqref{Beq:WTMGS-V(h2psih)=V(h2psi)}, 
we find that the contribution from graviton-exchanges
remains unchanged after the field redefinitions
\eqref{eq:transf-phi-psi}.\ 
Thus, we first deduce the following equality,
\beq 
\MM_{\rm{L}}^{h}[4\psi] = \MM_{\rm{L}}^{h}[4\psih]\,.
\eeq 
Using the new trilinear coupling of $\phih\psih\psih$
and the new quartic coupling of $\psih^4$ 
as given by Eq.\eqref{eq:V(phih-psih2)-V(4psi)},  
we compute the sub-amplitudes of $\MM_{\rm{L}}^{}[4\psih]$ 
and sum up the contributions from dilaton-exchanges and 
from the contact term for the field basis $(\phih,\psih )$,  
{\small 
\begin{align}
	\label{eq:ML(phi)+ML(c)}
	\hspace*{-5mm}
	\MM_{\rm{L}}^{\phih}[4\psih]+\MM_{\rm{L}}^{c}[4\psih]
	&= -\frac{\,\kappa^2 m_s^2
		\big[(7s^2\!-\!24m_s^2 s\!-\!16m_s^4)c_{2\theta}^{}
		\!-\!(7s^2\!-\!56m_s^2 s\!-\!16m_s^4)\big]\,}
	{8\hs s(s\!-\!4m_s^2)s_{\theta}^{2}} 
	\!-\!\kappa^2m_s^2 
	\nn\\
	&= \MM_{\rm{L}}^{\phi}[4\psi]+\MM_{\rm{L}}^{c}[4\psi]\,,
\end{align} 
}
\hspace*{-3mm}
which fully agrees with the dilaton-exchange contribution 
$\MM_{\rm{L}}^{\phi}[4\psi]$
in Eq.\eqref{eq:M[4psi]phi-sum-Landau}
and with the contact-term contribution $\MM_{\rm{L}}^{c}[4\psi]$
in Eq.\eqref{eq:M[4psi]-sum1-Landau}.\
In summary, the above analysis demonstrates that
after the field-redefinition transformations \eqref{eq:transf-phi-psi}, 
the full amplitude of the four-point physical scalar scattering
remains the same as before:  
\beq 
\MM_{\rm{L}}^{}[4\psi] = \MM_{\rm{L}}^{}[4\psih]\,.
\eeq 
This explicitly proves that the on-shell physical scalar amplitude
is invariant under the field-redefinition transformations
and verifies the consistency of our analysis.\


\section{\hspace*{-2mm}BRST Quantization of the WTMG Theory}
\label{app:C}
\label{app:D}

In this Appendix, we further present the BRST quantization of the
WTMG Lagrangian \eqref{eq:LTMG-phi}
and derive the relevant BRST identity for general LSZ reduction
(up to loop level) of the external line $\mathcal{F}_{2}^{}$
in the identity \eqref{eq:F2-GID}.

\vs

According to the gauge-fixing terms \eqref{eq:gaugefix},
we can write the following Faddeev-Popov ghost terms:
\beq
\mathcal{L}_{\rm{FP}}^{} = \bar{c}_{\mu}^{}\hs\ssh \mathcal{F}_{\rm{GF}1}^{\mu}
+ \bar{b}_{\mu}^{}\hs\ssh \mathcal{F}_{\rm{GF}2}^{\mu} \,,
\eeq
where the (anti-)ghost fields $(c_{\mu}^{},\,\bar{c}_{\mu}^{})$ and
$(b_{\mu}^{},\,\bar{b}_{\mu}^{})$ correspond to the gauge-fixing functions
$\mathcal{F}_{\rm{GF}1}^{\mu}$ and $\mathcal{F}_{\rm{GF}2}^{\mu}$ respectively.\

\vs

We drive the BRST transformations for the WTMG Lagrangian \eqref{eq:LTMG-phi}.\
We first consider the diffeomorphism transformation of the TMG.\
Let $\xi^{\mu}(x)$ be an infinitesimal coordinate transformation vector (normal parameter),
then the normal transformation of the metric tensor $g_{\mu \nu}$ can be written as follows:
\begin{equation}
\delta_{\xi}g_{\mu\nu}=\kappa\mathcal{L}_{\xi}g_{\mu\nu}
=-g_{\nu\lambda}\partial_{\mu}\xi^{\lambda}-g_{\mu\lambda}\partial_{\nu}\xi^{\lambda}
-\xi^{\lambda}\partial_{\lambda}^{}g_{\mu\nu}^{}\,.
\end{equation}
where $\mathcal{L}_{\xi}^{}$ is Lie derivative.\
Under weak field expansion $g_{\mn}^{}\!=\!\eta_{\mn}\!+\hsm\kappa h_{\mn}$, we obtain:
\begin{equation}
\kappa\delta_{\xi}h_{\mu\nu}
=-\partial_{\mu}\xi_{\nu}-\partial_{\nu}\xi_{\mu}-\kappa h_{\nu\lambda}\partial_{\mu}\xi^{\lambda}-\kappa h_{\mu\lambda}\partial_{\nu}\xi^{\lambda}-\kappa\xi^{\lambda}\partial_{\lambda}h_{\mu\nu}
\,.
\end{equation}
After the Weyl transformation \eqref{eq:CT-gmunu}, we derive the following
infinitesimal coordinate transformations on $\bar{h}_{\mn}^{}$ and $\phi\hs$, 
\begin{subequations}
\begin{align}
\kappa\delta_{\xi}h_{\mu\nu}
= &-\partial_{\mu}\xi_{\nu}-\partial_{\nu}\xi_{\mu}-\kappa h_{\nu\lambda}\partial_{\mu}\xi^{\lambda}-\kappa h_{\mu\lambda}\partial_{\nu}\xi^{\lambda}-\kappa\xi^{\lambda}\partial_{\lambda}h_{\mu\nu}
\nn\\
& +\frac{2}{3}(\eta_{\mu\nu}+\kappa h_{\mu\nu})(\partial_{\lambda}\xi^{\lambda}+\kappa h_{\alpha\beta}\partial^{\alpha}\xi^{\beta})\,,
\\
\kappa\delta_{\xi}\phi
= & -\frac{2}{3}\partial_{\lambda}\xi^{\lambda}-\frac{2}{3}\kappa h_{\alpha\beta}\partial^{\alpha}\xi^{\beta}-\kappa\xi^{\lambda}\partial_{\lambda}\phi\,,
\end{align}
\end{subequations}
where we have suppressed the ``bar'' on top of the graviton field $h_{\mn}^{}$ 
for notational simplicity.\ 
Then, we replace the normal parameter $\xi^{\mu}(x)$ by the Grassmannian ghost fields
$(c^\mu\hsm ,\,b^\mu)$.\
Denoting the BRST transformation operator by $\ssh\hs$
and considering the gauge-fixing function
$\mathcal{F}_{\rm{GF}2}^{\mu}$,
we express the BRST transformations of the graviton field $h_{\mn}^{}$ and the dilaton field $\phi$ as follows:
\begin{subequations}
\begin{align}
\hat{\tt s}h_{\mu\nu}
=& -\!\partial_{\mu}^{}b_{\nu}^{}\!-\!\partial_{\nu}b_{\mu}^{}
\!-\hsm\kappa\hs h_{\nu\lambda}^{}\partial_{\mu}^{}
b^{\lambda}  
\!-\!\kappa\hs h_{\mu\lambda}^{}\partial_{\nu}^{}b^{\lambda}
\!-\!\kappa\hs b^{\lambda}\partial_{\lambda}^{}h_{\mu\nu}^{}
\nn\\
& +\frac{2}{\hs 3\hs}(\eta_{\mn}^{}\!+\hsm\kappa\hs h_{\mn}^{})
(\partial_{\lambda}^{}b^{\lambda}
\!+\hsm\kappa\hs h_{\alpha\beta}^{}\partial^{\alpha}b^{\beta}) 
\hs,
\\
\hat{\tt s}\phi =&
-\!\frac{2}{\hs 3\hs}
\partial_{\lambda}^{}b^{\lambda}\!-\!\frac{2}{\hs 3\hs}
\kappa\hs h_{\alpha\beta}^{}\partial^{\alpha}b^{\beta}
\!-\hsm\kappa\hs b^{\lambda}\partial_{\lambda}^{}\phi \,.
\end{align}
\end{subequations}
Here ghost fields $c^\mu$ and $b^\mu$ represent 
diffeomorphism invariance and conformal symmetry, respectively.

\vs 

Accordingly, we write down the BRST transformations for the two types of
(anti-)ghost fields $(c^\mu\hsm ,\,\bar{c}^\mu)$
and $(b^{\mu},\,\bar{b}^{\mu})$:
\begin{equation}
\begin{split}
& \ssh\hs c^{\mu}=c^{\nu}\partial_{\nu}c^{\mu},   
\hspace*{7mm}
\ssh\hs \bar{c}^{\mu}=-\hs\xi^{-1}\mathcal{F}_{\rm{GF}1}^{\mu} ,
\\
& \ssh\hs b^{\mu}=b^{\nu}\partial_{\nu}b^{\mu},   
\hspace*{7mm}
\ssh\hs \bar{b}^{\mu} = -\hs \xi^{-1}\mathcal{F}_{\rm{GF}2}^{\mu} \,,
\end{split}
\end{equation}
where the gauge-fixing functions $\mathcal{F}_{\rm{GF}1}^{\mu}$ and
$\mathcal{F}_{\rm{GF}2}^{\mu}$
are given in Eq.\eqref{eq:gaugefix} of the main text
and we set $\xi=\zeta\hs$ for the current analysis.\

\vs

In Eq.\eqref{eq:F1-F3}, we have further introduced the 
contracted gauge-fixing function,
$\FF_2^{}\!=\hsm h\!-\hsm\xi\hs\phi\hs$,
by imposing the on-shell condition 
$p^{2}\!=\! -m^{2}${\hs}.\
Thus, we can reexpress $\FF_2^{}$ as follows:
\begin{equation}
\label{appeq:F2-K}
\begin{split}
& \FF_2^{}\hs
= h\!-\xi\hs\phi
= \mathbf{K}^{\text{T}}\mathbf{H} \,,
\\
& \mathbf{K}^{\rm{T}}\!\hsm
=\big(\eta_{\mn}^{},\,-\xi\hs\big), \quad
\mathbf{H}=(h^{\mu \nu}\hsm ,\,\phi)^{\hsm\rm{T}}\hs.
\end{split}
\end{equation}
Similar to the case of compactified KK gravity theories 
in 4d\,\cite{Hang:2024uny},
we derive the following BRST identity involving 
the gauge-fixing function
$\mathcal{F}_2^{}$ of Eq.\eqref{appeq:F2-K}:
\begin{equation}
\mathbf{K}^{\text{T}}\boldsymbol{\mathcal{D}}(p) =\langle0|T\FF_2^{}\mathbf{H}^{\text{T}}|0\rangle
\equiv -\mathbf{X}(p)^{\text{T}}\,.
\end{equation}
In the above,
the matrix propagator $\boldsymbol{\mathcal{D}}(p)$ 
and the matrix quantity $\mathbf{X}(p)$ are given by
\beqs
\begin{align}
\label{eq:D-X-S}
\mathcal{D}(p) &= \langle 0| T \mathbf{H}\mathbf{H}^{\text{T}}|0\rangle(p),\quad \mathbf{X}(p)=\bar{\mathbf{X}}(p)S(p),\quad
S(p)=\langle 0| T b\hs\bar{b}\hs |0\rangle(p)\hs,
\\[1.5mm]
\bar{\mathbf{X}}(p) & =\fr{\,2\hs\xi\,}{m}\!
\(\!
\begin{matrix}
\langle 0| T \ssh\hs h^{\mu \nu}|\bar{b}\rangle
\\[1mm]
\langle 0| T \ssh\hs \phi|\bar{b}\rangle
\end{matrix} \)
\!=\fr{\hs 4\hs\xi\hs}{\,3\,}\!\(\!
\begin{matrix}
\hs \eta^{\mu \nu}\!
\big[1\!+\!\Delta^{\!(1)}\hsm (p^{2})\big]
\\[1mm]
\hs \hs \hs- \hs \hs \hs[1\!+\!\Delta^{\!(2)}\hsm (p^{2})]
\end{matrix} \!\)
\!\equiv
\!\(\hsm
\begin{matrix}
\bar{X}^{\mn}_h\hsm (p)
\\[1mm]
\bar{X}_{\hsm\phi}^{}(p)
\end{matrix} \hsm\)
\!,
\hspace*{6mm}
\label{eq:barX}
\end{align}
\eeqs
where we have defined the notations
$b\!=\! \ii\hs \epsilon_{\mu}^{\rm{S}}{b^{\hs\mu}}$ and
$\bar{b}\!=\! 
 \ii\hs\epsilon_{\mu}^{\hs\rm{S}}\hs\bar{b}^{\mu}$.\
In Eq.\eqref{eq:barX},
$\Delta^{\!(1)}\hsm (p^{2})$ and $\Delta^{\!(2)}\hsm (p^{2})$ are loop quantities and generated by the non-linear terms 
of the BRST transformations.

\vs 

With the above, we apply the LSZ reduction to each external line $\mathcal{F}_{2}^{}$
of the Green function on the left-hand side of the identity \eqref{eq:F2-GID}
and derive the following:
\begin{align}
& \mathcal{G} \big[\mathcal{F}_2^{}(p),\cdots\big]
= \mathbf{K}^{\rm{T}}\DD (p)\MM\big[\mathbf{H}(p),\cdots\big]
= -\mathbf{X}(p)^{\hsm\rm{T}}\hsm
\MM\big[\mathbf{H}(p),\cdots\hsm\big]
\nn\\
& = -\mathcal{S}(p)\hs\bar{\mathbf{X}}(p)^{\hsm\rm{T}} \hsm\MM\big[\mathbf{H}(p),\cdots\big]
= -\fr{\hs{4\hs\xi}\hs}{\hs 3\hs}\mathcal{S}(p)
\!\left[1\!+\!\Delta^{\!(1)}\hsm (p^2)\right]\!
\MM\big[h\!-\hsm C\phi,\cdots\hsm\big],
%
%
\label{appeq:G-M-C}
\end{align}
where in the last step we have replaced the bare fields $(h_{\mn}^{},\,\phi)$
by the corresponding renormalized fields through the relations,
\beq
h_{\mn}^{}\!\longrightarrow\hsm Z_h^{\frac{1}{2}} h_{\mn}^{}\,,
\hspace*{8mm}
\phi \hsm\longrightarrow\hsm Z_{\phi}^{\frac{1}{2}} \phi \,,
\eeq
where $(Z_h^{},Z_{\phi}^{})$ denote wavefunction renormalization constants of the
(graviton,\,dilaton) fields.\  
In Eq.\eqref{appeq:G-M-C}, the radiative multiplicative modification factor $C$ is induced 
at loop level and takes the following form:
\begin{equation}
\label{appeq:C}
C =  \(\!\hsm\fr{\,Z_{\phi}^{}\,}{Z_h^{}}\hsm\!\)^{\!\!\hsm\frac{1}{2}}
\!\!
\fr{~1\!+\!\Delta^{\!(2)}\hsm (p^{2})\,}
{~1\!+\!\Delta^{\!(1)}\hsm (p^{2})\,} 
=1\!+\hsm {O}(\rm{loop})\hs .
\end{equation}
Thus, from Eq.\eqref{eq:F2-GID}, we can derive 
an LSZ-amputated identity as follows:
\beq
\MM\hsm \big[\FFB_2^{}(p),\cdots\big] = 0 \,,
\eeq
where we set the external momentum to be on-shell
($p^2\!=\!-m^2$) and the function $\FFB_2^{}$ is given by the following formula,
\beqs
\label{appeq:F2-Q}
\begin{align}
\label{appeq:F2-C}
\FFB_2^{} & = \hP\!-\hsm \frac{1}{\hs 2\hs}\big(\hat{C}\hs \phi \hsm +\! \htd_v^{}\big)
= \hP \! -\hsm \QQ\,,
\\
\label{appeq:Q-C}
\QQ & = \frac{1}{\hs 2\hs}\big(\hat{C}\hs \phi \hsm +\! \htd_v^{}\big) \hs,
\end{align}
\eeqs
In the above, we have defined the radiative modification factor
$\hat{C}\hsm\!=\!C\big|_{p^2=-m^2}^{}\hs$, 
and gaviton fields
$\hP\!=\!\epP^{\mn}h_{\mn}^{}$ and
$\tilde{h}_v\!=\!v^{\mn}h_{\mn}^{}$,
where the tensor factor $v^{\mn}\!=\!O(m/E)$.\

\vs

Extending the above procedure, we can make the LSZ reduction for all the external states of the identity \eqref{eq:F2-GID} and derive the following identity for the amputated $S$-matrix elements:
\begin{equation}
	\label{appeq:F-TGET-ID}
	\MM \big[{\FFB}_2^{}(p_1^{}),{\FFB}_2^{}(p_2^{}),
	\cdots \!,{\FFB}_2^{}(p_N^{}), \Phi \big]= 0 \,,
\end{equation}
where $\Phi$ denotes any LSZ-amputated external physical state.\
This provides the TGRET identity \eqref{eq:F2-MID2} in the main text.


\end{document}